\documentclass[Master,Submit,ngerman,UKenglish]{scrbook}
\usepackage[texlive=2014,physics]{ubonn-thesis}
\usepackage{ubonn-biblatex}

\usepackage[acronym,toc,nosuper]{glossaries}
\usepackage{thesis_defs}

\usepackage[ruled]{algorithm2e}

\usepackage{placeins}
\usepackage{verbatim}

\graphicspath{%
  {../figs/}%
  {../figs/cover/}%
  {../figs/graphics/}%
  {../feynmf/}%
  {figs/}%
}

\makeglossaries

\newglossaryentry{siunitx}{name=siunitx,
  description={the best package around for typesetting units}}
\newglossaryentry{csquotes}{name=csquotes,
  description={a very nice package for using consistent quotes that is
  language sensitive}}
\newglossaryentry{LaTeX}{name=\LaTeX,
  description={the typesetting program that is used for this guide}}

\newglossaryentry{FT}{name={Fourier transform},description={A Fourier transform converts functions between the time domain and the frequency domain.}}
\newglossaryentry{FrequencySamplingMethod}{name={frequency sampling method},description={A method for designing \glspl{FIRFilter} with a given frequency behavior (see chapter 6.7 of~\cite{DSP})}}

\newglossaryentry{ADCCore}{name={ADC-Core},description={single \gls{ADC} of a \gls{TIADC}}}

\newglossaryentry{BRAM}{name={BRAM},description={RAM integrated in \gls{FPGA}; storage location for larger datasets}}
\newglossaryentry{DSPSlice}{name={DSP-Slice},description={An integrated unit in \glspl{FPGA} to perform fast arithmetic operations}}
\newglossaryentry{gateware}{name={gateware},description={A file, which describes the configuration of an \gls{FPGA}}}

\newglossaryentry{MirrorSuppression}{name={Mirror Suppression},description={The ratio between the signal power and the power in mirror bands}}

\newglossaryentry{CooperPair}{name={Cooper pair},description={A weekly bound pair of electrons, which are responsible for super conductivity}}

\newglossaryentry{balun}{name={balun},description={A special transformer to convert a symmetric signal into an asymmetric signal and vice-versa}}

\newglossaryentry{FraunhoferLines}{name={Fraunhofer lines},description={A number of strong absorption lines in the optical solar spectrum}}

\newglossaryentry{CCD}{name={CCD},description={A charge-coupled device, which is typically a 2-dimensional array of light sensitive sensors}}

\newglossaryentry{IFHybrid}{name={IF-Hybrid},description={Splits an \gls{IF} signal into two \gls{IF} signals of equal power with a phase shift between the two signals (typically \SI{90}{\degree})}}

\newacronym{CCAT-p}{CCAT-prime}{Cerro Chajnantor Atacama Telescope-prime}
\newacronym{CHAI}{CHAI}{CCAT-prime Heterodyne Array Instrument}
\newacronym{APEX}{APEX}{Atacama Pathfinder Experiment}

\newacronym{RF}{RF}{radio frequency}
\newacronym{IF}{IF}{intermediate frequency}
\newacronym{LF}{LF}{low frequency}
\newacronym{LO}{LO}{local oscillator}
\newacronym{RFI}{RFI}{radio frequency interference}

\newacronym{SNR}{SNR}{signal/noise ratio}

\newacronym{ADC}{ADC}{analog digital converter}
\newacronym{DAC}{DAC}{digital analog converter}

\newacronym{TIADC}{TIADC}{time interleaved ADC}

\newacronym{ASIC}{ASIC}{application specific integrated circuit}
\newacronym{FPGA}{FPGA}{field programmable gate array}
\newacronym{LUT}{LUT}{look-up table}
\newacronym{FF}{FF}{flip-flop}
\newacronym{HDL}{HDL}{hardware description language}
\newacronym{VHDL}{VHDL}{very high speed integrated circuit hardware description language}

\newacronym{IC}{IC}{integrated circuit}
\newacronym{CMOS}{CMOS}{complementary metal-oxide-semiconductor}
\newacronym{SH}{S/H-stage}{sample and hold stage}

\newacronym{DSP}{DSP}{digital signal processing}

\newacronym{CPU}{CPU}{central processing unit}

\newacronym{FFT}{FFT}{fast Fourier transform}
\newacronym{IFFT}{IFFT}{inverse fast Fourier transform}

\newacronym{FFTS}{FFTS}{fast Fourier transform spectrometer}

\newacronym{DFT}{DFT}{discrete Fourier transform}

\newacronym{FIRFilter}{FIR-Filter}{finite impulse response filter}
\newacronym{IIRFilter}{IIR-Filter}{infinite impulse response filter}

\newacronym{MKID}{MKID}{microwave kinetic inductance detector}

\newacronym{COMCA}{COMCA}{COMplex CAlibration}

\newacronym{MPIfR}{MPIfR}{Max-Planck-Institut für Radioastronomie}

\newacronym{SSB}{SSB}{single-sideband}
\newacronym{2SB}{2SB}{two-single-sideband}

\ifthenelse{\equal{\ThesisVersion}{Draft}}{%
  \usepackage{background}
  \ifthenelse{\texlive < 2013}{%
    \SetBgContents{DRAFT}
    \SetBgColor{blue!30}
  }{%
    \backgroundsetup{contents=DRAFT, color=blue!30}
  }
}

\newcommand{\iu}{{i\mkern1mu}}
\newcommand{\FT}{\mathcal{F}}
\newcommand{\id}[1]{_\text{#1}}
\newcommand{\RealNumbers}{\mathbb{R}}
\newcommand{\NaturalNumbers}{\mathbb{N}}

\newcommand{\BigO}[1]{$\mathcal{O}\left( #1 \right)$}

\DeclareMathOperator{\atantwo}{atan2}
\DeclareMathOperator{\sinc}{sinc}
\DeclareMathOperator{\floor}{floor}
\DeclareMathOperator{\MirSup}{MirSup}
\DeclareMathOperator{\smod}{mod}

\DeclareSIUnit \sample {S}
\DeclareSIUnit \SPS {SPS}
\DeclareSIUnit \GSPS {GSPS}

\begin{document}

% Cover page of thesis - this is only needed for the printed final
% version to be submitted to the department library
% Do not use this page for thesis submission to the Prüfungsamt or Promotionsbüro!
\ifthenelse{\equal{\ThesisVersion}{PILibrary}}{%
  \typeout{Document \jobname, Info: PI library version of thesis}
  \input{cover/\ThesisType_Cover}
}{}

% Start counting pages from the title page
\frontmatter
% Dedication has to come before \maketitle
% \dedication{For ...}

% Select the correct title page(s)
\ifthenelse{\equal{\ThesisType}{Unknown}}{%
  \typeout{Document \jobname, Error: Unknown thesis type - no title page printed}
}{%
  % Bachelor thesis only has one title page
  \ifthenelse{\equal{\ThesisType}{Bachelor}}{%
    \typeout{Document \jobname, Info: Bachelor thesis}
    \input{cover/\ThesisType_Title}
  }{%
    \ifthenelse{\equal{\ThesisVersion}{Final} \OR \equal{\ThesisVersion}{PILibrary}}{%
      % Final and PI library versions
      \typeout{Document \jobname, Info: Final version of a \ThesisType  thesis}
      \input{cover/\ThesisType_Final_Title}
    }{% Submission and draft versions
      \input{cover/\ThesisType_Submit_Title}
      \typeout{Document \jobname, Info: Draft/submission version of a \ThesisType  thesis}
    }
  }
}

\pagestyle{scrplain}

%------------------------------------------------------------------------------
% You can add your acknowledgements here - don't forget to also add
% them to \includeonly above
%------------------------------------------------------------------------------
\chapter*{Acknowledgements}
\label{sec:ack}
%------------------------------------------------------------------------------
First of all I would like to thank Prof. Dr. Frank Bertoldi for being my first referee and giving me the opportunity to write this Master-thesis with a focus on instrumentation.
Secondly many thanks also to Prof. Dr. Bernd Klein for the possibility to work with the spectrometers developed at the DSP-division of the \gls{MPIfR}.

I have to thank the whole DSP division of the \gls{MPIfR} for making this experience an interesting one and providing a pleasant working atmosphere.
A very special word of thanks goes to Stefan Hochgürtel for being a superb supervisor, for always being willing to listen to my problems and for giving important feedback at all stages of this thesis.

Furthermore I want to thank my friends and fellow students for the unforgettable journey up to now and all the journeys still to come.
Zu guter Letzt möchte ich meinen Eltern danken für eure immerwährende Unterstützung in jeder Lebenslage.
Ohne euch wäre ich nicht die Person die ich bin und hätte sicher nicht diese Arbeit geschrieben.

%%% Local Variables: 
%%% mode: latex
%%% TeX-master: "../mythesis"
%%% End: 

\tableofcontents

\mainmatter
\pagestyle{scrheadings}

% Turn off DRAFT for the following pages
\ifthenelse{\equal{\ThesisVersion}{Draft}}{%
  \ifthenelse{\texlive < 2013}{%
    \SetBgContents{}
  }{%
    \backgroundsetup{contents={}}
  }
}{}

%------------------------------------------------------------------------------
% Add your chapters here - don't forget to also add them to \includeonly above
%==============================================================================
\chapter{Introduction}
\label{sec:intro}
%==============================================================================
Since the beginning of mankind the sky is a mystery and has a strong attraction for the human mind.
For a long time the sky was only observable in the optical regime (\SIrange{400}{800}{\nano\meter}).
The next window, which gave mankind a new picture of the sky, was the radio regime.
In 1931 the first radioastronomical observation was performed unintentionally by Karl Jansky.
He searched for the source of static noise in transatlantic telecommunication.
After observations with a directional antenna at \SI{20.5}{\mega\hertz} he discovered, that a part of the noise originates from Milky Way, especially from the center of our galaxy.
Inspired by the work of Jansky, Grote Reber built a parabolic dish with \SI{9}{\meter} in diameter by himself.
In 1938 he successfully observed the radio sky at \SI{160}{\mega\hertz}.
Since then the instruments became more and more powerful and new techniques were developed to observe at higher frequencies. 
Today observations in the sub-mm regime up to circa \SI{4.74}{\tera\hertz} are possible~\cite{history}.

Since the discovery of the \gls{FraunhoferLines} spectroscopy is a major technique in astronomy, not only in the optical regime but over all wavelengths.
An important aspect of spectroscopy is the identification of spectral lines and therefore the revelation of the chemical composition and physical conditions of an astronomical source.
In the optical regime spectral lines of atoms dominate.
But in the sub-mm regime molecular lines dominate the spectra.
There fore radioastronomical spectroscopy allows deep studies of molecular clouds, the birthplaces of stars.

Radioastronomical spectrometers can be implemented by a number of different technologies.
An analog filter bank is comparable to photometry with filters in the optical range.
%Each spectral channel has its own analog filter, but they operate in parallel as the analog signal is amplified and distributed to each filter.
This type of spectrometer is rather unstable as the analog components drift.
Furthermore the matching between the different spectral channels is challenging.
The usual number of spectral channels lies between $256$ and $512$ for analog filter banks.

Autocorrelation spectrometers form the power spectrum from the autocorrelation of the input signal.
The power spectrum is the \gls{FT} of the autocorrelation.
The autocorrelation can be implemented in the analog or digital regime.
Analog implementations achieve a greater bandwidth (\SI{6.5}{\giga\hertz} with $~200$ spectral channels~\cite{AACS}), while digital ones are more stable and offer more spectral channels (\SI{1}{\giga\hertz} with $1000$ spectral channels~\cite{DACS}).

The acousto-optical spectrometer uses acoustic waves in a crystal to diffract a laser beam, which then carries the information of the power spectrum.
The incoming radio signal is coupled into the crystal.
The laser beam is then projected onto a \gls{CCD}, which records the resulting spectrum. 
They provide up to \SI{3}{\giga\hertz} of bandwidth with a resolution of \SI{1}{\mega\hertz}~\cite{ASO}.
These spectrometers are especially sensitive to mechanical and thermal drifts due to the optics.

The chirp transform spectrometer uses sonic wave propagation in crystals to calculate the power spectrum with the help of the velocity dispersion of those waves.
These spectrometers are optimized for spectral resolution in the order of \SI{50}{\kilo\hertz}, but the bandwidth is limited to \SI{400}{\mega\hertz}~\cite{CTS}.
\nocite{ToolsOfRadioAstronomy}

\glspl{FFTS} calculate the power spectrum, using a \gls{FFT} to get the complex amplitude spectrum, which is squared to get the power spectrum.
\glspl{FFTS} achieve a bandwidth of \SI{4}{\giga\hertz} with up to $65536$ channels.

\Glspl{FFTS} are the newest type of spectrometer for radioastronomical receivers.
This thesis has been written at the DSP-devision of the \gls{MPIfR}, which is a leading developer of \glspl{FFTS}. 
The \glspl{FFTS} developed at the \gls{MPIfR} use \glspl{FPGA} to process digital signals from \glspl{ADC}.
The development of these \glspl{FFTS} was possible due to new generations of \glspl{ADC} with sample rates of multiple \si{\giga\hertz} and the increasing computational power of \glspl{FPGA}.
Due to a little number of analog components this type of spectrometer is quite stable in terms of drifts.

However there are also limiting components in \glspl{FFTS}, which can be optimized.
One is the \gls{ADC}, and its limiting characteristics shall be reduced in this thesis.
The \glspl{ADC} on the spectrometers are composed of a number of slower \glspl{ADC}, which slightly differ from each other (see chapter \ref{sec:tiadc}).
A calibration procedure has to correct these differences for an optimal performance of the spectrometers.
For small bandwidths these differences can be assumed to be frequency independent, but for larger bandwidths this assumption reduces the dynamical range of the spectrometers.
The current generation of \glspl{FFTS} calibrates these differences at a fixed frequency with a statistical approach (see chapter 4.2 of~\cite{prom}).
The goal of this thesis is to develop and implement a method in the \glspl{FFTS} to correct for these differences in a frequency dependent manner.
This correction shall enhance the dynamical range of the spectrometer and shall allow further advancement in the development of \glspl{FFTS}. 

This thesis starts with a short introduction to spectroscopic radioastronomy in chapter \ref{sec:radioastronomy}.
Chapter \ref{sec:theory} first describes the mathematical tools used in this thesis, then it describes the \glspl{ADC} used on the spectrometers.
Chapter \ref{sec:hardware} gives a brief overview over the hardware of used to measure the differences between different signal paths.
A method to characterize these differences is described in chapter \ref{sec:comma}.
Chapter \ref{sec:approaches} discusses different approaches to correct these differences.
In chapter \ref{sec:comca} the implementation details of the chosen approach in the spectrometers are discussed.
The implementation is tested and characterized, especially in terms of stability over time and temperature.
The conclusion in chapter \ref{sec:conclusion} summarizes the work and presents prospects for other application cases and further improvements.

%==============================================================================
\chapter{Fundamentals of spectroscopic radioastronomy}
\label{sec:radioastronomy}
%==============================================================================

\begin{figure}[htbp]
	\centering
	\includegraphics[width=0.95\textwidth]{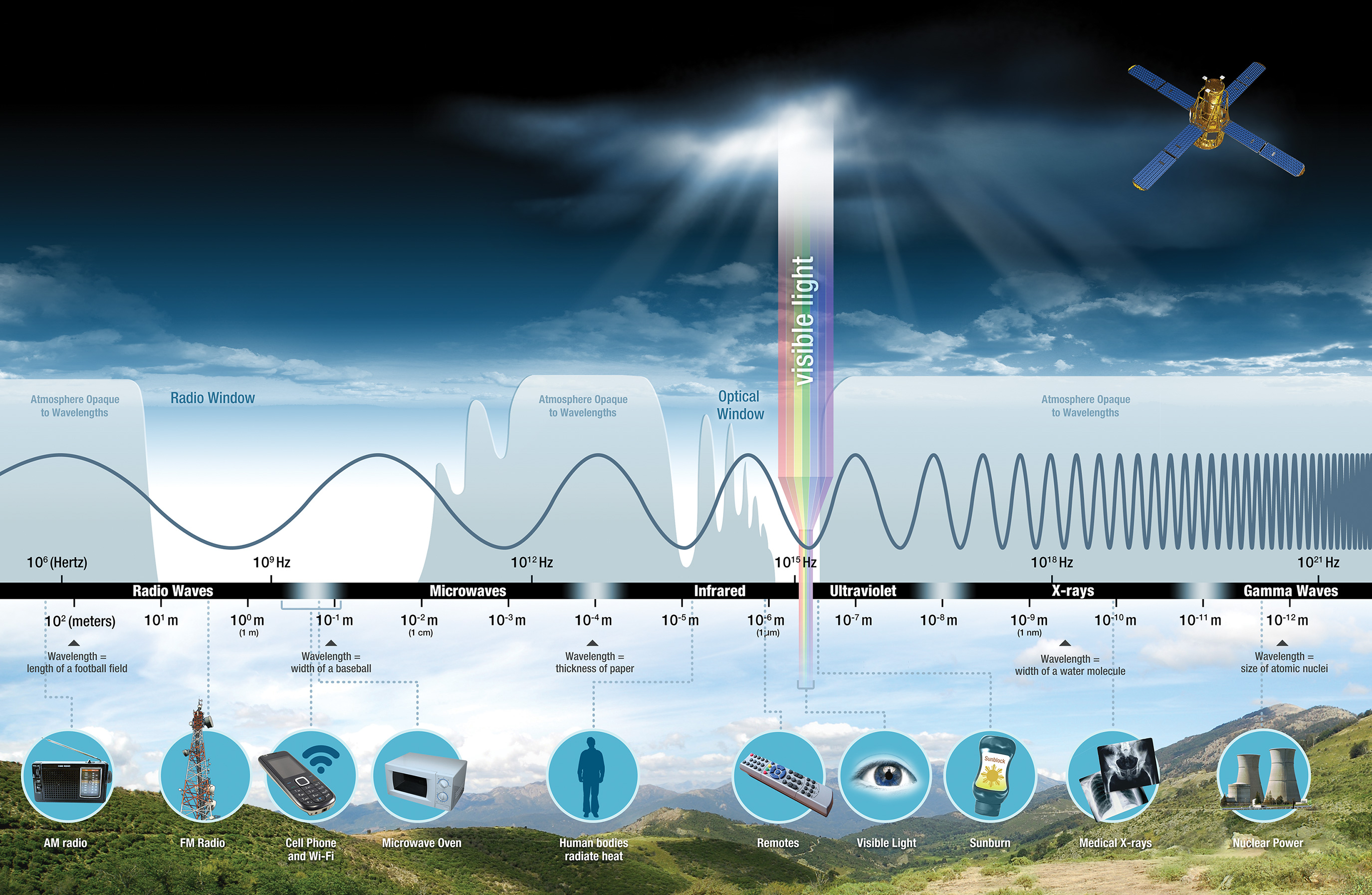}
	\caption{The atmospheric opacity for different electromagnetic regimes~\cite{AtmosphericOpacity}.}
	\label{fig:AtmosphericOpacity}
\end{figure}

Distant objects can only be investigated by messengers they emit or interact with. Only a few messengers can be received from these objects.  Currently these messengers are neutrinos, gravitational waves and electromagnetic radiation. The latter can be divided into different regimes, which corresponds to photons with distinct wavelengths respectively energies (see table \ref{tab:em_regims} for a short overview and figure \ref{fig:AtmosphericOpacity} for a graphically representation). Each regime reveals different information about the objects in the universe. For a complete understanding of the universe the information from all messengers and regimes are essential.
\begin{table}[htb]
	\centering
	\begin{tabular}{lllll}\toprule
		Regime 			& {Wavelength} & Energy & Observable from the ground &  \\\midrule
		Radio  			& \SI{10}{\meter} to \SI{1}{\centi\meter} & \SI{1.2}{\micro\eV} to \SI{1.2}{\milli\eV} & yes                        &  \\
		mm and sub-mm 	& \SI{1}{\centi\meter} to \SI{450}{\micro\meter} & \SIrange{1.2}{2.8}{\milli\eV} & only at high altitudes &  \\
		Far infrared 	& \SIrange{450}{30}{\micro\meter} & \SIrange{2.8}{40}{\milli\eV} & no &  \\
		Near infrared 	& \SI{30}{\micro\meter} to \SI{750}{\nano\meter} & \SI{40}{\milli\eV} to \SI{1.7}{\eV} & yes &  \\
		Optical 		& \SIrange{750}{320}{\nano\meter} & \SIrange{1.7}{4}{\eV} & yes &  \\
		Ultra violet 	& \SIrange{320}{10}{\nano\meter} & \SI{4}{\eV} to \SI{1.2}{\kilo\eV} & no &  \\
		X-ray 			& \SIrange{10}{0.01}{\nano\meter} & \SIrange{1.2}{120}{\kilo\eV} & no &  \\
		$\gamma$-ray 	& below \SI{0.01}{\nano\meter} & above \SI{120}{\kilo\eV} & not directly &  \\\bottomrule
	\end{tabular}
	\caption{Different electromagnetic regimes~\cite{ir1994}\cite{AtmosphericOpacity}\cite{Cowley2002}.}
	\label{tab:em_regims}
\end{table}

Each messenger and regime hold their own challenges.
For example neutrinos have a tiny cross section, hence the detection of neutrinos need massive detectors.
In the X-ray and $\gamma$-ray regime conventional mirrors have no effect, therefore mirrors based on total reflection are used.
The radio regime, mm and sub-mm have wavelengths from \SI{10}{\meter} to \SI{450}{\micro\meter}.
A single telescope in the radio regime has an angular resolution which is not comparable to the one of optical telescopes, because the maximal size of radio telescopes is limited due to constructive issues.
This lead to the development of interferometry, where the signals from different telescopes are combined to achieve a better angular resolution.
Furthermore the commercial and military use of frequencies in the radio regime causes \gls{RFI}.
The sub-mm regime is sensitive to the column of water in the atmosphere along the line of sight.
Therefore telescope for this regime are placed at high and dry location (e.g. Atacama Desert in Chile for the \gls{APEX}).
Another solutions are airborne or spaceborne telescopes.
Balloon experiments and plane based experiment give easier excess for maintenance and upgrade compared to satellites but suffer from the residual atmosphere.
Another challenge is the handling of signals at these high frequencies.
This becomes even more challenging in the regime of far infrared. 

\section{A simple radio-telescope}\label{sec:simpleRadiotelescope}
A radio-telescope is composed from different parts. In the following the most important parts are described in the order as they appear in the signal path. 

\begin{figure}[htbp]
	\centering
	\begin{subfigure}[t]{0.45\textwidth}
		\centering
		\includegraphics[width=0.9\textwidth]{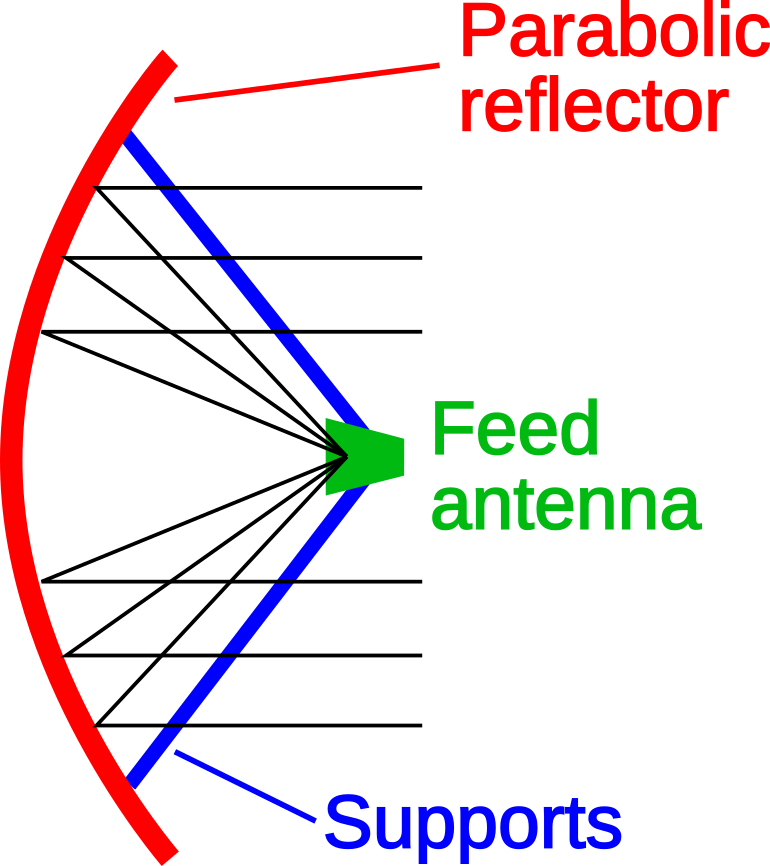}
		\caption{Parabolic reflector with feed antenna (modified version from~\cite{wiki:parabolic}). The feed is fixated with support structures.}
		\label{fig:ParabolicAntenna}
	\end{subfigure}
	\quad
	\begin{subfigure}[t]{0.45\textwidth}
		\centering
		\includegraphics[width=0.9\textwidth]{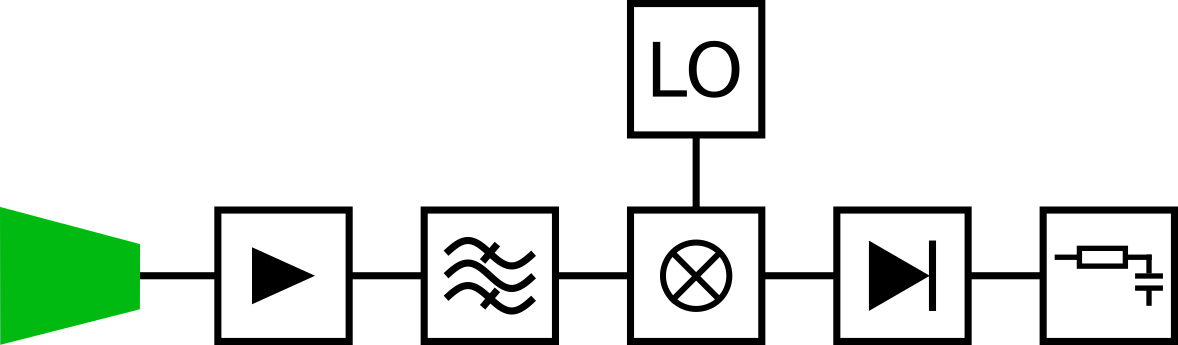}
		\caption{A simple total power radiometer, from left to right: 1. feed, 2. amplifier, 3. mixer with \gls{LO}, 4. diode with square characteristic, 5. integrator }
		\label{fig:radiometer}
	\end{subfigure}
\end{figure}

The first element is an antenna, which collects the signal and couples it into a transfer element (e.g. coax-cable or waveguide). A typical antenna configuration is a parabolic dish which focuses the incoming planar waves into the feed where the wave is picked up by a short wire. Figure \ref{fig:ParabolicAntenna} shows such an antenna configuration. The electromagnetic waves propagate in a coax cable or a wave-guide to an amplifier. This amplifier is used to increase the signal amplitude. The amplification decreases the influence of noise (e.g. thermal noise), which is introduced by the following components. The signal is then filtered to prevent contamination with \gls{RFI}. This filtered signal is then mixed with a \gls{LO} to a lower frequency range, where the further signal processing is easier. This part of the signal processing chain is the frontend, more precise it is a Heterodyne receiver. A Heterodyne receiver preserves the phase information of the received waves. Receivers, which preserves the phase information are called coherent. 
The further treatment of the signal, which is called backend, depends on the purpose of the observation. The simplest backend is a total power backend. This backend measures the total power of the received radiation in the frequency band of the receiver. Figure \ref{fig:radiometer} shows a simple total power receiver (consisting of a frontend and a total power backend). A spectrometer backend is similar to a total power backend. But instead of the power of the whole band, the spectrometer measures the power in a number of smaller bands (frequency channels). 

The phase information, which is preserved in coherent receivers, can be used for beam forming and interferometry. Both techniques use a number of antennas to achieve a better angular resolution. 

In contrast to the Heterodyne receiver bolometers are incoherent receivers. The basic idea is to detect photons based on their energy. The simplest bolometer is a black body connected to a heat reservoir (with a known thermal resistance between black body and reservoir). The temperature difference between black body and reservoir is then related to the absorbed radiation. Today bolometer technologies exists, which do not relay on the temperature (e.g. \gls{MKID} see \ref{sec:mkid}) but on other absorption mechanism of photons.

\section{Spectroscopy in the sub-mm regime}
\begin{figure}[htbp]
	\centering
	\includegraphics[width=0.45\textwidth]{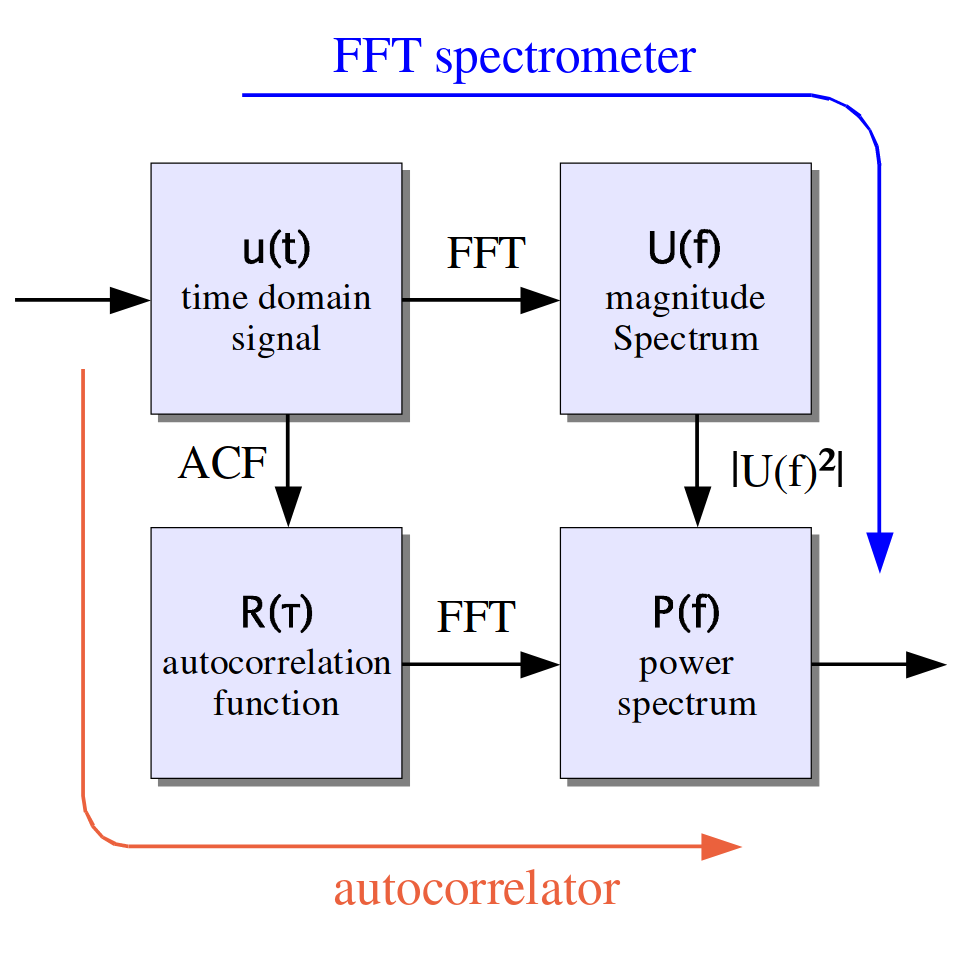}
	\caption{Visualization of the Wiener-Khinchin-Theorem. Show two different routes to generate the power spectrum $P(f)$ from the time domain signal $u(t)$~\cite{APEX}.}
	\label{fig:theorem}
\end{figure}
\Glspl{FFTS} are the preferred implementation to realize high spectral resolution over wide bands.
An \gls{FFTS} processes the mixed down signal in the digital regime.
Therefore an \gls{ADC} converts the analog \gls{IF} signal to a digital datastream.
An \gls{FFT} transforms this time-domain datastream into a complex spectrum, which is squared to get the power spectrum.
Finally the power spectra are summed up to perform an integration over time.
The basic principle of an \gls{FFTS} is shown in figure \ref{fig:theorem} in contrast to an autocorrelation spectrometer.
An autocorrelation spectrometer first computes the autocorrelation function of the time-domain signal, which is then integrated and finally converted into the power spectrum with an \gls{FFT}.

\begin{figure}[htbp]
	\centering
	\begin{subfigure}[t]{0.45\textwidth}
		\centering
		\includegraphics[scale=0.28]{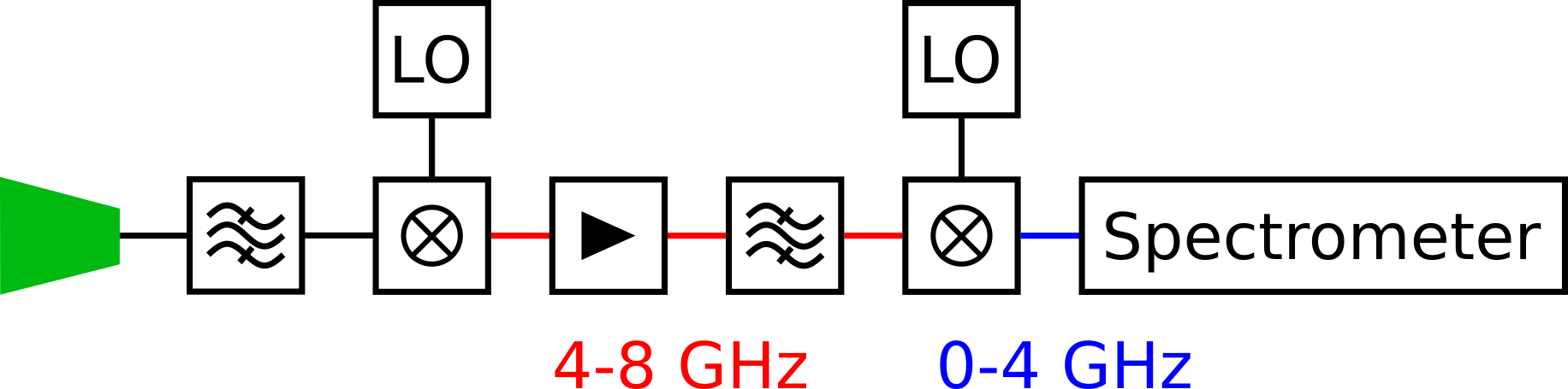}
		\caption{A heterodyne receiver for the sub-mm regime with \gls{IF} signal processing.}
		\label{fig:radiometerIF}
	\end{subfigure}
	\quad
	\begin{subfigure}[t]{0.45\textwidth}
		\centering
		\includegraphics[scale=0.28]{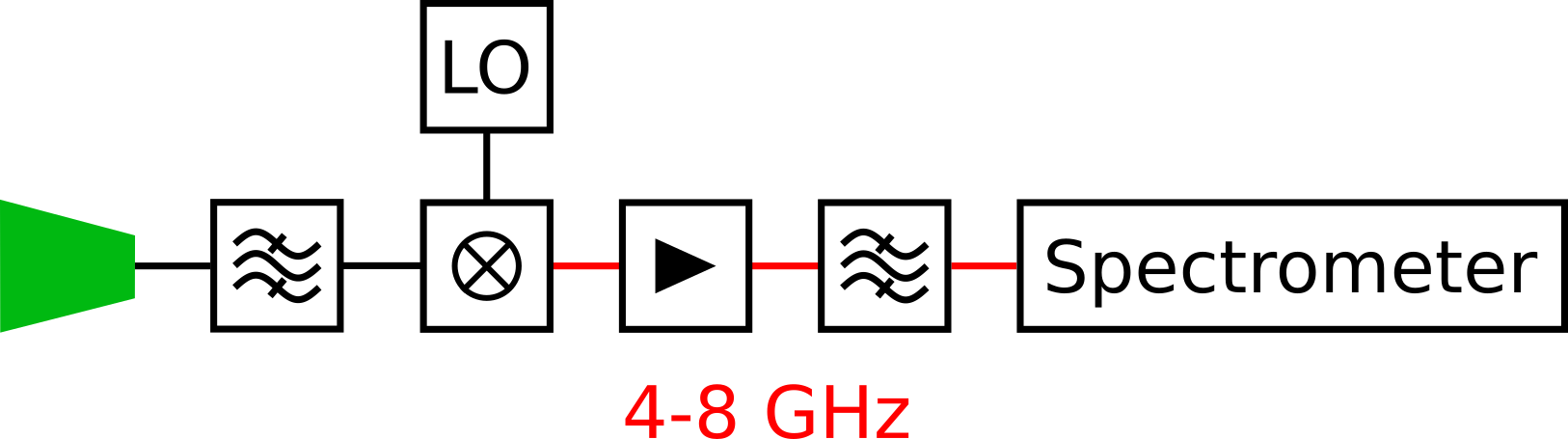}
		\caption{A heterodyne receiver for the sub-mm regime without \gls{IF} signal processing. The spectrometer is sampling directly the \gls{IF} frequency band.}
		\label{fig:radiometerWithoutIF}
	\end{subfigure}
	\caption{Receivers for the sub-mm regime.}
\end{figure}

The high frequency of the \gls{RF} signals in the sub-mm regime are technologically challenging, which requires a different setup for a receiver than the one given in chapter \ref{sec:simpleRadiotelescope}.
Currently no amplifiers are available for such frequencies.
Therefore the \gls{RF} signal is filtered and mixed down into the \gls{IF} band, which is typically \SI{4}{\giga\hertz} to \SI{8}{\giga\hertz}.
The \gls{RF} cannot be mixed into the band between \SI{0}{\giga\hertz} and \SI{4}{\giga\hertz} as the technologies of these mixers do not support to couple out low frequency signals.
The \gls{IF} signal is then amplified and filtered further.
In a second mixing stage the \gls{IF} signal is translated into the baseband (\SI{0}{\giga\hertz} to \SI{4}{\giga\hertz}).
This baseband signal is then sampled by the spectrometer. 
Figure \ref{fig:radiometerIF} shows the schematics of such a receiver.

The bandwidths of \glspl{FFTS} are increasing, due to the advances of \gls{ADC} development. 
For example the \gls{ADC} \verb|ADC12DJ3200| has an analog input bandwidth of \SI{8}{\giga\hertz}~\cite{ADC12DJ3200}.
This increasing analog input bandwidth means, that the \gls{FFTS} is not only sensitive in the baseband (\SIrange{0}{4}{\giga\hertz}) but also in the second Nyquist-band (\SIrange{4}{8}{\giga\hertz}).
Therefore the second mixing stage can be skipped and the \gls{FFTS} can directly sample the \gls{IF} signal.
Such a receiver is shown in figure \ref{fig:radiometerWithoutIF}.
This simplification of the receiver chain, leads to a more stable system.
Furthermore it consumes less power, space and money.
 
\subsection{Multi-beam heterodyne receivers}
In optical observations different positions at the sky are observed simultaneously due to a number of pixels in the focal plane. A similar concept is developing for heterodyne receivers. For this an array of feeds is placed in the focal plane. The output of each feed is treated separately in the signal processing (filtering, amplifying, mixing and digitalization). 

An example for such a design is the \gls{CHAI} at the \gls{CCAT-p} telescope, which is located at the Atacama desert of northern Chile and is conceived as survey telescope. 
It is a ground based instrument operating at submillimeter to millimeter wavelengths~\cite{CCAT} with \SI{6}{\meter} opening.
\gls{CHAI} is composed of two separate receivers.
One operating at a lower frequency range of \SIrange{455}{495}{\giga\hertz} and one at a higher frequency range of \SIrange{800}{820}{\giga\hertz}.
The primary science targets is the star forming in the interstellar medium.
This is covered by the observation of CI fine-structure transitions $2-1$ and $1-0$, as well as the rotational transitions $4-3$ and $7-6$ of CO.
Each of the two receivers consists of $64$ pixels at first light.
A possible extension to $128$ pixels each is planned.
The Multi-beam approach increases the mapping speed of the instrument, which will lead to a very powerful and fast survey instrument.
The limiting factor for the number of pixels is the cost per pixel and the space available at the telescope site~\cite{CCATPresentation}.
Each pixel needs the complete signal processing chain.
Therefore the costs of each part of the chain has to be optimized, including the used spectrometers. 
Furthermore the complexity of the system has to be minimized to make such a project feasible, luckily both requirements go hand in hand most of the time.
On important point of improvement is the elimination of the \gls{IF} system, as this reduces the number of needed mixer and removes one complete stage of the signal chain.
Therefore the backends have to operate between \SI{4}{\giga\hertz} and \SI{8}{\giga\hertz} instead of \SIrange{0}{4}{\giga\hertz}.
Another requirement is the available space at the telescope site, which is limited.
With the elimination of the \gls{IF} system, its space can be used by other components of the receiver.

All this requirements have to be balanced to successfully implement a project like this.
The current planed solution to this situation is a \texttt{qFFTS-8G} with four inputs per spectrometer, which can operate between \SIrange{0}{4}{\giga\hertz} and from \SIrange{4}{8}{\giga\hertz}.
With a \texttt{qFFTS-8G} between $32$ and $64$ spectrometer are needed.

\section{Microwave kinetic inductance detectors}\label{sec:mkid}
\begin{figure}[htbp]
	\centering
	\includegraphics[width=0.40\textwidth]{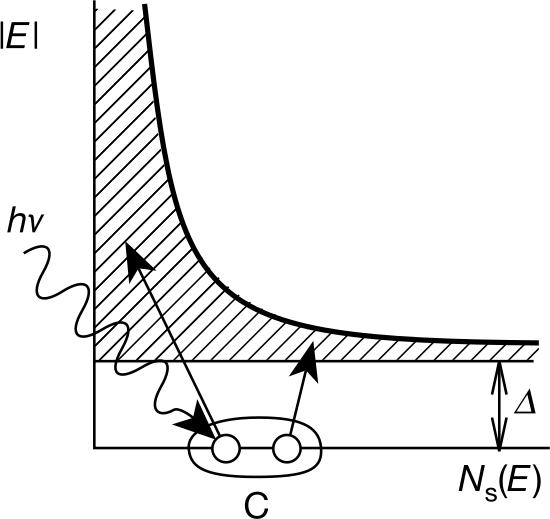}
	\caption{A \gls{CooperPair} is broken up by a photon into quasi-particles, which block some energy states. This leads to a change in the impedance~\cite{Day2003}.}
	\label{fig:MKIDpair}
\end{figure}
\noindent
\Glspl{MKID} are superconducting devices to detect photons.
Their surface impedance changes when they absorb a photon.
This effect is based on the creation of quasi-particles by incident radiation.
This quasi-particles block energy states, which would be occupied with \glspl{CooperPair} otherwise (due to the exclusion principle).
This blocking leads to a change in the impedance.
Figure \ref{fig:MKIDpair} shows the energy states of such a system, where a \gls{CooperPair} is broken up by a photon.

\begin{figure}[htbp]
	\centering
	\begin{subfigure}[t]{0.45\textwidth}
		\centering
		\includegraphics[width=0.90\textwidth]{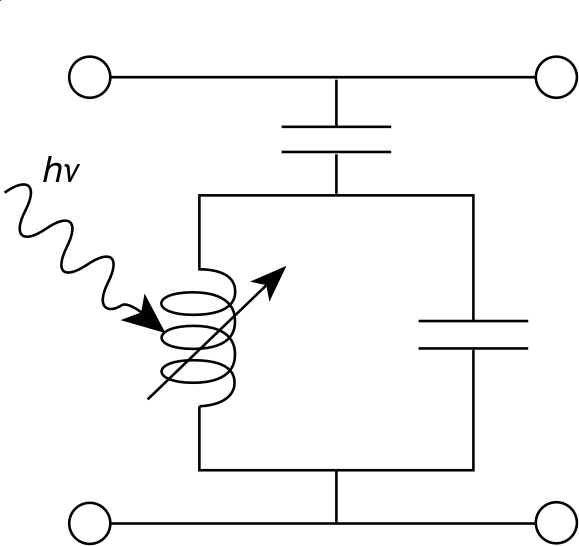}
		\caption{A resonator with a \gls{MKID} as frequency changing element. The resonator is coupled to a readout line~\cite{Day2003}.}
		\label{fig:MKIDcircuit}
	\end{subfigure}
	\quad
	\begin{subfigure}[t]{0.45\textwidth}
		\centering
		\includegraphics[width=0.90\textwidth]{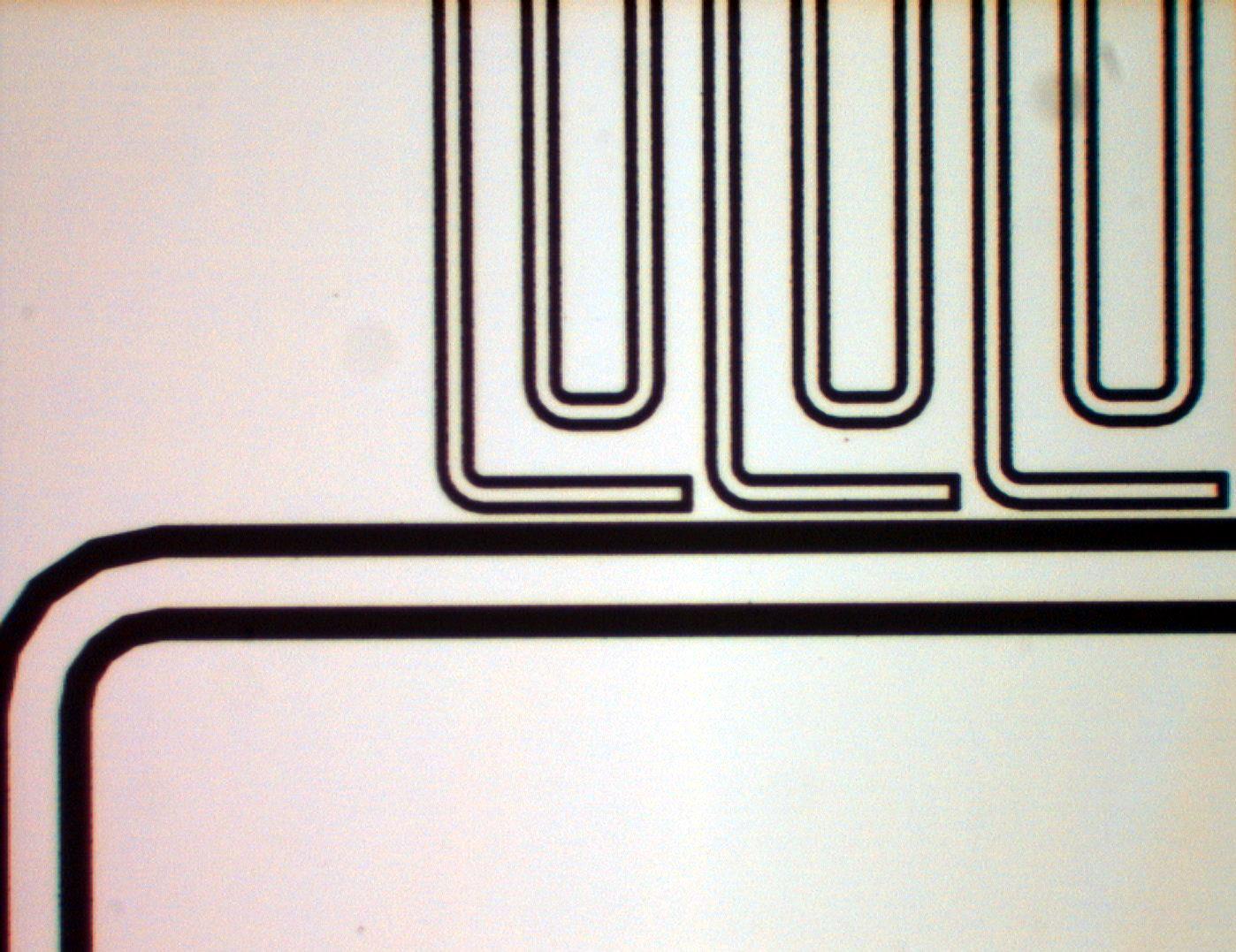}
		\caption{A number of \glspl{MKID} (upper part) coupled to one readout line (lower part)~\cite{MKID}.}
		\label{fig:MKIDcoupled}
	\end{subfigure}
	\caption{\gls{MKID} circuit as schematic and real implementation.}
\end{figure}

\Glspl{MKID} are non coherent receivers as the phase of the photons is lost in the receiving process. The sensitive bandwidth is given by the optics and filter in front of the \glspl{MKID} and the minimum energy to create a quasi-particle $2\Delta$ ($\Delta$ is the superconducting gap energy)~\cite{2009AIPC.1185..135M}. Consequently, these detectors are sensitive to a larger bandwidth than heterodyne receivers; but the spectral resolution, if an energy measurement of the photons is implemented, is worse than a heterodyne receiver combined with a spectrometer.

\begin{figure}[htbp]
	\centering
	\begin{subfigure}[t]{0.45\textwidth}
		\centering
		\includegraphics[width=0.9\textwidth]{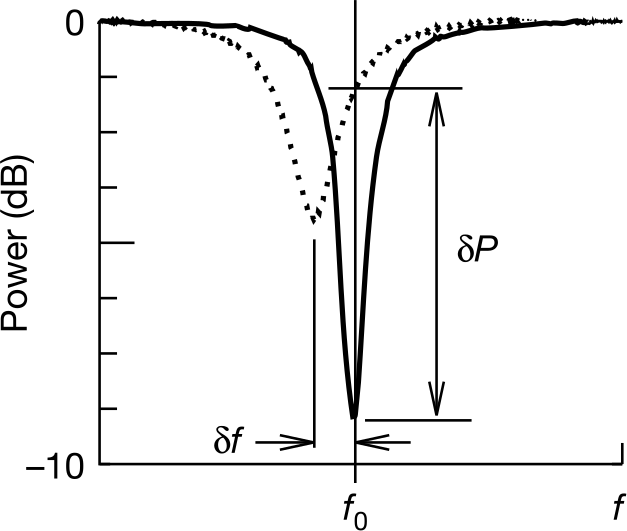}
		\caption{The amplitude change $\delta P$ of a resonant \gls{MKID} readout, due to a changed resonance frqeuncy~\cite{Day2003}.}
		\label{fig:MKIDamplitude}
	\end{subfigure}
	\quad
	\begin{subfigure}[t]{0.45\textwidth}
		\centering
		\includegraphics[width=0.9\textwidth]{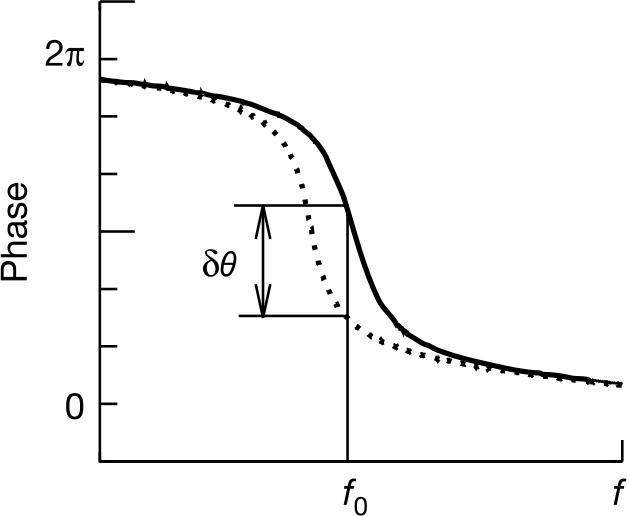}
		\caption{The phase change $\delta \theta$ of a resonant \gls{MKID} readout, due to a changed resonance frqeuncy~\cite{Day2003}.}
		\label{fig:MKIDphase}
	\end{subfigure}
	\caption{The two possibles readout methods for \glspl{MKID}.}
\end{figure}

The surface impedance of \glspl{MKID} is mostly inductive. The measurement of the changing surface impedance is done with a superconducting resonant circuit.
The \gls{MKID} is connected to a capacity, the resonance frequency of the circuit consisting of the capacity and \gls{MKID} changes with the surface impedance of the \gls{MKID}.
Figure \ref{fig:MKIDcircuit} shows a schematic representation of such a resonant circuit, while figure \ref{fig:MKIDcoupled} presents a real implementation of \glspl{MKID}, which are coupled to one readout line.
The typical readout of a \gls{MKID} is done via the phase or amplitude change of an applied sinusoid-signal at the reference frequency. Figures \ref{fig:MKIDamplitude} and \ref{fig:MKIDphase} show the amplitude change $\delta P$ and phase change $\delta \theta$ due to a resonance change.
Compared to the measurement of the position of the resonance frequency this is more sensitive to changes and only needs one applied frequency instead of a frequency sweep. 

\Glspl{MKID} can be produced, together with the capacities, in thin-film lithography.
Therefore an array of \gls{MKID} pixels can be produced with conventional optical lithography, which reduces the costs per pixel compared to other pixel technologies.
The different pixels are multiplexed in the the frequency domain. This means that each pixel has a different resonance frequency.
Up to $1000$ pixels can be read out via one readout line in figure \ref{fig:MKIDcircuit}.
A \gls{DAC} generates a signal with a frequency component for each pixel.
This signal is applied to one side of the readout line, on the other side a spectrometer is connected, which measures the amplitude and phase of the frequency components.
From these information the resonance frequency of the circuits can be derived and from there the absorbed radiation. 

The increasing bandwidths of \glspl{FFTS} allow to increase the number of pixels per readout line or to reduce the number of spectrometer per readout line.
This simplifies the overall design and reduces space, power and cost requirements.

\nocite{2009AIPC.1185..135M} 
\nocite{MKID}
%==============================================================================
\chapter{Digital signal processing theory}
\label{sec:theory} \nocite{Seo2005}
%==============================================================================
\Gls{DSP} allows the transformation and analysis of digital signals. In this context a digital signal is a signal which is discrete in time and amplitude. These can be implemented in digital circuits like an \gls{FPGA} or a \gls{CPU}. Compared to analog signal processing, the digital counterpart has different advantages:
\begin{itemize}
	\item More stable:
Analog filters vary over time due to different environmental parameters. After the digitalization the signals and transformations are not sensitive to such drifts any more. 

	\item Changeable after deployment:
It is possible to program most processing units, like \glspl{FPGA} or \glspl{CPU}, after an instrument is deployed.

	\item Better footprint scaling:
The footprint of \gls{DSP} circuits scales better than the footprint of analog circuits. E.g. generating amplitude spectrum needs an sophisticated analog filter for each channel; while the digital counterpart is composed of a few \glspl{IC}.  
\end{itemize}
The downside is the need of high processing power and the possible loss of information due to discretization. But due to the exponential development of the processing power in the last decades, the processing power is on hand. By designing the signal processing chain carefully, in both regimes the analog and digital, an unwanted information loss can be constrained. 

In the digital domain clocks are used to synchronize different components and their processes. In most cases these signals are square wave to archive a low errors in the synchronization between the components. Either the rising edge, falling edge or both are used for clocking. In the figures of this thesis the the rising edge will be used. 

A function, continuous with respect to its argument, is represented by the notation $a(t)$. $a_k$ is instead used for a function discrete in terms of $k$, where $k \in \NaturalNumbers$ is the index of the sample.

\section{Complex numbers}
The complex numbers are an extension of the real numbers. This is done by defining a solution for negative square roots $\iu = \sqrt{-1}$, the imaginary unit. Less abstract the one dimensional number line of the real numbers is extended by a second dimension. A complex number $a = b + c\iu$ can be represented by two real numbers, the real part $\real(a) = b$ and the imaginary part $\imaginary(a) = c$. 

The complex plain, in which the complex numbers are living, is typically described by rectangular coordinates ($a = b + c\iu$) or by polar coordinates ($a = g e^{\iu\phi}$). The transformation from rectangular coordinates to polar ones is given as follows:
\begin{align}
	g &= \abs{a} \\
	\phi &= \atantwo\left(\imaginary\left(a\right), \real\left(a\right)\right)
\end{align}
The transformation other way round is given as follows:
\begin{align}
	a &= g \cdot e^{\iu\phi}
\end{align}
In terms of \gls{DSP} the angle $\phi$ of a complex number corresponds to the phase of a signal and the absolute value $g$ corresponds to the amplitude of the signal. 

\section{Fourier transform}\label{sec:ft}
The continuous \gls{FT} of a function $a(t)$ is defined as:
\begin{align}
	\FT a(f) &= \int e^{-2\pi\iu f \cdot t} \cdot a(t) \dd{t} = \int e^{-\iu \omega \cdot t} \cdot a(t) \dd{t}
	\label{eq:ft}
\end{align}
The \gls{FT} can be used to transform functions between a representation in time $a(t)$ and in frequency $A(f)$. In general both representations live in the complex plain. In this thesis the Fourier pairs are notated by lower case and capital letters, for example $\FT a(f) = A(f)$ as the Fourier pair of $a(t)$. 
A shift in time corresponds to a phase shift in the frequency domain. For a function $a(t)$ that is shifted in time by $\Delta t$, the following applies: 
\begin{align}
	a'(t) 		&= a(t - \Delta t) \\ 
	\FT a'(f) 	&= \int e^{-2\pi\iu f \cdot t} \cdot a(t - \Delta t) \dd{t} \label{eq:ftShift} \\
	         	&= \int e^{-2\pi\iu f \cdot (t' + \Delta t)} \cdot a(t') \dd{t'}\hspace{-1.5mm}\text{, }t'= t - \Delta t \nonumber \\
				&= e^{-2\pi\iu f \cdot \Delta t} \cdot \FT a(f) \nonumber
\end{align}
Hence a shift in time corresponds to a phase shift, which depends on the frequency of the component. 

In the regime of \gls{DSP} there are no continuous signals but time discrete signals, which are a sequence of single samples. Therefore the \gls{DFT} is used instead of the continuous Fourier transformation. A \gls{DFT} on $N$ samples leads to a spectrum of length $N$. The $k$-th channel of the \gls{DFT} $\hat{a}_k$ is defined similar to the continuous one: 
\begin{align}
	\hat{a}_k &= \sum_{j = 0}^{N - 1} e^{-2\pi\iu \frac{jk}{N}} \cdot a_j
	\label{eq:df}
\end{align}
The notation $\hat{a}_k$ is used in this thesis for a \gls{DFT} of the sequence $a_n$. 
The calculation of a spectrum of length $N$ needs $N^2-N$ operation with the \gls{DFT} algorithm.
The complexity of the \gls{DFT} algorithm is \BigO{N^2}. Therefore the number of operations scales as $N^2$, relative to the length $N$ of a spectrum.

From the definition of the \gls{DFT} follows its linearity. The periodicity of the complex exponential function leads to a periodicity of the \gls{DFT}, each $N$ channels:
\begin{align}
	\hat{a}_{k+N} &= \sum_{j = 0}^{N - 1} e^{-2\pi\iu \frac{j(k+N)}{N}} \cdot a_j \\
				  &= \sum_{j = 0}^{N - 1} e^{-2\pi\iu \frac{jN}{N}} \cdot e^{-2\pi\iu \frac{jk}{N}} \cdot a_j \nonumber \\
				  &= \sum_{j = 0}^{N - 1} e^{-2\pi\iu j N} \cdot e^{-2\pi\iu \frac{jk}{N}} \cdot a_j \nonumber \\
				  &= \sum_{j = 0}^{N - 1} e^{-2\pi\iu \frac{jk}{N}} \cdot a_j \nonumber \\
				  &= \hat{a}_{k} \nonumber
\end{align}
A typical digitized signal has no imaginary contribution, therefore it is real. This leads to another symmetry of the \gls{DFT}. When $a_j \in \RealNumbers$ ($a_j = a_j^*$):
\begin{align}
	\hat{a}_{N-k} 	&= \sum_{j = 0}^{N - 1} e^{-2\pi\iu \frac{j(N-k)}{N}} \cdot a_j \\
					&= \sum_{j = 0}^{N - 1} e^{-2\pi\iu j} \cdot e^{2\pi\iu \frac{jk}{N}} \cdot a_j \nonumber \\
					&= \sum_{j = 0}^{N - 1} \left(e^{-2\pi\iu \frac{jk}{N}}\right)^* \cdot a_j^* \nonumber \\
					&= \left( \sum_{j = 0}^{N - 1} e^{2\pi\iu \frac{jk}{N}} \cdot a_j \right)^* \nonumber \\
					&= \hat{a}_k^* \nonumber
\end{align}
The calculation of a \gls{DFT} can be sped up by some optimizations. A popular algorithm is the \gls{FFT} for \glspl{DFT} with a length of $N = 2^n$. That allows to split up the calculation into two parts. The even samples and the odd ones,
\begin{align}
	\hat{a}_k &= \sum_{j = 0}^{N - 1} e^{-2\pi\iu \frac{jk}{N}} \cdot a_j \\
	&= \sum_{j = 0}^{N/2 - 1} e^{-2\pi\iu \frac{2jk}{N}} \cdot a_{2j} + \sum_{j = 0}^{N/2 - 1} e^{-2\pi\iu \frac{(2j+1)k}{N}} \cdot a_{2j+1} \nonumber \\
	&= \sum_{j = 0}^{N/2 - 1} e^{-2\pi\iu \frac{jk}{N/2}} \cdot a_{2j} + e^{-2\pi\iu \frac{k}{N}} \cdot \sum_{j = 0}^{N/2 - 1} e^{-2\pi\iu \frac{jk}{N/2}} \cdot a_{2j+1} \nonumber \\
	&= \sum_{j = 0}^{N/2 - 1} e^{-2\pi\iu \frac{jk}{N/2}} \cdot a'_{j} + e^{-2\pi\iu \frac{k}{N}} \cdot \sum_{j = 0}^{N/2 - 1} e^{-2\pi\iu \frac{jk}{N/2}} \cdot a''_{j} \nonumber \\
	&= \hat{a}'_{k} + e^{-2\pi\iu \frac{k}{N}} \cdot \hat{a}''_{k} \nonumber
\end{align} where $\hat{a}'$ results from the \gls{DFT} of the even samples and $\hat{a}''$ from the \gls{DFT} of the odd samples. 

Using the periodicity of the \gls{DFT} of length $N/2$, the second half of the spectrum can be calculated ($N/2 \le k < N$): 
\begin{align}
	\hat{a}'_k  &= \hat{a}'_{k-N/2} \\
	\hat{a}''_k &= \hat{a}''_{k-N/2}
\end{align}
The smaller \glspl{DFT} $\hat{a}'$ and $\hat{a}''$ can be calculated the same way, down to a \gls{DFT} of length $1$:
\begin{align}
	\hat{a}_k &= \sum_{j = 0}^{0} e^{-2\pi\iu jk} \cdot a_j \\
			  &= a_j \nonumber
\end{align}
The number of needed operations for a \gls{FFT} of size $N$ without the calculation of the smaller spectra is $T(N) = N + 2T(N/2)$. The whole calculation needs $\log_2 (N)$ iterations to reach a spectrum length of $1$ which is trivial to calculate. This  leads to a complexity of \BigO{N \log(N)} for the \gls{FFT}.

\nocite{DSPSciEng}

\section{Convolution}\label{sec:convolution}
A continuous convolution is given by
\begin{align}
	(f*g)(t) = \int f(\tau) \cdot g(t-\tau) \dd{\tau}\,.
	\label{eq:convolution}
\end{align}

Under a \gls{FT} a convolution transforms as follows:
\begin{align}
	\FT \left((f*g)(t)\right) &= \FT \left(  \int f(\tau) \cdot g(t-\tau) \dd{\tau} \right) \\
							  &= \int e^{-2\pi\iu f \cdot t} \int f(\tau) \cdot g(t-\tau) \dd{\tau} \dd{t} \nonumber \\
							  &= \int e^{-2\pi\iu f \cdot t} \int e^{-2\pi\iu f \cdot \tau} e^{2\pi\iu f \cdot \tau} f(\tau) \cdot g(t-\tau) \dd{\tau} \dd{t} \nonumber \\
							  &= \int \int  e^{-2\pi\iu f \cdot \tau} \cdot  f(\tau) \cdot e^{-2\pi\iu f \cdot (t-\tau)} \cdot g(t-\tau) \dd{\tau} \dd{t} \nonumber \\	      
							  &= \int e^{-2\pi\iu f \cdot \tau} \cdot  f(\tau)  \dd{\tau} \cdot \int  e^{-2\pi\iu f \cdot t'} \cdot g(t') \dd{t'}\hspace{-1.5mm}\text{, }t' = t - \tau \nonumber \\    
	&= \FT f(t) \cdot \FT g(t) \nonumber
\end{align}
The linear discrete convolution corresponds to the continuous convolution.
For two signals $f_n$ of length $N$ and $g_n$ of length $M$ the linear discrete convolution is defined as follows:
\begin{align}
	(f*g)_n = \sum^{N-1}_{k=0} f_k \cdot g_{n-k}
\end{align}
$g_{n-k}$ has the value $0$ for $n-k < 0$ and $n-k \ge M$. The resulting signal $(f*g)_n$ has a length of $N + M -1$.
If the the series $g_n$ is $M$-periodic ($g_n = g_{n + M}$) then the linear discrete convolution $(f*g)_n$ is also $M$-periodic:
\begin{align}
	(f*g)_n = \sum^{N-1}_{k=0} f_k \cdot g_{n-k} = \sum^{N-1}_{k=0} f_k \cdot g_{n+M-k} = (f*g)_{n+M}
\end{align}
Furthermore a circular discrete convolution exists.
Its definition, with two signals $f_n$ of length $N$ and $g_n$ of length $M$, is as follows:
\begin{align}
	(f \circledast g)_n = \sum^{N-1}_{k=0} f_k \cdot g_{(n-k)\smod N}
\end{align}
The result $(f \circledast g)_n$ has again a length of $N$. Due to the modulus operation the additional samples, which occur at the linear convolution, are summed up with the other ones.
The circular discrete convolution is connected to the \gls{DFT}, similar to the relationship between \gls{FT} and continuous convolution:
\begin{align}
	\widehat{(f \circledast g)}_k &= \sum_{j = 0}^{N - 1} e^{-2\pi\iu \frac{jk}{N}} \sum^{N-1}_{n=0} f_n \cdot g_{(j-n)\smod N} \\
								&= \sum_{j = 0}^{N - 1} \sum^{N-1}_{n=0} e^{-2\pi\iu \frac{jk}{N}}  f_n \cdot g_{(j-n)\smod N} \cdot e^{-2\pi\iu \frac{nk}{N}} \cdot e^{+2\pi\iu \frac{nk}{N}} \nonumber \\
								&= \sum_{j = 0}^{N - 1} \sum^{N-1}_{n=0} e^{-2\pi\iu \frac{(j-n)k}{N}} \cdot g_{(j-n)\smod N} \cdot e^{-2\pi\iu \frac{nk}{N}} \cdot   f_n  \nonumber \\
    							&= \sum_{m = 0}^{N - 1}  e^{-2\pi\iu \frac{mk}{N}} \cdot g_{m \smod N} \cdot  \sum^{N-1}_{n=0} e^{-2\pi\iu \frac{nk}{N}} \cdot   f_n \nonumber \\
								& = \left( \hat{f} \cdot \hat{g} \right)_{k} \nonumber
\end{align}
Due to the low complexity of the \gls{FFT} this enables fast circular convolutions for long input series.
In general the circular discrete convolution and the linear discrete convolution generate different output series for the same input.
But in the case of a series $f_n$ of length $N$ and an $N$-periodic series $g_n$ ($g_n = g_{n + N}$) both, the linear discrete convolution and the circular discrete convolution, yield the same result:
\begin{align}
	(f \circledast g)_n = \sum^{N-1}_{k=0} f_k \cdot g_{(n-k)\smod N} = \sum^{N-1}_{k=0} f_k \cdot g_{n-k} = (f*g)_n\label{eq:lcConvEq}
\end{align}
Furthermore the circular discrete convolution can be used to compute the linear discrete convolution between a kernel $f$ of length $N$ and a longer signal $g$ of length $M$, $N < M$~\cite{overlapsave}.
The basic idea is to split up the signal in overlapping chunks and use the following:
\begin{align}
	(f*g)_n = (f \circledast g)_n \text{ for } N - 1 \le n < M 
\end{align}
Hence the calculation of linear convolutions with long input series can be sped up with the use of an \gls{FFT}, too. 

\section{Sampling theorem}\label{sec:samplingTheorem}
\begin{figure}[htbp]
	\centering
	\includegraphics[width=0.95\textwidth]{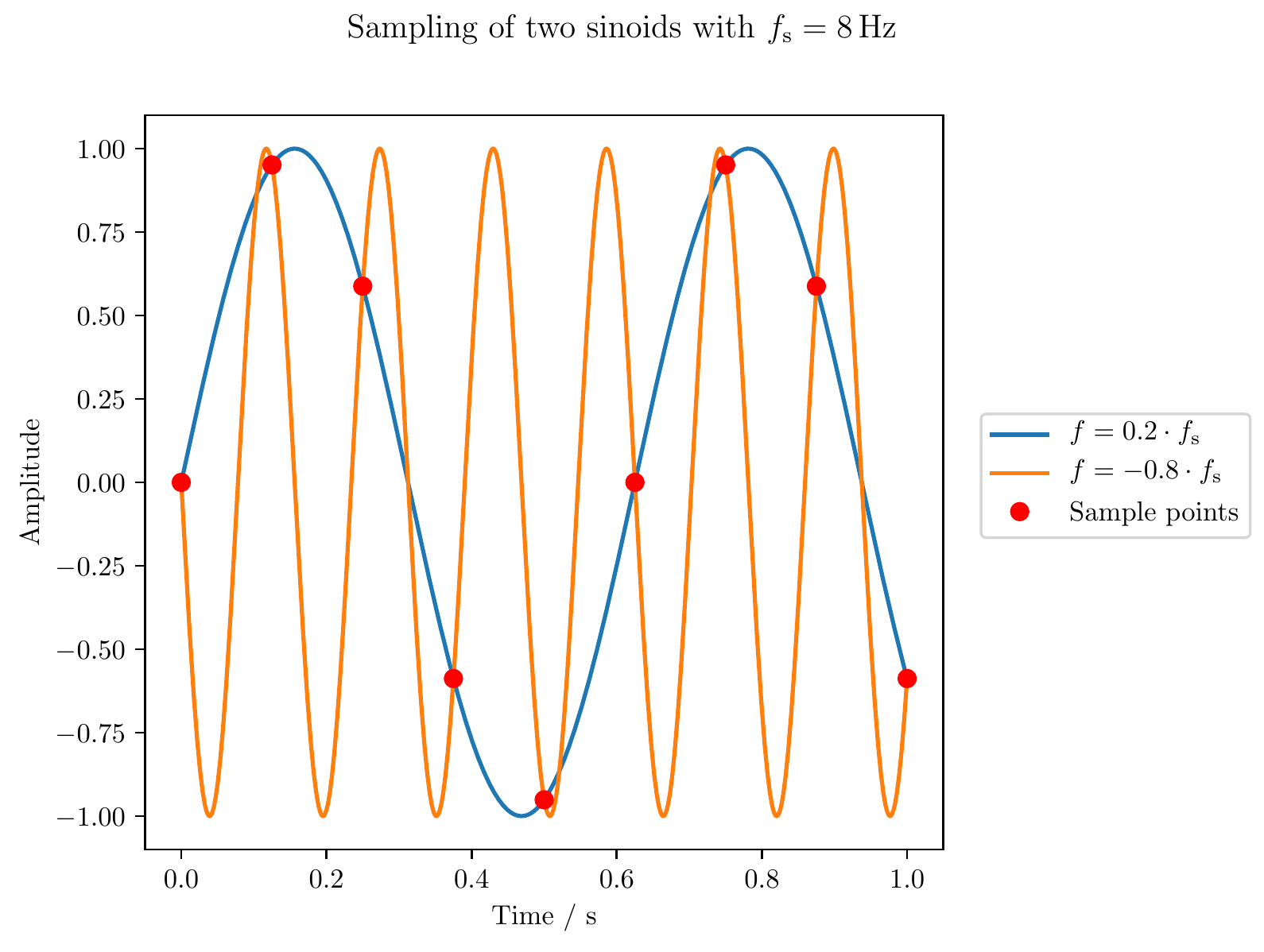}
	\caption{Two different sinusoidal signals sampled, which result in the same samples.}
	\label{fig:sampling}
\end{figure}
\noindent
The sampling converts a time-continuous signal into a time-discrete series. It is one of the premises for \gls{DSP} as it only operates on time-discrete series. The sampling can be represented by a multiplication with an impulse train $p(t)$ (for the transform of $p(t)$ see chapter 2.2 of~\cite{comb}): 
\begin{align}
	p(t) &= \sum_{n = -\infty}^{\infty} \delta\left( t - \frac{n}{f\id{s}} \right) \\
	\FT p(f) &= f\id{s} \sum_{k = -\infty}^{\infty} \delta\left(f - k f\id{s} \right) \label{eq:FTimpulseTrain}
\end{align}
The time-continuous input signal is $a(t)$. Therefore the signal after the sampling, $b(t)$, is
\begin{align}
	b(t) &= a(t) \cdot p(t) \\
	\FT b(f) &= \left(\FT{a} * \FT{p}\right)(f) \\
	&= \int \FT{a}(f-u) \cdot \FT{p}(u) \dd{u} \nonumber \\ 
	&= f\id{s} \sum_{k = -\infty}^{\infty} \int \FT{a}(f-u) \cdot \delta\left(u - k f\id{s} \right) \dd{u} \nonumber \\
	&= f\id{s} \sum_{k = -\infty}^{\infty} \FT{a}(f- k f\id{s})  \,. \nonumber
\end{align} 
\begin{figure}[htbp]
	\centering
	\includegraphics[width=0.95\textwidth]{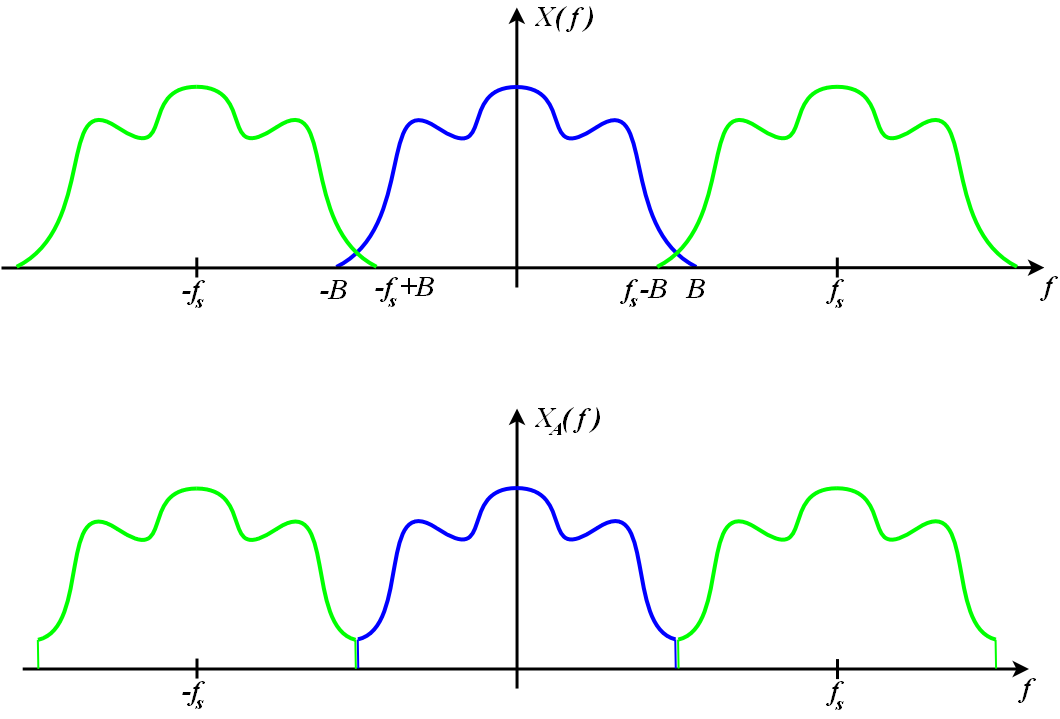}
	\caption{The power spectrum of two functions after sampling. The sampling frequency is given by $f\id{s}$ and the bandwidth of the signal by $B$. The upper one is not bandlimited, hence there are alias artifacts. The lower one is bandlimited ($B$ < $f\id{s}/2$), no alias artifacts appear~\cite{wiki:aliased}.}
	\label{fig:alisedSpectrum}
\end{figure}
From the last equation it is obvious that $\FT b(f)$ can contain an infinite number of frequency components from the time-continuous signal $a(t)$. Figure \ref{fig:sampling} shows this effect for a sampling with \SI{8}{\hertz} of two sinusoidal signals with \SI{1.6}{\hertz} and \SI{6.4}{\hertz}. At the sample points both signals have the same value and are not distinguishable after the sampling. 
To get an unambiguous spectrum the input signal has to be confirmed in the frequency region $(n-1) \cdot \frac{f\id{s}}{2}$ to $n \cdot \frac{f\id{s}}{2}$; this region is called $n$-th Nyquist-band. If the signal is not limited to a bandwidth of $\frac{f\id{s}}{2}$, different frequency components of the continuous signal are summed up in one frequency of the time discrete signal. Figure \ref{fig:alisedSpectrum} illustrates this in the frequency domain. If the Nyquist-sampling is not fulfilled the (upper part) different frequency components are summed up. This effect is called aliasing. The unaliased signal cannot be restored from the sampled data. To get a not aliased sampled signal, analog filters are needed before the sampling process (Anti-Aliasing-Filter) to limit the bandwidth of the continuous signal.

\section{Linear digital filter}
\label{sec:filters}
\begin{figure}[htbp]
	\centering
	\includegraphics[width=0.40\textwidth]{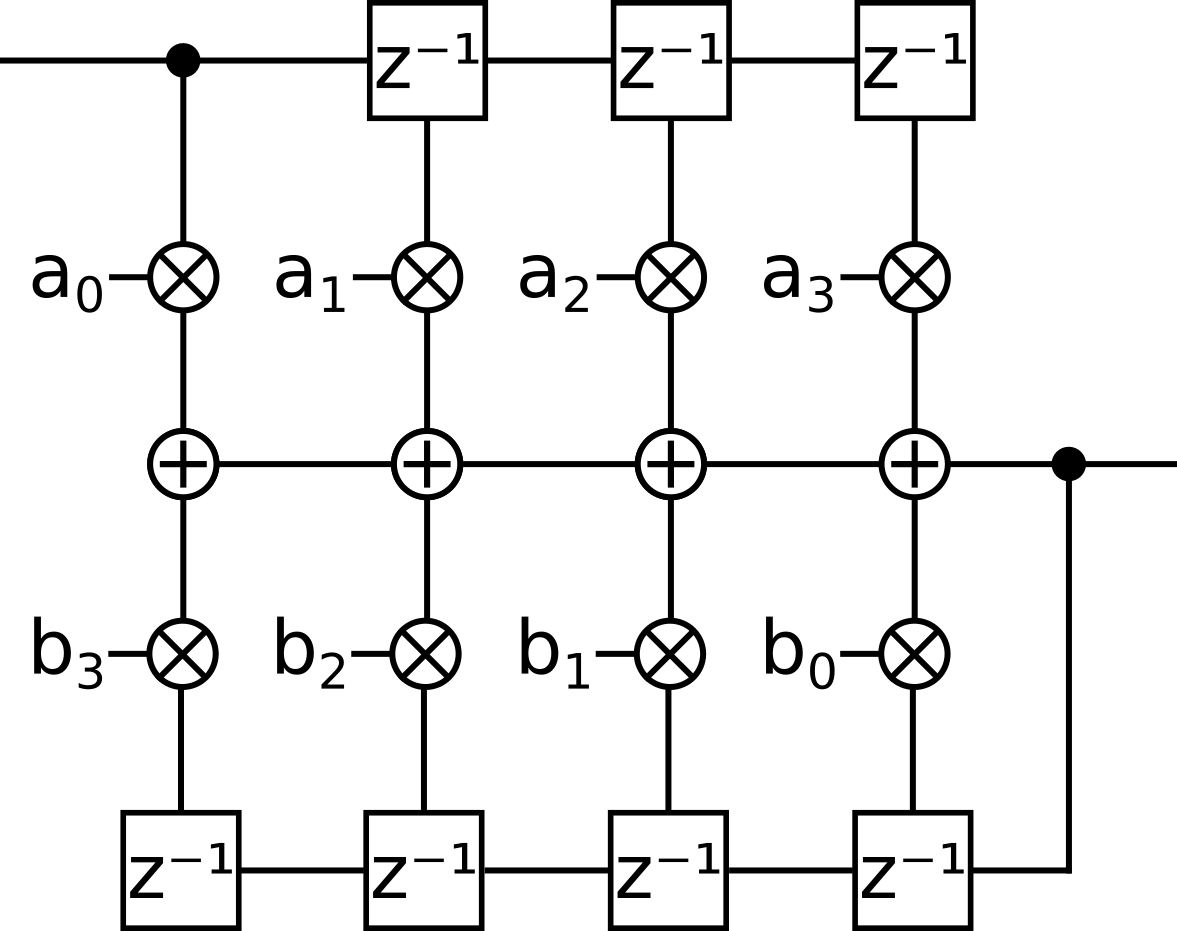}
	\caption{An \gls{IIRFilter} of length $M = N = 4$ with the coefficients $a_k$ and $b_j$.}
	\label{fig:IIR}
\end{figure}

Digital filters are the analogue to analog filter. They can be characterized by their impulse response. The impulse response is the output of a filter after a Dirac-Impulse $\delta_{0,n}$. Two groups of linear digital filters are distinguished
\begin{itemize}
	\item \Gls{FIRFilter}
	\item \Gls{IIRFilter}
\end{itemize}
They differ in the length of their impulse response. An \gls{FIRFilter} has a finite impulse response. Therefore its output only depends on a finite number of $N$ input samples and is $0$ after $n > N$ samples of value $0$. 
These $N$ samples are saved in memory. As there is no feedback in this architecture, \gls{FIRFilter} are stable under all conditions. An \gls{FIRFilter} of length $N$ with input $x_n$, output $y_n$ and coefficients $h_n$ can be defined as follows:
\begin{align}
	y_n &= \sum_{k=0}^{N-1} x_{n-k} \cdot h_k
\end{align}
\begin{figure}[htbp]
	\centering
	\includegraphics[width=0.40\textwidth]{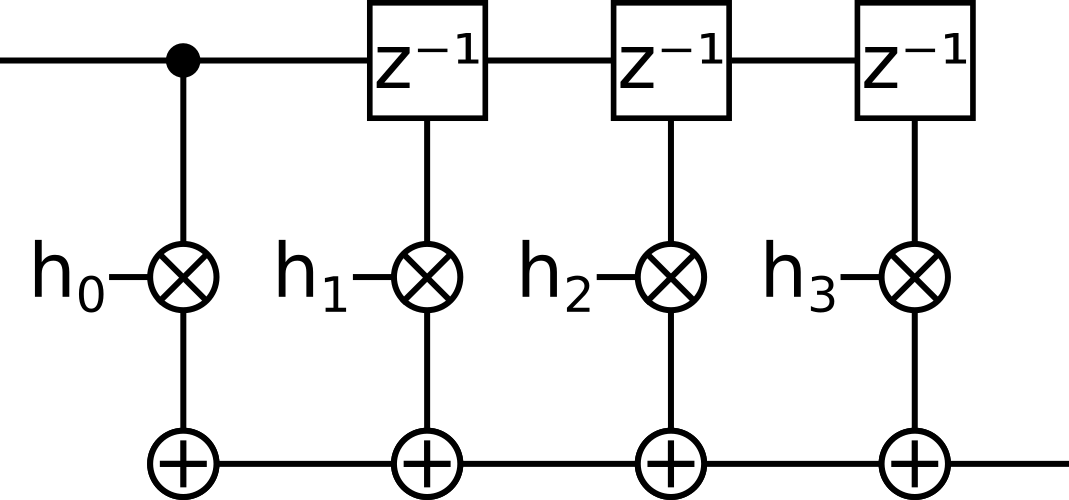}
	\caption{A \gls{FIRFilter} of length $4$ with the coefficients $h_k$.}
	\label{fig:FIR}
\end{figure}
Figure \ref{fig:FIR} represents a \gls{FIRFilter} of length $4$. The length of a \gls{FIRFilter} is sometimes also called number of taps. An \gls{FIRFilter} with $4$ taps is a filter with a length of $4$. 
A \gls{FIRFilter} corresponds to a linear convolution: 
\begin{align}
	y_n &= (h * x)_n
\end{align}
An \gls{IIRFilter} has an infinite impulse response, the output of the filter depends on all inputs samples, which have been processed by the filter. As the memory to save samples is limited, \gls{IIRFilter} use their output as second input (next to the sample which is to be processed). This feedback can lead to instabilities of the filter (e.g. behavior which depends strongly on the initial conditions and can lead to periodic behavior). There are algorithms to generate coefficients for stable \glspl{IIRFilter}, but the quantization of these coefficients can lead again to instable filters. Therefore the coefficients have to be checked after the quantization. 
An \gls{IIRFilter} with input $x_n$, output $y_n$ and $N$ coefficients $a_n$, $M$ coefficients $b_m$ can be defined as follows:
\begin{align}
	y_n &= \sum_{k=0}^{N-1} x_{n-k} \cdot a_k + \sum_{j=0}^{M-1} y_{n-j-1} \cdot b_j 
\end{align}
Figure \ref{fig:IIR} shows an \gls{IIRFilter} with $M = N = 4$.  

The transfer function $H(f)$ describes the changes in phase and amplitude of a signal component with frequency $f$ due to a filter. For digital filters the coefficients of a filter determine the corresponding transfer function. The transfer function at frequency bin $f_n$ of a a \gls{FIRFilter} with coefficients $h_n$ is given by the following:
\begin{align}
	H(f_n) &= \hat{h}_n
\end{align}
The transfer function between the points $H(f_n)$ is given by a $\sinc$-interpolation. 

\section{Analog digital converter}
\begin{figure}[htbp]
	\centering
	\begin{subfigure}[t]{0.45\textwidth}
		\centering
		\includegraphics[scale=0.8]{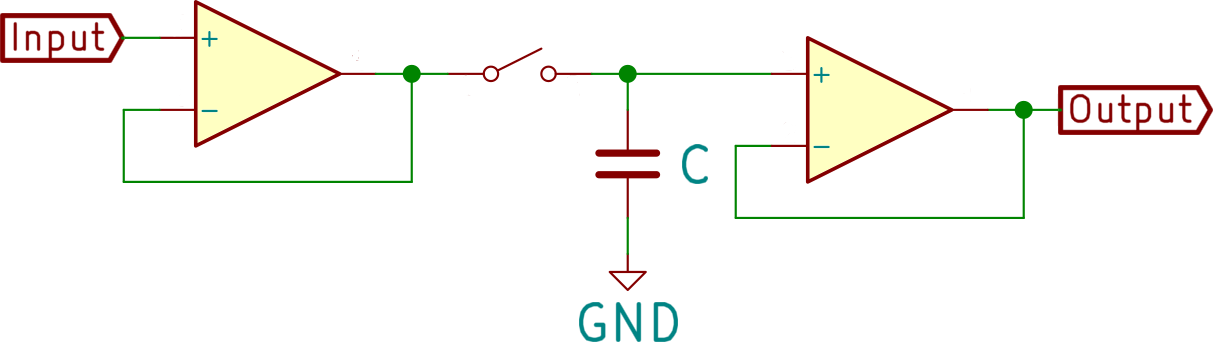}
		\caption{A \gls{SH} circuit, which samples the \texttt{Inupt} when the switch is closed. The sampled voltage is stored in the capacity and transferred to the \texttt{Output}. The buffer amplifiers decouple the \texttt{Input} from the capacitor \texttt{C} and the capacitor \texttt{C} from the \texttt{Output}.}
		\label{fig:SH}
	\end{subfigure}
	\quad
	\begin{subfigure}[t]{0.45\textwidth}
		\centering
		\includegraphics[scale=0.8]{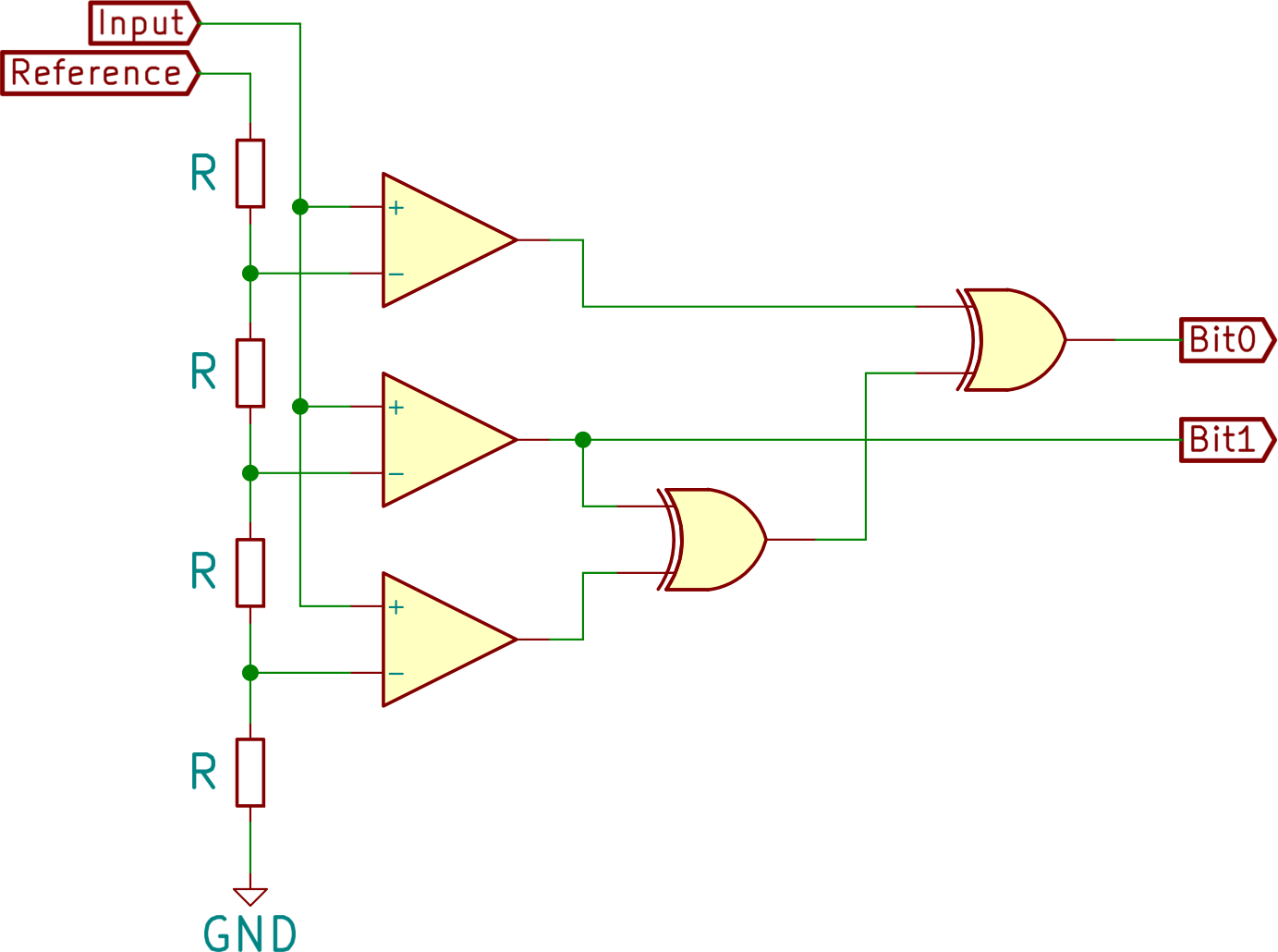}
		\caption{A simple $2$-bit FLASH-\gls{ADC} with a simple encoder logic composed of three comparators and two exclusive or gates. The signal which should be converted is applied to \texttt{Input}, the upper reference voltage is applied to \texttt{Reference}, whereas the lower one is given by \texttt{GND}.}
		\label{fig:ADC}
	\end{subfigure}
	\caption{The basic building blocks of an \gls{ADC}.}
\end{figure}

The \gls{ADC} is an electronic device to convert an analog signal (continuous in time and value) into a digital signal (discrete in time and value). An \gls{ADC} consists of a \gls{SH} followed by a converter unit.
The \gls{SH} stores the amplitude of the analog signal at one time; this element is responsible for the discretization in time.
Figure \ref{fig:SH} shows a simple \gls{SH} circuit.
The sampling frequency is given in \si{\SPS} (samples per second). 
This stored amplitude is then converted into a digital value of a certain bit-width by the converter unit; this element is responsible for the discretization in amplitude. This is always relative to an upper and lower reference voltage (e.g. \SI{0}{\volt} and \SI{5}{\volt}). Only amplitudes between these two are valid. 

A common type of \gls{ADC} is a FLASH-\gls{ADC}. Its converter is implemented by an array of comparators with different threshold voltages. These are followed by an encoding logic which converts the comparator outputs to a binary number. A simple $2$-bit version of this type is shown in figure \ref{fig:ADC}. 

An \gls{ADC} can be represented by a filter, with a transfer function, followed by a \gls{SH}. An input signal $a(t)$ is changed due to the transfer function $h$ of the \gls{ADC} as follows:
\begin{align}
	a'(t) &= \left(a * h\right)(t) \\
	\FT{a}'(f) &= \FT{a}(f) \cdot H(f)
\end{align}
The \gls{SH} is responsible for the sampling of the time-continuous signal. It corresponds to a multiplication of $a'(t)$ with an impulse train $p(t)$. The combination of both stages, the filter and the \gls{SH}, results in a digital signal as follows: 
\begin{align}
	b(t) &= a'(t) \cdot p(t) \\
	\FT{b}(f) &= \left(\FT{a}' * \FT{p}\right)(f) \\
	&= f\id{s} \sum_{k = -\infty}^{\infty} \FT{a}'(f- k f\id{s}) \nonumber \\
	&= f\id{s} \sum_{k = -\infty}^{\infty} \FT{a}(f- k f\id{s}) \cdot H(f- k f\id{s}) \nonumber
\end{align} 
The gain $g(f)$ and phase $\phi(f)$ alterations of the sampled signal with frequency $f$ is connected to the transfer function $h$ of the \gls{ADC}:
\begin{align}
	g(f) &= \abs{\FT{h}(f)}\\
	\phi(f) &= \arctan\left(\frac{\imaginary(\FT{h}(f))}{\real(\FT{h}(f))}\right) \\
	H(f) = \FT{h}(f) &= g(f) \cdot e^{\iu \phi(f)}
\end{align}

\subsection{Quantization error}\label{sec:quantization}
The quantization maps a range of continuous values $]x - \frac{\Delta}{2}, x + \frac{\Delta}{2}]$ to one discrete value $x$. This may cause a difference $\delta$ between discrete and continuous value, which is called quantization error. 
The statistical properties of this error can be derived under the assumption that the difference is uniformly distributed. The probability density $p(\delta) = const$ of each difference $\delta$ is determined by the normalization of the probability.  
\begin{align}
	1 &= \int_{-\frac{\Delta}{2}}^{+\frac{\Delta}{2}} p(\delta) \dd{\delta} \\
	  &= \Delta \cdot p \nonumber \\
	\Rightarrow p &= \frac{1}{\Delta}
\end{align}
The mean of this error is obviously zero ($p \cdot \int_{-\frac{\Delta}{2}}^{+\frac{\Delta}{2}} \delta \dd{\delta} = 0$). The standard deviation is derived as follows:
\begin{align}
	p \cdot \int_{-\frac{\Delta}{2}}^{+\frac{\Delta}{2}} \delta^2 \dd{\delta} &= \frac{1}{3} \cdot p \cdot \left[ \delta^3 \right]_{-\frac{\Delta}{2}}^{+\frac{\Delta}{2}} \\
	&= \frac{1}{3\Delta} \cdot \frac{2\Delta^3}{8} \nonumber \\
	&= \frac{\Delta^2}{12} \nonumber
\end{align}
Hence the error (or the noise) due to the quantization depends quadratically on the step size of the quantization. 

\section{Time interleaved ADC}\label{sec:tiadc}
\begin{figure}[htbp]
	\centering
	\includegraphics[height=0.40\textheight]{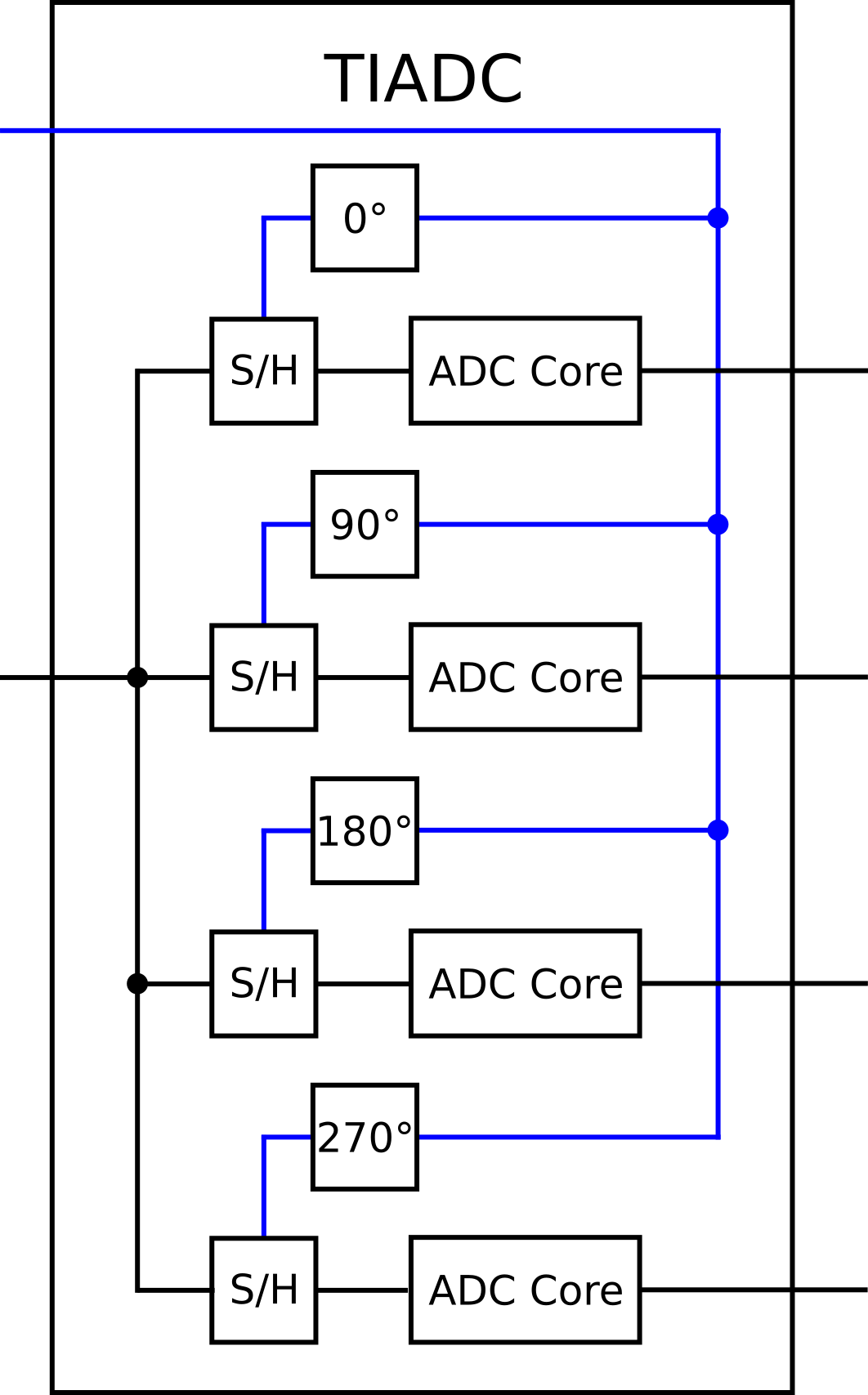}
	\caption{Schematic of an ideal \gls{TIADC} with $M=4$. The blue wires are the main clock of the \gls{TIADC}, which is then phase shifted before the \glspl{SH}.}
	\label{fig:tiadc}
\end{figure}

The sampling rate and the conversion rate of \glspl{ADC} are limited.
To achieve higher rates $M$ \glspl{ADC} ($M > 1$) can be combined into a \gls{TIADC}, where each \gls{ADC} samples with a frequency of $\frac{f\id{s}}{M}$.
A single \gls{ADC} within a \gls{TIADC} is called \gls{ADCCore}.
The clock for each \gls{ADCCore} $m$ ($m \in \mathbb{N}$, $m < M$) is shifted by $\frac{m}{f\id{s}}$ or $\frac{m}{M} \cdot \SI{360}{\degree}$ with respect to the phase of the input clock of the \gls{TIADC}. 
Figure \ref{fig:tiadc} shows this schematically for an ideal \gls{TIADC} with $4$ \glspl{ADCCore}.
Figure \ref{fig:tiadcSignals} shows the input signals (clock and signal) for such a \gls{TIADC} together with the phase shifted clock for each \gls{ADCCore} and the output of the corresponding \glspl{SH}.
The \glspl{SH} in front of the \glspl{ADCCore} have to be fast enough to sample at the total sample rate of the \gls{TIADC} $f\id{s}$.
Yet the following parts of each \gls{ADCCore} only run at $\frac{f\id{s}}{M}$.

\begin{figure}[htbp]
	\centering
	\includegraphics[width=0.95\textwidth]{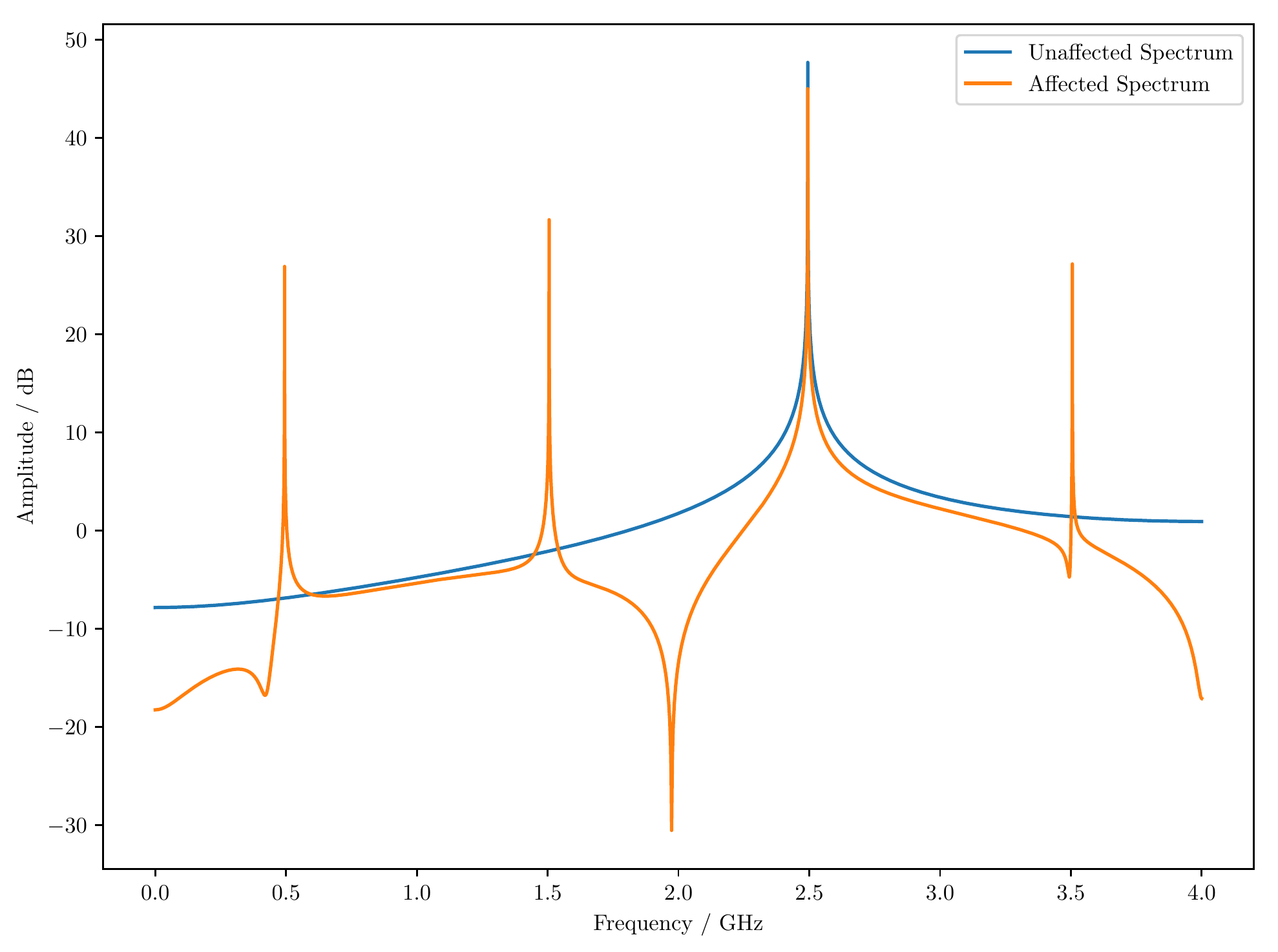}
	\caption{The spectrum of a sinusoidal signal with \SI{2.4944}{\giga\hertz} (unaffected spectrum) and the spectrum after the sampling with a \gls{TIADC} ($M=4$, no offset mismatches; affected spectrum).}	
	\label{fig:interleavingSpectrum}
\end{figure}

Each \gls{ADCCore} differs from the others (e.g. gain, offset, phase shift of input signal).
An offset of an \gls{ADCCore} shifts the whole signal by a constant voltage.
As the offset is constant it does not depend on the frequency of the input signal.
Hence offsets between the \glspl{ADCCore} are frequency independent mismatches.

Gain and phase mismatches can most easily be explained by the example of a sinusoidal signal, which is applied to the \gls{TIADC}.
In the case of gain mismatches the different \glspl{ADCCore} sample sinusoidal signals with different amplitudes.
In the case of phase mismatches the different \glspl{ADCCore} sample sinusoidal signals, which are shifted against each other in time.
The phase and gain mismatches between \glspl{ADCCore} can change with the frequency of the applied sinusoidal signal.
Therefore these are frequency dependent mismatches.
For a general signal, each frequency component of the signal is affected differently, depending on the frequency behavior of the gain and phase mismatches.
These two mismatches can be described by one complex number, as the gain and phase are the representation in polar coordinates.

An offset mismatch between \glspl{ADCCore} leads to spurs at the frequencies $\frac{m}{M} f\id{s}$.
Whereas gain and phase mismatches create mirror signals of a signal with frequency $f\id{in}$ at the frequencies $\frac{m}{M} f\id{s} \pm f\id{in}$~\cite{Seo2005}.
Figure \ref{fig:interleavingSpectrum} shows a spectrum with spurs due to phase and gain mismatches in comparison to the unaffected spectrum. 

\Glspl{ADCCore} within a \gls{TIADC} often allow to trim their offset, gain and phase.
But a single setting to trim these may not be sufficient for all frequencies; especially if high bandwidths or higher Nyquist-bands are needed.
Therefore a frequency dependent correction of phase and gain is highly advisable. 

\Glspl{TIADC} can be implemented in one \gls{IC} (e.g. \verb|ADC12DJ3200|~\cite{ADC12DJ3200}) or by using more than one \gls{IC} (e.g. \verb|dFFTS-4G| spectrometer~\cite{bklein}).
In the first case the differences between the \glspl{ADC} are smaller than in the latter one.
This is due to the lower variance in the production process for high proximities on a wafer and the longer shared signal flow. 

In the following section a model is developed for the frequency dependent mismatches and their impact on the spectrum of the signal.

\subsection{Model of a TIADC}
\begin{figure}[htbp]
	\centering
	\includegraphics[height=0.40\textheight]{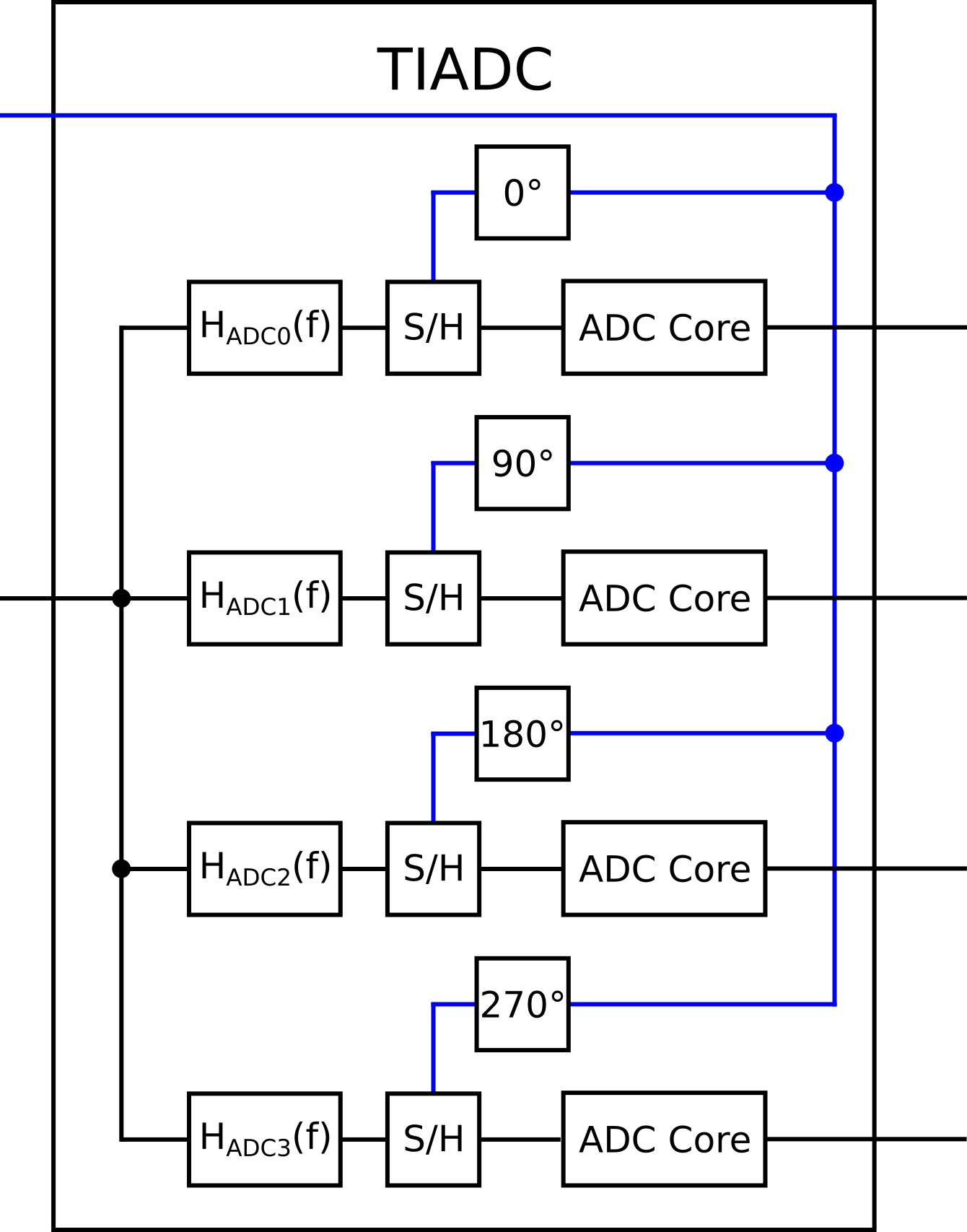}
	\caption{Schematic of a real \gls{TIADC} with $M=4$. The blue wires are the main clock of the \gls{TIADC}, which is then phase shifted before the \glspl{SH}.}
	\label{fig:tiadc_filter}
\end{figure}

The following model takes into account only the phase and gain mismatches.
The offset is assumed to be $0$.
The \gls{TIADC} consists of $M$ \glspl{ADCCore}, where each \gls{ADCCore} samples with $\frac{f\id{s}}{M}$.
The input signal $x(t)$ is band-limited to $\frac{f\id{s}}{2}$.
Hence the spectrum $X(f)$ is limited between $0$ and $f\id{s}/2$ and a sampling with $f\id{s}$ leads to no aliasing.
Each \gls{ADCCore} $m$ has its own phase-shifted impulse train $p_m(t)$ and transfer function $H_{\text{ADC}m} = \FT h_{\text{ADC}m}$, which represent the mismatches between the \glspl{ADCCore}.
Figure \ref{fig:tiadc_filter} shows a schematic view of a real \gls{TIADC}, where the filters $H_{\text{ADC}m}$ are included to model the mismatches.

The spectrum of each impulse train $p_m(t)$ is given by $P_m(f)$ and calculates as follows:
\begin{align}
	p_m(t) &= p\left(t - \frac{m}{f\id{s}}\right) \\
	P_m(f) &= \FT{p}_m(f) = \FT{p}(f) \cdot e^{-2\pi\iu f \frac{m}{f\id{s}}} = P(f) \cdot e^{-2\pi\iu f \frac{m}{f\id{s}}}
\end{align}
The sampled, recombined waveform is given by $b(t)$, which is the sum of all \glspl{ADCCore}:
\begin{align}
	b_m(t) &= (x * h_{\text{ADC}m})(t) \cdot p_m(t) \\
	b(t) &= \sum_{m = 0}^{M-1} b_m(t)
\end{align}
The \gls{FT} of $b_m(t)$ is then given by:
\begin{align}
	B_m(f) &= \FT(b_m)  \\
	       &= \left(\left(X\cdot H_{\text{ADC}m}\right) * P_m\right)(f) \nonumber \\
	       &= \int X(f-\tau) \cdot H_{\text{ADC}m}(f-\tau) \cdot P_m(\tau) \dd{\tau} \nonumber \\
	       &= \int e^{-2\pi\iu \tau \frac{m}{f\id{s}}} P(\tau) \cdot X(f-\tau) \cdot H_{\text{ADC}m}(f-\tau) \dd{\tau} \nonumber \\
	       &= \frac{f\id{s}}{M} \sum_{k = -\infty}^{\infty} \int e^{-2\pi\iu \tau \frac{m}{f\id{s}}} \delta\left(\tau - k \frac{f\id{s}}{M}\right) \cdot X(f-\tau) \cdot H_{\text{ADC}m}(f-\tau) \dd{\tau} \nonumber \\
	       &= \frac{f\id{s}}{M} \sum_{k = -\infty}^{\infty} e^{-2\pi\iu k \frac{f\id{s}}{M} \frac{m}{f\id{s}}} \cdot X\left(f - k \frac{f\id{s}}{M}\right) \cdot H_{\text{ADC}m}\left(f - k \frac{f\id{s}}{M}\right) \nonumber \\
	       &= \frac{f\id{s}}{M} \sum_{k = -\infty}^{\infty} e^{-2\pi\iu \frac{k\cdot m}{M}} \cdot X\left(f - k \frac{f\id{s}}{M}\right) \cdot H_{\text{ADC}m}\left(f - k \frac{f\id{s}}{M}\right) \nonumber
\end{align}
As the input signal $x(t)$ is band-limited, the spectrum $X(f)$ is only non zero for $0 \le f < \frac{f\id{s}}{2}$:
\begin{align}
	B_m(t) &= \frac{f\id{s}}{M} \sum_{k = 0}^{M-1} e^{-2\pi\iu \frac{k\cdot m}{M}} \cdot X\left(f - k \frac{f\id{s}}{M}\right) \cdot H_{\text{ADC}m}\left(f - k \frac{f\id{s}}{M}\right)
\end{align}
The combined signal is
\begin{align}
	B(f) &= \sum_{m = 0}^{M-1} \frac{f\id{s}}{M} \sum_{k = 0}^{M-1} e^{-2\pi\iu \frac{k\cdot m}{M}} \cdot X\left(f - k \frac{f\id{s}}{M}\right) \cdot H_{\text{ADC}m}\left(f - k \frac{f\id{s}}{M}\right) \\
		 &= \frac{f\id{s}}{M} \sum_{k = 0}^{M-1} X\left(f - k \frac{f\id{s}}{M}\right) \sum_{m = 0}^{M-1} e^{-2\pi\iu \frac{k\cdot m}{M}} \cdot H_{\text{ADC}m}\left(f - k \frac{f\id{s}}{M}\right) \nonumber \\
		 &= f\id{s} \sum_{k = 0}^{M-1} X\left(f - k \frac{f\id{s}}{M}\right) \cdot c_k\left(f - k \frac{f\id{s}}{M}\right) \nonumber
\end{align}
with 
\begin{align}
	c_k(f) &= \frac{1}{M} \sum_{m = 0}^{M-1} e^{-2\pi\iu \frac{k\cdot m}{M}} \cdot H_{\text{ADC}m}\left(f\right)\,.
\end{align}
For an ideal \gls{TIADC}, the filter characteristics in front of the \glspl{ADCCore} are given by $H_{\text{ADC}m}(f) = 1$.
This leads to $c_k(f) = \delta_{0,k}$ and $B(f) = \frac{f\id{s}}{M} X\left(f\right)$, which corresponds to a spurious-free spectrum as it is expected for this case.

\subsection{Corrections of mismatching transfer functions}\label{subsec:correction}
\begin{figure}[htbp]
	\centering
	\includegraphics[height=0.40\textheight]{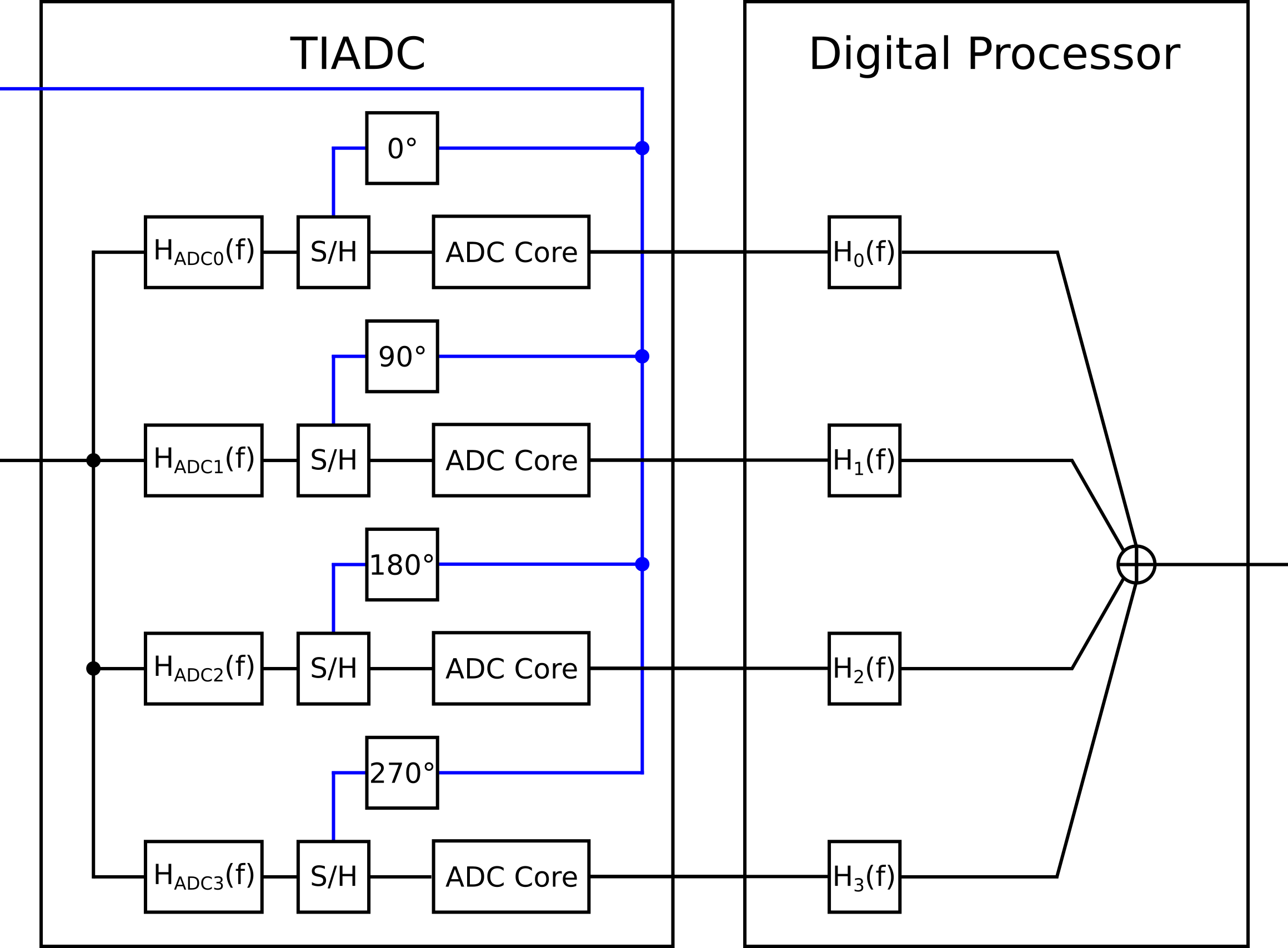}
	\caption{Schematics of a \gls{TIADC} with $M=4$ and a digital circuit to correct the mismatches with four filters; one for each \gls{ADCCore}.}
	\label{fig:tiadc_fir}
\end{figure}

The model of a \gls{TIADC} can now be used to find a scheme to compensate for the mismatches. To achieve this, the filters $H_{\text{ADC}m}(f)$ have to be compensated. This can be done with a digital filter $H_m(f)$ after each \gls{ADCCore} (see figure \ref{fig:tiadc_fir} for a schematic).
The added filter $H_m(f)$ changes the model of a \gls{TIADC} as follows:
\begin{align}
	\tilde{B}_m(f) &= \frac{f\id{s}}{M} \sum_{k = -\infty}^{\infty} e^{-2\pi\iu \frac{k\cdot k}{M}} \cdot X\left(f - k \frac{f\id{s}}{M}\right) \cdot H_{\text{ADC}m}\left(f - k \frac{f\id{s}}{M}\right) \cdot H_m(f)
\end{align}
This leads to the equation
\begin{align}
	\tilde{B}(f) &= f\id{s} \sum_{k = 0}^{M-1} X\left(f - k \frac{f\id{s}}{M}\right) \cdot \tilde{c}_k\left(f - k \frac{f\id{s}}{M}\right)
\end{align}
with 
\begin{align}
	\tilde{c}_k(f) &= \frac{1}{M} \sum_{m = 0}^{M-1} e^{-2\pi\iu \frac{k\cdot m}{M}} \cdot H_{\text{ADC}m}\left(f\right) \cdot H_m\left(f + k \frac{f\id{s}}{M}\right)\,.
\end{align}
To achieve a spur free spectrum, $\tilde{c}_k$ has to be $0$ for $k \ne 0$ and for all frequencies. In this case $\tilde{B}(f)$ simplifies to:
\begin{align}
	\tilde{B}(f) &= f\id{s} X(f) \cdot \tilde{c}_0\left(f\right)
\end{align}
Therefore the corrected spectrum contains no artifacts due to the mismatches except a disturbed bandpass behavior of the whole system. To also achieve an undisturbed bandpass assure $\tilde{c}_0 = 1$ for all frequencies. 

In the next step a scheme to calculate $H_m(f)$ from a given mismatch set $H_{\text{ADC}m}(f)$ is developed. At first the special case $M=2$ is solved, to get a rough idea of how to approach the problem. Second this approach is generalized to $M$ \glspl{ADCCore}. 

To achieve a spectrum that is free from mirror signals and has a flat bandpass behavior the following conditions have to be matched:
\begin{itemize}
	\item $\tilde{c}_0(f) = 1$
	\item $\tilde{c}_k(f) = 0$ for $k  \ne 0$  
\end{itemize}
This leads to the equation:
\begin{align}
	\delta_{m,0} = \tilde{c}_m(f) \label{eq:correction_equations}
\end{align}
For $M=2$ the following system of equations can be found:
\begin{align}
	2 &= H_{\text{ADC}0}(f) \cdot H_0(f) + H_{\text{ADC}1}(f) \cdot H_1(f) \\
	0 &= H_{\text{ADC}0}(f) \cdot H_0\left(f + \frac{1}{2} f\id{s}\right) - H_{\text{ADC}1}(f) \cdot H_1\left(f + \frac{1}{2} f\id{s}\right) 
\end{align}
The frequency components $f$ of the signal are bandlimited to $0 \le f < f\id{s}/2$. Consequently the transfer function of the filters $H_m$ and \glspl{ADCCore} $H_{\text{ADC}k}$ are also limited to this range. Therefore the following properties of the \gls{DFT} are needed to transform $H_m$ and $H_{\text{ADC}k}$ into the frequency range:
\begin{itemize}
	\item The symmetry of the spectrum of a sampled, real valued function: $H(f) = H^*(f\id{s}-f)$
	\item The periodicity of the spectrum of a sampled function: $H(f + f\id{s}) = H(f)$
\end{itemize}
As a consequence of the first property, the filter kernels $h_m$ is restricted to be real ($h_m \in \RealNumbers$). With the use of the symmetry the system of equations transforms as follows:
\begin{align}
	2 &= H_{\text{ADC}0}(f) \cdot H_0(f) + H_{\text{ADC}1}(f) \cdot H_1(f) \\
	0 &= H_{\text{ADC}0}(f) \cdot H^*_0\left(\frac{1}{2} f\id{s} - f\right) - H_{\text{ADC}1}(f) \cdot H^*_1\left(\frac{1}{2} f\id{s} - f\right) 
\end{align}
The system of equations consists of $2$ equations but $4$ unknowns. Therefore another $2$ equations with the same $4$ unknowns are needed. The start point is a second instance of the system of equations with a different frequency $f' = \frac{1}{2}f\id{s} - f$:
\begin{align}
	2 &= H_{\text{ADC}0}\left(\frac{1}{2} f\id{s} - f\right) \cdot H_0\left(\frac{1}{2} f\id{s} - f\right) + H_{\text{ADC}1}\left(\frac{1}{2} f\id{s} - f\right) \cdot H_1\left(\frac{1}{2} f\id{s} - f\right) \\
	0 &= H_{\text{ADC}0}\left(\frac{1}{2} f\id{s} - f\right) \cdot H_0\left(f\id{s} - f\right) - H_{\text{ADC}1}\left(\frac{1}{2} f\id{s} - f\right) \cdot H_1\left(f\id{s} - f\right) 
\end{align}
Applying the symmetry onto $H_0\left(\frac{1}{2} f\id{s} - f\right)$ and $H_1\left(\frac{1}{2} f\id{s} - f\right)$ leads to:
\begin{align}
	2 &= H_{\text{ADC}0}\left(\frac{1}{2} f\id{s} - f\right) \cdot H_0\left(\frac{1}{2} f\id{s} - f\right) + H_{\text{ADC}1}\left(\frac{1}{2} f\id{s} - f\right) \cdot H_1\left(\frac{1}{2} f\id{s} - f\right) \\
	0 &= H_{\text{ADC}0}\left(\frac{1}{2} f\id{s} - f\right) \cdot H^*_0\left(f\right) - H_{\text{ADC}1}\left(\frac{1}{2} f\id{s} - f\right) \cdot H^*_1\left(f\right) 
\end{align}
Combining these two systems of equations and complex conjugating the latter two equations leads to:
\begin{align}
	2 &= H_{\text{ADC}0}(f) \cdot H_0(f) + H_{\text{ADC}1}(f) \cdot H_1(f) \\
	2 &= H_{\text{ADC}0}\left(\frac{1}{2} f\id{s} - f\right) \cdot H_0\left(\frac{1}{2} f\id{s} - f\right) + H_{\text{ADC}1}\left(\frac{1}{2} f\id{s} - f\right) \cdot H_1\left(\frac{1}{2} f\id{s} - f\right) \\
	0 &= H^*_{\text{ADC}0}(f) \cdot H_0\left(\frac{1}{2} f\id{s} - f\right) - H^*_{\text{ADC}1}(f) \cdot H_1\left(\frac{1}{2} f\id{s} - f\right) \\
	0 &= H^*_{\text{ADC}0}\left(\frac{1}{2} f\id{s} - f\right) \cdot H_0\left(f\right) - H^*_{\text{ADC}1}\left(\frac{1}{2} f\id{s} - f\right) \cdot H_1\left(f\right) 
\end{align}

For the general case of $M$ \glspl{ADCCore} the initial equation system consists of $M$ equations where the $k$-th equation ($0 \le k < M$) is (see equation \ref{eq:correction_equations}):
\begin{align}
	\delta_{k,0} = \frac{1}{M} \sum_{m = 0}^{M-1} e^{-2\pi\iu m \frac{k}{M}} \cdot H_{\text{ADC}m}(f) \cdot H_m\left(f + \frac{k}{M} f\id{s} \right) 
\end{align}
This system of equations has $M\cdot M$ unknowns $H_m\left(f + \frac{k}{M} f\id{s} \right), 0 \le k, m < M$. Therefore another $M^2 - M = M\cdot(M-1)$ equations are needed to solve this system. These equations have the same structure as the ones above  (see equation \ref{eq:correction_equations}), but with a different frequencies $f_j' = f + \frac{j}{M}f\id{s}$ with $0 < j < M$. 
The full system of equations is consequently the collection of the following equation with $k$ and $j$ running independently from $0$ to $M$. The $(k,j)$-th equation of the system is given by:
\begin{align}
	\delta_{k,0} = \frac{1}{M} \sum_{m = 0}^{M-1} e^{-2\pi\iu m \frac{k}{M}} \cdot H_{\text{ADC}m}\left(f + \frac{j}{M}f\id{s}\right) \cdot H_m\left(f + \frac{k+j}{M} f\id{s} \right) 
	\label{eq:generateFilter}
\end{align}
This system has $2M^2-M$ unknowns $H_m\left(f + \frac{k+j}{M} f\id{s} \right)$, but with the use of the following two properties of the \gls{DFT} the number of unknowns can be reduced to $M^2$:
\begin{itemize}
	\item The symmetry of the spectrum of a sampled, real valued function: $H(f) = H^*(f\id{s}-f)$
	\item The periodicity of the spectrum of a sampled function: $H(f + f\id{s}) = H(f)$
\end{itemize}
Which property has to be applied to which equation depends on $f + \frac{j}{M}f\id{s}$ and $f + \frac{j+k}{M}f\id{s}$; it has to be chosen in such a way that the frequency $f'$ after the transformation $0 \le f' < f\id{s}/2$. After applying the properties the system has a unique solution. The generated $H_m(f)$ corresponds to \glspl{FIRFilter} with real filter coefficients and a length of $N$. If the signal is not in the first Nyquist-band, but still limited to a band of width $f\id{s}/2$, $f$ can be simply replaced by the aliased frequency. 

\section{Field programmable gate array}
\Glspl{FPGA} are \glspl{IC} that allow repeated (re)configuration. They consist of different blocks which are implemented in silicon:
\begin{itemize}
	\item \Glspl{LUT} with \glspl{FF}
	\item Clock management units
	\item Input/Output blocks
	\item \glspl{BRAM} memory cells
	\item \glspl{DSPSlice}
	\item Configurable connections between blocks
\end{itemize} 
To connect all this different blocks a configurable interconnect matrix exists in-between the blocks. The \glspl{LUT} are used to implement binary functions. The associated \glspl{FF} can store the output of the \glspl{LUT}. The \glspl{FF} are controlled by a clock, which allows the creation of clocked and therefore time-dependent functions. The clocks are controlled by a clock management unit, which allows functions, such as the multiplication of the frequency by a constant factor. The Input/Output blocks connect the \gls{FPGA} to the other components of the circuit. This blocks transforms the signals between the different signal standards used inside of an \gls{FPGA} and in the circuit (e.g. single ended/differential or the voltage level). 

Early \glspl{FPGA} contained only of simple elements, like \glspl{LUT}, \glspl{FF} and Input/Output blocks. Modern \glspl{FPGA} have further blocks which are more specialized to one purpose. While the \glspl{BRAM} are used to save larger sets of data, the \glspl{DSPSlice} are used to perform fast arithmetic operations (especially multiplications and accumulations). 

The logic to be implemented in the \gls{FPGA} is described in a \gls{HDL}. A synthesizer compiles the described logic into a list of used blocks and their connection. Then this block instances are placed in the \gls{FPGA} and the connection between them are routed. This is called implementation of a design.  
In this thesis the used \gls{HDL} is \gls{VHDL}.

In contrast to \glspl{CPU}, which operates sequentially, an \gls{FPGA} operates in parallel. So all described operations are carried out in each clock-cycle. This leads to a deterministic data-flow. Therefore they are well suited for high-throughput systems, in combination with a high computational power due to the \glspl{DSPSlice}. The next step in computational power is delivered by \glspl{ASIC}, which are custom \gls{IC}. The workflow is comparable to the one for an \gls{FPGA}. But the blocks can be placed freely on the chip because later they are etched into the silicium wafer. The costs for setting up an \gls{ASIC} production are high but a single \gls{IC} is cheap. 
Hence \glspl{ASIC} are cheaper for very large quantities of equal \glspl{IC}. But the are not reconfigurable, which reduces the flexibility and possibility to fix bugs.

%==============================================================================
\chapter{Hardware}
\label{sec:hardware}
%==============================================================================
\section{Spectrometer}
\begin{figure}[htbp]
	\centering
	\includegraphics[width=0.55\textwidth]{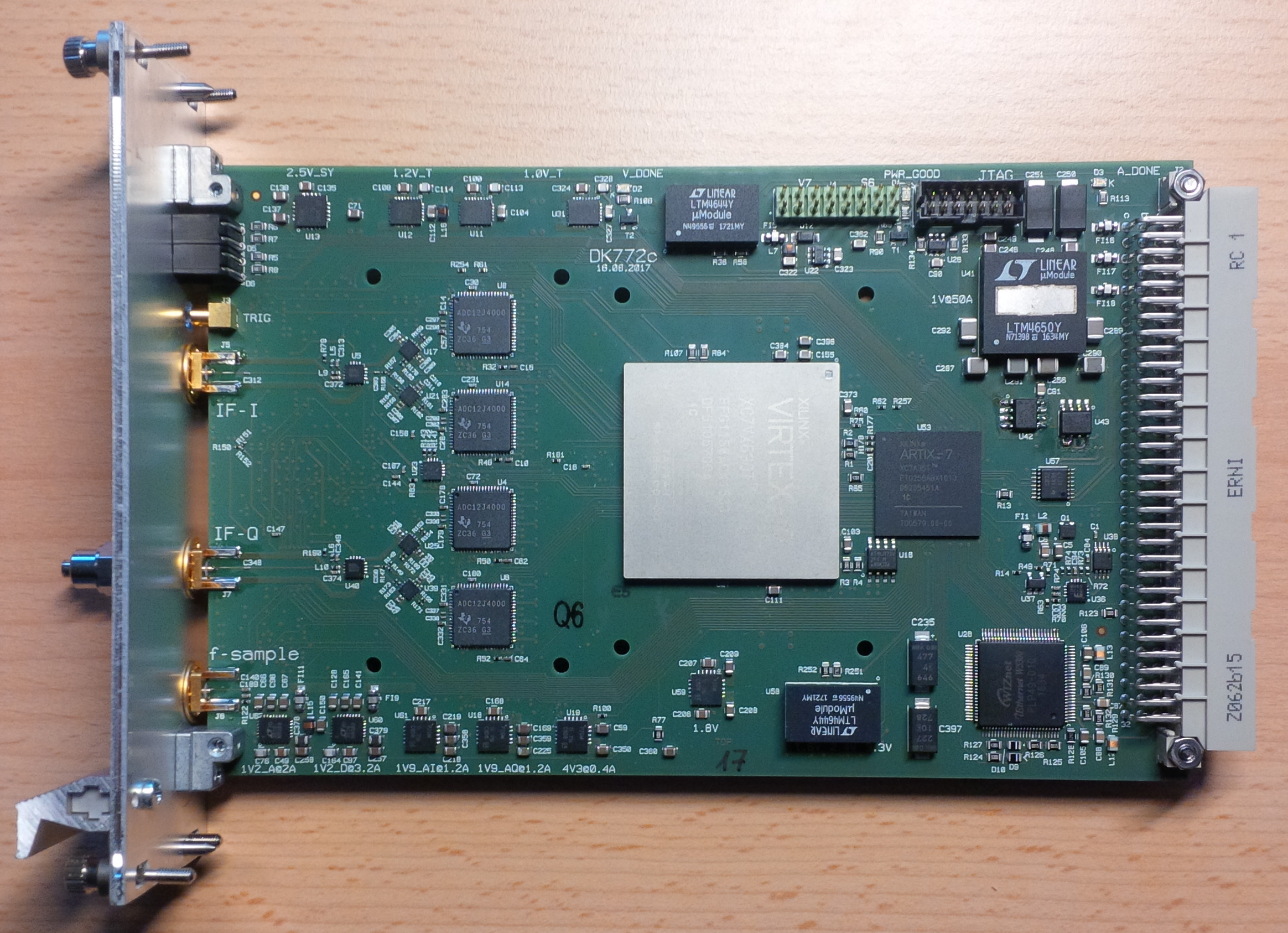}
	\caption{The current generation \texttt{dFFTS-4G} of spectrometers without heatspreader. Modified version from~\cite{bklein}.}
	\label{fig:dffts}
\end{figure}
\begin{figure}[htbp]
	\centering
	\includegraphics[width=0.95\textwidth]{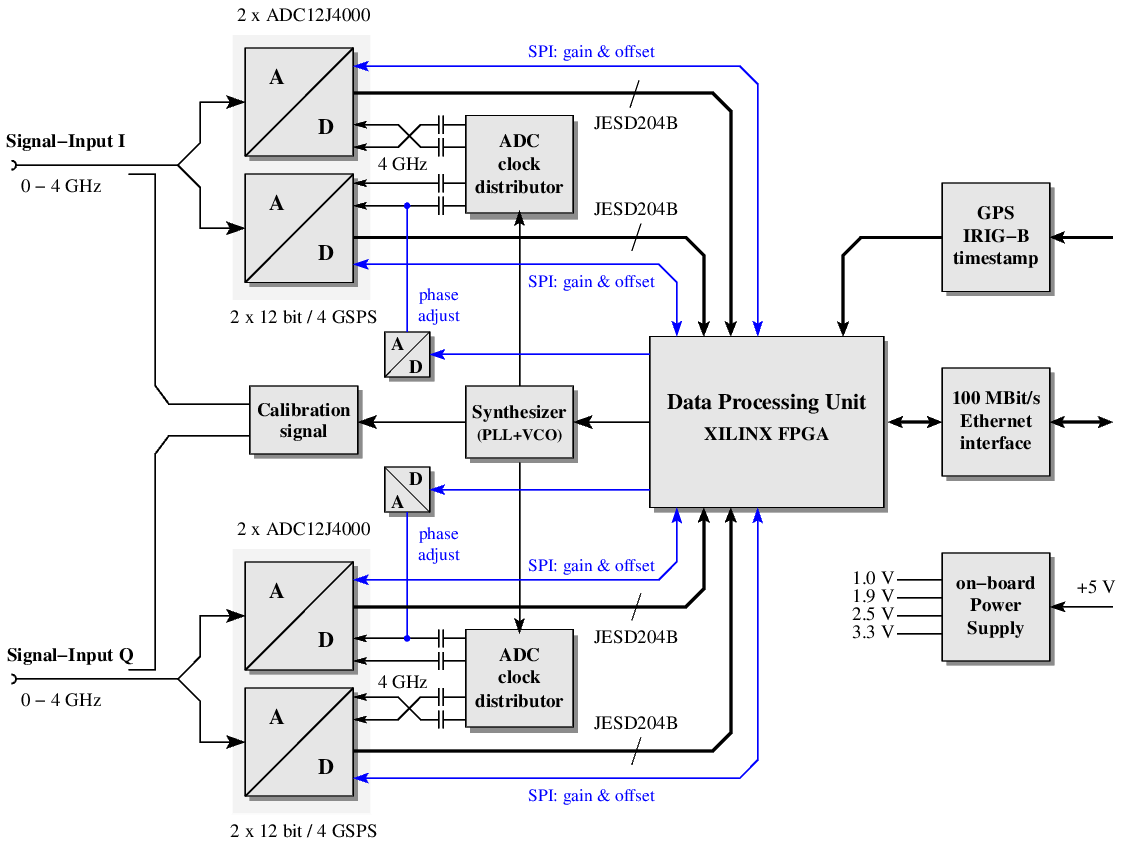}
	\caption{The current generation \texttt{dFFTS-4G} of spectrometers~\cite{bklein}.}
	\label{fig:schematics_dffts}
\end{figure}

\noindent
The latest generation of spectrometer developed by the \gls{DSP}-division is the \verb|dFFTS-4G|. Figure \ref{fig:dffts} pictures the \verb|dFFTS-4G|, while figure \ref{fig:schematics_dffts} shows a schematic view of the different componnets of the spectrometer. This spectrometer has two independent inputs \verb|I| and \verb|Q|, each of them is sampled with \SI{8}{\GSPS}, resulting in a Nyquist-frequency of \SI{4}{\giga\hertz}. It is optimized for observations in the first Nyquist-band. To achieve this samplerate, two \glspl{ADC} of type \verb|ADC12DJ4000| are externaly interleaved for each input. Each \gls{ADC} consists of $4$ \glspl{ADCCore}, resulting in $8$ \glspl{ADCCore} per input. The analog bandwidth (defined by the \SI{-3}{\dB} point in the frequency domain) of the \glspl{ADC} is at \SI{3.2}{\giga\hertz}~\cite{ADC12J4000}. But the spectrometer can still sample the second Nyquist-band directly, but with a worse sensitivity. The number of spectral channels varies between \SI{16}{k} and \SI{64}{k} per input, resulting in a channel width of \SIrange{61.04}{244.10}{\kilo\hertz}. The spectrometer achieves a frequency resolution of up to \SI{70.8}{\kilo\hertz}~\cite{bklein}. An onboard synthesizer can be used to generate a periodic signal for the calibration of the two \glspl{TIADC}.
The \glspl{ADC} are connected to a \verb|Xilinx®| \verb|XC7VX690T|~\cite{Virtex7} \gls{FPGA} that receives and processes the samples. The \gls{FPGA} belongs to the \verb|Virtex-7| family, which is the high-performance family in a \SI{28}{\nano\meter} process from \verb|Xilinx®|. The \gls{FPGA} has $3600$ \glspl{DSPSlice} and $1470$ \glspl{BRAM} with a total size of \SI{52920}{\kibi\bit}. In the typicall applications of the spectromter the two inputs are treated independently in terms of signal processing. Therefore this is also done in this thesis. The methods developed in this thesis are also applicable if they are not treated independently. 

The spectrometer communicates via a \SI{100}{\mega\bit} Ethernet connection with the controlling computer. This Ethernet connection is controlled by a second smaller \gls{FPGA}, which communicates with the \gls{FPGA} used for \gls{DSP}. The readout frequency of the spectrometer is limited due to the Ethernet connection. If the readout frequency is chosen to high, spectra are lost.

The current implementation already features a calibration of the \glspl{ADCCore}.
A rectangular test signal of \SI{901}{\mega\hertz} is injected into the signal path.
The harmonics of the rectangular signal are filtered to achieve a more pure sinusoidal signal.
The gain mismatches of the \glspl{ADCCore} are estimated by the standard deviation between the samples of from them.
After the gain is corrected, the phase mismatches can be estimated by the integral over the absolute differences between the samples of the \glspl{ADCCore}. More information about this calibration can be found in chapter 4.2 of ~\cite{prom}.
With phase and gain adjustments in the \gls{ADC}-\glspl{IC} this mismatches are then minimized.
Given by first principles this solution can only minimize the mismatches at one frequency.

\section{Measurement setup}
\begin{figure}[htbp]
	\centering
	\includegraphics[width=0.55\textwidth]{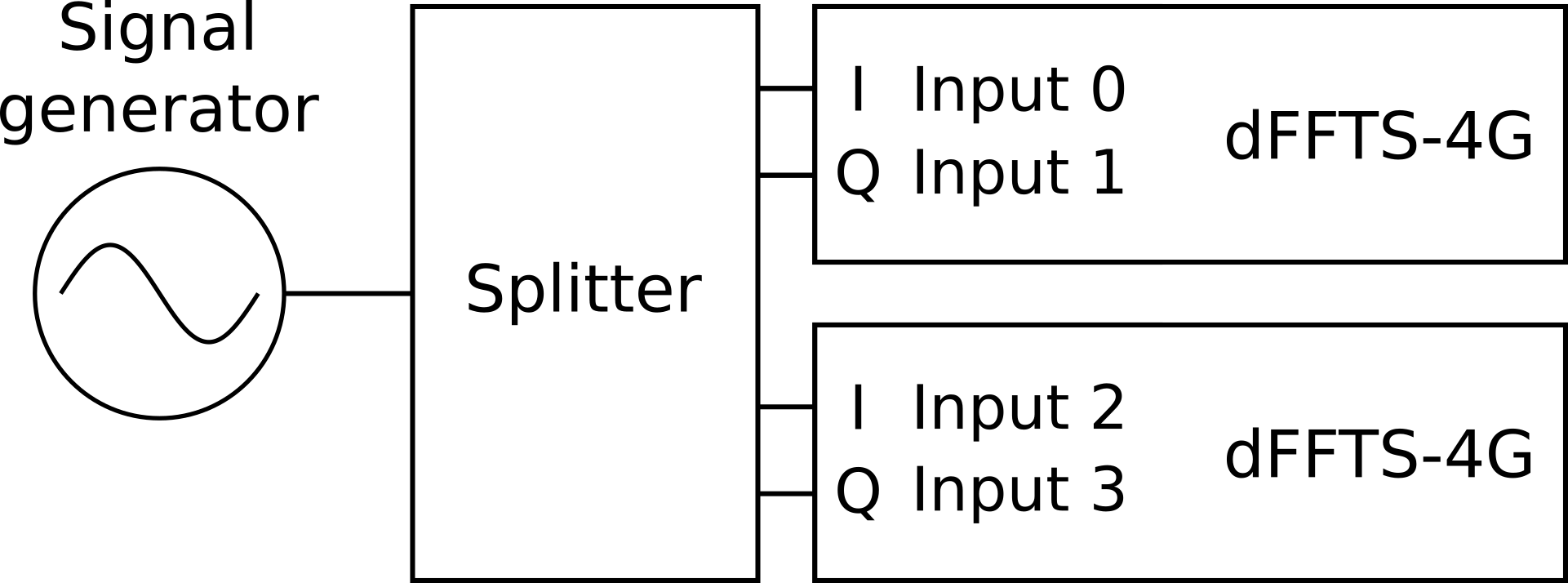}
	\caption{Schematic of the wiring of the test setup. The signal generator is an \texttt{R\&{}S\textsuperscript{\textregistered} SMB 100A}~\cite{SMB100A} and the splitter is a \texttt{ZFRSC-4-842-S+}~\cite{Splitter}.}
	\label{fig:hardware}
\end{figure}

\begin{figure}[htbp]
	\centering
	\includegraphics[width=0.95\textwidth]{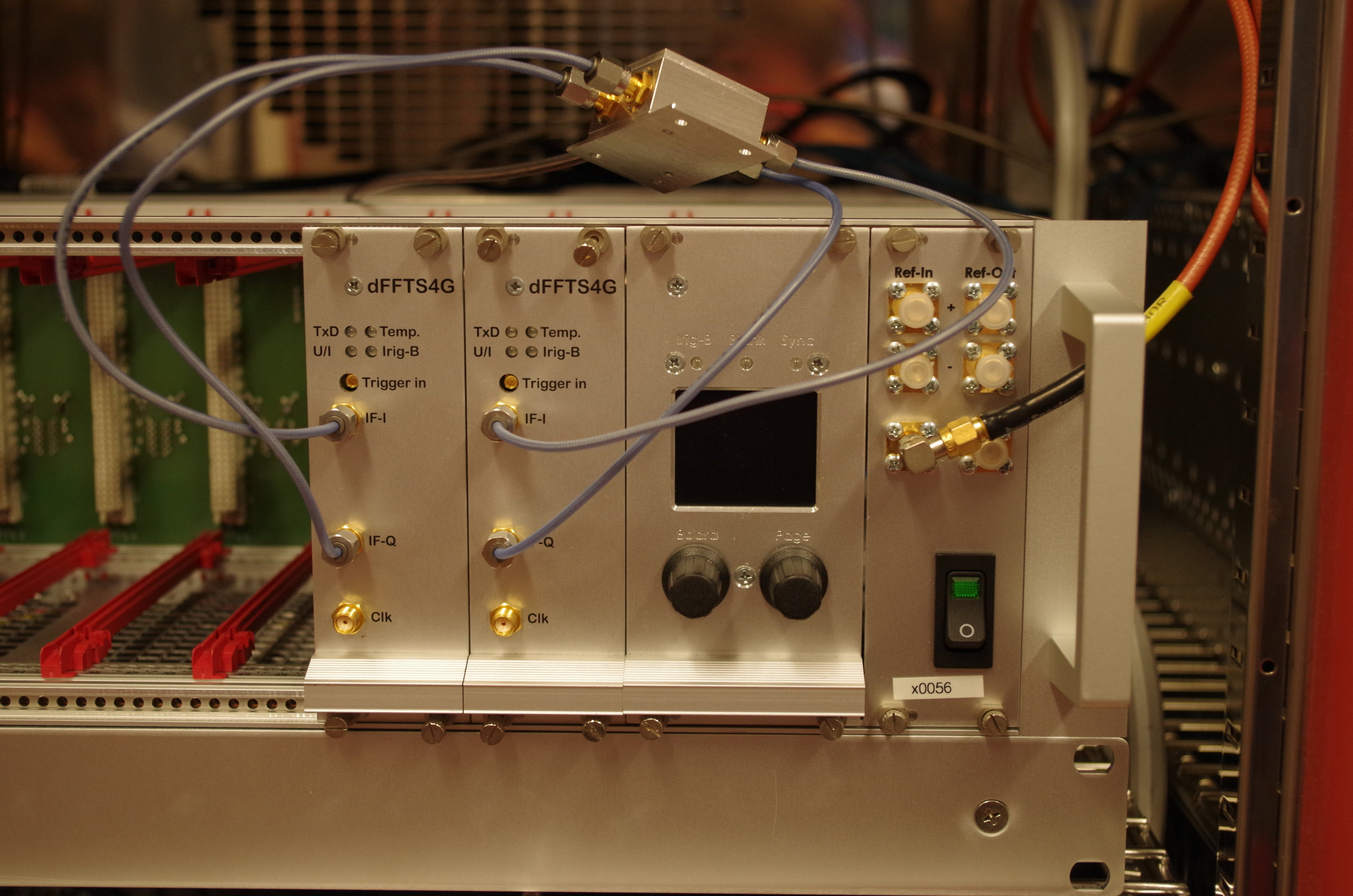}
	\caption{The setup placed in the temperature chamber. On the left the two \texttt{dFFTS4G} are visible, followed by a controller which has administrative tasks. The splitter is placed on top of the crate, the cable going from the splitter to the back is connected to the signal generator. This is not visible on the picture, as it is not placed in the temperature chamber, due to place and temperature stability reasons.}
	\label{fig:hardware_picture}
\end{figure}

The signal generation is performed with a \texttt{R\&{}S\textsuperscript{\textregistered} SMB 100A}~\cite{SMB100A}. This generator can generate sinusoidal signals with a frequency between \SI{100}{\kilo\hertz} and \SI{40}{\giga\hertz}. The output of the generator is split up into $4$ coax cables with a \verb|ZFRSC-4-842-S+|~\cite{Splitter} from \texttt{Mini Circuits\textsuperscript{\textregistered}}. The splitter is a resistive splitter and has a bandwidth between DC and \SI{8.4}{\giga\hertz}. It is used to distributed the input signal form the signal generator equally to the spectrometer inputs.
These outputs are connected to the inputs of the spectrometers. Figure \ref{fig:hardware} shows a schematic wiring plan of the test setup, while figure \ref{fig:hardware_picture} shows a picture of the setup, which is placed in the temperature chamber.

To achieve a controlled environment temperature, the spectrometers are placed in a temperature chamber \verb|VT 4021|. The temperature stability over time is better than \SI{0.5}{\kelvin} and temperature homogenity in the test chamber is better than \SI{1.5}{\kelvin}~\cite{KlimaSchrank}.
%==============================================================================
\chapter{Measurement of transfer functions}
\label{sec:comma}
%==============================================================================
Before the mismatches of different \glspl{ADCCore} can be corrected, they have to be measured. In this chapter a measurement scheme is developed and optimized. Of course the measurement can only be performed at a limited number of frequencies. Hence the mismatches $\hat{h}_{m,n}$ at discrete frequencies $f_n$ are used to approximate the continuous mismatches $H_{\text{ADC}m}(f_n)$.   
To achieve a good \gls{SNR}, an integration over more than one measurement is advisable. A continuous readout is not possible due to the limited readout-rate.  Therefore the method should allow an integration in the \gls{FPGA} to achieve a short calibration time.

\section{Measurement of mismatching transfer functions}
A straightforward way of measuring the transfer functions of the \glspl{ADCCore} is the use of a sinusoidal test-signal.
To measure the whole frequency range, the frequency of the signal is changed for each measurement.
The advantage of a sinus is the sharp peak in spectrum, therefore the energy of such a signal is distributed only to a few spectral channels.
Hence the \gls{SNR} is high compared to other signals with the same amplitude, which is limited by the \glspl{ADCCore}.
As the analog bandwidth of the \glspl{ADCCore} is around \SI{3.2}{\giga\hertz} a test-signal with an \gls{SNR} as high as possible is highly recommended.
Therefore this test-signal is used to implement and test the measurement.
Possible other test-signals are discussed later in this chapter. 

The basic idea is to compare the phase and amplitude of this signal after the digitalization between the different \glspl{ADCCore}. A $2048$ point \gls{FFT} is calculated for the samples of each \gls{ADCCore}, $\hat{h}_{m,n}$ is the $n$-th frequency channel of the \gls{ADCCore} $m$. From the complex spectrum the amplitude $g$ and phase $\phi$ of the signal at each \gls{ADCCore} can be extracted  via transformation from rectangular coordinates to polar coordinates:
\begin{align}
	g_{m,n} &= \abs{\hat{h}_{m,n}} \\
	\phi_{m,n} &= \atantwo\left(\imaginary\left(\hat{h}_{m,n}\right), \real\left(\hat{h}_{m,n}\right)\right) \\
	\hat{h}_{m,n} &= g_{m,n} \cdot e^{\iu \phi_{m,n}}
\end{align}

There is no fixed phase relation between the sinusoidal signal and the sampling clock. Hence the phase of the signal from the \gls{FFT} differs for each \gls{FFT}-calculation. An integration of the complex spectra leads to partial or complete cancellation of the signal. Consequently no integration is possible without further processing the data. To allow an integration the phases of the \glspl{FFT} have to be normalized, which can be achieved in possible ways. On the one hand the average of all \glspl{ADCCore} can be used as reference. On the other hand one of the \glspl{ADCCore} can be used. To achieve a low consumption of \gls{FPGA} resources the first \gls{ADCCore} is used as reference:
\begin{align}
	\hat{h}'_{m,n} = \frac{\hat{h}_{m,n}}{\hat{h}_{0,n}} &= \frac{\hat{h}_{m,n} \cdot \hat{h}_{0,n}^*}{\abs{\hat{h}_{0,n}}^2} \\
										&= \frac{\abs{\hat{h}_{m,n}}}{\abs{\hat{h}_{0,n}}} e^{\iu(\phi_{m,n} - \phi_{0,n})} \nonumber
\label{eq:complexDiv}
\end{align}
A complex division needs in total $6$ real multiplications and one real division. By choosing a different normalization this can be improved. If equation \ref{eq:complexDiv} is multiplied with $\abs{\hat{h}_{0,n}}^2$ only $4$ real multiplications and no divisions are needed:
\begin{align}
	\hat{h}''_{m,n} &= \abs{\hat{h}_{0,n}}^2 \cdot \hat{h}'_{m,n} \\
										&= \abs{\hat{h}_{0,n}}^2 \cdot \frac{\hat{h}_{m,n}}{\hat{h}_{0,n}} \nonumber \\
										&= \hat{h}_{m,n} \cdot \hat{h}_{0,n}^* \nonumber \\
										&= \abs{\hat{h}_{m,n}}\cdot\abs{\hat{h}_{0,n}} e^{\iu(\phi_{m,n} - \phi_{0,n})}  \nonumber
\end{align}
This is especially important for the implementation of the integration in an \gls{FPGA} as the \glspl{DSPSlice} do not support division. Furthermore their number is limited, hence a minimal number of multiplications is preferred.  

\begin{figure}[htbp]
	\centering
	\includegraphics[height=0.40\textheight]{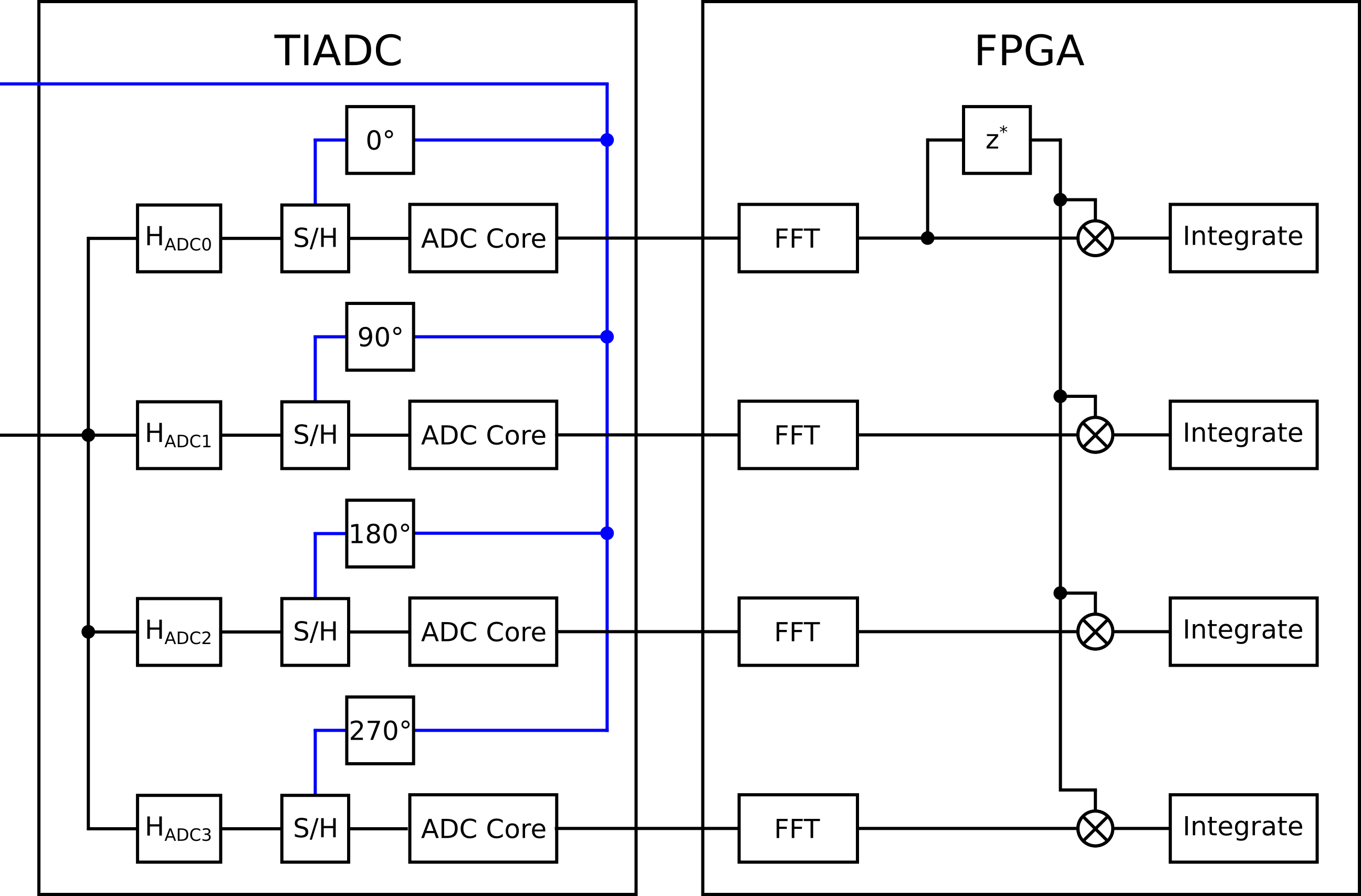}
	\caption{Schematics of a \gls{TIADC} with $M=4$ and a digital processor to measure the mismatches (with integration in the digital processor).}
	\label{fig:tiadcMeasure}
\end{figure}

The implementation in the \gls{FPGA} consists of a standard spectrometer configured in such  way, that each \gls{ADCCore} is handled as an independent input. After the \gls{FFT} the first \gls{ADCCore} is complex conjugated and then multiplied with the other \glspl{ADCCore}. These complex numbers are then summed up while the spectrometer is integrating. Figure \ref{fig:tiadcMeasure} shows the schematics of such an implementation. Including this summation all operations are performed on fixed point numbers, but before the spectra are read from a PC, these summed values are converted to floating point numbers. 
After the data is read, it is further processed with \verb|Python|. Only the frequency bins belonging to the applied calibration frequency are analyzed by the following procedure. Each $\hat{h}''_{m,n}$ is divided by $\hat{h}''_{0,n}$ to reconstruct $h'_{m,n}$:
\begin{align}
	\frac{\hat{h}''_{m,n}}{\hat{h}''_{0,n}} &= \frac{\hat{h}_{m,n} \cdot \hat{h}_{0,n}^*}{\hat{h}_{0,n} \cdot \hat{h}_{0,n}^*} \\
											&= \frac{\hat{h}_{m,n}}{\hat{h}_{0,n}} \nonumber \\
											&= \hat{h'}_{m,n} \nonumber
\end{align}
To estimate the error of the measurement a number of integrations can be done at each frequency.
The transfer functions $\hat{h}'_{m,n}$ at the corresponding frequency are compute from the spectra.
From these transfer functions the mean and standard deviation are computed.
This ensures, that the highest possible \gls{SNR} is achieved. 
The next step depends on the Nyquist-band of the calibration frequency (with respect to $\frac{f\id{s}}{M}$). The signal components in the even Nyquist-bands are mirrored and therefore their phase is also mirrored. This is compensated with the complex conjugation of $\hat{h'}_{m,n}$ if it is an even Nyquist-band (e.g. second or fourth Nyquist-band).

In the next step the inherent phase of the different \glspl{ADCCore} due to the shifted clocks has to be calculated and then corrected. The sampling point for \gls{ADCCore} $m$ is shifted in time compared to \gls{ADCCore} $0$ by
\begin{align}
	\Delta t'_{m} &= \frac{m}{f\id{s}}\,.
\end{align}
Therefore the sampled signal from \gls{ADCCore} $m$ is shifted by
\begin{align}
	\Delta t_m &= -\frac{m}{f\id{s}}
\end{align}
with respect to sampled signal from \gls{ADCCore} 0.
According to the shift theorem (s. equation \ref{eq:ftShift}) of the \gls{FT} this corresponds to a factor of
\begin{align}
	e^{2\pi\iu \cdot m \frac{f}{f\id{s}}}
\end{align} 
in the frequency domain or a phase of
\begin{align}
	\phi = 2\pi \cdot m \frac{f}{f\id{s}}\,.
\end{align}
The normalized spectrum is then corrected by:
\begin{align}
	\hat{h}'''_{m,n} 	&= \frac{\hat{h}'_{m,n}}{e^{2\pi\iu \cdot m \frac{f}{f\id{s}}  }} \label{eq:phaseInterleaving}   \\
					&= \hat{h}'_{m,n} \cdot e^{-2\pi\iu \cdot m \frac{f}{f\id{s}}} \nonumber
\end{align}
The same applies to $\hat{h}''_{m,n}$. After the last step, the correction of the inherent phase due to the interleaving by applying equation \ref{eq:phaseInterleaving}, the resulting $\hat{h}'''_{m,n}$ corresponds to the transfer function of \gls{ADCCore} $m$ with respect to the first \gls{ADCCore}.  

A perfect sinusoidal signal has only one frequency component $f$. In reality such a signal does not exist. Due to the imperfections of the signal it contains harmonics. The $n$-th harmonic has a frequency of $(n+1)\cdot f, n \in \NaturalNumbers$. Typically higher harmonics are weaker than lower harmonics. The calibration frequencies $f$ have been chosen such, that $f$ and $(n+1)\cdot f$ belong to different frequency bins of the \gls{FFT} for a sufficient range of $n$. To achieve this the calibration frequencies $f$ and the bandwidth $f\id{s}/2$ should be relatively prime. 

\begin{figure}[htbp]
	\centering
	\includegraphics[width=0.95\textwidth]{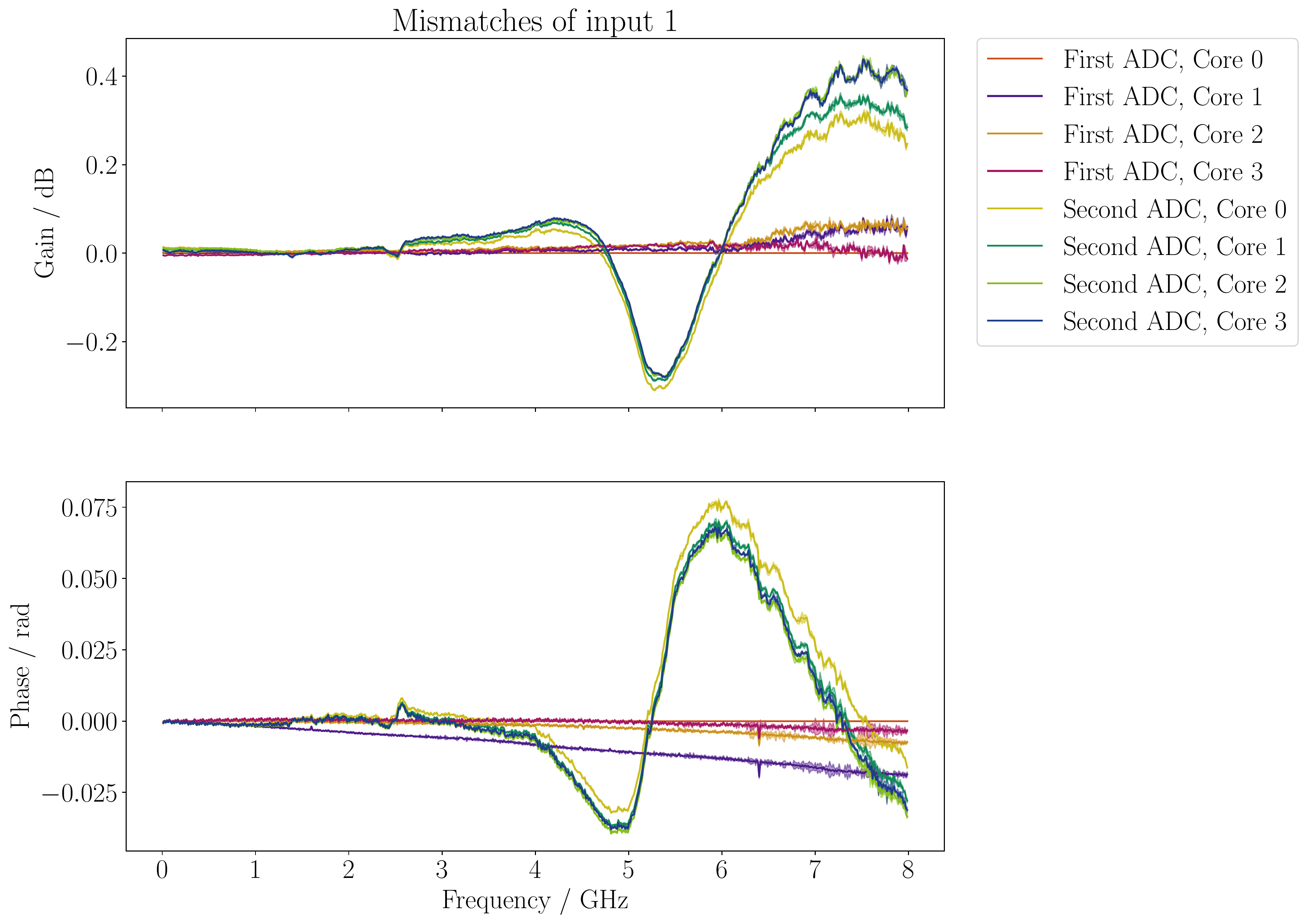}
	\caption{Typical example of the mismatches of a \texttt{dFFTS-4G} (input \texttt{Q} of the first \texttt{dFFTS-4G} board). The figure shows the gain given by $\left| \hat{h}'''_{m,n} \right|$ and phase given by $\atantwo\left(\imaginary\left(\hat{h}'''_{m,n}\right), \real\left(\hat{h}'''_{m,n}\right)\right)$ of the \glspl{ADCCore} with respect to \gls{ADCCore} $0$ of the first \gls{ADC}. The standard deviations of the measurements are shown as colored bands around the mean.}
	\label{fig:mismatches}
\end{figure}

A typical example of the mismatches of a \verb|dFFTS-4G| spectrometer is shown in figure \ref{fig:mismatches}.
The upper plot shows the gain of the different \glspl{ADCCore} with changing frequency relative to \gls{ADCCore} $0$.
The lower one shows the phase of the different \glspl{ADCCore} over frequency also with respect to \gls{ADCCore} $0$.
The measured points have a spacing of \SI{12.8}{\mega\hertz} and span the first and second Nyquist-Band.
The two distinct \glspl{IC} (first and second \gls{ADC} with each four \glspl{ADCCore}) are clearly visible as the mismatches of the \glspl{ADCCore} of one \gls{IC} are close to each other. 
The \gls{SNR} decreases for higher frequencies, as the applied sinusoidal signal is more attenuated at higher frequencies, in the analog part of the spectrometer in-front of the \glspl{ADCCore}.

\section{Optimizations}
\begin{figure}[htbp]
	\centering
	\includegraphics[width=0.75\textwidth]{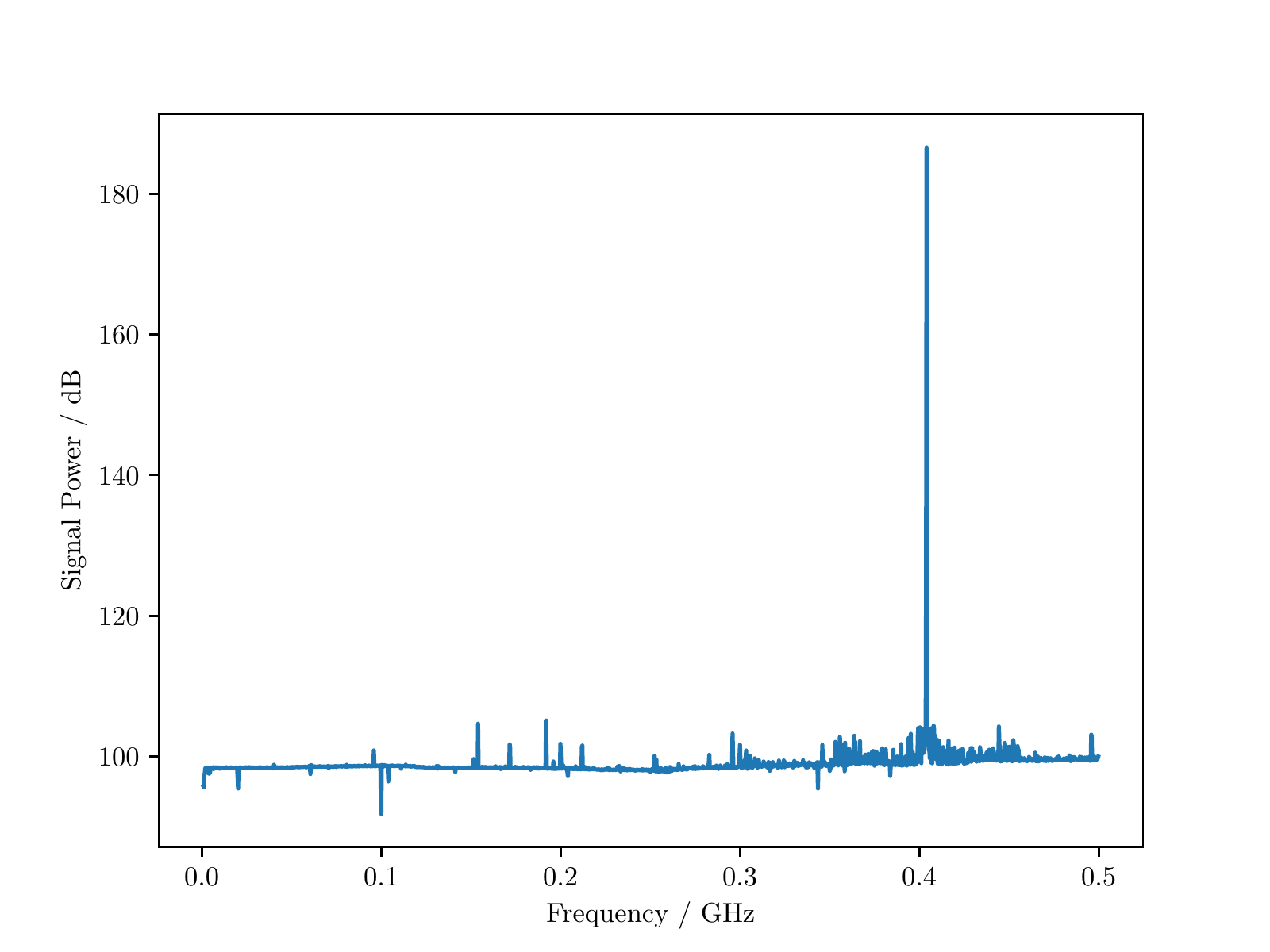}
	\caption{The amplitude spectrum of a \SI{2.404}{\giga\hertz} signal from the onboard calibrator sampled with a single \gls{ADCCore} ($f\id{s} = \SI{1}{\giga\hertz}$). The aliased ground wave is at \SI{404}{\mega\hertz} and the second harmonic at \SI{212}{\mega\hertz}. }
	\label{fig:adccore}
\end{figure}

The spectrometer features an onboard synthesizer for calibration, but it is designed only for the first Nyquist-band. In the following the usability of it for a mismatch measurment in the second Nyquist-band is discussed. The synthesizer generates a square wave with a frequency of up to \SI{4.4}{\giga\hertz}~\cite{ADF4351}. As a square wave has only even harmonics, the strongest harmonic is the second one. A \gls{balun} with a bandwidth of \SI{4}{\giga\hertz}~\cite{TCM2-43} is used to transform the the symmetric output of the synthesizer into an asymmetric signal. A lowpass of third order with a cutoff frequency of \SI{2.375}{\giga\hertz} filters the asymmetric signal. The filtered asymmetrical signal is the injected into the signal path of the spectrometer. Therefore the second harmonic should be used for the calibration to reach a good \gls{SNR}. As an example figure \ref{fig:adccore} shows the power spectrum from one \gls{ADCCore}, while the synthesizer generates a square wave with \SI{2.404}{\giga\hertz}. The second harmonic of this signal is \SI{7.212}{\giga\hertz}. The aliased ground wave is located at \SI{404}{\mega\hertz} and the aliased second harmonic at \SI{212}{\mega\hertz}. The suppression of the second harmonic relative to the ground wave is \SI{85(2)}{\dB}. The signal path of the spectrometer including the \glspl{ADCCore} has a lose of \SI{17(1)}{\dB} at \SI{7.212}{\giga\hertz} relative to a signal at \SI{2.404}{\giga\hertz}~\cite{shochguertel}. The lowpass filter has an attenuation of \SI{29}{\dB} between the two frequencies. In total this leads to a relative attenuation of \SI{46(1)}{\dB}. The missing \SI{39(1)}{\dB} arise one the one hand from the attenuation of the \gls{balun}, which has a bandwidth of \SI{4}{\giga\hertz} with a unknown attenuation behavior above \SI{4}{\giga\hertz}. On the other hand from the synthesizer, which also has a limited output bandwidth. Consequently the current onboard synthesizer is not usable for the second Nyquist-band. 

As described in chapter \ref{sec:hardware} the readout rate is limited by the Ethernet connection. To achieve a faster calibration more than one calibration frequency per integration is needed or the speed of the Ethernet connection has to be increased. A number of non-zero frequency components in spectrum requires a non-sinusoidal signal. One possibility is to change the frequency of a sinusoidal signal while integrating. The abrupt change in frequency creates further frequency components. It has to be ensured that these components do not disturb the calibration. 

A second possibility is to use a non sinusoidal periodic signal with different frequency components. An interesting choice is a signal of form $\sum_{n = -\infty}^{\infty} \delta\left( t - \frac{n}{f} \right)$, an impulse train, as this signal has evenly spaced frequency components with constant amplitude (see equation \ref{eq:FTimpulseTrain}). Depending on the bandwidth of the generator the signal should be filtered to restrict the frequency components to the used Nyquist-Band. This signal has less energy ($E = \int x^2(t) \dd{t}$) compared to the first one. A signal with a comparable amplitude spectrum is a sawtooth wave. In contrast to the impulse train the amplitude of such a signal decreases for the higher frequency components. 

Furthermore, the current measurement is implemented in an own \gls{gateware}. For a fast determination of the mismatches, in an observation run, the measurement should be integrated in the standard \gls{gateware} of the spectrometer. This eliminates the time needed to reconfigure the \gls{FPGA}.
%==============================================================================
\chapter{Approaches to calibration}
\label{sec:approaches}
%==============================================================================
The basic idea for calibration is to apply a digital filter to the samples from each \gls{ADCCore}. The possible implementations of this filter are discussed in this chapter. Especially the needs for an implementation in an \gls{FPGA} are considered. The number of \glspl{DSPSlice} needed for each approach will be estimated, as the calibration is computationally intensive. 

\section{Finite impulse response filter}\label{sec:approachesFIR}
As discussed in \ref{sec:filters} an \gls{FIRFilter} is a linear, non-recursive filter of the form
\begin{align}
	y_n &= \sum_{k=0}^{N-1} x_{n+k} \cdot c_k\,.
\end{align}
A generic filter of length $N$ needs $N$ multiplications and $N-1$ summations. 

One filter per \gls{ADCCore} can be used to correct the mismatches.
This filter structure has to work at the full sample rate (in this case \SI{8}{\GSPS}) to reach unambiguity for all frequency components.
If the filter would work at a lower sample rate different frequency components of the signal would be mixed indistinguishable (see \ref{sec:samplingTheorem}); e.g. at \SI{2}{\GSPS} sample rate (the sample rate of each \gls{ADCCore}) a signal with frequency of \SI{0.5}{\giga\hertz} would not distinguishable from a signal of \SI{1.5}{\giga\hertz}.
Hence the filter can not apply the right corrections to this signal. 
A sufficient sampling rate is achieved either by the use of the samples from all \glspl{ADCCore} or by padding with zeros~\cite{Seo2005}\cite{6587074}.
A \gls{DFT} of the latter with length $M \cdot N$ is closely related to a \gls{DFT} of the unpadded signal.
This relation between the unpadded signal $a_j$ and the padded signal $a'_j$ is given by:
\begin{align}
	a'_j &= 
	\begin{cases}
		a_j, & \text{if } j \smod M = 0 \\
		0, & \text{otherwise}
	\end{cases}
\end{align}
\begin{align}
	\sum_{j = 0}^{M \cdot N - 1} e^{-2\pi\iu \frac{jk}{M \cdot N}} \cdot a'_j 
	= \sum_{j = 0}^{N - 1} e^{-2\pi\iu \frac{(M\cdot j)k}{M \cdot N}} \cdot a'_{(M\cdot j)} 
	= \sum_{j = 0}^{N - 1} e^{-2\pi\iu \frac{jk}{N}} \cdot a_j\label{eq:paddedSpectrum}
\end{align}
Therefore the zero-padding does not alter the spectrum of the signal.

\begin{figure}[htbp]
	\centering
	\includegraphics[width=0.75\textwidth]{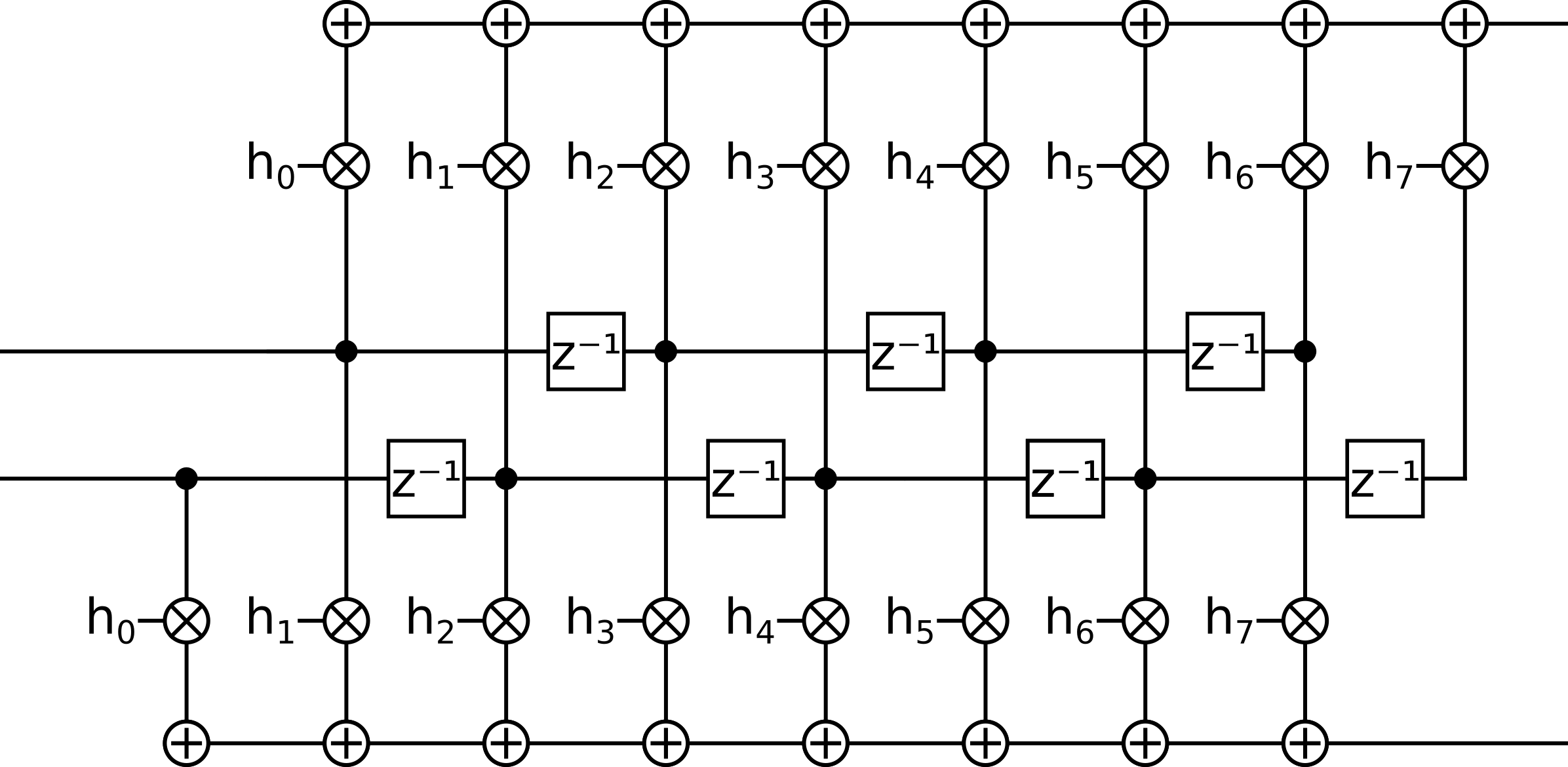}
	\caption{A \gls{FIRFilter} of length $8$ with two parallel data-streams.}
	\label{fig:parallelFIR}
\end{figure}

\Glspl{FPGA} do not support clock frequencies as high as the sampling frequency of the spectrometer.
Therefore the data-stream has to be processed in parallel. The first parallelism is the split up into individual streams from each \gls{ADCCore}.
But this is not sufficient, furthermore each stream of an \gls{ADCCore} has to be split up into $P$ parallel streams ($P$ digitized values per \gls{ADCCore} and clock cycle).
Therefore the filter receives $M \cdot P$ samples in each clock cycle and generates $M \cdot P$ new samples in each clock cycle, as the filter is running at the full sample rate to distinguish between frequency components.
Figure \ref{fig:parallelFIR} shows an \gls{FIRFilter} of length $8$ with two parallel data-streams.
As each \gls{FIRFilter} needs $N$ multiplications, this leads to $N\cdot P \cdot M$ multiplications per clock cycle per \gls{ADCCore}.
Hence the number of multiplications for the correction per clock cycle is $N \cdot P \cdot M^2$.
If the upsampling is done by inserting zeros, the multiplication and summation of these samples can be skipped.
This leads to a minimum number of $N \cdot P \cdot M$ multiplications per clock cycle. 
$P$ and $M$ are fixed by the hardware, only $N$ can be varied. A typical spectrometer configuration consists of $P = 4$ and $M = 8$ and two independent inputs.
The number of free \glspl{DSPSlice} equals $2200$ for a configuration with \SI{16}{k} spectral channels, consequently there are $1100$ \glspl{DSPSlice} per input.
Hence the filter length is limited to $N = \frac{1100}{P \cdot M} \approx 32$.

To explore this calibration method a simulation was developed using \verb|Python|.
It consists of a model of a \gls{TIADC}, where each \gls{ADCCore} is parameterized separately (gain, phase and bandpass behavior).
Each of the  $8$ \glspl{ADCCore} has a simulated sampling rate of \SI{1}{\GSPS}, consequently the \gls{TIADC} has a sampling rate of \SI{8}{\GSPS} and a bandwidth of \SI{4}{\giga\hertz}.
The output data of the \gls{TIADC} is then split up into data-streams from each \gls{ADCCore}.
These are upsampled to the sample rate of the whole \gls{TIADC} by inserting zeros.
In the next step each of these upsampled streams is processed by an \gls{FIRFilter}.
The transfer functions of these filters are calculated by the method given in chapter \ref{subsec:correction} and then converted into coefficents using the \gls{FrequencySamplingMethod}~\cite{DSPSciEng}.
This method generates the coefficients from a truncated \gls{IFFT} of the transfer functions.
In the last step the filtered data from all \gls{ADCCore} are summed up to form the calibrated data-stream.
Figure \ref{fig:mirrorbandAll} presents the \gls{MirrorSuppression} which is achieved with \glspl{FIRFilter} of different length in comparison to the \gls{MirrorSuppression} without a correction. 

\section{FIR filters by convolution}
\begin{figure}[htbp]
	\centering
	\includegraphics[width=0.75\textwidth]{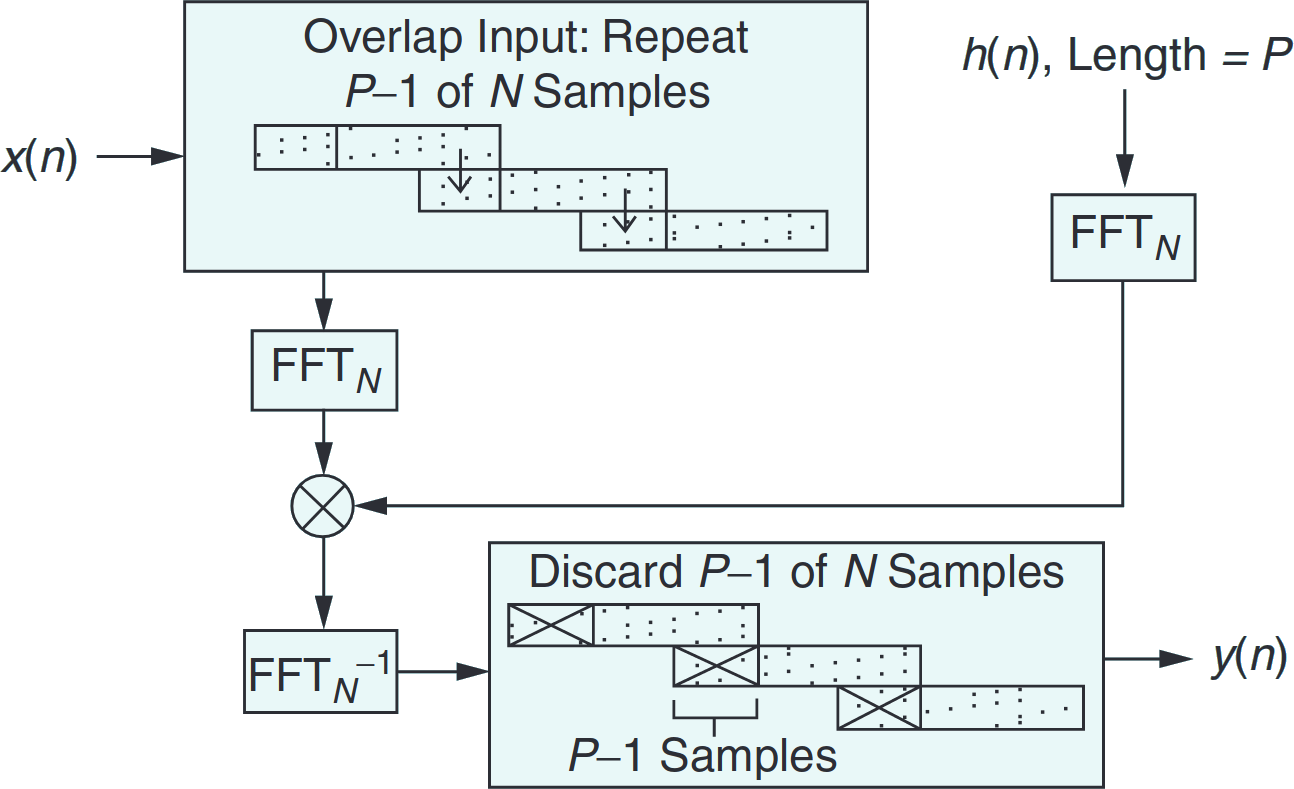}
	\caption{A linear convolution ($y(n) = x(n) * h(n)$) implemented with the overlap-save method (modified version from~\cite{overlapsave}). $h(n)$ has a length of $P$ and $x(x)$ has a length of $N$.}
	\label{fig:overlapsave}
\end{figure}

\Glspl{FIRFilter} are mathematically identical to linear convolutions as described in chapters \ref{sec:convolution} and \ref{sec:filters}.
The linear convolution of a signal with a shorter kernel can be split up into a number of circular convolutions (see chapter \ref{sec:convolution} and~\cite{overlapsave}), which can be implemented in the complex spectrum.
Therefore the linear convolution can be implemented with an \gls{FFT}, complex multiplications and an \gls{IFFT}.
But the input signals of the \gls{FFT} have to overlap as described in chapter \ref{sec:filters}.
Figure \ref{fig:overlapsave} shows an implementation of a linear convolution with \gls{FFT} and \gls{IFFT}, also called overlap-save method~\cite{overlapsave}.

A lower limit for the number of \glspl{DSPSlice} needed for the overlap-save method can be estimated by the amount needed for the \gls{FFT} of length $N$, the \gls{IFFT} of length $N$ and the multiplication in the frequency domain.
The filter must run at the full sample rate of the \gls{TIADC} and not just the sample rate of a single \gls{ADCCore}.
This can be done via inserting zeros (as described in chapter \ref{sec:approachesFIR}), which does not alter the spectrum, as shown in equation \ref{eq:paddedSpectrum}.
Therefore the \gls{FFT} only needs a size of $\frac{N}{M}$ and the padding can be done without computation in the frequency domain by periodic continuation of the spectrum.
The total number of complex multiplications for an \gls{FFT} of length $\frac{N}{M}$ can be estimated by 
\begin{align}
\frac{N}{M} \log \left( \frac{N}{M} \right)\,.
\end{align}
To calculate this \gls{FFT} in realtime, it has to be completed in
\begin{align}
\frac{N}{P\cdot M}
\end{align} clock cycles ($P$ the number of samples per clock cycle and per \gls{ADCCore}).
Therefore in each clock cycle 
\begin{align}
P \cdot \log\left(\frac{N}{M}\right)
\end{align} complex multiplications have to be performed for the \gls{FFT}.
The \gls{FFT} is followed by the $N$ complex multiplications in the frequency domain in 
\begin{align}
\frac{N}{P\cdot M}
\end{align} clock cycles.
Hence $P\cdot M$ complex multiplications per clock cycle are needed for this.
The \gls{IFFT} handles the full sample rate ($N\cdot log(N)$ complex multiplications in $\frac{N}{P\cdot M}$ clock cycles).
Therefore the \gls{IFFT} needs 
\begin{align}
	P\cdot M \cdot \log(N)
\end{align} complex multiplications per clock cycle.
Consequently the total number of complex multiplications per clock cycle per \gls{ADCCore} is
\begin{align}
	P \cdot \left( M + M\cdot\log(N) + \log\left(\frac{N}{M}\right)\right)\,.
\end{align}
For a straight forward complex multiplication $C = 4$ real multiplications are needed.
This can be reduced to $C = 3$ real multiplications but a higher number of additions are needed.
The complete count of real multiplications per clock cycle, which is an estimator for the count of needed \glspl{DSPSlice}, is therefore 
\begin{align}
	C \cdot M \cdot P \cdot \left( M + M\cdot\log(N) + \log\left(\frac{N}{M}\right)\right)\,.
\end{align}
The $1100$ available \glspl{DSPSlice} per input limit the filter length to below $N=2$, as it is a lower limit estimation, with the other given hardware constrains ($P = 4$, $M=8$). 
For this scenario this solution is worse than a direct implementation of a linear convolution, as described in chapter \ref{sec:approachesFIR}.

\begin{figure}[htbp]
	\centering
	\includegraphics[width=0.4\textwidth]{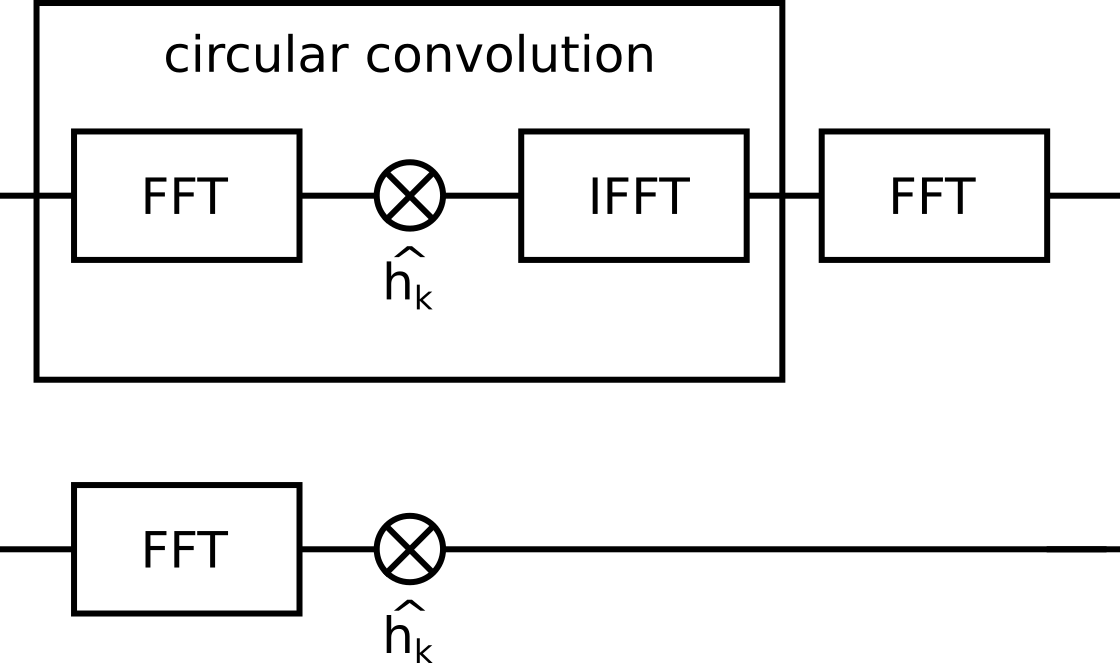}
	\caption{The upper datapath is a circular convolution followed by an \gls{FFT}. In the lower one the \gls{FFT} and \gls{IFFT} cancel each other out. Both datapaths generate the same result.}
	\label{fig:circularOptimization}
\end{figure}

For the case of an $N$-periodic signal and a filter of length $N$ the linear convolution and circular convolution are equal (s. equation \ref{eq:lcConvEq}).
The basic assumption of a \gls{DFT} and therefore also of an \gls{FFT} is that the signal is periodic.
Consequently the circular convolution can be used instead of the linear convolution without inaccuracies for a \gls{FFTS} as the \gls{FFT} already assumes periodic signals.
Of course the length of the filter has to be matched to the length, and therefore the periodicity, of the \gls{FFT}.
Furthermore the circular convolution can be integrated into the existing \gls{DSP} pipeline as illustrated in figure \ref{fig:circularOptimization}, which decreases the amount of needed resources.
This reuse of the existing \gls{FFT} reduces the number of additional \glspl{DSPSlice} to these used for the complex multiplications in the frequency domain.
The estimation for the number of \glspl{DSPSlice} is given by $C\cdot M^2\cdot P = 1024$ with $C = 4$.

\begin{figure}[htbp]
	\centering
	\includegraphics[width=0.75\textwidth]{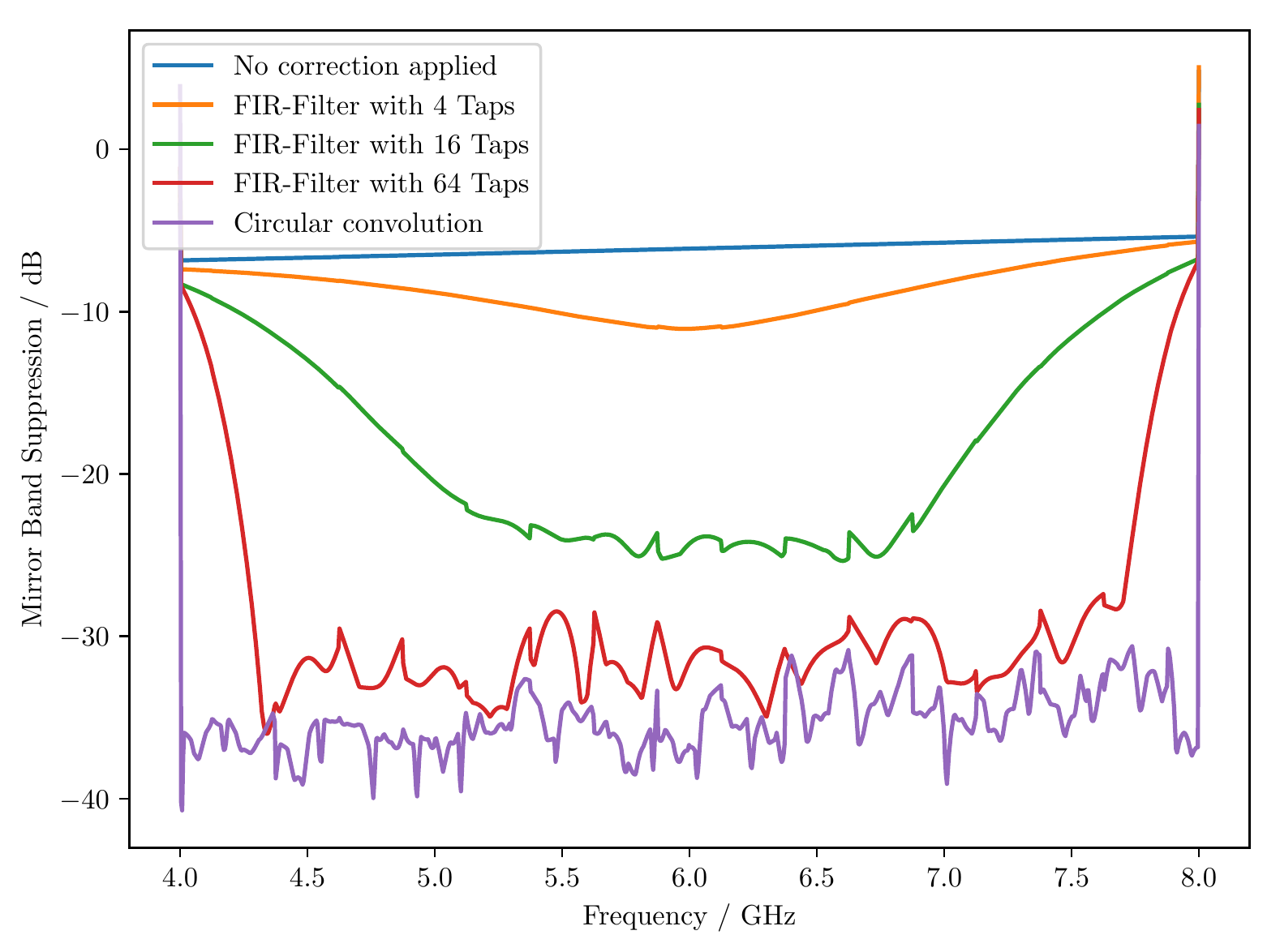}
	\caption{Simulation of the \gls{MirrorSuppression} of the different calibration approaches. The \gls{TIADC} consists of $8$ \glspl{ADCCore}. Each \gls{ADCCore} has a different gain and phase. Furthermore each \gls{ADCCore} has a different bandpass behavior.}
	\label{fig:mirrorbandAll}
\end{figure}

The circular convolution method was simulated similarly to the one in chapter \ref{sec:approachesFIR}.
The simulation separated the data-streams of each \gls{ADCCore} of the \gls{TIADC} and applied an \gls{FFT}.
With the periodicity of the \gls{DFT} the complex spectrum is extended from $f\id{s}/M$ to $f\id{s}$ (which corresponds to padding with zeros in the time domain).
This spectrum is multiplied with the complex transfer function of the filters $H_m(f)$ as given in chapter \ref{subsec:correction}.
The result is transferred back into the time domain with an \gls{IFFT} and summed up there.
That sum is the corrected data-stream of the \gls{TIADC}. Figure \ref{fig:mirrorbandAll} shows the \gls{MirrorSuppression} after such a correction compared to the original \gls{MirrorSuppression}.
The \glspl{FFT} of the data-streams from the \gls{ADCCore} are of length $4096$.

\section{Conclusion}
	\begin{figure}[htbp]
		\centering
		\includegraphics[width=0.75\textwidth]{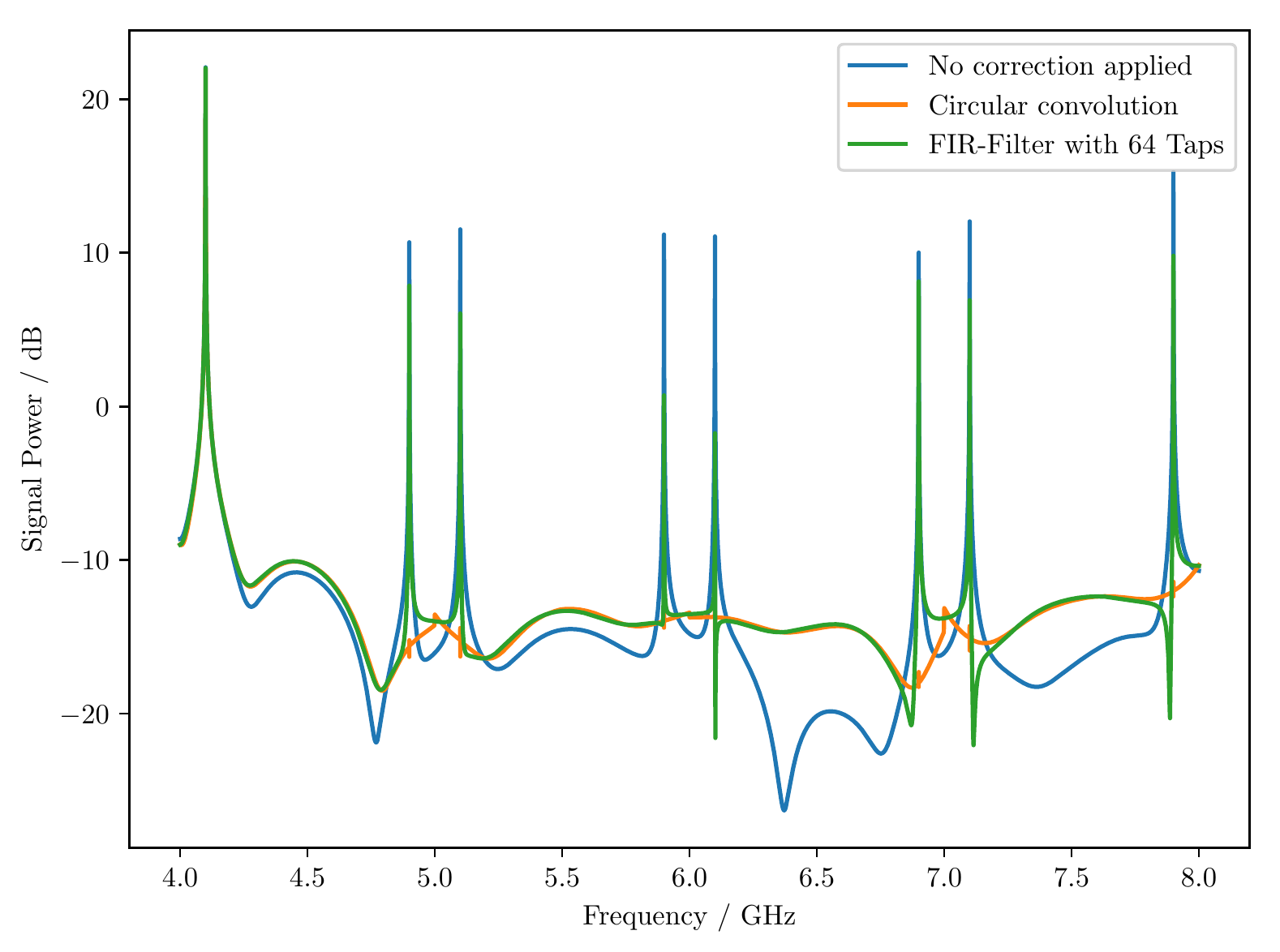}
		\caption{A simulated spectrum without calibration, with \glspl{FIRFilter} and with circular convolution. The figure shows the spectrum of a sinusoidal signal of \SI{4.1}{\giga\hertz}. The \gls{TIADC} consists of $8$ \glspl{ADCCore}. Each \gls{ADCCore} has a different gain and phase. Furthermore each \gls{ADCCore} has a different bandpass behavior.}
		\label{fig:spectra2}
	\end{figure}
Figure \ref{fig:mirrorbandAll} compares the \gls{MirrorSuppression} of calibration with \glspl{FIRFilter} and circular convolution.
The linear convolution is not shown as it is mathematically identical to the approach with \glspl{FIRFilter}.
The figure shows that the calibration with \glspl{FIRFilter} is approaching the performance of the calibration with circular convolution for increasing filter-lengths, as the circular convolution is equal to \glspl{FIRFilter} with $4096$ taps (length of \gls{FFT} after the \glspl{ADCCore}).
Especially at the edges of the band the circular convolutions shows a better \gls{MirrorSuppression} than \glspl{FIRFilter} with $64$ taps.
Furthermore the circular convolution needs the least amount of resources, when it is integrated into the existing \gls{DSP} pipeline.

Figure \ref{fig:spectra2} show the spectra of a sinusoidal signal with and without calibration.
The baseline behavior of the spectrum without correction, calibration with \glspl{FIRFilter} and with circular convolution is comparable.
The exact values differ, but the general behavior is similar. 

Therefore the circular convolution is the best approach to correct for mismatches between \glspl{ADCCore} in the \gls{FFTS} \texttt{dFFTS4G}.
For these reasons a calibration based on circular convolutions is developed and optimized in the following chapter.
%==============================================================================
\chapter{Calibration of the complex amplitude spectrum}
\label{sec:comca}
%==============================================================================
The idea of the calibration is to correct each \gls{ADCCore} in the frequency domain using the filter, defined in chapter \ref{subsec:correction}.
It uses the circular convolution, as this approach yields a good \gls{MirrorSuppression} with the resources available compared to other filter implementations.
This allows to correct frequency dependent mismatches.
The approach corresponds to a calibration of the complex amplitude spectrum (\verb|COMCA| for COMplex CAlibration).
To optimize the use of resources the last stages of the \gls{FFT} are absorbed into the correction and the stages before are also integrated into the calibration. 

In the first step the measured mismatches are converted into a correction filter $H_k$.
Then these coefficients are applied to the data in the signal processing chain in the \gls{FPGA}.
The method is developed for the first Nyquist-band, but can be used for all Nyquist-bands.
Instead of the real frequency $f, f < \frac{f\id{s}}{2}$ the aliased frequency $f'$ has to be used in the following.

The calculation of the filter coefficients is done in software, which is implemented in \verb|Python 3|
The package \verb|pandas| is used for data storage, while \verb|numpy| and \verb|scipy| are used to perform the calculations.

\section{Calculation of filter coefficients to compensate mismatching transfer functions}\label{sec:comcaCali}
Equation \ref{eq:generateFilter} is used to calculate the filter coefficients from the measured mismatches $H_{\text{ADC}m}$:
\begin{align}
	\delta_{k,0} = \frac{1}{M} \sum_{m = 0}^{M-1} e^{-2\pi\iu m \frac{k}{M}} \cdot H_{\text{ADC}m}\left(f + \frac{j}{M}f\id{s}\right) \cdot H_m\left(f + \frac{k+j}{M} f\id{s} \right), 0 \le k, j < M, k, j \in \NaturalNumbers
\end{align} 
Furthermore the periodicity of the spectrum of a sampled function, $H(f) = H(f+f\id{s})$, and the symmetry of the spectrum of a real valued, sampled function, $H(f\id{s}-f) = H^*(f)$, are needed to reduce the $2M^2-M$ unknowns to $M^2$, as described in chapter \ref{subsec:correction}.
These properties have to be applied in such a way, that the frequencies $f + \frac{k+j}{M}\cdot f\id{s}$ and $f + \frac{j}{M}\cdot f\id{s}$ lie between $0$ and $\frac{f\id{s}}{2}$.
Then the resulting system of equations can be solved.

To solve the equation system in \ref{eq:generateFilter} of size $M^2$, it is expressed with a matrix $\boldsymbol{A} = (a_{nu})$ and two vectors $\vec{H} = (H'_n)$ and $\vec{s} = (s_n)$ as $\boldsymbol{A}\vec{H} = \vec{s}$. The elements of the matrix are given by the algorithm explained below. The vector elements $s_n$ are given by:
\begin{align}
	s_n &= 
	\begin{cases}
		M, & \text{if } n < M \\
		0, & \text{otherwise}
	\end{cases}\label{eq:svector}	
\end{align}
The vector elements $H_n$ are given by:
\begin{align}
	n' &= \floor\left(\frac{n}{M}\right) \\
	H'_n &= 
	\begin{cases}
		H_{(n\smod M)}\left(f + \frac{n'}{M} f\id{s} \right), & \text{if } n < \frac{M^2}{2} \\
		H_{(n\smod M)}\left(\frac{M-n'}{M} f\id{s} - f \right), & \text{otherwise}
	\end{cases}\label{eq:frequencies}
\end{align}
The algorithm \ref{alg:cali} is used to generate one row of the matrix $\boldsymbol{A}$.
This algorithm transforms the frequencies into the frequency range between $0$ and $f\id{s}/2$ with the symmetry and periodicity given above.
Due to the structure of the system of equations a frequency $f$ is connected to another $M-1$ different frequencies (see equation \ref{eq:frequencies}).
Hence $f$ can be limited to $0 < f < \frac{f\id{s}}{2M}$, while the other frequencies cover the remaining part of the Nyquist-band ($\frac{f\id{s}}{2M} < f' < \frac{f\id{s}}{2}$).
Furthermore this limit simplifies the algorithm as this limits the frequency of the term $H_{\text{ADC}m}\left(f + \frac{j}{M}f\id{s}\right)$ to be below $f\id{s}$.
The resulting matrix equation is then solved with the function \verb|linalg.solve| provided by \verb|numpy|.  
\begin{algorithm}
	\DontPrintSemicolon
	\caption{Generation of matrix coefficients $a_{nu}$, to solve the system of equations given by equation \ref{eq:generateFilter}.}
	\label{alg:cali}
	
	\KwData{frequency $f < \frac{f\id{s}}{2M}$, index $j$ and $k$, mismatches $H_{\text{ADC}m}$}
	\KwResult{row $n = k\cdot M + j$ of $\boldsymbol{A}$}
	
	\BlankLine
		$row$ = ArrayWithZeros($M^2$)\;
		\For{$m=0$ \KwTo $M-1$}{
			\eIf{$f + f\id{s}\frac{j}{M} \ge \frac{f\id{s}}{2}$}{
				$H = H^*_{\text{ADC}m}\left(\frac{M-j}{M}f\id{s} - f\right)$\;			
			}{
				$H = H_{\text{ADC}m}\left(f + f\id{s}\frac{j}{M}\right)$\;	
			}	
			$H = H \cdot \exp\left(-2\iu\pi \frac{k\cdot m}{M}\right)$\;
			\eIf{$\frac{f\id{s}}{2} \le f + \frac{k+j}{M} f\id{s} < f\id{s} $}{
				$row[(k + j)\cdot M + m] = H$\; 
			}{
				$row[(k + j)\cdot M + m] = H^*$\; 
			}	
		}
		\Return $row$\;
\end{algorithm}

The structure of this calibration requires a factor for each frequency bin and \gls{ADCCore}.
But the number of measured $H_{\text{ADC}m}$ is limited, due to time constraints of the measurement process and the selection of suitable calibration frequencies.
Therefore the measured $H_{\text{ADC}m}$ are interpolated within the grid given by the frequency bins.
The interpolation should approximate the real, but unknown, frequency behavior as good as possible.
On the other hand a short calibration time is preferred to a longer one.
Therefor the interpolation should be light in terms of computational complexity.
A simple interpolation is the linear interpolation between the grid points. 
It has been chosen due to its lightness and good results.

\section{Applying of the calibration onto data}\label{sec:scheme}
The calibration is applied to the data based on a split up \gls{FFT}, where each \gls{ADCCore} is treated independently in the beginning (it is assumed that $M$ is a factor of $N$). 
The total $N$-point spectrum is given by (s. equation \ref{eq:df}):
\begin{align}
	\hat{a}_k &= \sum_{j = 0}^{N - 1} e^{-2\pi\iu \frac{jk}{N}} \cdot a_j
\end{align}
This is then split into sub-spectra $\hat{a}_{k,m}$ of size $\frac{N}{M}$ for each \gls{ADCCore} $m$:
\begin{align}
	\hat{a}_k &= \sum_{m=0}^{M-1} \sum_{j = 0}^{\frac{N}{M} - 1} e^{-2\pi\iu \frac{(j\cdot M + m)k}{N}} \cdot a_{j\cdot M + m} \\
			  &= \sum_{m=0}^{M-1} e^{-2\pi\iu \frac{mk}{N}} \sum_{j = 0}^{\frac{N}{M} - 1} e^{-2\pi\iu \frac{j Mk}{N}} \cdot a_{j\cdot M + m} \nonumber \\
			  &= \sum_{m=0}^{M-1} e^{-2\pi\iu \frac{mk}{N}} \hat{a}_{k,m} \nonumber 
\end{align}
The spectra $\hat{a}_{k,m}$ of each \gls{ADCCore} are calculated in the first step.
Then the calibration factor $H_{k,m}$ for this frequency and \gls{ADCCore} is applied to each sub-spectrum, leading to the corrected spectrum $\hat{b}_k$:
\begin{align}
	\hat{b}_k &= \sum_{m=0}^{M-1} e^{-2\pi\iu \frac{mk}{N}} \hat{a}_{k,m} \cdot H_{k,m}\label{eq:applyCali}
\end{align}
The steps to apply the calibration to an $N$-point \gls{DFT} are the following (with $M$ \glspl{ADCCore}):
\begin{enumerate}
	\item Calculate an $\frac{N}{M}$-bin \gls{DFT} $\hat{a}_{k,m}$ from the samples $a_{j\cdot M + m}$ of each \gls{ADCCore}.
	\item Multiply each bin of the $M$ \glspl{DFT} $\hat{a}_{k,m}$ with the corresponding complex number $H_k(f\id{bin}) = H_{k,f\id{bin}}$, which corrects phase and gain mismatches at that frequency bin.
	\item Calculate the sum given by equation \ref{eq:applyCali}, which is similar to an $M$-point \gls{DFT}. The result is the corrected complex $N$-point spectrum.
\end{enumerate}
Equation \ref{eq:applyCali} suggest a simple optimization.
The twiddle factor $e^{-2\pi\iu \frac{mk}{N}}$ can be absorbed into the calibration factor $H_{k,m}$, which is generated on the computer during the calibration.
Without this optimization a \gls{DFT} of size $M$ has to be applied to the spectra $\hat{a}_{k,m} \cdot H_{k,m}$ of the \glspl{ADCCore}.
Note that an \gls{FFT} can not be used instad of a \gls{DFT}, due to the lack of periodicity:
\begin{align}
	\hat{a}_{k,m} \cdot H_{k,m} \neq  \hat{a}_{k+N/M,m} \cdot H_{k+N/M,m}
\end{align}

\section{Implementation}
\begin{figure}[htbp]
	\centering
	\includegraphics[width=0.9\textwidth]{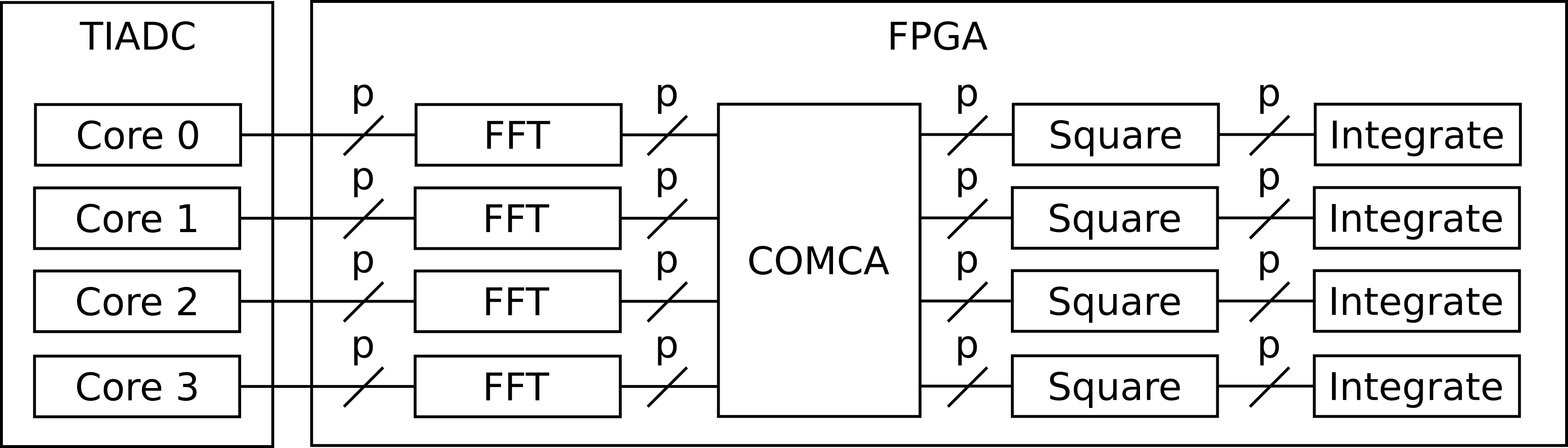}
	\caption{The simplified data-flow in the spectrometer for $4$ \glspl{ADCCore}.}
	\label{fig:spectrometerflow}
\end{figure}

\begin{figure}[htbp]
	\centering
	\begin{subfigure}[t]{0.45\textwidth}
		\centering
		\includegraphics[width=0.9\textwidth]{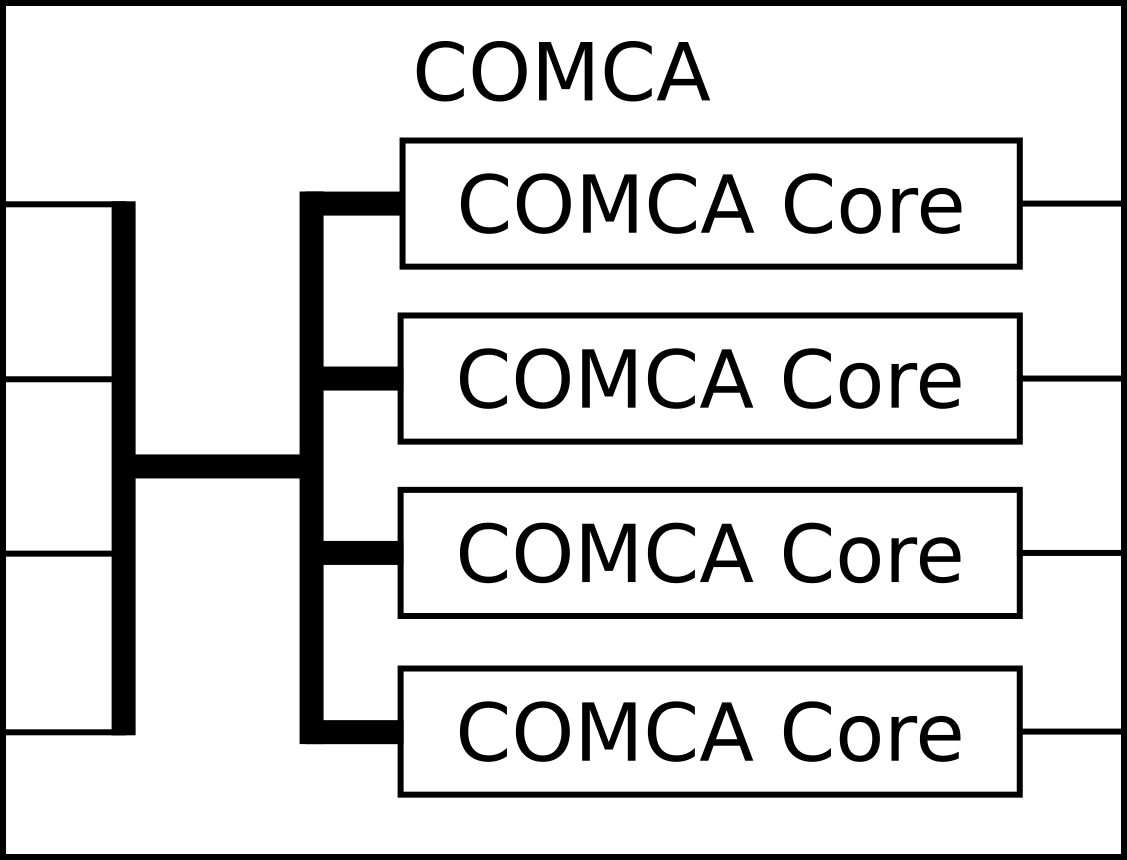}
		\caption{The simplified data-flow in the COMCA component for $4$ \glspl{ADCCore} and one parallel datastream $P=1$. In case of $P > 1$ each COMCA Core is instantiated $P$ times. The thick lines correspond to the aggregate of all input-data.}
		\label{fig:comcainflow}
	\end{subfigure}
	\quad
	\begin{subfigure}[t]{0.45\textwidth}
		\centering
		\includegraphics[width=0.9\textwidth]{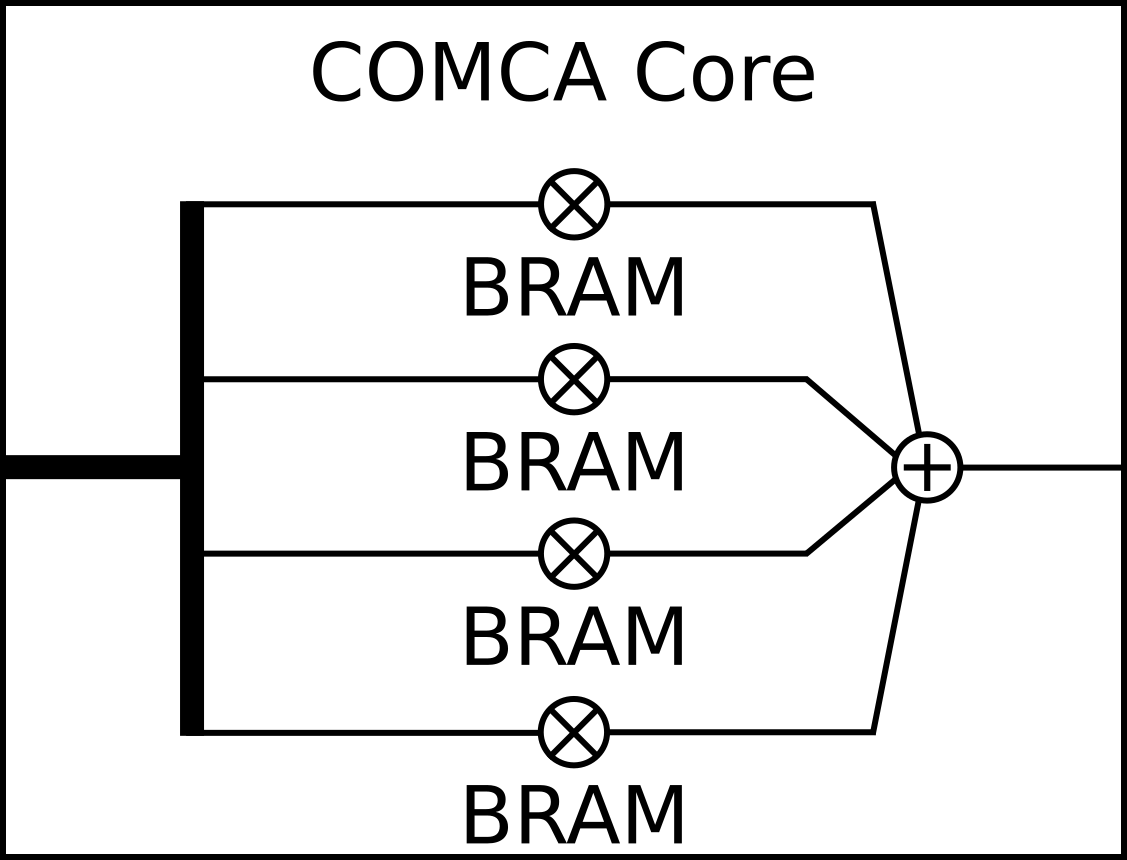}
		\caption{The data-flow in the COMCA Core component for $4$ \glspl{ADCCore}. The thick lines correspond to the aggregate of all input-data.}
		\label{fig:comcacoreflow}
	\end{subfigure}
	\caption{The data-flow in the implemented calibration.}
	\label{fig:comcaflow}
\end{figure}

The implementation in the real spectrometer follows the scheme from chapter \ref{sec:scheme}.
As the spectrometer is implemented in an \gls{FPGA} its signal processing has to be adopted to the parallel data-flow.
The simplified data-flow, including the calibration, in the spectrometer is shown in figure \ref{fig:spectrometerflow}.
Figure \ref{fig:comcaflow} shows the hierarchy of the developed components.
In the following the calibration of a spectrometer with $N/2$ spectral channels is described.

Each \gls{ADCCore} samples with \SI{1}{\GSPS}, while the signal processing in the \gls{FPGA} runs at \SI{250}{\mega\hertz}.
Therefore the \gls{FPGA} is not fast enough to process each sample sequential.
Hence a number of samples from each \gls{ADCCore} are processed per clock cycle, resulting in a parallel processing of the data.  

The coefficients $e^{-2\pi\iu \frac{mk}{N}} \cdot H_{k,m}$ are saved in \gls{BRAM} instances on the \gls{FPGA}.
The coefficients are loaded into the \glspl{BRAM} via an \SI{8}{\bit} register.
The register is accessible from the network, which is connected to the controlling computer.
Inside the \gls{FPGA} the communication from the register to the \gls{BRAM} is realized with an \SI{8}{\bit} wide data-bus with a \verb|data ready| signal and a \verb|reset| signal.
\verb|data ready| indicates a new data-word on the bus, if it is set the data is stored in the \gls{BRAM}'s current pointer position and the pointer is incremented for the next data-word.
\verb|reset| signal resets the pointer to the beginning of the \gls{BRAM}.
The filter coefficients $H_{k,m}$ are calculated on a computer, using the algorithm from chapter \ref{sec:comcaCali}.
Furthermore the filter coefficients $H_{k,m}$ are extended by the twiddle factor $e^{-2\pi\iu \frac{mk}{N}}$ to $c_{k,m} = H_{k,m} \cdot e^{-2\pi\iu \frac{mk}{N}}$.
This saves memory for the twiddle factors and \glspl{DSPSlice} for the multipliers on the \gls{FPGA}.

The first step of the signal processing (see figure \ref{fig:spectrometerflow}) uses the usual \gls{FFT} implementation of the spectrometer with $N' = N/M$ channels to calculate $\hat{a}_{k',m}$, $k' = k \smod N'$ for each \gls{ADCCore} $m$ (for further information see chapter 4.3 and 4.4 in~\cite{prom}).
The typical input stream of the spectrometer is composed of real numbers.
Therefore the second half of the spectrum contains the same information as the first half.
So only the first part of the spectrum is calculated ($ 0 \le k' < N'/2$).
Consequently the second half has to be restored from the first one via $\hat{a}_{k',m} = \hat{a}^*_{N'-k',m}$.
The channel $N'/2$ is lost in this process as $\hat{a}_{N'/2,m} = \hat{a}^*_{N'/2,m}$.

The spectral channels of the \glspl{FFT} are the input of the \verb|COMCA| component.
This is composed of $M \cdot P$ independent \verb|COMCA Cores|, each of them is responsible for $\frac{N}{P \cdot M}$ channels of the final spectrum.
Each \verb|COMCA Core| requires all of the $M$ \gls{FFT} outputs (see equation \ref{eq:applyCali}).
Figure \ref{fig:comcainflow} shows this data-flow.
The \verb|COMCA Core| fetches the $M$ ($0 \le m < M$) complex coefficients $c_{k,m}$ for the current spectral channel $k$ from the \gls{BRAM} and multiplies them with the spectral channels $a_{k',m}$ from the \glspl{FFT}.
Finally the products are added up to obtain the corrected spectral channel.
Figure \ref{fig:comcacoreflow} shows this for one \verb|COMCA Core|. 

The channel $k'$ from the \glspl{FFT} of the \glspl{ADCCore} is connected to $2\cdot M$ channels of the final spectrum.
Due to the symmetry of the \gls{DFT} for real input signals only the first $M$ channels are unique (see chapter \ref{sec:ft}). 
With equation \ref{eq:applyCali} and the periodicity of the \gls{DFT} the relationships between $k'$ and the final channel $k$ can be found:
\begin{itemize}
	\item The channels directly given by $k'$: $k' \rightarrow k =  k' + N'\cdot m$ \\
	\item The channels given by the reconstructed channels $N'-k'$: $k' \rightarrow k = N' - k' + N' \cdot m$
\end{itemize}
Therefore a part of the channels have to be reordered.
This can be simply implemented by flipping and then shifting to the right of the corresponding channels.
Table \ref{tab:reordering} illustrates the reordering.
Furthermore the not reconstructable channels can be found from this relations, as $k$ is limited $0 \le k' < N'/2$.
Consequently the channels $k = N'/2 + N' \cdot m$ cannot be reconstructed.
This is also shown in table \ref{tab:reordering}. 
\begin{table}[htbp]
	\centering
	\begin{tabular}{@{}lllllllllllllllll@{}}
		\toprule
		k                   & 0 & 1 & 2 & 3 & 4 & 5 & 6 & 7 & 8 & 9 & a & b & c & d & e & f \\ \midrule
		k'                  & 0 & 1 & 2 & 3 & - & 3 & 2 & 1 & 0 & 1 & 2 & 3 & - & 3 & 2 & 1 \\
		linear              & 0 & 1 & 2 & 3 & 0 & 1 & 2 & 3 & 0 & 1 & 2 & 3 & 0 & 1 & 2 & 3 \\
		flipped             & 0 & 1 & 2 & 3 & 3 & 2 & 1 & 0 & 0 & 1 & 2 & 3 & 3 & 2 & 1 & 0 \\
		flipped and shifted & 0 & 1 & 2 & 3 & 0 & 3 & 2 & 1 & 0 & 1 & 2 & 3 & 0 & 3 & 2 & 1 \\ \bottomrule
	\end{tabular}
	\caption{The channel reordering for $N = 32$, $N' = 8$ and $M = 4$. $k$ represents the correct and full channel sequence, $k'$ the correct channel sequence if channels are reconstructed, linearly in a not reordered sequence, the last two rows represent the reordered sequence with flipping and shifting. The not reconstructable channels are marked with a -.}
	\label{tab:reordering}
\end{table}
The reordering can be done after the readout of the spectrometer in software or when the integrated spectrum is copied to a buffer for sending it to the computer via Ethernet. 

\section{Consumption of resources in the FPGA and their optimization}
The calibration requires a significant amount of the resources of the \gls{FPGA}.
Due to the high utilization the used resources are spread over the chip, which leads to a high number of long connections and causes congestion and violations of timing requirements.
The utilization of resources limits the size of the \glspl{FFT}.
Therefore an optimization of these resources can lead to a higher spectral resolution with the same \gls{FPGA} or the possibility to use a smaller and cheaper \gls{FPGA}.

\subsection{Block RAM}
The \gls{BRAM} is used to store the coefficients $H_{k,m} \cdot e^{-2\pi\iu \frac{mk}{N}}$.
It can be written on the fly via the network connection.
The \gls{BRAM} size that is required to correct an $N$-point \gls{FFT} with $M$ \glspl{ADCCore} and coefficients that are \SI[number-math-rm = \mathnormal, parse-numbers = false]{B}{\bit} corresponds to:
\begin{align}
	 n\id{BRAM} = 2 \cdot \SI[number-math-rm = \mathnormal, parse-numbers = false]{M \cdot N \cdot B}{\bit}
\end{align}
For two \glspl{FFT} with $16$k spectral channels and $8$ \glspl{ADCCore} and $\SI{16}{bit}$ coefficients \SI{8192}{\kibi\bit} of \gls{BRAM} are needed. This corresponds to \SI{17.4}{\percent} of the \gls{BRAM} resources of a Virtex 7 \verb|XC7VX690T|.

The \gls{BRAM} is the limiting resource in the spectrometer design.
The \gls{BRAM} utilization of the calibration increases linearly with the size of the \gls{FFT}.
Therefore an optimization of the \gls{BRAM} usage is highly advisable.
The usage can be reduced if not every coefficient is stored, but only every $n$-th coefficient.
The missing coefficients have to be interpolated from the otherss.
The linear interpolation is well suited for an \gls{FPGA} implementation as only summations and a multiplication of small width are needed.
Higher order interpolations need bigger multiplication widths, which then require \glspl{DSPSlice}.  
The downside of the interpolation is the need of higher bandwidth from the \gls{BRAM} as per resulting coefficient two values from the \gls{BRAM} are required.
This bandwidth can be achieved by a wider \gls{BRAM} or a higher readout clock.
Another possibility is to reduce the size of the coefficient, but this has an negative impact on the \gls{SNR}~\cite{4263314}.
The increased bandwidth is satisfied by doubling the width of the \glspl{BRAM}.
Because doubling the clock of the \glspl{BRAM} is impossible, since their speed is limited to \SI{458.09}{\mega\hertz}~\cite{Switching} in the used \gls{FPGA} and the current spectrometer design runs at \SI{250}{\mega\hertz}.
A fractional ratio between the clocks of \glspl{BRAM} and the \gls{DSP} part of the design would require buffering and therefore additional \glspl{BRAM}.

\subsubsection{Implementation of the interpolation}
The implementation of the interpolation uses two \glspl{BRAM}.
One stores the even coefficients, one stores the odd ones.
Hence the effective width of the \glspl{BRAM} is doubled.
Depending on the needed, interpolated coefficient $c'_{k,m}$ the coefficients $c_{k_0(k),m}$ and $c_{k_1(k),m}$ are fetched from the \glspl{BRAM} and interpolated $I$ times with:
\begin{align}
	k_0(k)		&= 2I \cdot \floor\left( \frac{k}{2I} + 0.5 \right) \\
	k_1(k) 		&= 2I \cdot \floor\left( \frac{k}{2I} \right) + I \\
	K'			&= \left|I - k \smod 2I \right| \\
	K''			&= I - K' \\
	c'_{k,m} 	&= \frac{1}{I} \left(K' \cdot c_{k_0(k),m} + K'' \cdot c_{k_1(k),m} \right)
\end{align}
The indexing given by this is shown in table \ref{tab:interpolation} for $I = 4$.
$I$ should be a power of $2$ to simplify the division to a shift operation to optimize the calculation complexity.  
The number of stored complex coefficients without interpolation is $N\cdot M$; with $I$ times interpolation it is $M \cdot \left( \frac{N}{I} + 1 \right)$.
Hence the \gls{BRAM} size that is required to correct an $N$-point \gls{FFT} with $M$ \glspl{ADCCore} and coefficients that are \SI[number-math-rm = \mathnormal, parse-numbers = false]{B}{\bit} using an $I$ times interpolation corresponds to:
\begin{align}
	 n\id{BRAM} = 2 \cdot M \cdot \left( \frac{N}{I} + 1 \right) \cdot \SI[number-math-rm = \mathnormal, parse-numbers = false]{B}{\bit}
\end{align}

\begin{table}[htbp]
	\centering
	\begin{tabular}{@{}llllllllllllllllll@{}}
		\toprule
		$k$                   & 0 & 1 & 2 & 3 & 4 & 5 & 6 & 7 & 8 & 9 & 10 & 11 & 12 & 13 & 14 & 15 \\ \midrule
		$k_0(k)$              & 0 & 0 & 0 & 0 & 8 & 8 & 8 & 8 & 8 & 8 & 8 & 8 & 16 & 16 & 16 & 16 \\
		$k_1(k)$              & 4 & 4 & 4 & 4 & 4 & 4 & 4 & 4 & 12 & 12 & 12 & 12 & 12 & 12 & 12 & 12 \\
		$K'$             	  & 4 & 3 & 2 & 1 & 0 & 1 & 2 & 3 & 4 & 3 & 2 & 1 & 0 & 1 & 2 & 3 \\
		$K''$ 				  & 0 & 1 & 2 & 3 & 4 & 3 & 2 & 1 & 0 & 1 & 2 & 3 & 4 & 3 & 2 & 1 \\ \bottomrule
	\end{tabular}
	\caption{The indexing for a $4$ times interpolation. $c'_{k,m}$ is the requested coefficient, $c_{k_0(k),m}$ and $c_{k_1(k),m}$ are the coefficients stored in the \glspl{BRAM}, while $K'$ and $K''$ are the scaling factors for $c_{k_0(k),m}$ and $c_{k_1(k),m}$. }
	\label{tab:interpolation}
\end{table}

\subsubsection{Error estimation of the interpolation}
The linear interpolation $P_1(k)$ introduces an error between the interpolated coefficient and the non interpolated.
The error $\Delta\id{I}$ due to linear interpolation of a function $f(k)$ based on $k_0$ and $k_1$ at $k \in [k_0, k_1]$ is given by~\cite{Schwarz2004}:
\begin{align}
	\Delta\id{I} &= \left| f(k) - P_1(k) \right| \leq \frac{\left| (k-k_0)(k-k_1) \right|}{2} \max \left| f''(k) \right|
\end{align}
The absolute value of the filter coefficients $\left| H_{k,m} \right|$ is close to $1$.

The value of the filter coefficients $\left| H_{k,m} \right|$ will be approximated by $1$ in the following error estimation, as their course is not known.
Furthermore their value only differs slightly from $1$.

Hence they will be neglected in the following error estimation. Therefore $f(k) = e^{-2\pi\iu \frac{mk}{N}}$ and $f''(k) = -4\pi^2 \frac{m^2}{N^2} e^{-2\pi\iu \frac{mk}{N}}$. The absolute value of $f''(k)$ is $\left| f''(k) \right| = 4\pi^2 \frac{m^2}{N^2}$. 
\begin{align}
	\Delta\id{I} 	&\leq \frac{\left| (k-k_0)(k-k_1) \right|}{2} \max \left| f''(k) \right| \\
			&= 2\pi^2 \frac{m^2}{N^2} \left| (k-k_0)(k-k_1) \right| 
\end{align}
The term $\left| (k-k_0)(k-k_1) \right|$ becomes maximal within the interval $k \in \left[k_0, k_1\right]$ for $k = \frac{k_0+k_1}{2}$:
\begin{align}
	\Delta\id{I}	&\leq \frac{\pi^2}{2} \frac{m^2}{N^2} (k_1-k_0)^2
\end{align}
The difference $k_1-k_0$ without interpolation is given by:
\begin{align}
	k_1-k_0 &= 1
\end{align}
When the coefficients are interpolated $I$ times:
\begin{align}
	k_1-k_0 &= I + 1
\end{align}
This leads to 
\begin{align}
	\Delta\id{I} &\leq \frac{\pi^2}{2} \frac{m^2}{N^2} (I + 1)^2 \,.
\end{align}
To estimate the maximum error $m$ is set to $M-1$:
\begin{align}
	\Delta\id{I} &\leq \frac{\pi^2}{2} \frac{(M-1)^2}{N^2} (I + 1)^2
\end{align}
The quantization error is given by $\frac{\Delta^2}{12}$ (see chapter \ref{sec:quantization}).
The quantization of a \SI{16}{\bit} twiddle factor, which is normalized to $1$, is $\frac{1}{2^16} \approx 1.5 \cdot 10^{-5}$.
Therefore the error due to the quantization is $1.9 \cdot 10^{-11}$.
The interpolation error $\Delta\id{I}$ in the case of the spectrometer ($M = 8$, $N = \SI{16}{k}$) for $I=2$ is $\Delta\id{I} = 8.1 \cdot 10^{-6}$ and for $I = 4$ $\Delta\id{I} = 2.2 \cdot 10^{-5}$. 
The error due toe interpolation is around $5$ to $6$ orders of magnitude bigger as the one due to quantization.
But the interpolation is only applied at one stage of the computation of the spectrum, while the twiddles are applied $\log_2(N)$ times.
The influence of the interpolation on the \gls{MirrorSuppression} is investigated in chapter \ref{sec:impactInterpolation}.

\subsection{DSP-Slices}
The \glspl{DSPSlice} are used to perform fast multiplications. Each complex multiplication needs at least $3$ \glspl{DSPSlice}; depending on the bit-width of the complex-numbers it can increase as one \gls{DSPSlice} is limited to a multiplication of \SI{25}{\bit} by \SI{18}{\bit} (see~\cite{DSP48E1}). The number of needed \glspl{DSPSlice} depends on the number of \glspl{ADCCore} $M$, the number of \glspl{DSPSlice} per complex multiplication $C$ and the number of parallel data-streams per \gls{ADCCore} $P$:
\begin{align}
	n\id{DSP} = M \cdot M \cdot P \cdot C
\end{align}
For two \glspl{FFT} with $16$k spectral channels and $8$ \glspl{ADCCore}, $2$ parallel data-streams per \gls{ADCCore} and $\SI{28}{bit}$ complex number from the \gls{FFT} ($C = 8$) $2048$ \glspl{DSPSlice} are needed.
This corresponds to \SI{56.9}{\percent} of the \glspl{DSPSlice} of a Virtex 7 \verb|XC7VX690T|.
The number of \glspl{DSPSlice} per complex multiplication can be reduced to $C = 6$ for $\SI{28}{bit}$ complex numbers at a cost of a higher number of summations.
For the same setup as above this leads to $1536$ \glspl{DSPSlice} and a utilization of \SI{42.7}{\percent} for the calibration.

The further reduction of \gls{DSPSlice} would require to minimize $C$ by a reduced input bit-width, which equals a loss in dynamical range.
A multiplication with \glspl{FF} and \glspl{LUT} would also reduce \gls{DSPSlice} utilization, but would need a disproportionately larger amount of logic and connections. 

\section{Measurements}
The implementation of the calibration is tested and characterized in this section.
At first the \gls{MirrorSuppression} of the enhanced calibration and the current calibration is compared.
Then the stability of the calibration and the influence of the interpolation of the filter coefficients in the \gls{FPGA} are characterized.

The measurements are done with two \verb|dFFTS-4G| spectrometers (see chapter \ref{sec:hardware} for more information).
Each spectrometer has two independent inputs, which results in total a number of $4$ spectra.
The spectrometers are placed in a temperature chamber, while the calibration and the measurements are performed, to ensure a controlled ambient temperature. 
After changing the temperature setting, the measured temperature has to settle within \SI{0.2}{\celsius} around the set temperature before starting a measurement.
The signal generator is used to apply sinusoidal signals to the spectrometers.
This setup and its components are described in more detail in chapter \ref{sec:hardware}.
All measurements are performed with $16$k spectral channels per input.
This corresponds to a $32$k \gls{FFT} where only the first half of the spectrum is calculated, as the inputs are real valued.
Unless stated otherwise the measurements are performed without interpolating the coefficients in the \gls{FPGA}.
The transfer functions of the \glspl{ADCCore} are measured with the method described in chapter \ref{sec:comma}.
The measured points are spaced in steps of \SI{64}{\mega\hertz} from \SI{64}{\mega\hertz} to \SI{8064}{\mega\hertz}.
In this section only one of the four independent inputs is shown.
The figures for all four inputs can be found in appendix \ref{sec:app}.

The \gls{MirrorSuppression} is measured in \SI{12.8}{\mega\hertz} steps, therefore a sinusoidal signal of each frequency from the signal generator is applied to the inputs. 
In the first Nyquist-band the measured points are ranged from \SI{12.8}{\mega\hertz} to \SI{3993.6}{\mega\hertz}, in the second one from \SI{4006.4}{\mega\hertz} to \SI{7987.2}{\mega\hertz}.
This selection of frequencies ensures, that not each measured frequency coincides with a calibration frequency, which tests the linear interpolation of the transfer functions of the \glspl{ADCCore}.
Furthermore this selection of frequencies reduces the influence of harmonic frequencies on the measurement, as only negligible harmonics share the same frequency bins with the signals of interest (input signal and mirror signals) for this selection.

\subsection{Extraction of the Mirror Suppression from a spectrum}
The integrated spectra are saved along with the frequency of the sinusoidal signal applied to the inputs.
If the input signal frequency $f\id{in}$ does not belong to the first Nyquist-band, it is aliased into the first Nyquist-band; the aliased frequency is $f'\id{in}$.
With the relation $f_{\text{mirror},m} = \pm f'\id{in} + \frac{m}{M} f\id{s}, 0 \le m < M$ the frequencies of the mirror signals can be determined.
These mirror frequencies $f_{\text{mirror},m}$ are also aliased into the first Nyquist-band; the aliased mirror frequencies are $f'_{\text{mirror},m}$.
The spectral channel $n$ is connected to the center frequency $f\id{c}$ of this channel via the bandwidth $f\id{b}$ and the number of spectral channels $N$ as:
\begin{align}
	n = \frac{f\id{c}}{f\id{b}} \cdot N 
\end{align}
The \gls{MirrorSuppression} is then given by 
\begin{align}
	\MirSup(f\id{in}) = 10 \cdot \log_{10} \left( \frac{\max\limits_{0 \le m < M} \left(p\left(\frac{f'_{\text{mirror},m}}{f\id{b}} \cdot N \right)\right)}{p\left(\frac{f'\id{in}}{f\id{b}} \cdot N\right)} \right) \si{\dB}
\end{align} with $p(n)$ the spectral power in the spectral channel $n$. 

To quantify the uncertainty of the \gls{MirrorSuppression} a number of spectra is recorded instead of a single, longer integrated one. 
Therefore six spectra with \SI{200}{\milli\second} integration time each are recorded at each frequency.
The \gls{MirrorSuppression} of each spectrum is calculated, as well as their average \gls{MirrorSuppression} and its standard deviation.
The latter one measures the uncertainty of the \gls{MirrorSuppression}.

%Hence at each frequency six spectra are recorded with \SI{200}{\milli\second} integration time for each spectrum. 
%From these spectra the \glspl{MirrorSuppression} are calculated. 
%This \glspl{MirrorSuppression} are then combined on the linear scale into their average and standard deviation. 
%The latter one measures the uncertainty of the \gls{MirrorSuppression}.

\begin{figure}[htbp]
	\centering
	\includegraphics[width=0.95\textwidth]{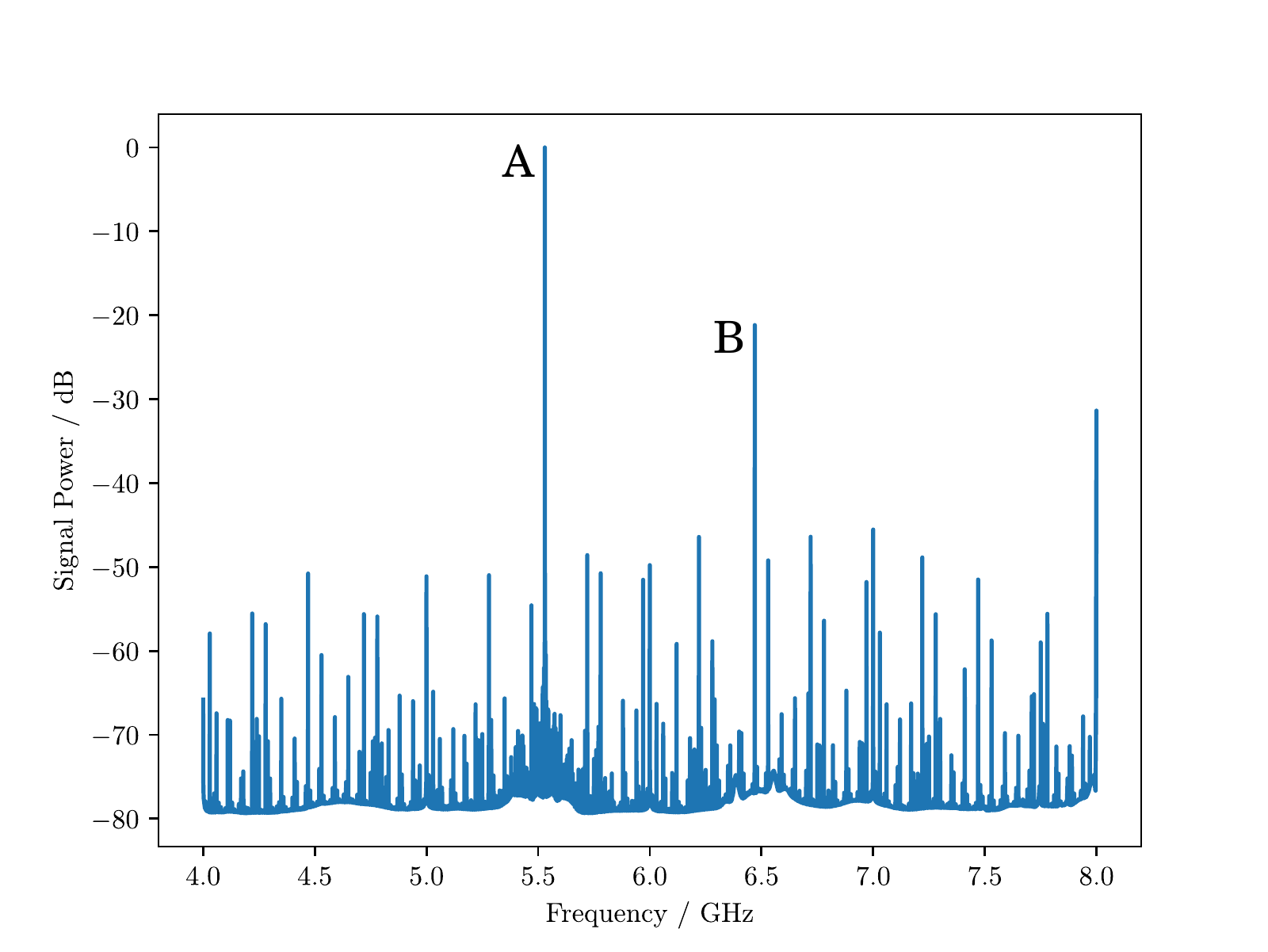}
	\caption{An example spectrum from the spectrometers after a frequency independent calibration. The input signal is a sinusoidal signal at \SI{5.53}{\giga\hertz} (component \texttt{A}). The strongest mirror signal is located at \SI{6.47}{\giga\hertz} (component \texttt{B}).}
	\label{fig:mirrorsuppression}
\end{figure}

Figure \ref{fig:mirrorsuppression} shows an example spectrum from a spectrometer after a frequency independent calibration, recording a sinusoidal signal at \SI{5.53}{\giga\hertz}.
The strongest peak in the spectrum (component \verb|A|) is the signal itself, but the second strongest one (component \verb|B|) is a mirror signal of the applied signal, due to mismatches between the \glspl{ADCCore}. 

Particularly important is the worst \gls{MirrorSuppression} in a Nyquist-band, as this may limit the dynamical range of the spectrometer.
In the following the worst \gls{MirrorSuppression} in the $n$-th Nyquist-band ($f\id{start} = \frac{f\id{s}}{2} \cdot (n-1)$, $f\id{end} = \frac{f\id{s}}{2} \cdot n$) is defined as 
\begin{align}
	\max\limits_{f\id{start} \le f\id{in} < f\id{end}} \MirSup(f\id{in}) \, .
\end{align}
The estimation of the uncertainties of the worst \gls{MirrorSuppression} is given by the uncertainty in \gls{MirrorSuppression}  of the worst point $\max\limits_{f\id{start} \le f\id{in} < f\id{end}} \MirSup(f\id{in})$. 
This approach is chosen as no general underlining model of the \gls{MirrorSuppression} in the region around the worst point can be found with ease.
In some cases this point is given by a single, dominant peak (e.g. figure \ref{fig:interpolation_adc0}). 
In other cases it is a strong peak in a noisy region (e.g. figure \ref{fig:interpolation_adc1}).

\subsection{Impact of the calibration on the Mirror Suppression}
To quantify the impact of the calibration on the \gls{MirrorSuppression}, it is measured for the first two Nyquist-bands with a frequency independent calibration and with a frequency dependent calibration. 
The frequency independent calibration is carried out at \SI{901}{\mega\hertz}. 
The frequency dependent calibration is split up into one calibration from \SIrange{0}{4}{\giga\hertz} and one from \SIrange{4}{8}{\giga\hertz}.
The filter coefficients corresponding to the measured Nyquist-band are loaded accordingly.
While calibration and measurement the spectrometers operate at a constant ambient temperature of \SI{25}{\celsius} to eliminate drift due to temperature changes. 

\begin{figure}[htbp]
	\centering
	\includegraphics[width=0.95\textwidth]{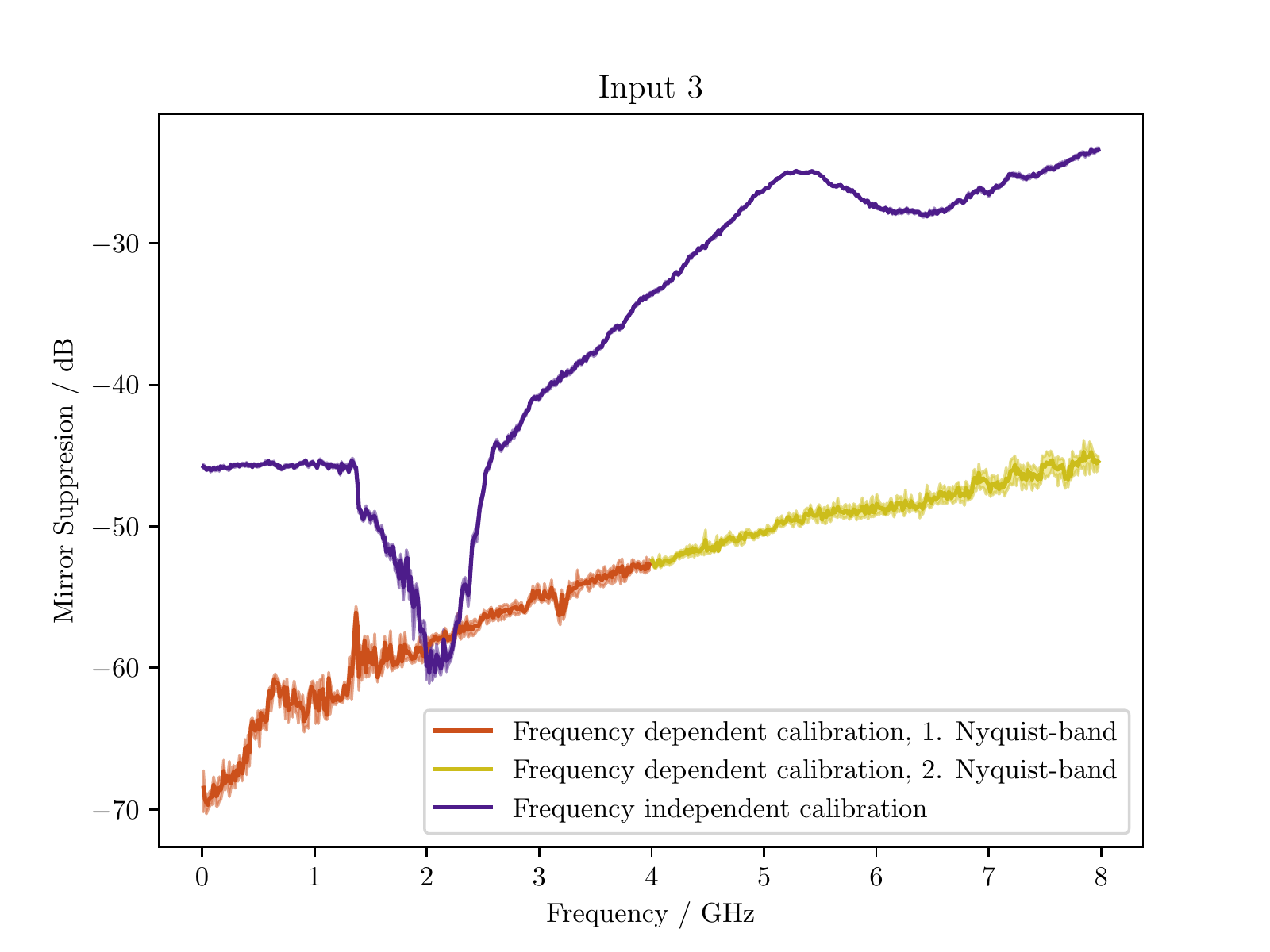}
	\caption{Comparison of \gls{MirrorSuppression} between calibration at one frequency and frequency dependent calibration for input \texttt{Q} of the second \texttt{dFFTS-4G} board.}
	\label{fig:comparsion_adc3_in}
\end{figure}

Figure \ref{fig:comparsion_adc3_in} shows the resulting \glspl{MirrorSuppression} for input \verb|Q| of the second \verb|dFFTS-4G| board.
The \glspl{MirrorSuppression} of all boards can be found in figures \ref{fig:comparsion_adc0}, \ref{fig:comparsion_adc1}, \ref{fig:comparsion_adc2} and \ref{fig:comparsion_adc3}.
Table \ref{tab:impactCalibration} summarizes the worst \gls{MirrorSuppression} for the two different calibrations and Nyquist-bands.
The frequency dependent calibration improves the \gls{MirrorSuppression} compared to the frequency independent calibration in both Nyquist-bands. 
In the first Nyquist-band the frequency dependent calibration improves the worst \gls{MirrorSuppression} for at least \SI{-10}{\dB}.
In the second Nyquist-band the improvements are better than \SI{-17.5}{\dB}.
The improvements are better in the second Nyquist-band as the mismatches are bigger and change stronger than in the first one (see chapter \ref{sec:comma}). 
\begin{table}[htbp]
	\centering
	\begin{tabular}{p{5cm}llll}
	\toprule
                                      					& Input 0 & Input 1 & Input 2 & Input 3 \\ \midrule
1. Nyquist-band frequency independent calibration 	& \SI[parse-numbers=false]{-28.50^{+0.7}_{-0.7}}{\dB} & \SI[parse-numbers=false]{-43.2^{+0.2}_{-0.2}}{\dB} & \SI[parse-numbers=false]{-33.70^{+0.09}_{-0.07}}{\dB} & \SI[parse-numbers=false]{-33.6^{+0.1}_{-0.1}}{\dB} \\
1. Nyquist-band frequency dependent calibration		& \SI[parse-numbers=false]{-49.3^{+0.1}_{-0.2}}{\dB} & \SI[parse-numbers=false]{-53.2^{+0.7}_{-0.9}}{\dB} & \SI[parse-numbers=false]{-52.8^{+0.3}_{-0.5}}{\dB} & \SI[parse-numbers=false]{-52.7^{+0.5}_{-0.4}}{\dB} \\
2. Nyquist-band frequency independent calibration	& \SI[parse-numbers=false]{-13.15^{+0.01}_{-0.02}}{\dB} & \SI[parse-numbers=false]{-27.8^{+0.1}_{-0.1}}{\dB} & \SI[parse-numbers=false]{-20.22^{+0.04}_{-0.03}}{\dB} & \SI[parse-numbers=false]{-23.4^{+0.1}_{-0.1}}{\dB} \\
2. Nyquist-band frequency dependent calibration		& \SI[parse-numbers=false]{-40.5^{+0.7}_{-0.7}}{\dB} & \SI[parse-numbers=false]{-45.3^{+0.4}_{-0.5}}{\dB} & \SI[parse-numbers=false]{-44.2^{+0.7}_{-0.9}}{\dB} & \SI[parse-numbers=false]{-44.7^{+0.8}_{-1.1}}{\dB} \\ \bottomrule
	\end{tabular}
	\caption{The worst \gls{MirrorSuppression} in the first two Nyquist-bands for the different spectrometer inputs. }
	\label{tab:impactCalibration}
\end{table}

The \gls{MirrorSuppression} shows significant structure over the band with the frequency independent calibration.
With the frequency dependent calibration the \gls{MirrorSuppression} rather follows a general trend towards lower \gls{MirrorSuppression} at high frequencies.
In the case of the frequency independent calibration the \gls{MirrorSuppression} is dominated by the mismatches between the \glspl{ADCCore}. 
In the case of the frequency dependent calibration the it is mainly created by statistical fluctuations.
One example for such fluctuations is the aperture jitter (typically \SI{100}{\femto\second}~\cite{ADC12J4000}) of the two distinct \gls{ADC}-\glspl{IC}.
This jitter is the random variation of the sampling points from sample to sample. 
It has a similar effect on the \gls{MirrorSuppression} as a phase mismatch between \glspl{ADCCore}, but as it is statistically instead of systematically, it cannot be corrected.

\subsection{Time stability}
The time stability of the \gls{MirrorSuppression} after a calibration is important to schedule the calibrations during observation runs.
There should be as few calibrations as possible, to maximize the time on target, but enough calibrations to achieve a \gls{MirrorSuppression} which does not affect the measurements.
In the following the stability of the \gls{MirrorSuppression} is characterized over three days after an initial calibration. 

\begin{figure}[htbp]
	\centering
	\includegraphics[width=0.95\textwidth]{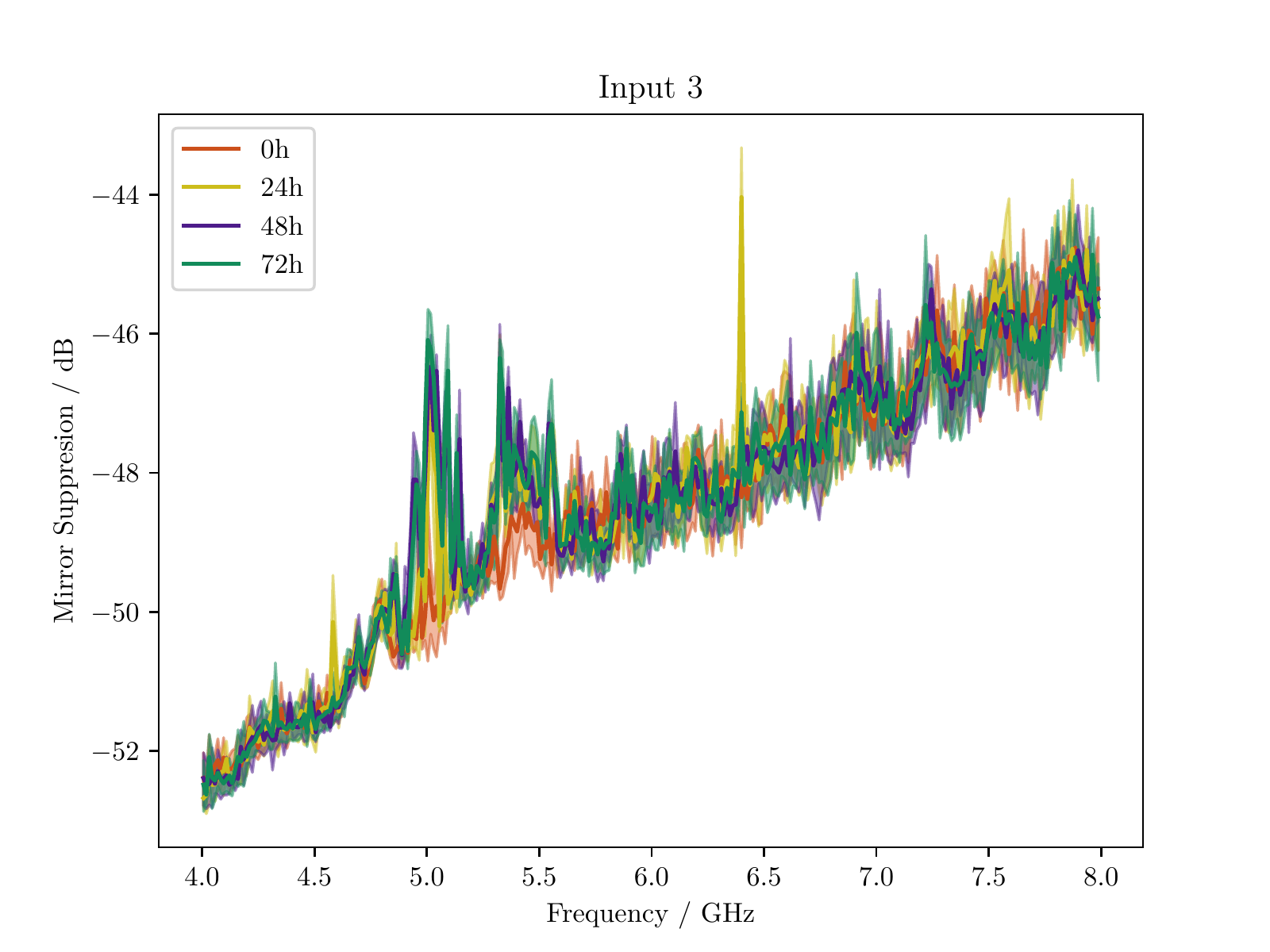}
	\caption{The time stability of the \gls{MirrorSuppression} with constant ambient temperature ($T = \SI{25}{\celsius}$) for input \texttt{Q} of the second \texttt{dFFTS-4G} board.}
	\label{fig:time_adc3_in}
\end{figure}

\begin{figure}[htbp]
	\centering
	\includegraphics[width=0.95\textwidth]{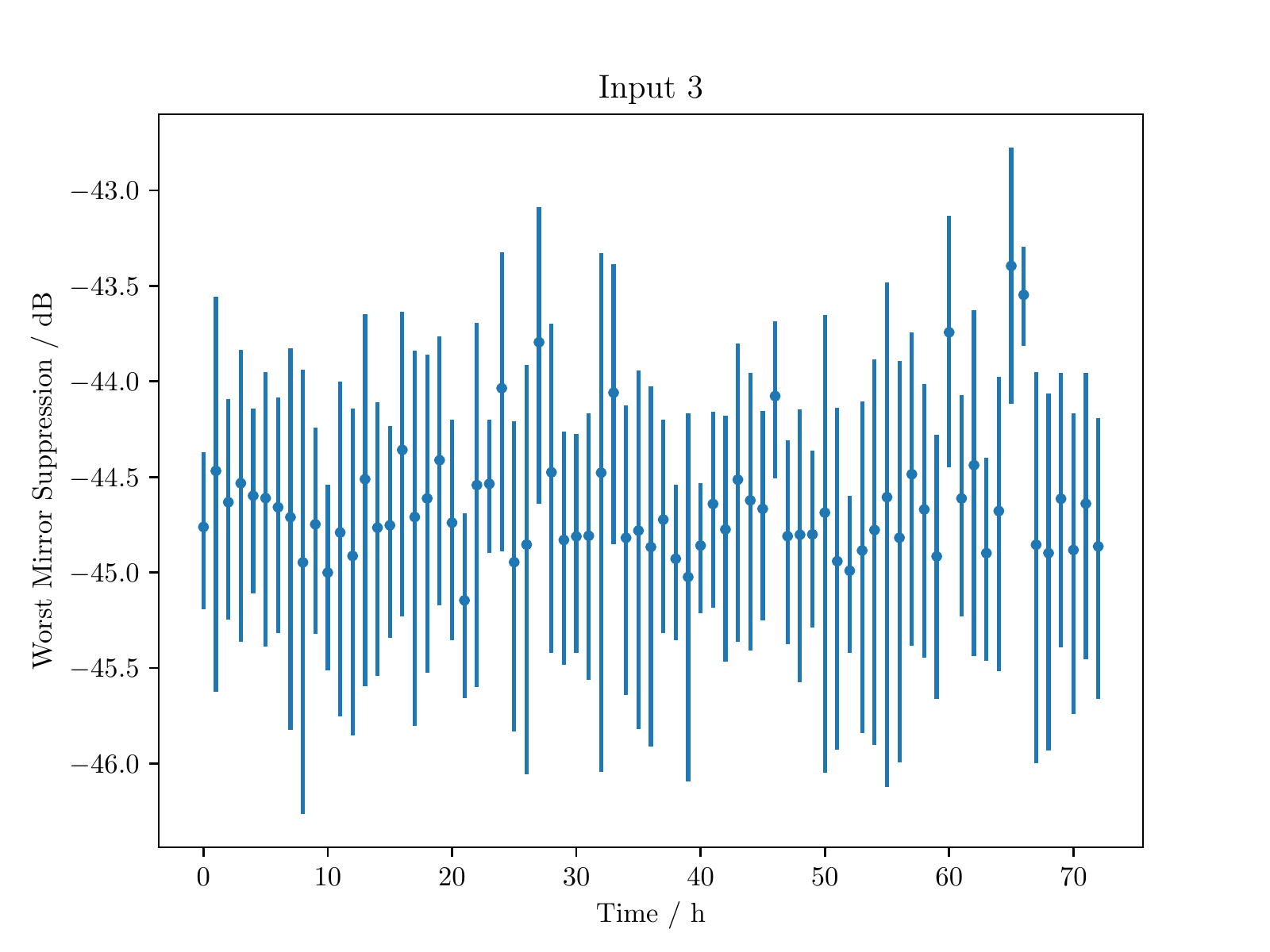}
	\caption{The time stability of the worst \gls{MirrorSuppression} with constant ambient temperature ($T = \SI{25}{\celsius}$) for input \texttt{Q} of the second \texttt{dFFTS-4G} board. The uncertainties of the worst \gls{MirrorSuppression} are given by the error bars.}
	\label{fig:time_adc3_worst_in}
\end{figure}

To eliminate temperature influences the temperature chamber is set to \SI{25}{\degreeCelsius}.
At the beginning of the measurement the transfer functions of the \glspl{ADCCore} are measured, the filter coefficients are generated and loaded into the spectrometers.
A frequency sweep through the second Nyquist-band is performed every hour and the \gls{MirrorSuppression} is extracted from the spectra.
The second Nyquist-band is selected as it has stronger variations between the \glspl{ADCCore} and hence tends to drift more easily than the first one. 
Figure \ref{fig:time_adc3_in} displays the \gls{MirrorSuppression} for input \verb|Q| of the second \verb|dFFTS-4G| board in \SI{24}{\hour} steps. 
Figure \ref{fig:time_adc3_worst_in} characterizes the worst \gls{MirrorSuppression} in the second Nyquist-band over time for this input. 
The figures \ref{fig:time_adc0_worst} to \ref{fig:time_adc3} show the time stability of all $4$ inputs of the two spectrometers.
Three of the four tested inputs show only negligible deterioration of the \gls{MirrorSuppression} over time, but input \verb|I| of the first spectrometer shows a deterioration of roughly \SI{11(1)}{\dB} over \SI{30}{\hour} (see figure \ref{fig:time_adc0_worst} and \ref{fig:time_adc0}). 
After this time the \gls{MirrorSuppression} is stable. 
The cause for this drift could not be determined from the data, especially as the second input of the same spectrometer does not show such a drift. 
To clarify if this is an abnormality or a typical behavior of some of the spectrometers a higher number of boards has to be tested. 

The time stability of the \gls{MirrorSuppression} under constant ambient temperature of the majority of the inputs is sufficient to calibrate them once a day.

\subsection{Temperature stability}
Additionally to time stability after a calibration the sensitivity on temperature changes affects the \gls{MirrorSuppression}.
Both can limit the time between calibrations while observing.
It is expected, that a temperature shift has a strong influence as the geometry on the \gls{IC} changes with it, which leads to changes of the inherent capacities and inductivities in the \gls{IC}.
This effect is important as the ambient temperature of the spectrometers is typically not stabilized at the telescope sites. Hence it can limit the time between calibrations. 

\begin{figure}[htbp]
	\centering
	\includegraphics[width=0.95\textwidth]{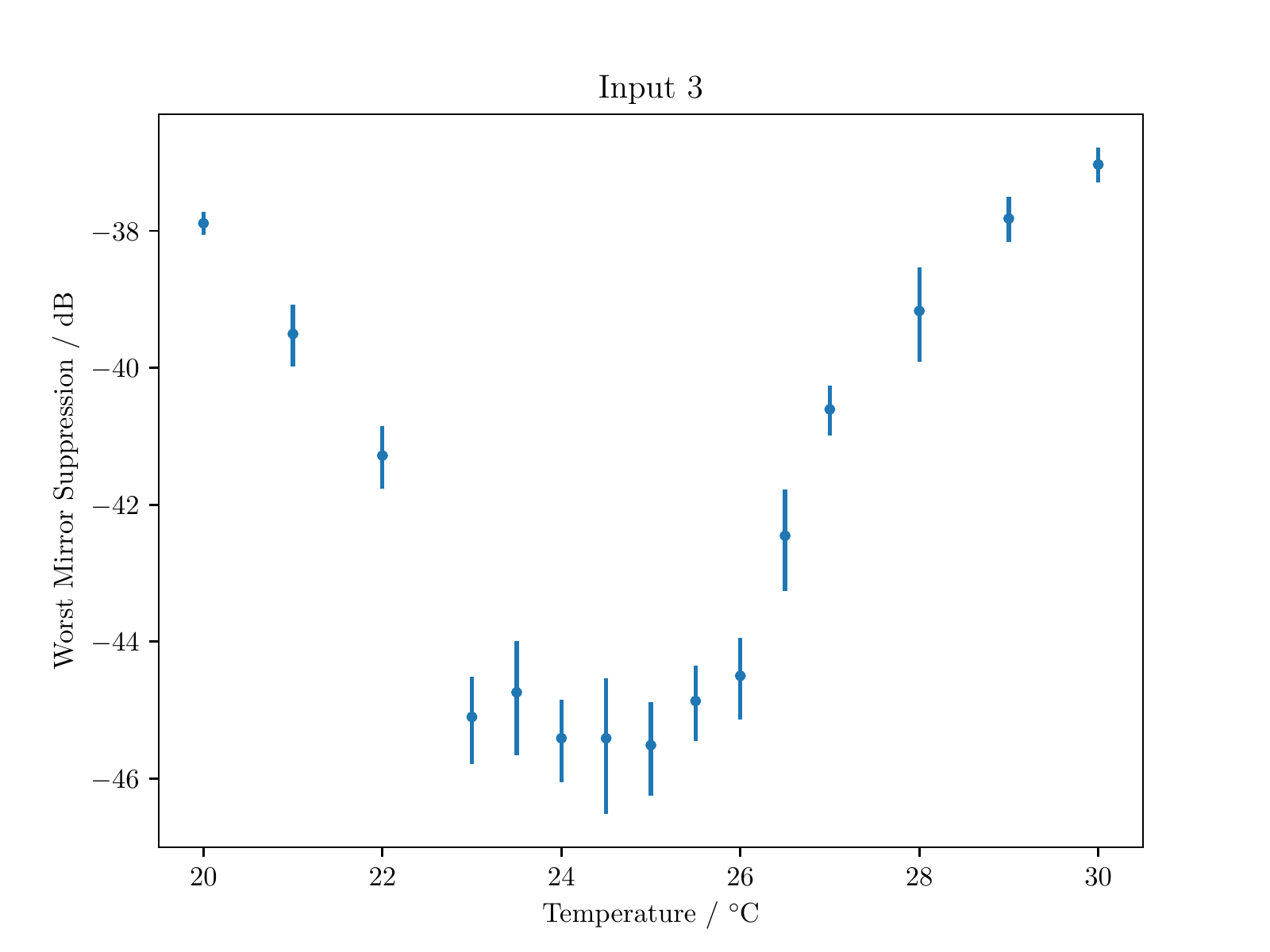}
	\caption{The temperature stability of the worst \gls{MirrorSuppression} in the second Nyquist-band for input \texttt{Q} of the second \texttt{dFFTS-4G} board. The calibration is performed at \SI{25}{\celsius}.}
	\label{fig:temperature_adc3_worst_in}
\end{figure}

\begin{figure}[htbp]
	\centering
	\includegraphics[width=0.95\textwidth]{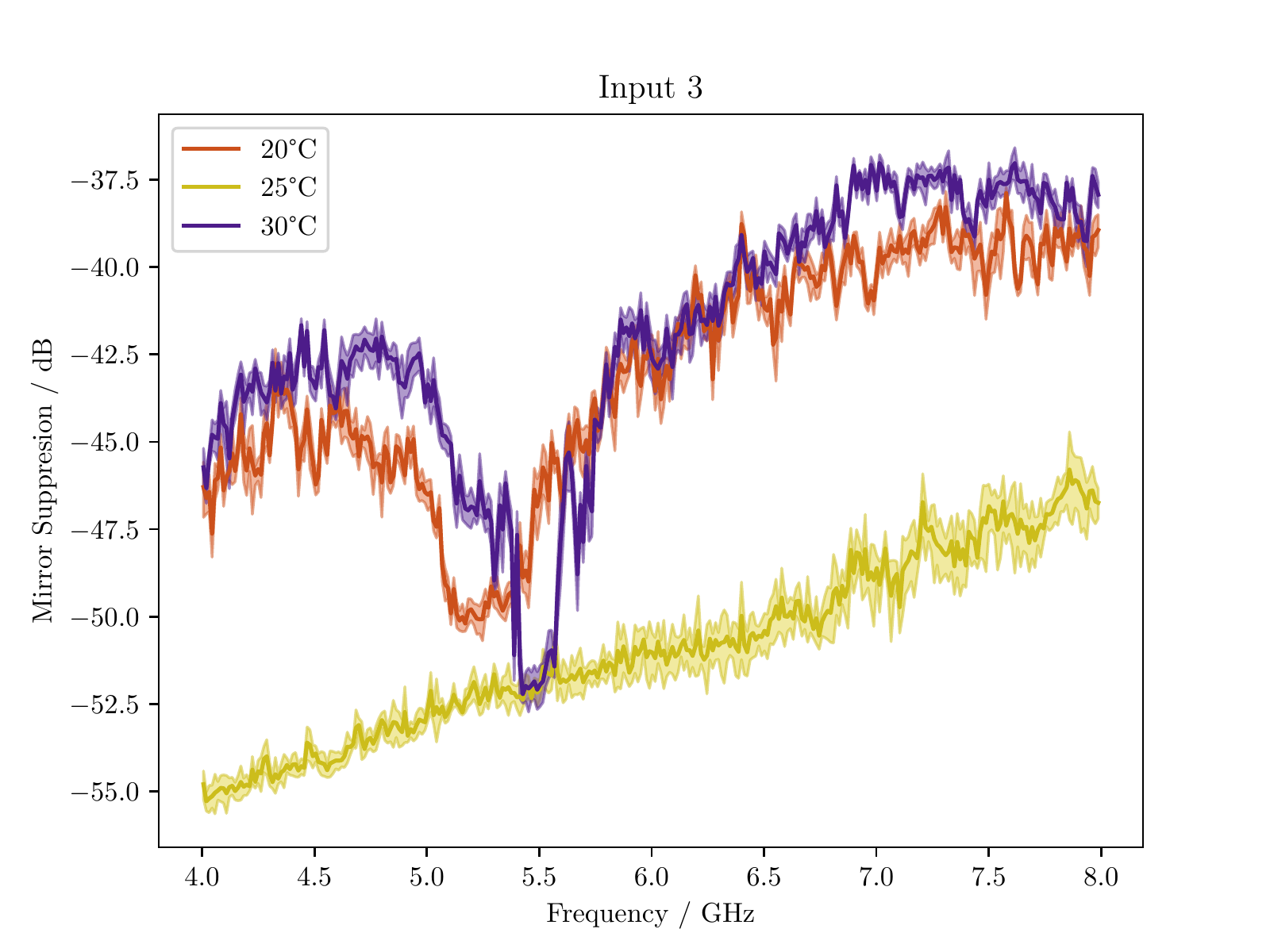}
	\caption{The temperature stability of the \gls{MirrorSuppression} in the second Nyquist-band for input \texttt{Q} of the second \texttt{dFFTS-4G} board. The calibration is performed at \SI{25}{\celsius}. The uncertainties of the worst \gls{MirrorSuppression} are given by the error bars.}
	\label{fig:temperature_adc3_in}
\end{figure}

The transfer functions of the \glspl{ADCCore} are measured at an ambient temperature of \SI{25}{\celsius} and the calibration coefficients generated from them are loaded into the spectrometers.
A frequency sweep over the second Nyquist-band is performed at different temperatures within \SI{\pm 5}{\celsius}.
Figure \ref{fig:temperature_adc3_in} displays the resulting \glspl{MirrorSuppression} for input \verb|Q| of the second spectrometer, while figures \ref{fig:time_adc0}, \ref{fig:time_adc1} and \ref{fig:time_adc2} display the \gls{MirrorSuppression} of the remaining inputs.
Figure \ref{fig:temperature_adc3_worst_in} shows, how the worst \gls{MirrorSuppression} of input \verb|Q| of the second spectrometer develops over temperature (the remaining three inputs can be found in figures \ref{fig:temperature_adc0_worst}, \ref{fig:temperature_adc1_worst} and \ref{fig:temperature_adc2_worst}).
A temperature dependence of the \gls{MirrorSuppression} is visible, as the transfer function of the \glspl{ADCCore} change with temperature. 
The exact behavior differs from spectrometer to spectrometer. 
Some show a broad plateau with changing temperature (e.g. in figure \ref{fig:temperature_adc3_worst_in}), while other show an instant deterioration over temperature (e.g. in figure \ref{fig:temperature_adc0_worst}).
Therefore the temperature of the \glspl{ADCCore} on the spectrometers should not change significantly (more than \SI{\pm 1}{\celsius}) to limit the deterioration of the \gls{MirrorSuppression} to \SI{3}{\dB}.
As no active temperature control is implemented in the spectrometers, the ambient temperature should be kept as constant as possible, for example with a temperature controlled ventilation system. 

\subsection{Stability over power cycles}
The stability over power cycles is not just interesting in case a hard reset is required, but it also evaluates two possible drift sources.
The ramp down and ramp up of the supply voltages itself, as well as temperature cycle of the \glspl{ADCCore} perform due to a shutdown.

\begin{figure}[htbp]
	\centering
	\includegraphics[width=0.95\textwidth]{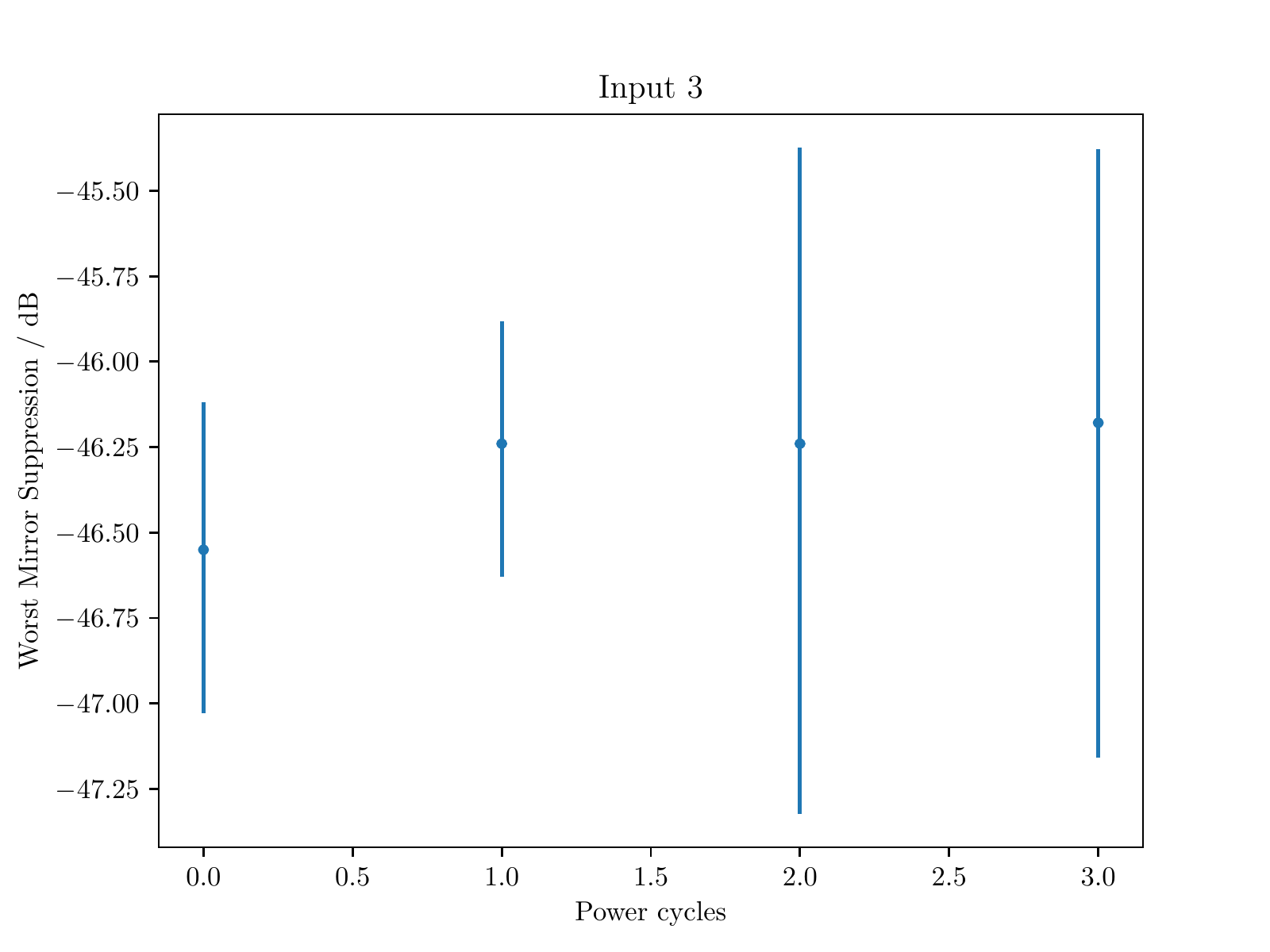}
	\caption{The stability of the worst \gls{MirrorSuppression} over power cycles in the second Nyquist-band for input \texttt{Q} of the second \texttt{dFFTS-4G} board. The calibration and the measurement are performed at \SI{25}{\celsius}. The uncertainties of the worst \gls{MirrorSuppression} are given by the error bars.}
	\label{fig:power_cycles_worst3_in}
\end{figure}

\begin{figure}[htbp]
	\centering
	\includegraphics[width=0.95\textwidth]{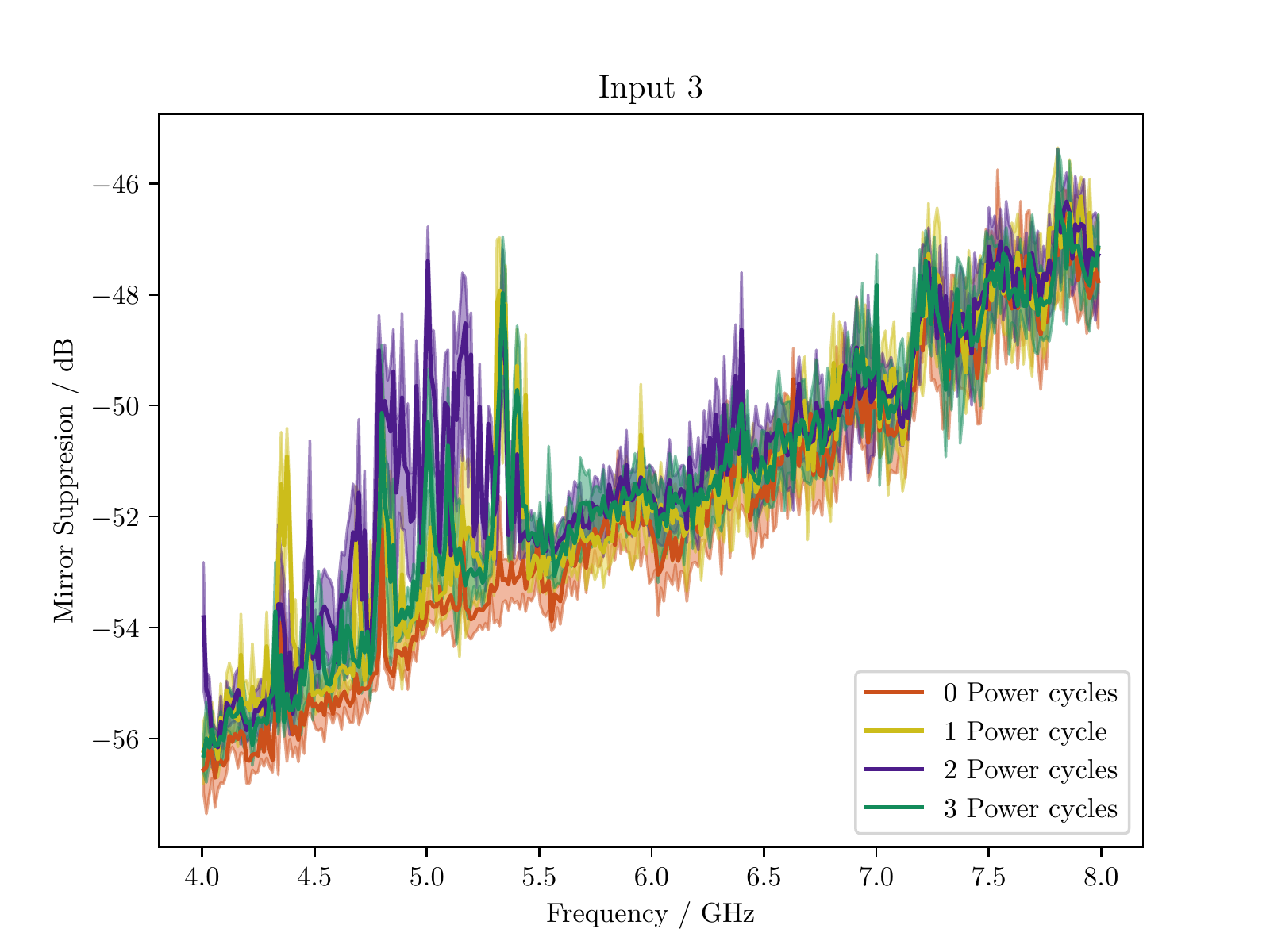}
	\caption{The stability of the \gls{MirrorSuppression} over power cycles in the second Nyquist-band for input \texttt{Q} of the second \texttt{dFFTS-4G} board. The calibration and the measurement are performed at \SI{25}{\celsius}.}
	\label{fig:power_cycles_adc3_in}
\end{figure}

In the first step the transfer functions of the \glspl{ADCCore} are measured and a set of calibration coefficients are generated.
These coefficients are used to carry out a frequency sweep through the second Nyquist-band.
Then a power cycle of the spectrometers is performed and both \gls{gateware} and the calibration coefficients are transferred to the \glspl{FPGA} again.
Finally the frequency sweep is repeated.
With this procedure the \gls{MirrorSuppression} is measured over $3$ power cycles. 
The ambient temperature is held constant at \SI{25}{\celsius} during this procedure.
Figure \ref{fig:power_cycles_adc3_in} displays the \glspl{MirrorSuppression} for the input \verb|Q| of the second spectrometer, figure \ref{fig:power_cycles_worst3_in} displays the change of the  worst \gls{MirrorSuppression} over the power cycles for this input. 
The same is shown for all 4 inputs in figures \ref{fig:power_cycles_worst0} to \ref{fig:power_cycles_adc3}.
The worst \gls{MirrorSuppression} does not change significantly over the power cycles expect for input \texttt{I} of the first spectrometer.
This input shows a change of \SI{3.5(6)}{\dB}.
Hence the \gls{MirrorSuppression} after a calibration is stable over power cycles for the majority of the inputs. 

\subsection{Impact of the interpolation on the calibration}\label{sec:impactInterpolation}
The interpolation should not corrupt the measurements.
Therefore it is characterized by measuring the \gls{MirrorSuppression} for a calibration without interpolation, with 2 times interpolation in the \gls{FPGA} and with 4 times interpolation in the \gls{FPGA}.

\begin{figure}[htbp]
	\centering
	\includegraphics[width=0.95\textwidth]{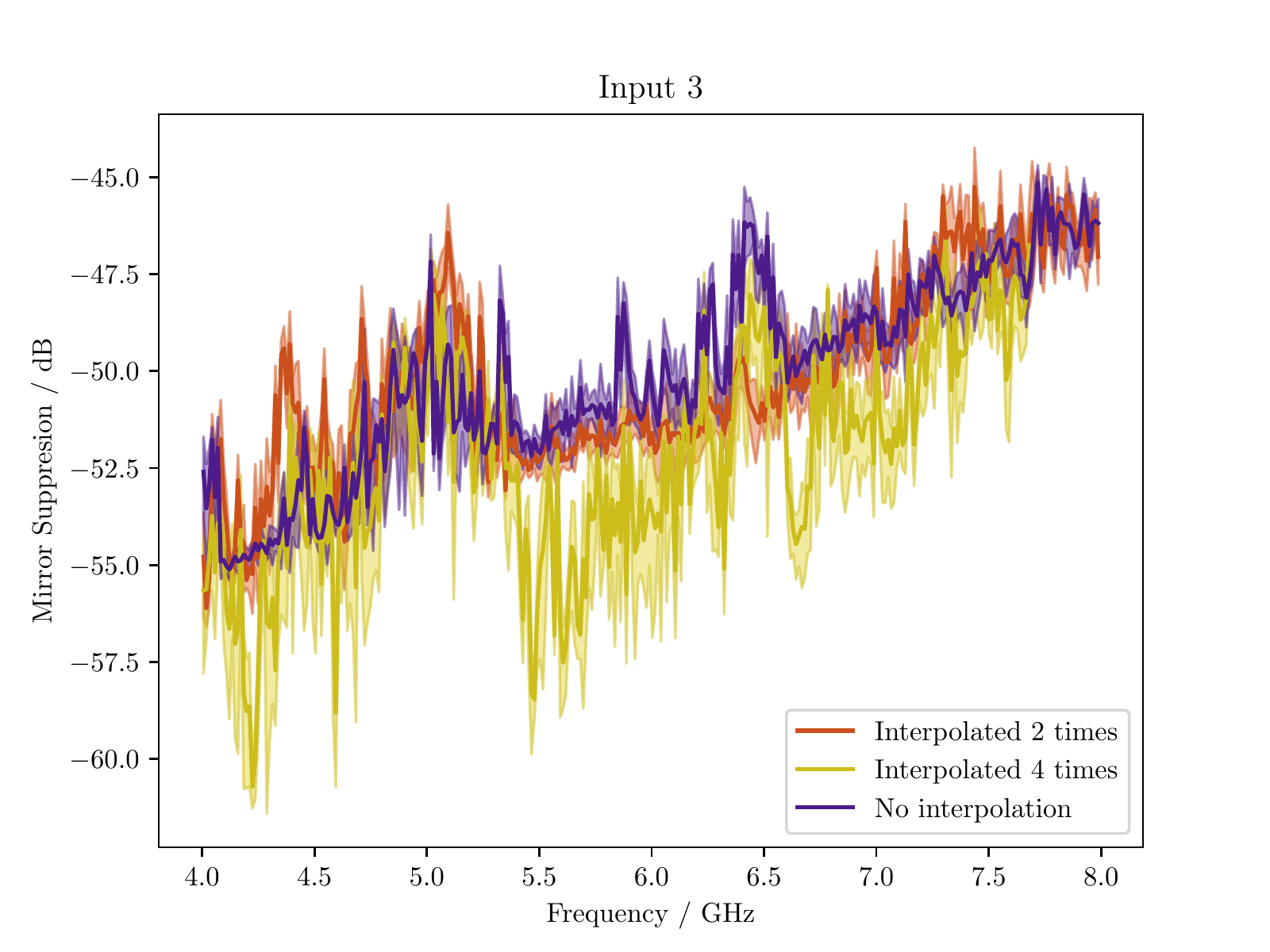}
	\caption{The \gls{MirrorSuppression} in the second Nyquist-band for different interpolations for input \texttt{Q} of the second \texttt{dFFTS-4G} board. The calibration and measurements are performed at an ambient temperature of \SI{25}{\celsius}.}
	\label{fig:interpolation_adc3_in}
\end{figure}

The calibration and measurements of the \gls{MirrorSuppression} are performed at a constant ambient temperature of \SI{25}{\celsius}.
At first the mismatches of the \glspl{ADCCore} are measured.
Then calibrations coefficients are generated, one set without interpolation, one with an interpolation factor of 2 and one with an interpolation factor of 4.
The \gls{MirrorSuppression} is measured for the second Nyquist-band for each of these three cases with the corresponding \gls{gateware} and coefficients.
Figure \ref{fig:interpolation_adc3_in} displays the \gls{MirrorSuppression} for the input \verb|Q| of the second spectrometer for the different interpolations.
Figures \ref{fig:interpolation_adc0} to \ref{fig:interpolation_adc3} show the same for all four inputs.
The worst points of the \gls{MirrorSuppression} depending on the interpolation are listed in table \ref{tab:worstInterpolation}. 
\begin{table}[htbp]
	\centering
	\begin{tabular}{p{5cm}llll}
	\toprule
                                      					& Input 0 & Input 1 & Input 2 & Input 3 \\ \midrule
Interpolated 2 times 	& \SI[parse-numbers=false]{-41.8^{+0.5}_{-0.5}}{\dB} & \SI[parse-numbers=false]{-46.2^{+0.7}_{-0.8}}{\dB} & \SI[parse-numbers=false]{-43.9^{+0.8}_{-1.1}}{\dB} & \SI[parse-numbers=false]{-45^{+1}_{-1}}{\dB} \\
Interpolated 4 times		& \SI[parse-numbers=false]{-43.7^{+0.4}_{-0.5}}{\dB} & \SI[parse-numbers=false]{-46.4^{+0.4}_{-0.5}}{\dB} & \SI[parse-numbers=false]{-45.1^{+0.8}_{-0.9}}{\dB} & \SI[parse-numbers=false]{-46.7^{+0.9}_{-1.1}}{\dB} \\
No interpolation	& \SI[parse-numbers=false]{-42.0^{+0.3}_{-0.4}}{\dB} & \SI[parse-numbers=false]{-41.7^{+0.3}_{-0.4}}{\dB} & \SI[parse-numbers=false]{-44.0^{+0.4}_{-0.5}}{\dB} & \SI[parse-numbers=false]{-45.1^{+0.4}_{-0.5}}{\dB} \\ \bottomrule
	\end{tabular}
	\caption{The worst \gls{MirrorSuppression} in the second Nyquist-band for different interpolations. }
	\label{tab:worstInterpolation}
\end{table}

The linear interpolation inside of the \gls{FPGA} causes to no significant deterioration of the \gls{MirrorSuppression}. 
Note, that the worst \gls{MirrorSuppression} is best for the 4 times interpolation inside of the \gls{FPGA} for all inputs.

\subsection{Power consumption}
The \gls{FPGA} on the spectrometer boards is based on the \gls{CMOS} technology.
The logic gates are controlled by charge. 
Some of these gates have to be charged and discharged each clock cycle, the \gls{FPGA} consumes different amounts of power depending on clock frequency, utilization and the processed data.
The toolchain from \verb|Xilinx®| allows to estimate the power consumption of a design.
The \gls{FFTS} design with $2 \cross 16$k spectral channels and a calibration for $8$ \glspl{ADCCore} has an estimated power consumption of \SI{42}{\ampere} for the \SI{1.0}{\volt} power rail, which powers the logic and therefore carries most of the current.

The consumption can be measured with an onboard circuit on the spectrometer board.
In the following the current measurement circuit and its accuracy are briefly described.
At first a hall-effect sensor converts the current to a voltage, which is then digitized with an \gls{ADC}.
The current-voltage-converter has a typical error of $\pm\SI{3}{\percent}$~\cite{ACS723} and the \gls{ADC} has a gain error of $\pm\SI{0.5}{\percent}$, an offset error of $\pm\SI{0.2}{\percent}$ and an error due to nonlinearity of $\pm\SI{0.05}{\percent}$~\cite{SwitchingArtix}, which adds up to a total systematic error of $\pm\SI{3.75}{\percent}$.
Figures \ref{fig:current_noInterpolation}, \ref{fig:current_2Interpolation} and \ref{fig:current_4Interpolation} compare the electric current for \glspl{gateware} without interpolation, two times interpolated and 4 times interpolated calibration coefficients. Each \gls{gateware} contains two \glspl{FFT} with $16$k spectral channels each and processes sinusoidal signals with different frequencies for the measurement.
The \glspl{gateware} with interpolation consume more power, because more logic is needed to generate the interpolated coefficients.
The maximum power consumption is as high as \SI{50(2)}{\ampere} for the \SI{1}{\volt} power rail.
A \gls{gateware} with more spectral channels would consume even more.
The \SI{1}{\volt} power rail uses a power supply module with a maximum current of \SI{50}{\ampere}~\cite{LTM4650}.
Therefore \glspl{gateware} with more spectral channels can overload the power rail, which could lead to malfunction of the spectrometers.

%==============================================================================
\chapter{Conclusion}
\label{sec:conclusion}
%==============================================================================
\Glspl{FFTS} sample their input signal and compute the power spectrum from the sampled signal.
One of the limiting components of an \glspl{FFTS} is the \gls{ADC}.
Its limiting characteristics have been successfully reduced in this thesis.
The \glspl{ADC} on the spectrometers are composed of a number of slower \glspl{ADC}, which slightly differ from each other.
A calibration procedure has to correct these mismatches for an optimal performance of the spectrometers.
For small bandwidths these mismatches can be assumed to be frequency independent, but for larger bandwidths this assumption reduces the dynamical range of the spectrometers.
The current generation of \glspl{FFTS} calibrates these differences at a fixed frequency with a statistical approach (see chapter 4.2 of~\cite{prom}).
The goal of this thesis was to develop and implement a method in the \glspl{FFTS} to correct for these differences in a frequency dependent manner.
Indeed, it was found, that the newly implemented frequency dependent calibration enhances the dynamical range of the spectrometer by increasing the \gls{MirrorSuppression}.

The mismatches were measured with a series of sinusoidal signals with different frequencies.
These mismatches were then converted into filter coefficients based on a model of the imperfect gains and phases of a \gls{TIADC} using \glspl{FIRFilter}. 
Multiplications on the complex amplitude spectrum, which correspond to circular convolutions in time domain, were used to implement these filters.
These circular convolutions equal linear convolutions in this special case as an \gls{FFT} assumes periodic signals.

The correction of mismatches between \glspl{ADCCore} on the complex amplitude spectrum in a frequency dependent manner improved the \gls{MirrorSuppression} of the \texttt{dFFTS-4G} up to \SI{20}{\dB} compared to a frequency independent correction.
The mismatches vary with temperature, therefore the \gls{MirrorSuppression} deteriorates if the temperature of the spectrometer changes between calibration and observation.
The deterioration could be limited to \SI{3}{\dB} when the temperature is stabilized to \SI{\pm 1}{\celsius}.
The deterioration of the \gls{MirrorSuppression} over time and power cycles was negligible for three of the four tested inputs.

The input \verb|I| of the first \texttt{dFFTS-4G} spectrometer (input 0) showed anomalies.
Its \gls{MirrorSuppression} was systematically worse and it had significantly higher mismatches (up to \textasciitilde{} \SI{2.5}{\dB} and \textasciitilde{} \SI{0.4}{\radian}) compared to the other inputs (up to \textasciitilde{} \SI{1}{\dB} and \textasciitilde{} \SI{0.2}{\radian}).
It is also notable, that this input had a loose connection, which had been resoldered.
However, additional measurements with more spectrometers have to clarify the reason for this behavior and whether it is an individual problem.

With respect to the complexity of the new calibration its implementation consumes a significant amount of the resources of the used \gls{FPGA}.
The amount of \glspl{DSPSlice} depends on the samplerate of the \glspl{ADCCore} and the number of \glspl{ADCCore}, but it is independent of the number of spectral channels.
In contrast to that, the amount of \gls{BRAM} depends linearly on the number of spectral channels and the number of \glspl{ADCCore}.
Furthermore the \gls{BRAM} is already the limiting resource for \gls{FFTS} designs without frequency dependent corrections.
Therefore an interpolation of the coefficients was developed to reduce the amount of used \gls{BRAM}, at the cost of a higher amount of other logic to implement the interpolation.
It was found, that an interpolation of factor 2 or 4 showed no significant deterioration of the \gls{MirrorSuppression} compared to exact coefficients.
An interpolation of factor $4$ was successfully implemented as part of the calibration in the \gls{FPGA} and reduced the \gls{BRAM} size from \SI{8192}{\kibi\bit} to \SI{2048.5}{\kibi\bit}.

The calibration procedure was performed with a sinusoidal signal, sweeped through the Nyquist-band.
A signal with more frequency components would speed up the calibration as different spectral channels are probed simultaneously.
A candidate for such a signal would be an impulse train.
Furthermore the speed of loading the coefficients into the \gls{BRAM} is limited by the \SI{100}{\mega\bit} Ethernet connection.
An interface with higher datarate, for example a \SI{1}{\giga\bit} Ethernet connection, would therefore speed up the calibration cycle.

This new calibration not only improved the dynamical range of the spectrometers, but can also be used as a basis for further development of the spectrometer.
The implemented method is not limited to the transfer functions of the \glspl{ADCCore}, but may also correct a more general transfer function of other components in front of the \glspl{SH} or even outside the spectrometer itself. 
\Gls{SSB} and \gls{2SB} receivers with wide bandwidths suffer from similar problems as \glspl{TIADC}.
The side band suppression is limited due to the imperfections of the \SI{90}{\degree} \gls{IFHybrid}.
These \glspl{IFHybrid} neither have a \SI{90}{\degree} phase shift over the whole bandwidth, nor do they split the input power into equal parts over the whole bandwidth. 
Therefore the sideband suppression is not constant over the bandwidth and can be improved if these divergent phases and amplitudes are corrected. 
The calibration developed in this thesis is a promising possibility to also correct these imperfections.
To do so, the calibration would be applied to the $2$ inputs of the spectrometer instead of applying it to each \gls{ADCCore}. 
Going one step further, the \SI{90}{\degree} phase shift can be realized with calibration alone.
Therefore eliminating the need for an \gls{IFHybrid} and simplifying the hardware of \gls{SSB} and \gls{2SB} receivers, while providing a better sideband suppression than traditional receivers with \gls{IFHybrid}. 

% Uncomment the following command to get references per chapter.
% Put it inside the file or change \include to \input if you do not want the references
% on a separate page
% \printbibliography[heading=subbibliography]

%------------------------------------------------------------------------------
% Use biblatex for the bibliography
% Add bibliography to Table of Contents
% Comment out this command if your references are printed for each chapter.
\printbibliography[heading=bibintoc]

%------------------------------------------------------------------------------
% Include the following lines and comment out \printbibliography if
% you use BiBTeX for the bibliography.
% If you use biblatex package the files should be specified in the preamble.
% \KOMAoptions{toc=bibliography}
% {\raggedright
%   \bibliographystyle{../refs/atlasBibStyleWithTitle.bst}
%   % \bibliographystyle{unsrt}
%   \bibliography{./thesis_refs,../refs/standard_refs-bibtex}
% }

%------------------------------------------------------------------------------
\appendix
% \part*{Appendix}
% Add your appendices here - don't forget to also add them to \includeonly above
%------------------------------------------------------------------------------
\chapter{Appendix}
\label{sec:app}
%------------------------------------------------------------------------------

\section{The internal signals of a TIADC}
\begin{figure}[htbp!]
	\centering
	\includegraphics[height=0.99\textheight]{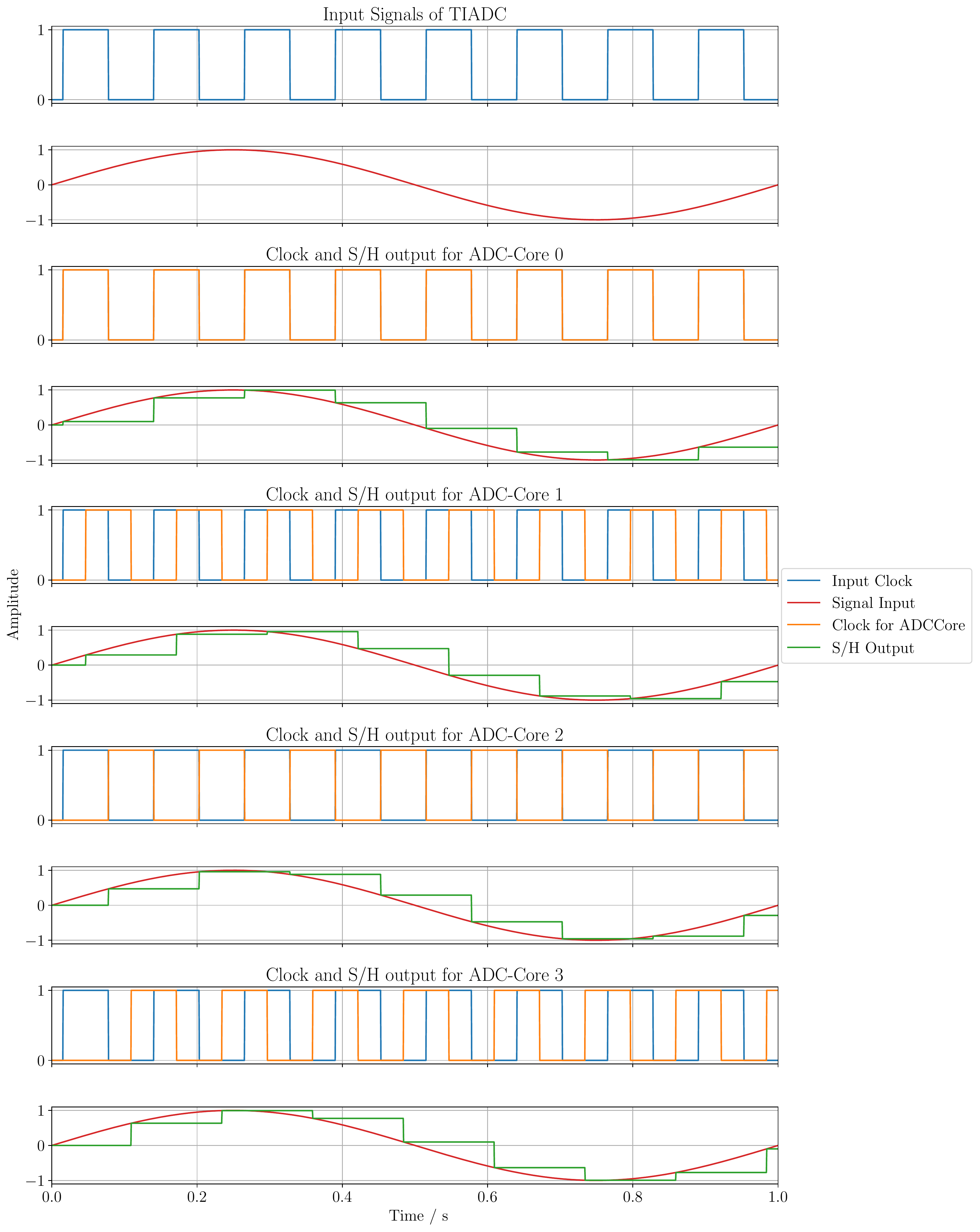}
	\caption{Signals of a \gls{TIADC} with $M=4$, the input clock, input signal and the phase shifted clocks for each \gls{ADCCore} with the outputs of the corresponding \glspl{SH}.}
	\label{fig:tiadcSignals}
\end{figure}

\FloatBarrier
\section{Mismatches}
%The measured mismatches of the used \verb|FFTS-4G| boards are shown in figure \ref{fig:mismatches0}, \ref{fig:mismatches1}, \ref{fig:mismatches2} and \ref{fig:mismatches3}. They are measured according to the procedure described in chapter \ref{sec:comma}. The measurement points have a spacing of \SI{12.8}{\mega\hertz} and span the first and second Nyquist-band. The standard derivation of the measurements are displayed as colored bands around the means of the measurements. 
\begin{figure}[htbp]
	\centering
	\includegraphics[width=0.95\textwidth]{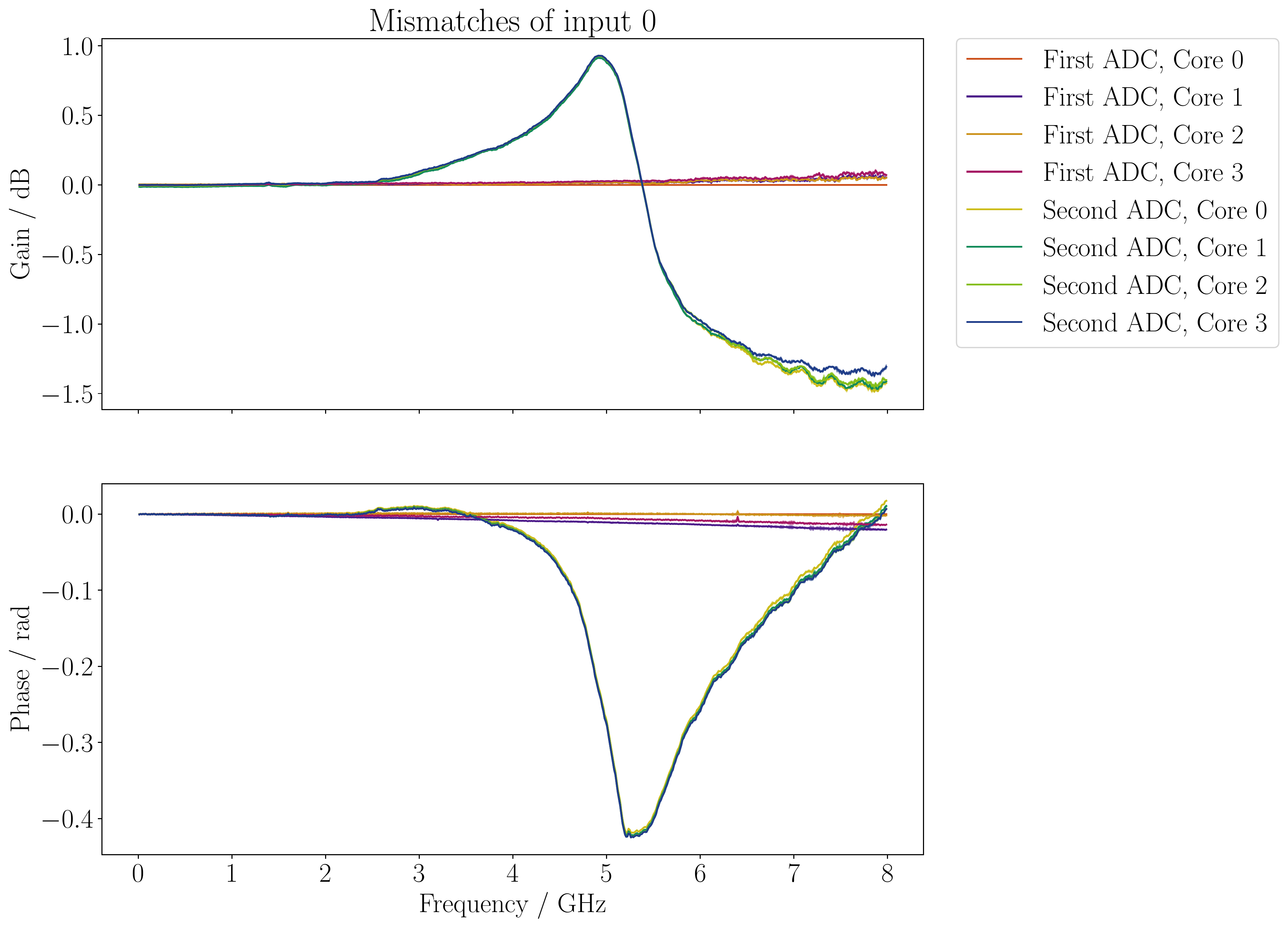}
	\caption{The mismatches of input \texttt{I} of the first \texttt{dFFTS-4G} board. The figure shows the gain given by $\left| \hat{h}'''_{m,n} \right|$ and phase given by $\atantwo\left(\imaginary\left(\hat{h}'''_{m,n}\right), \real\left(\hat{h}'''_{m,n}\right)\right)$ of the \glspl{ADCCore} with respect to \gls{ADCCore} $0$ of the first \gls{ADC}. The standard deviations of the measurements are shown as colored bands around the mean.}
	\label{fig:mismatches0}
\end{figure}
\begin{figure}[htbp]
	\centering
	\includegraphics[width=0.95\textwidth]{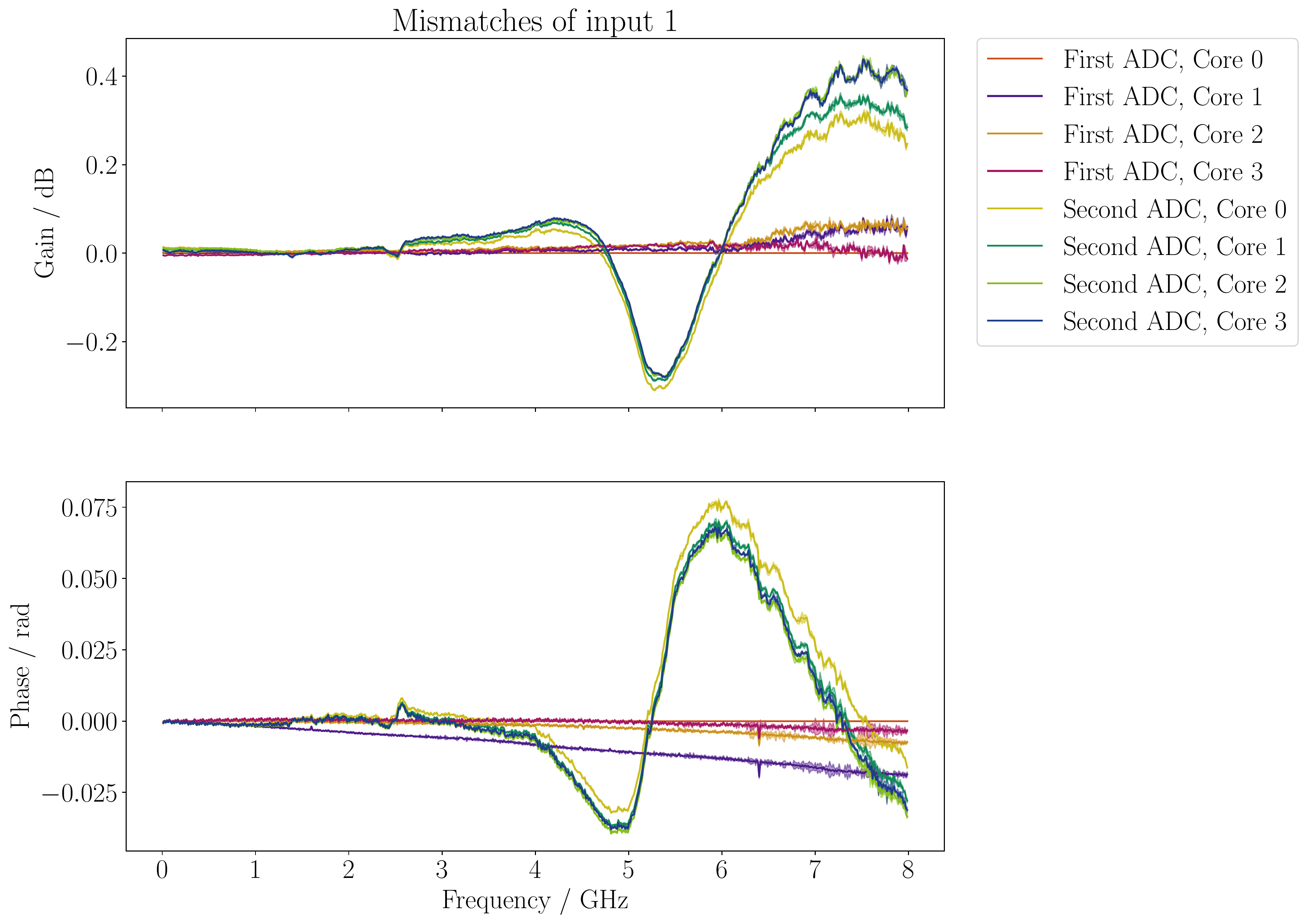}
	\caption{The mismatches of input \texttt{Q} of the first \texttt{dFFTS-4G} board. The figure shows the gain given by $\left| \hat{h}'''_{m,n} \right|$ and phase given by $\atantwo\left(\imaginary\left(\hat{h}'''_{m,n}\right), \real\left(\hat{h}'''_{m,n}\right)\right)$ of the \glspl{ADCCore} with respect to \gls{ADCCore} $0$ of the first \gls{ADC}. The standard deviations of the measurements are shown as colored bands around the mean.}
	\label{fig:mismatches1}
\end{figure}

\begin{figure}[htbp]
	\centering
	\includegraphics[width=0.95\textwidth]{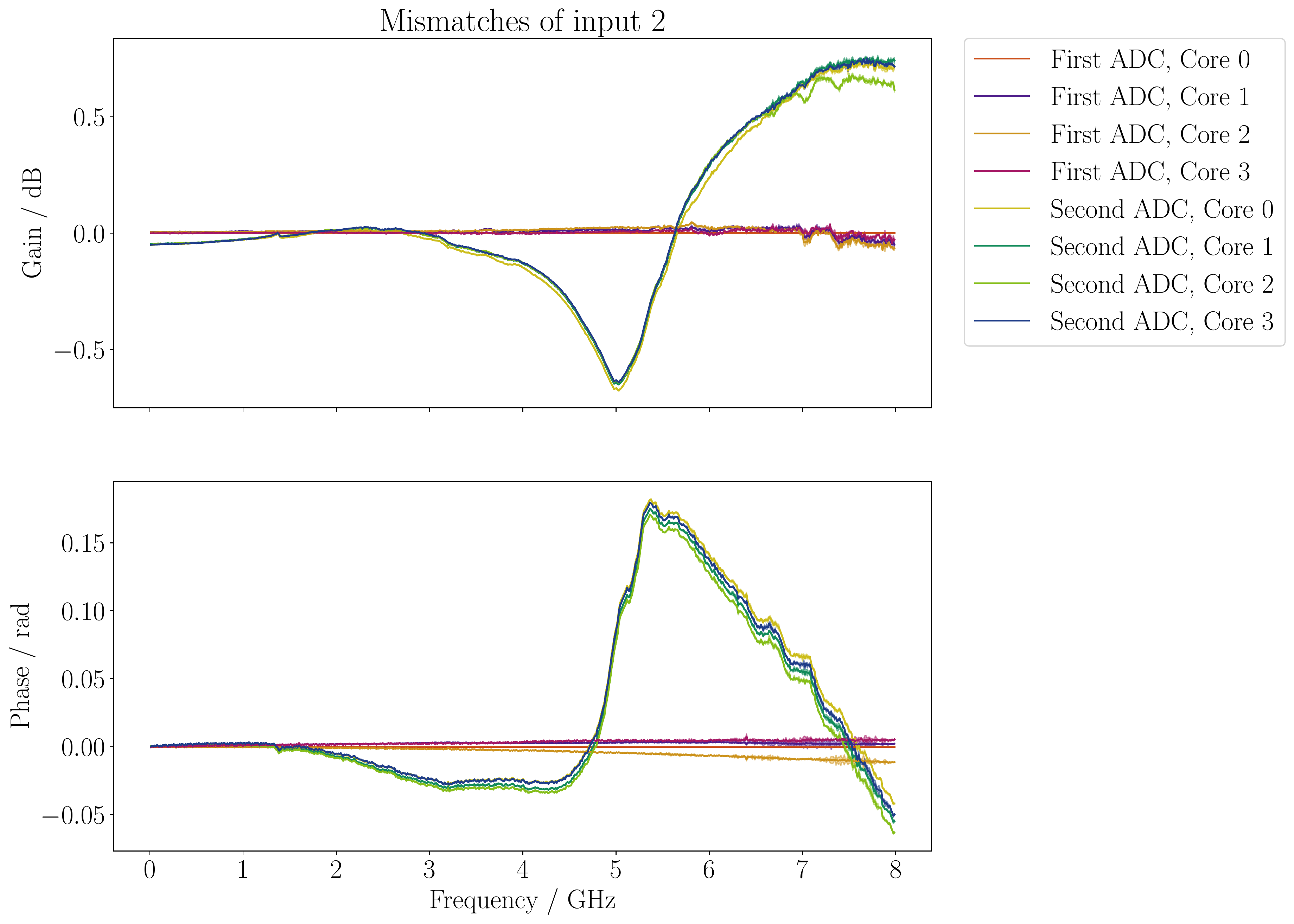}
	\caption{The mismatches of input \texttt{I} of the second \texttt{dFFTS-4G} board. The figure shows the gain given by $\left| \hat{h}'''_{m,n} \right|$ and phase given by $\atantwo\left(\imaginary\left(\hat{h}'''_{m,n}\right), \real\left(\hat{h}'''_{m,n}\right)\right)$ of the \glspl{ADCCore} with respect to \gls{ADCCore} $0$ of the first \gls{ADC}. The standard deviations of the measurements are shown as colored bands around the mean.}
	\label{fig:mismatches2}
\end{figure}
\begin{figure}[htbp]
	\centering
	\includegraphics[width=0.95\textwidth]{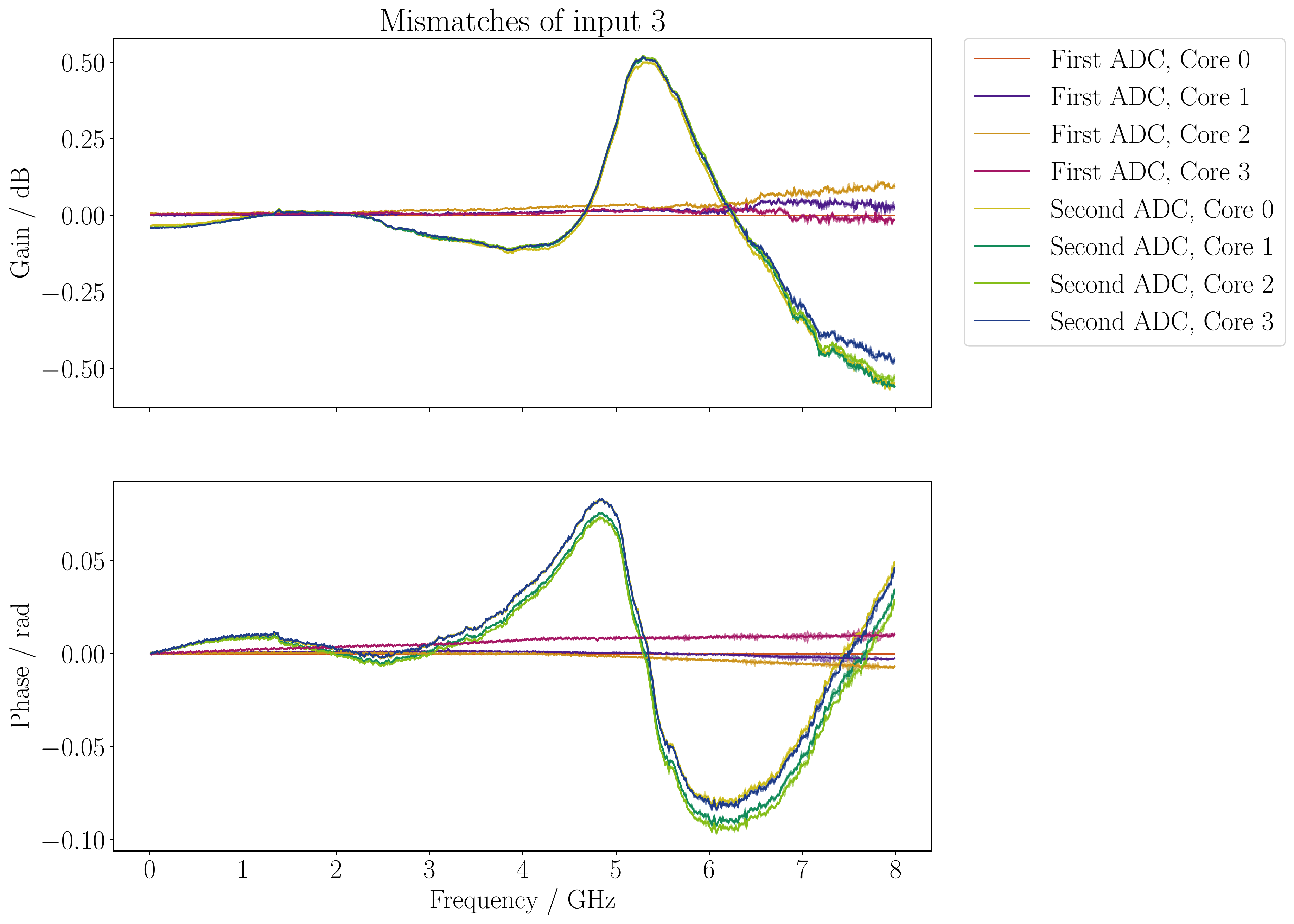}
	\caption{The mismatches of input \texttt{Q} of the second \texttt{dFFTS-4G} board. The figure shows the gain given by $\left| \hat{h}'''_{m,n} \right|$ and phase given by $\atantwo\left(\imaginary\left(\hat{h}'''_{m,n}\right), \real\left(\hat{h}'''_{m,n}\right)\right)$ of the \glspl{ADCCore} with respect to \gls{ADCCore} $0$ of the first \gls{ADC}. The standard deviations of the measurements are shown as colored bands around the mean.}
	\label{fig:mismatches3}
\end{figure}

\FloatBarrier
\section{Impact of calibration}
\begin{figure}[htbp]
	\centering
	\includegraphics[width=0.95\textwidth]{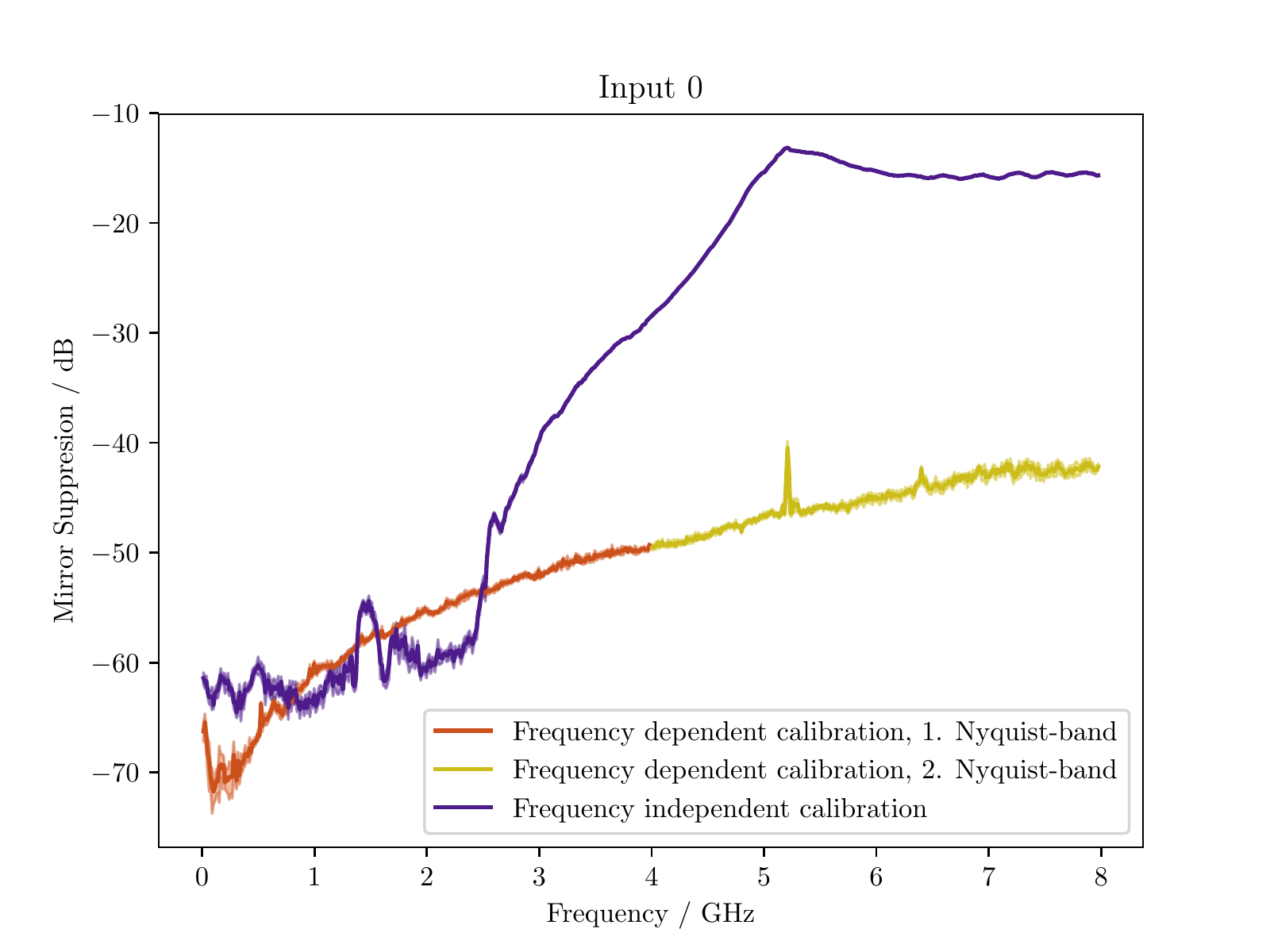}
	\caption{Comparison of \gls{MirrorSuppression} between calibration at one frequency and frequency dependent calibration for input \texttt{I} of the first \texttt{dFFTS-4G} board.}
	\label{fig:comparsion_adc0}
\end{figure}
\begin{figure}[htbp]
	\centering
	\includegraphics[width=0.95\textwidth]{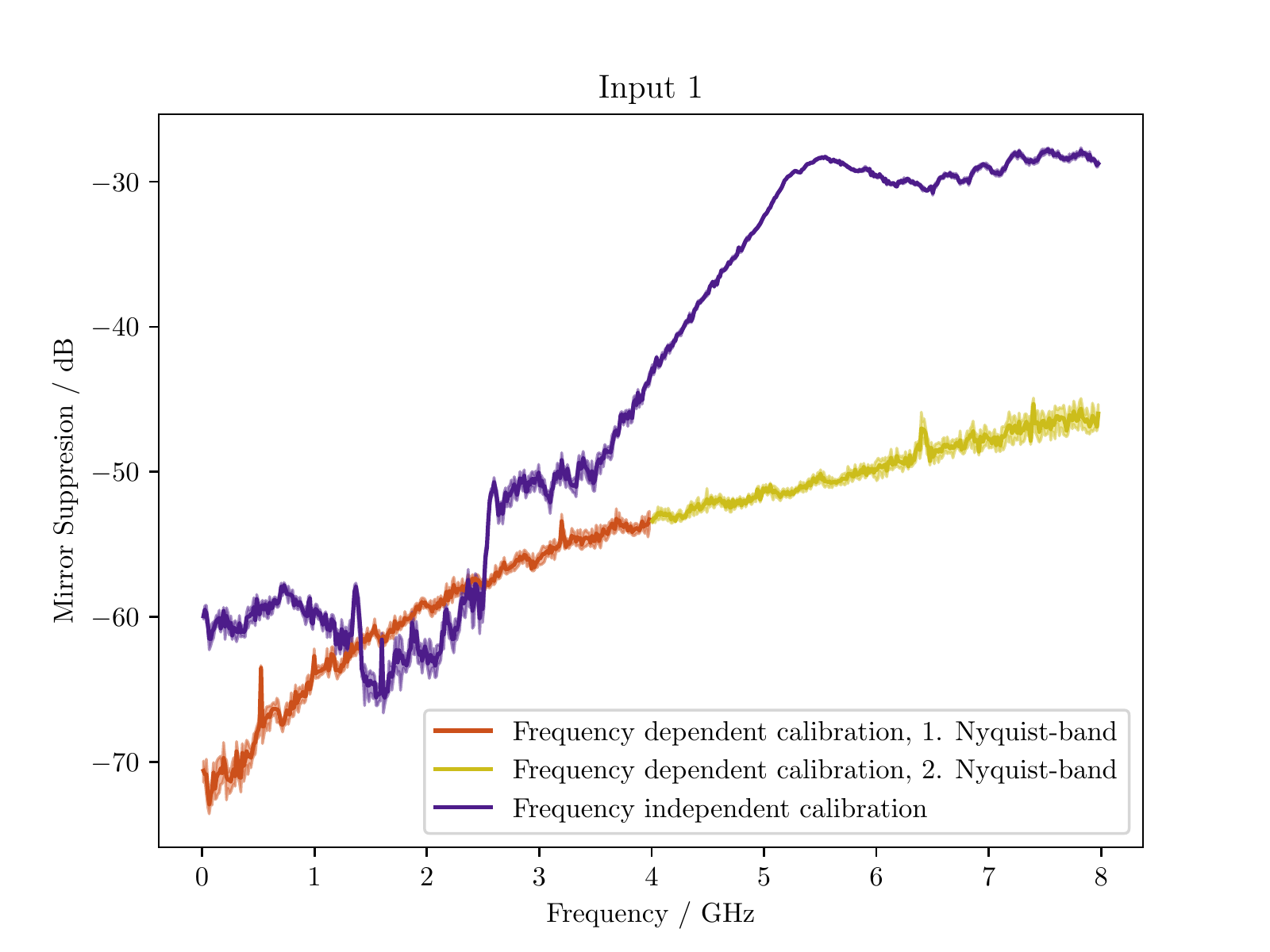}
	\caption{Comparison of \gls{MirrorSuppression} between calibration at one frequency and frequency dependent calibration for input \texttt{Q} of the first \texttt{dFFTS-4G} board.}
	\label{fig:comparsion_adc1}
\end{figure}
\begin{figure}[htbp]
	\centering
	\includegraphics[width=0.95\textwidth]{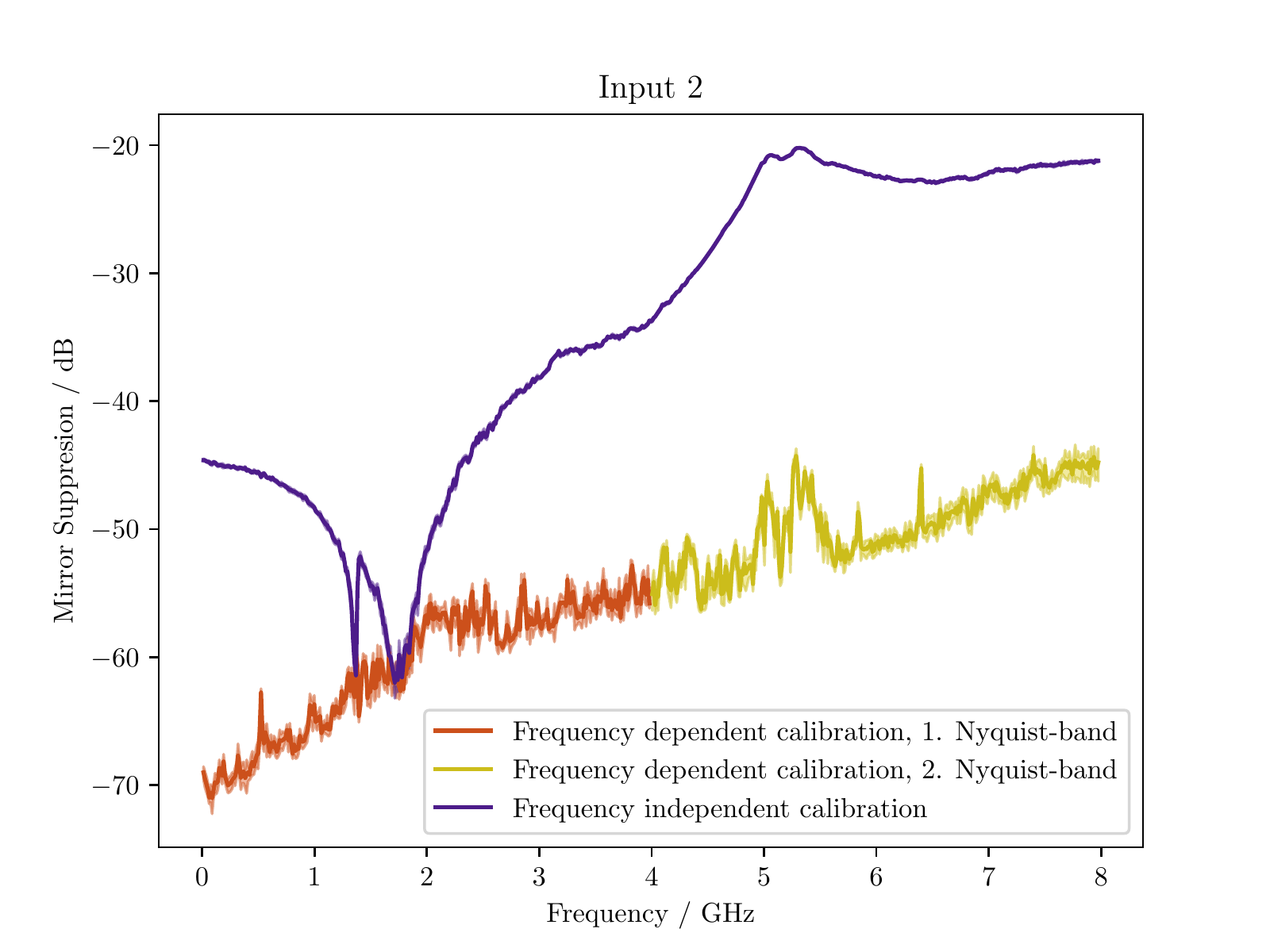}
	\caption{Comparison of \gls{MirrorSuppression} between calibration at one frequency and frequency dependent calibration for input \texttt{I} of the second \texttt{dFFTS-4G} board.}
	\label{fig:comparsion_adc2}
\end{figure}
\begin{figure}[htbp]
	\centering
	\includegraphics[width=0.95\textwidth]{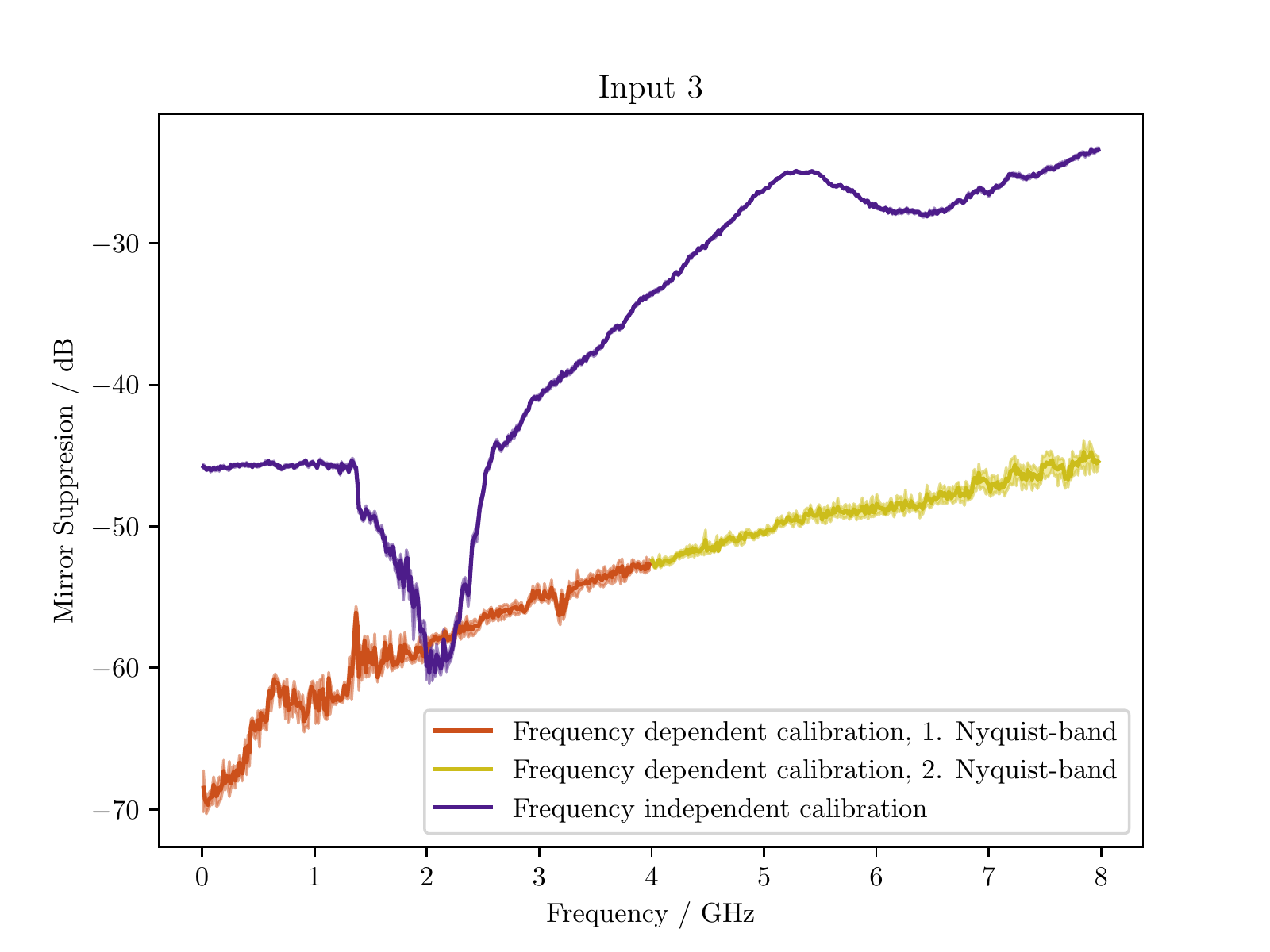}
	\caption{Comparison of \gls{MirrorSuppression} between calibration at one frequency and frequency dependent calibration for input \texttt{Q} of the second \texttt{dFFTS-4G} board.}
	\label{fig:comparsion_adc3}
\end{figure}

\FloatBarrier
\section{Time stability}
\begin{figure}[htbp]
	\centering
	\includegraphics[width=0.95\textwidth]{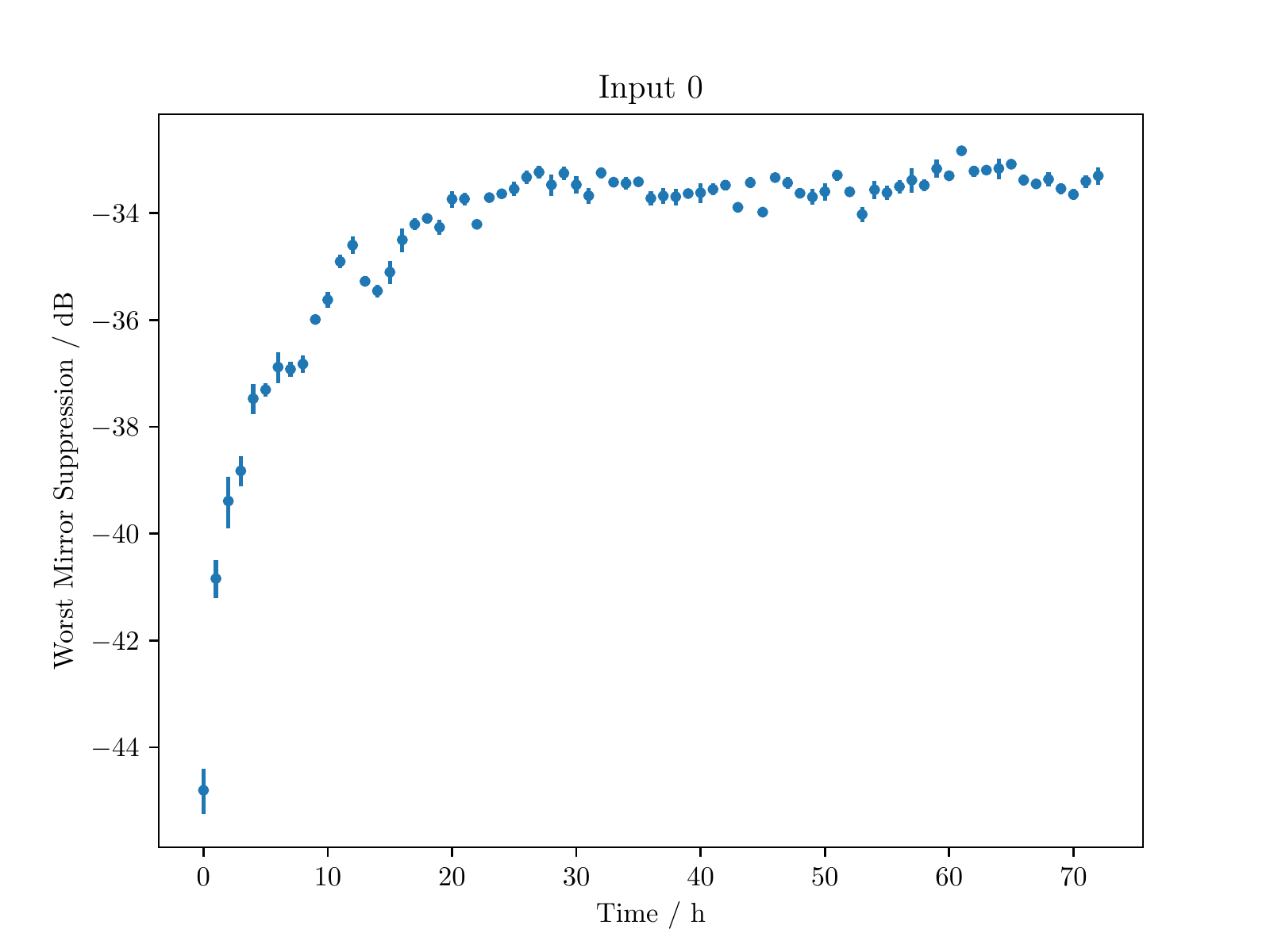}
	\caption{The time stability of the worst \gls{MirrorSuppression} with constant ambient temperature ($T = \SI{25}{\celsius}$) for input \texttt{I} of the first \texttt{dFFTS-4G} board. The uncertainties of the worst \gls{MirrorSuppression} are given by the error bars.}
	\label{fig:time_adc0_worst}
\end{figure}
\begin{figure}[htbp]
	\centering
	\includegraphics[width=0.95\textwidth]{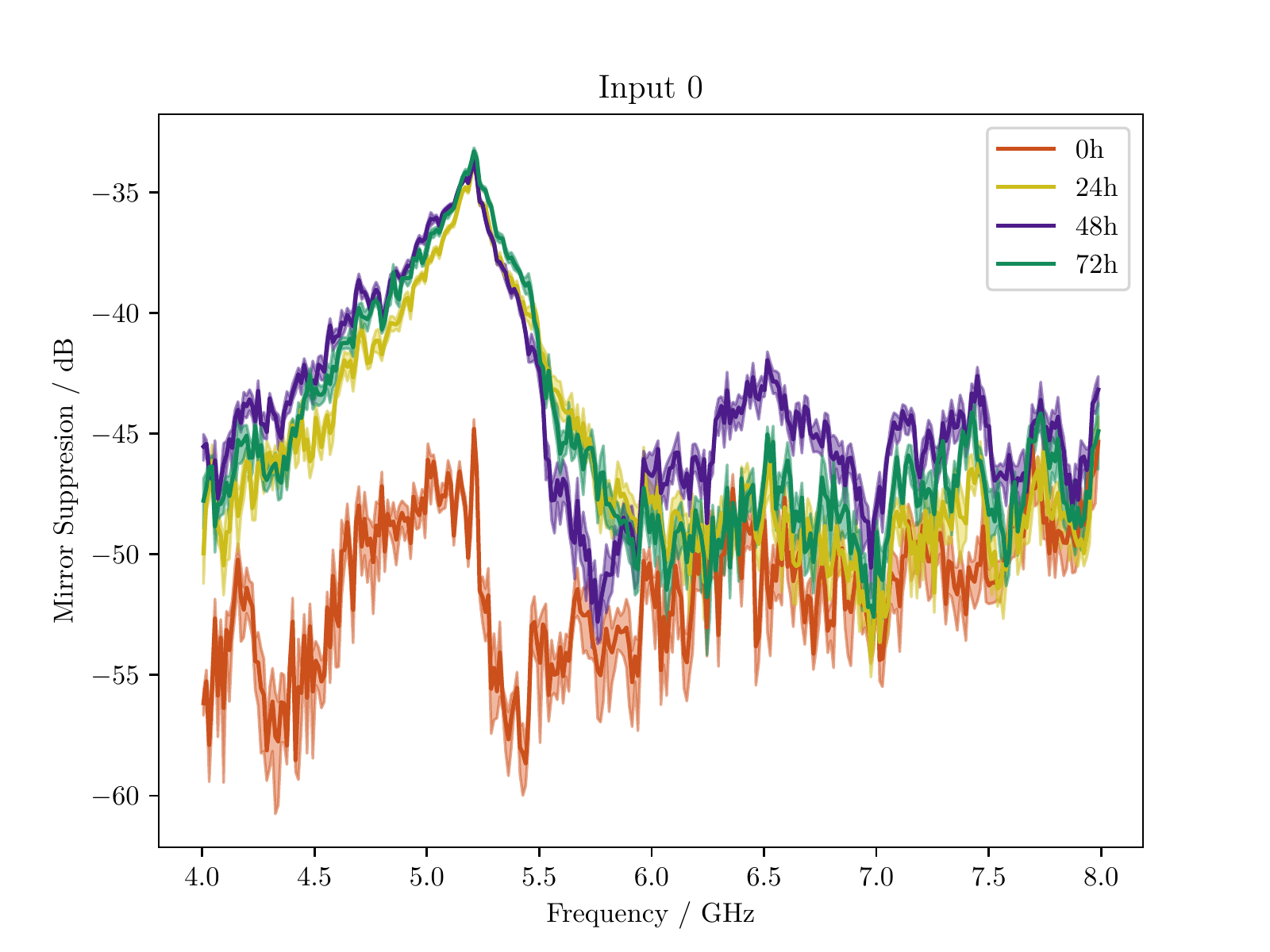}
	\caption{The time stability of the \gls{MirrorSuppression} with constant ambient temperature ($T = \SI{25}{\celsius}$) for input \texttt{I} of the first \texttt{dFFTS-4G} board.}
	\label{fig:time_adc0}
\end{figure}
\begin{figure}[htbp]
	\centering
	\includegraphics[width=0.95\textwidth]{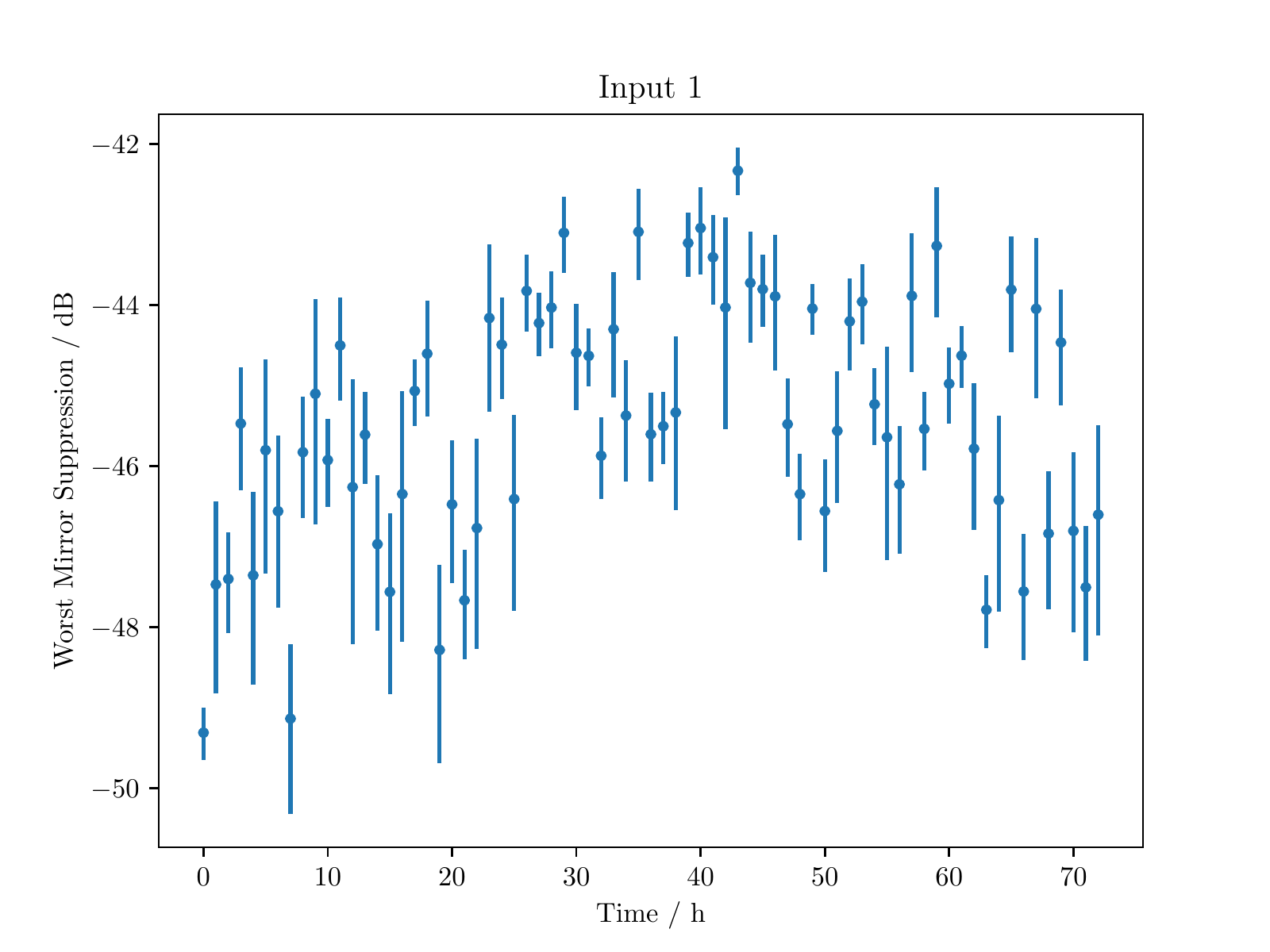}
	\caption{The time stability of the worst \gls{MirrorSuppression} with constant ambient temperature ($T = \SI{25}{\celsius}$) for input \texttt{Q} of the first \texttt{dFFTS-4G} board. The uncertainties of the worst \gls{MirrorSuppression} are given by the error bars.}
	\label{fig:time_adc1_worst}
\end{figure}
\begin{figure}[htbp]
	\centering
	\includegraphics[width=0.95\textwidth]{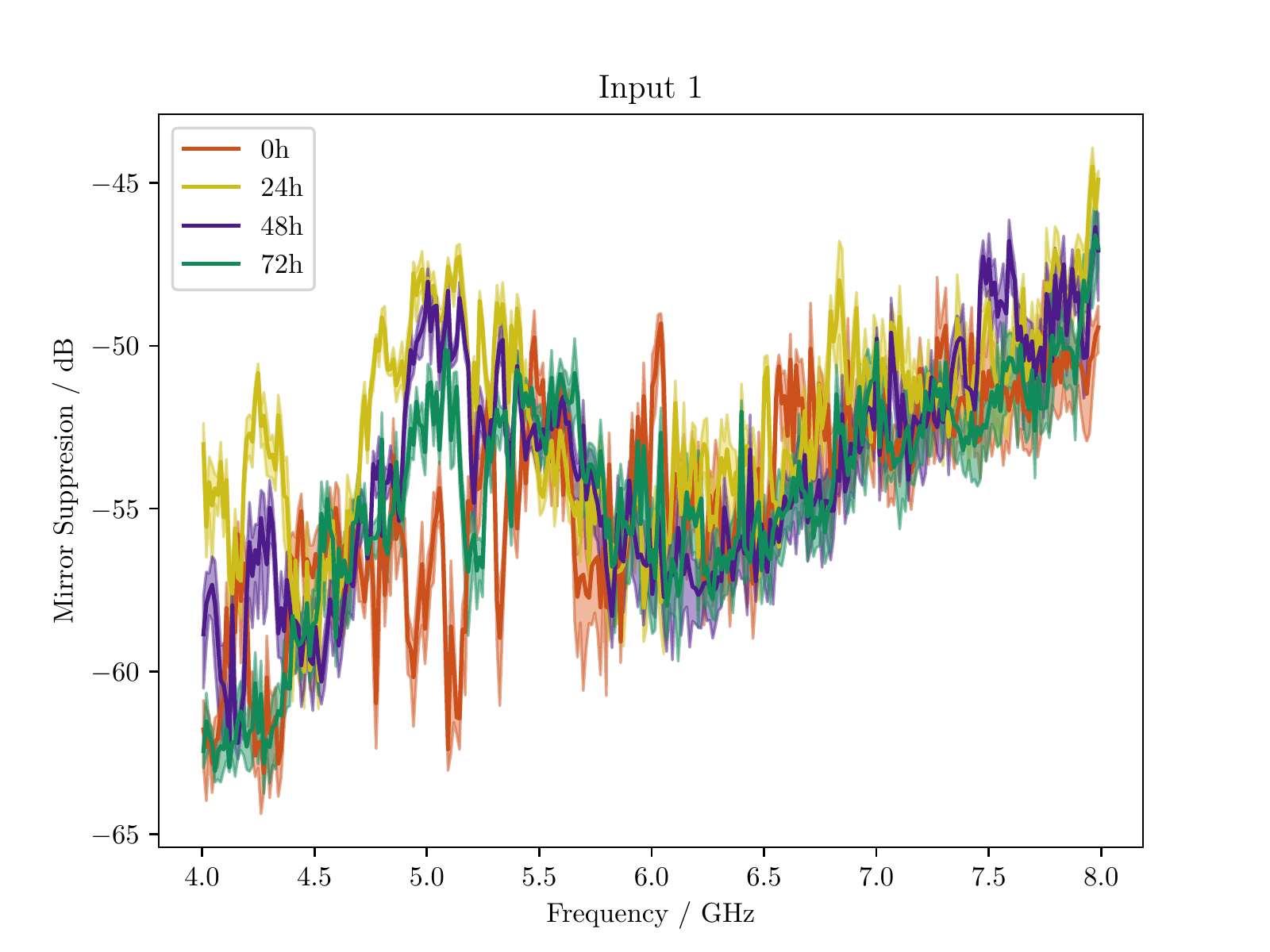}
	\caption{The time stability of the \gls{MirrorSuppression} with constant ambient temperature ($T = \SI{25}{\celsius}$) for input \texttt{Q} of the first \texttt{dFFTS-4G} board.}
	\label{fig:time_adc1}
\end{figure}
\begin{figure}[htbp]
	\centering
	\includegraphics[width=0.95\textwidth]{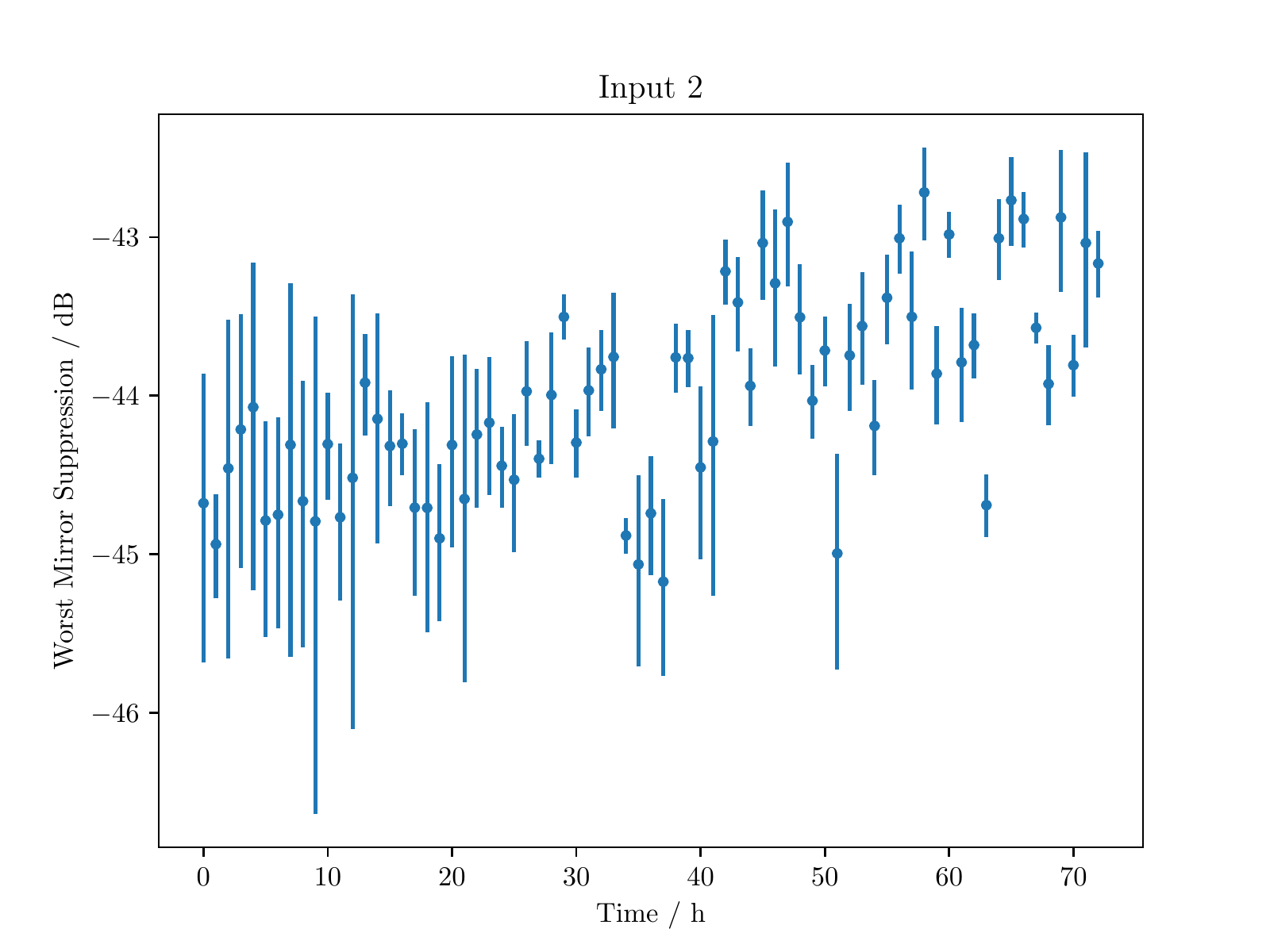}
	\caption{The time stability of the worst \gls{MirrorSuppression} with constant ambient temperature ($T = \SI{25}{\celsius}$) for input \texttt{I} of the second \texttt{dFFTS-4G} board. The uncertainties of the worst \gls{MirrorSuppression} are given by the error bars.}
	\label{fig:time_adc2_worst}
\end{figure}
\begin{figure}[htbp]
	\centering
	\includegraphics[width=0.95\textwidth]{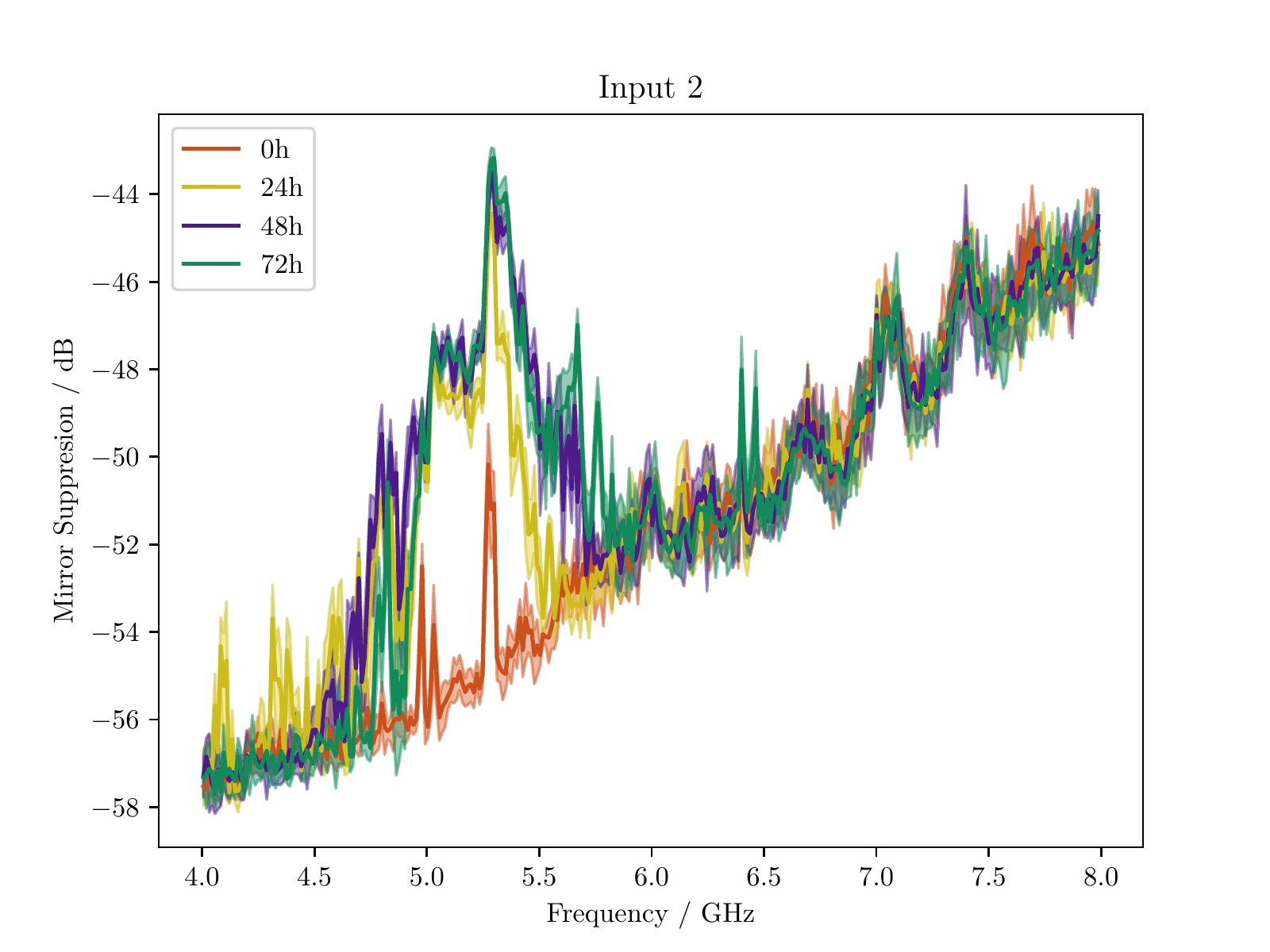}
	\caption{The time stability of the \gls{MirrorSuppression} with constant ambient temperature ($T = \SI{25}{\celsius}$) for input \texttt{I} of the second \texttt{dFFTS-4G} board.}
	\label{fig:time_adc2}
\end{figure}
\begin{figure}[htbp]
	\centering
	\includegraphics[width=0.95\textwidth]{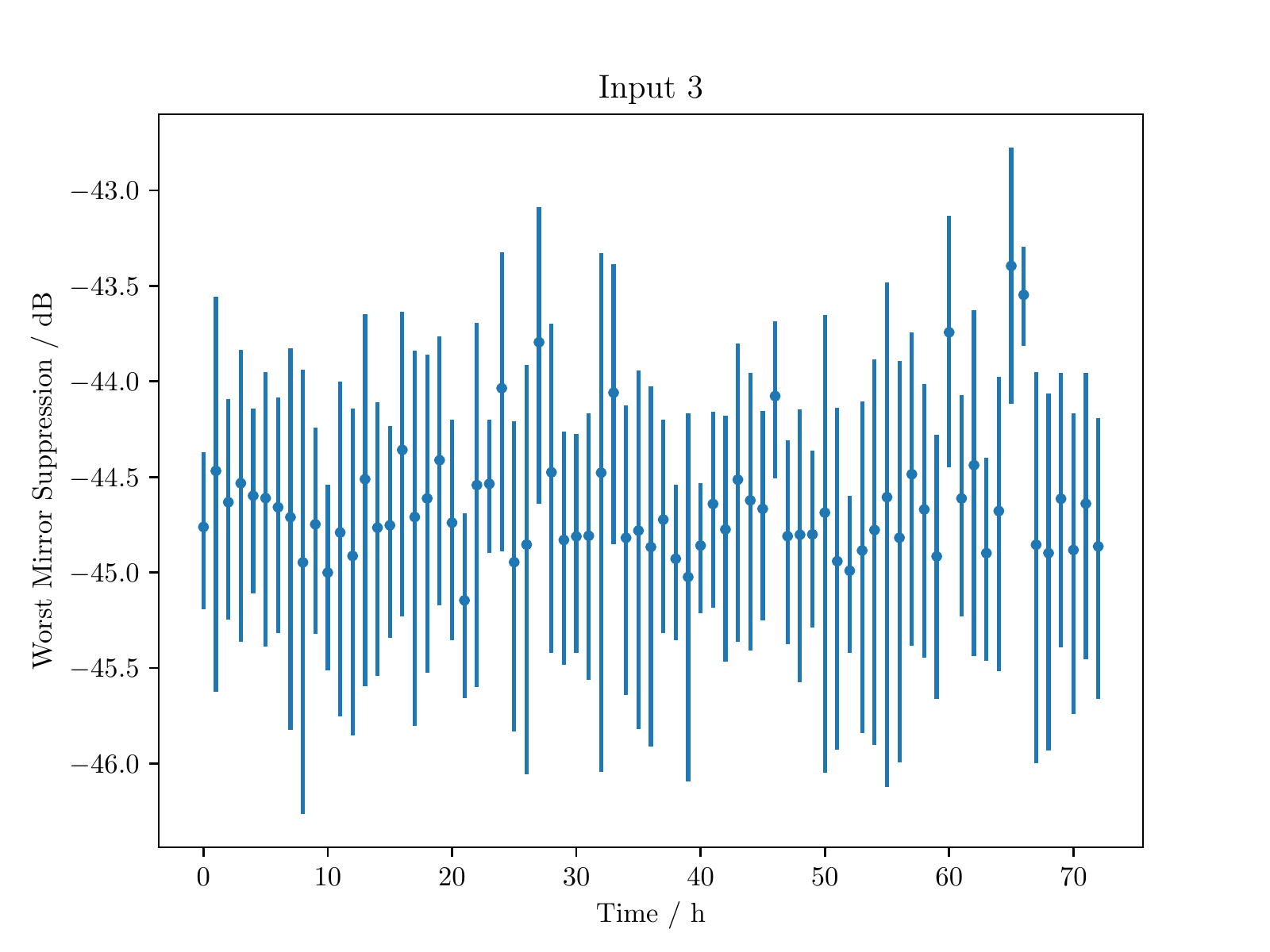}
	\caption{The time stability of the worst \gls{MirrorSuppression} with constant ambient temperature ($T = \SI{25}{\celsius}$) for input \texttt{Q} of the second \texttt{dFFTS-4G} board. The uncertainties of the worst \gls{MirrorSuppression} are given by the error bars.}
	\label{fig:time_adc3_worst}
\end{figure}
\begin{figure}[htbp]
	\centering
	\includegraphics[width=0.95\textwidth]{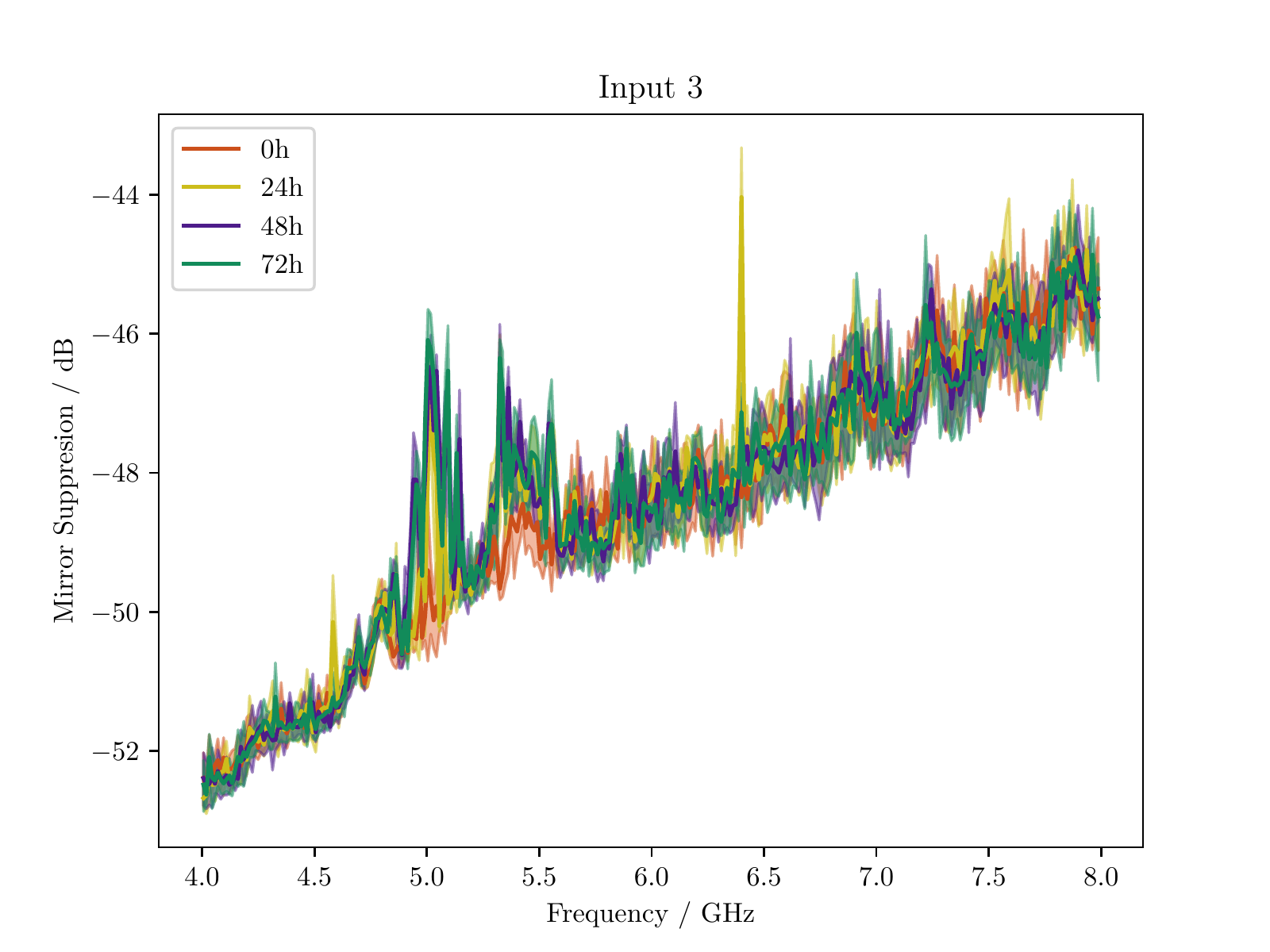}
	\caption{The time stability of the \gls{MirrorSuppression} with constant ambient temperature ($T = \SI{25}{\celsius}$) for input \texttt{Q} of the second \texttt{dFFTS-4G} board.}
	\label{fig:time_adc3}
\end{figure}

\FloatBarrier
\section{Temperature stability}
\begin{figure}[htbp]
	\centering
	\includegraphics[width=0.95\textwidth]{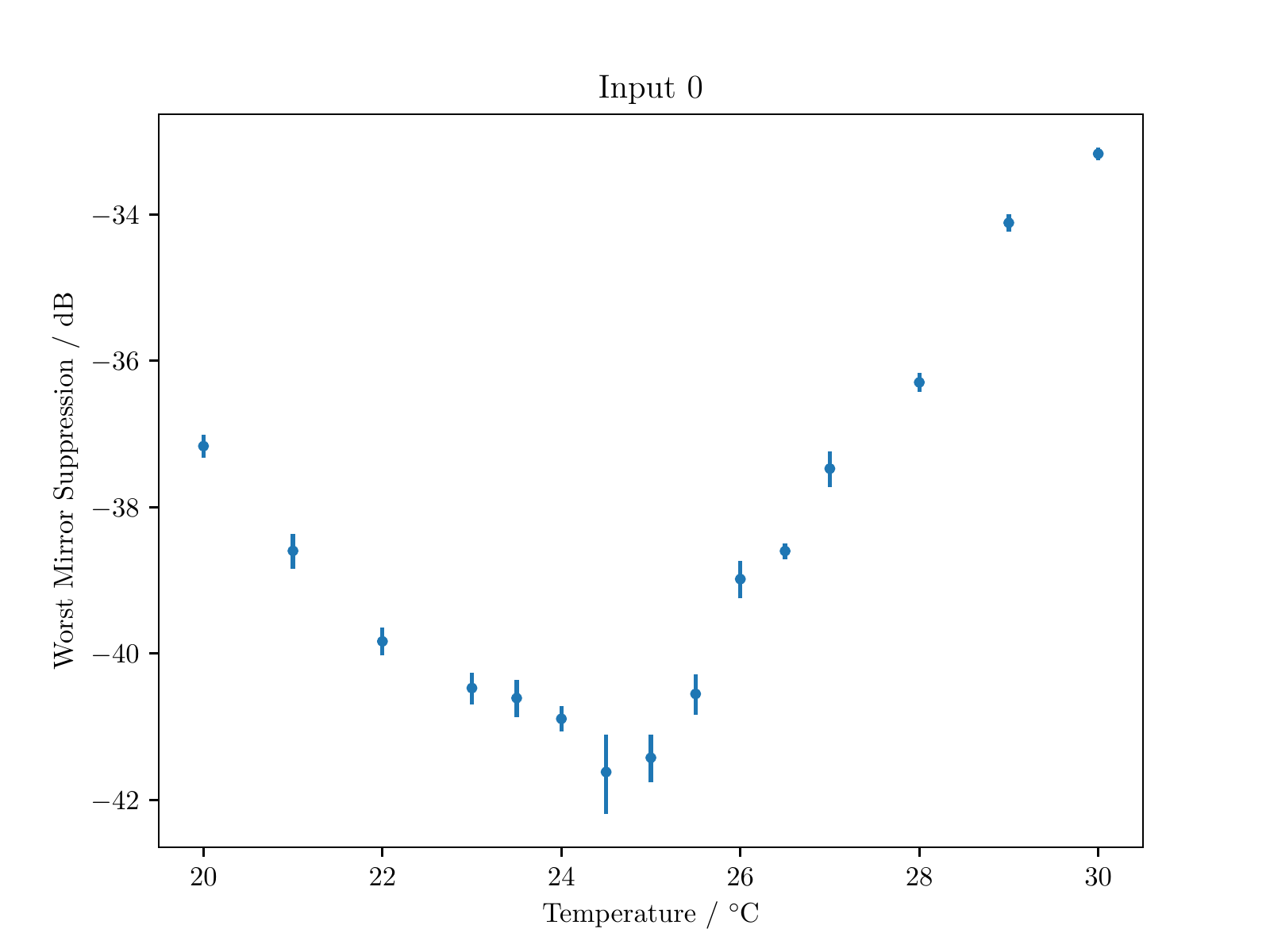}
	\caption{The temperature stability of the worst \gls{MirrorSuppression} in the second Nyquist-band for input \texttt{I} of the first \texttt{dFFTS-4G} board. The calibration is performed at \SI{25}{\celsius}. The uncertainties of the worst \gls{MirrorSuppression} are given by the error bars.}
	\label{fig:temperature_adc0_worst}
\end{figure}
\begin{figure}[htbp]
	\centering
	\includegraphics[width=0.95\textwidth]{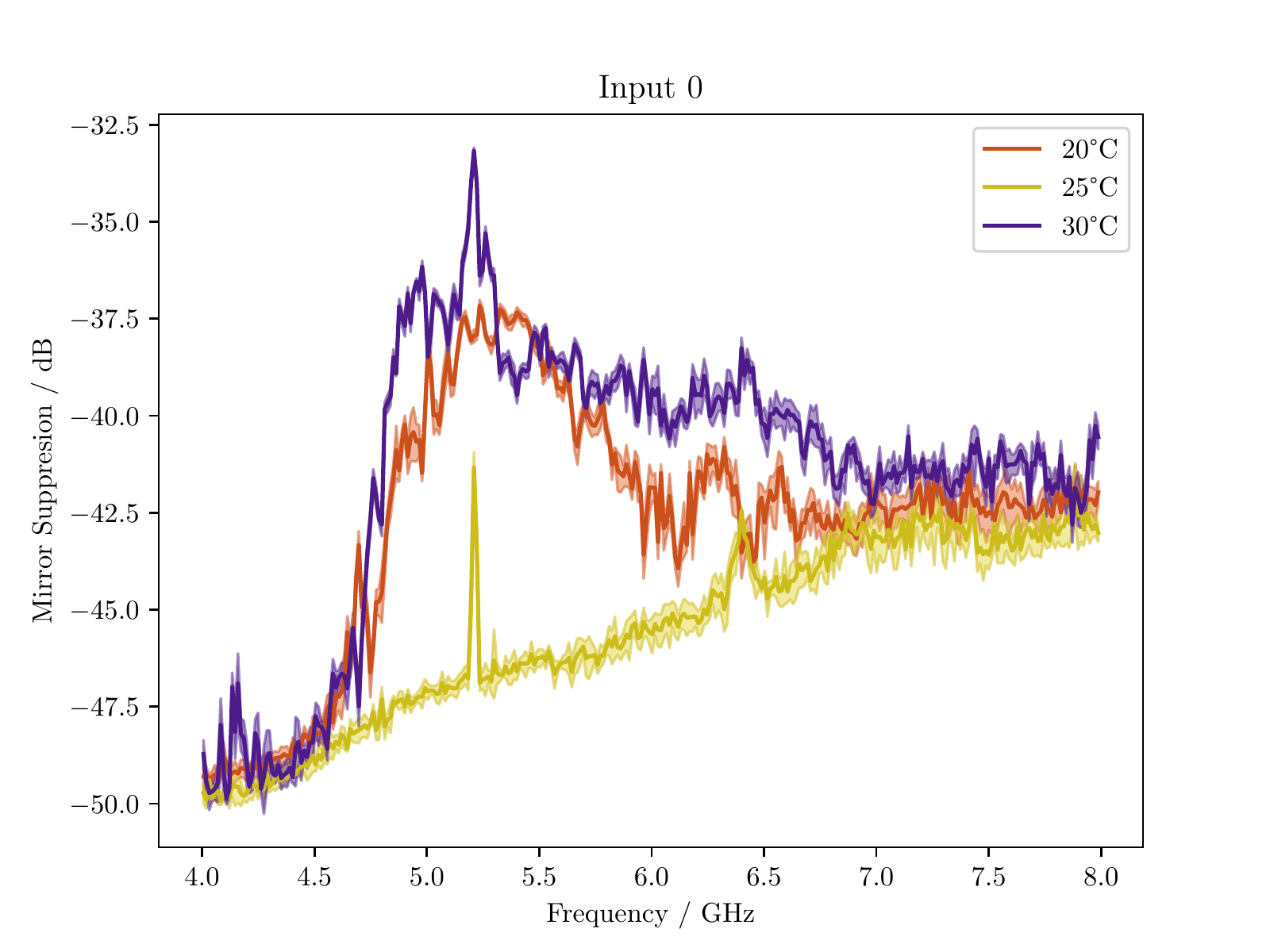}
	\caption{The temperature stability of the \gls{MirrorSuppression} in the second Nyquist-band for input \texttt{I} of the first \texttt{dFFTS-4G} board. The calibration is performed at \SI{25}{\celsius}. To qualify drifts due to time, a measurement at \SI{25}{\celsius}  is taken at the beginning and end of the measurement series.}
	\label{fig:temperature_adc0}
\end{figure}
\begin{figure}[htbp]
	\centering
	\includegraphics[width=0.95\textwidth]{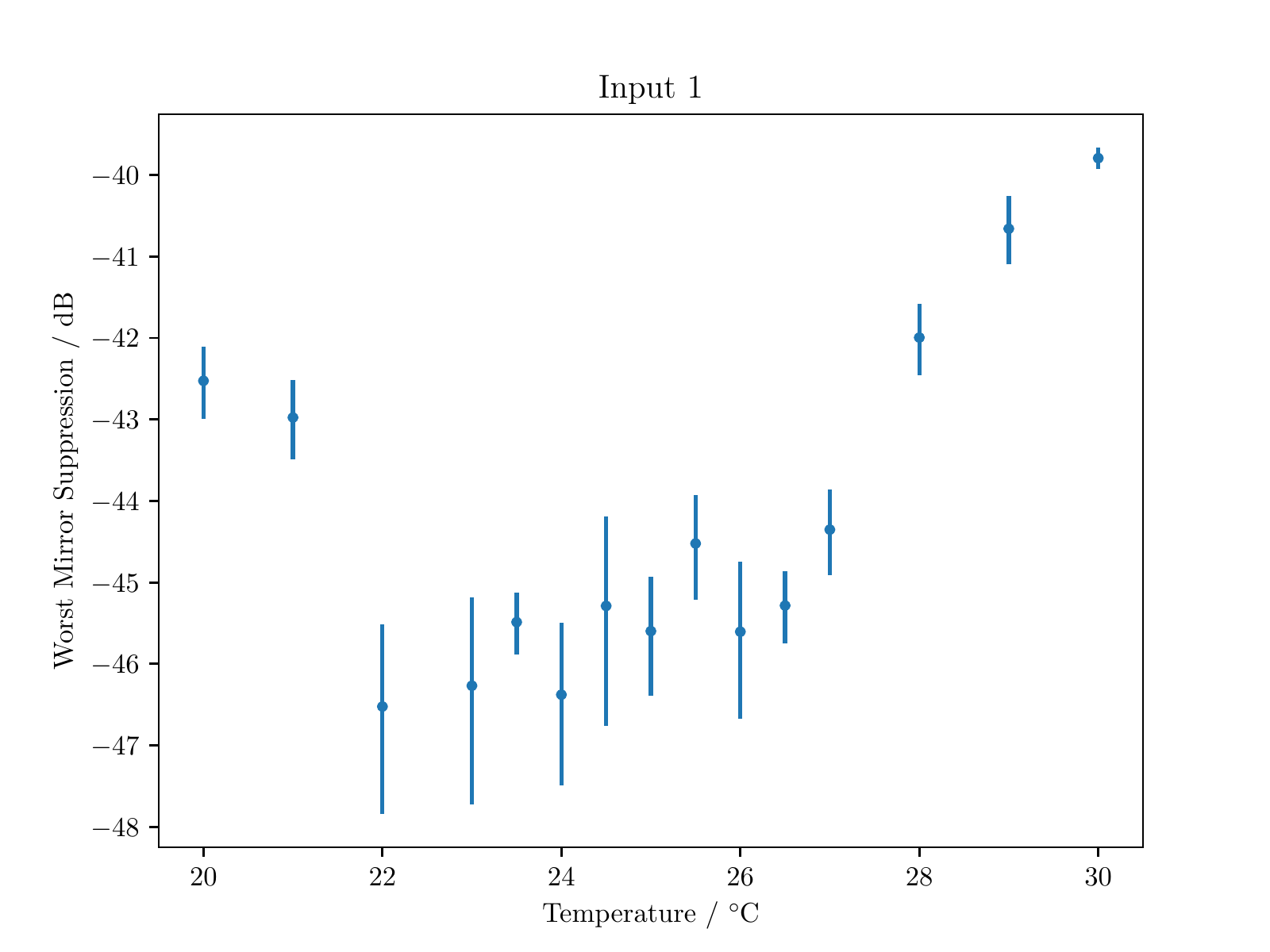}
	\caption{The temperature stability of the worst \gls{MirrorSuppression} in the second Nyquist-band for input \texttt{Q} of the first \texttt{dFFTS-4G} board. The calibration is performed at \SI{25}{\celsius}. The uncertainties of the worst \gls{MirrorSuppression} are given by the error bars.}
	\label{fig:temperature_adc1_worst}
\end{figure}
\begin{figure}[htbp]
	\centering
	\includegraphics[width=0.95\textwidth]{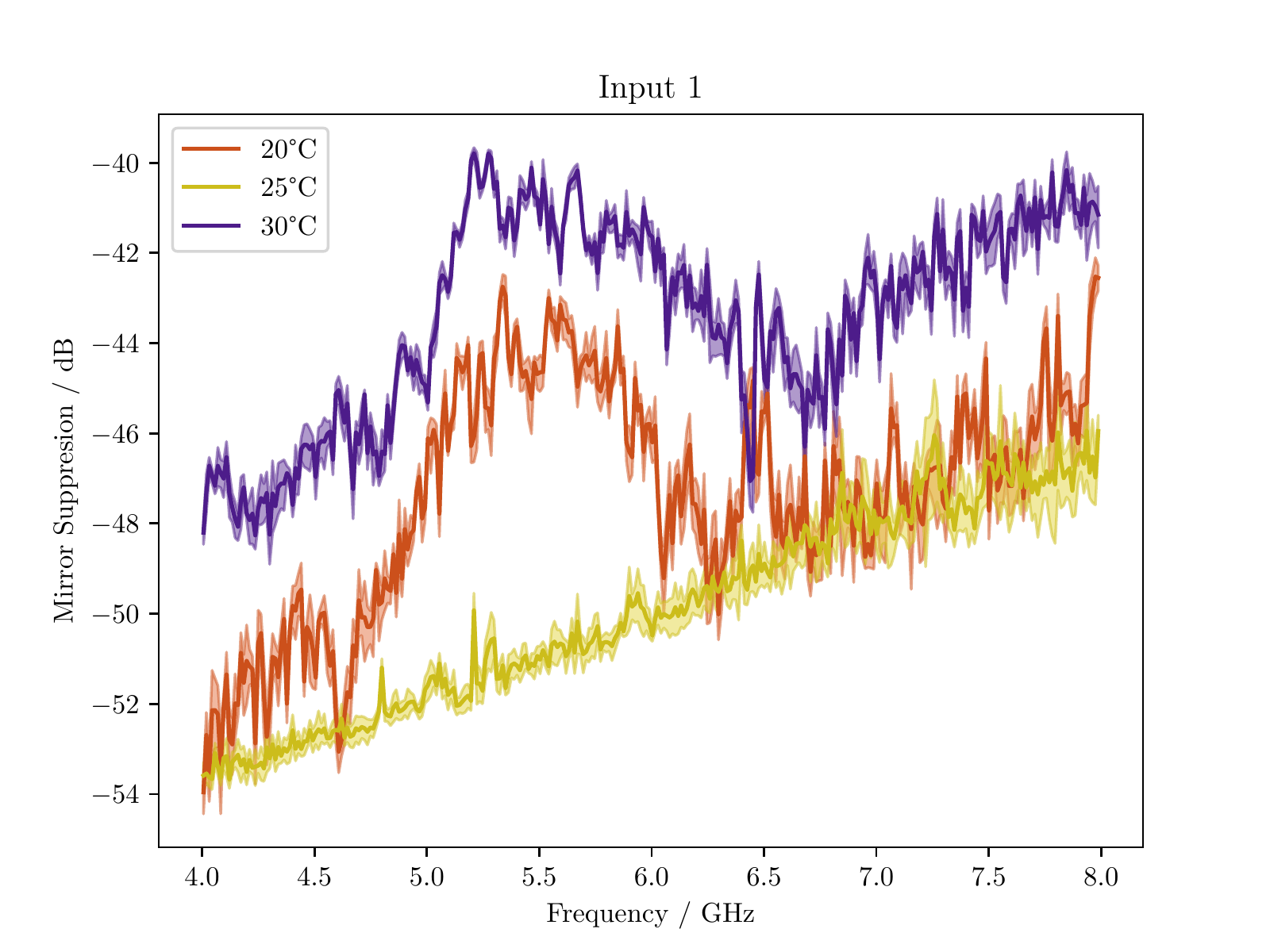}
	\caption{The temperature stability of the \gls{MirrorSuppression} in the second Nyquist-band for input \texttt{Q} of the first \texttt{dFFTS-4G} board. The calibration is performed at \SI{25}{\celsius}. To qualify drifts due to time, a measurement at \SI{25}{\celsius} is taken at the beginning and end of the measurement series.}
	\label{fig:temperature_adc1}
\end{figure}
\begin{figure}[htbp]
	\centering
	\includegraphics[width=0.95\textwidth]{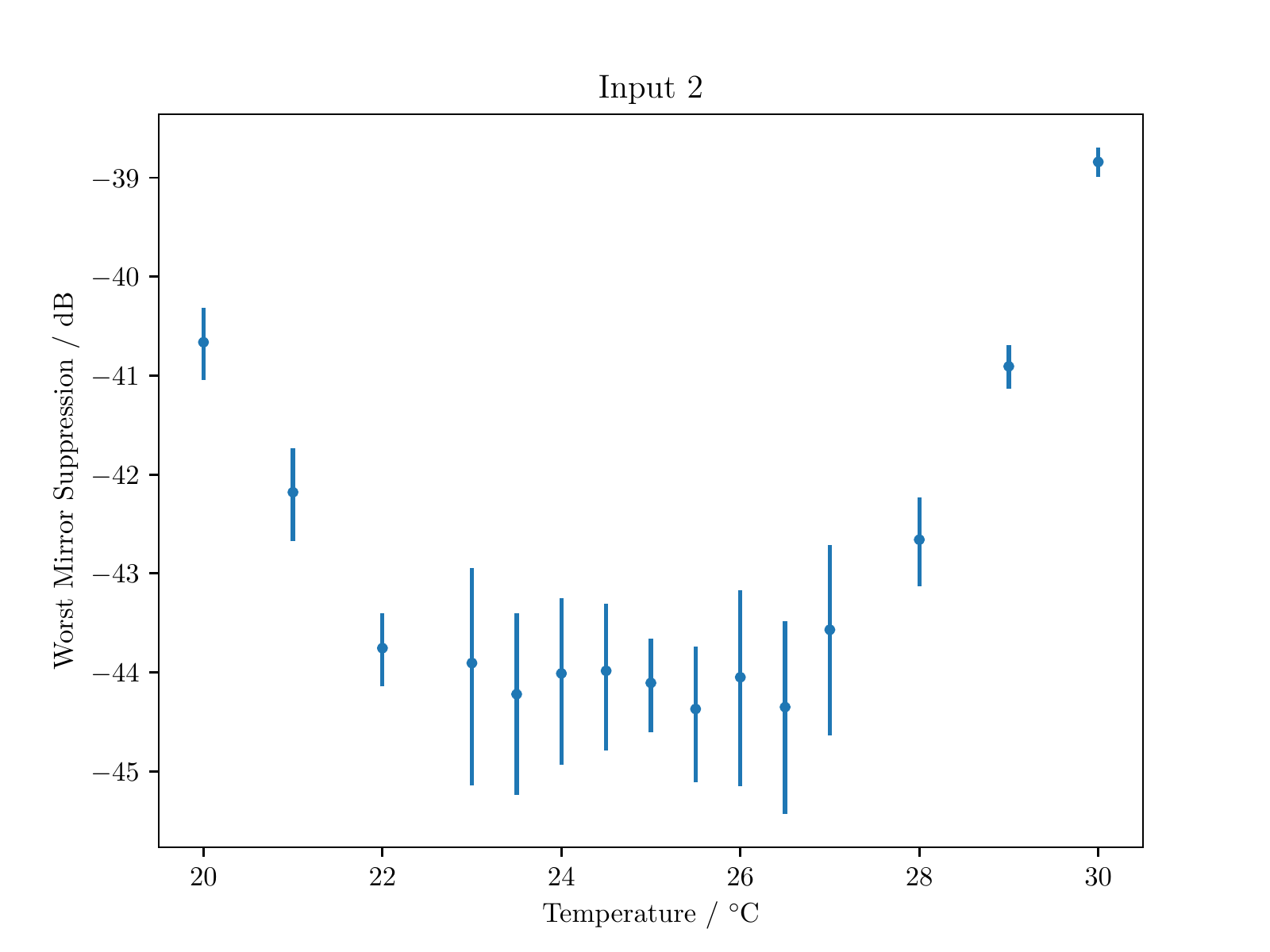}
	\caption{The temperature stability of the worst \gls{MirrorSuppression} in the second Nyquist-band for input \texttt{I} of the second \texttt{dFFTS-4G} board. The calibration is performed at \SI{25}{\celsius}. The uncertainties of the worst \gls{MirrorSuppression} are given by the error bars.}
	\label{fig:temperature_adc2_worst}
\end{figure}
\begin{figure}[htbp]
	\centering
	\includegraphics[width=0.95\textwidth]{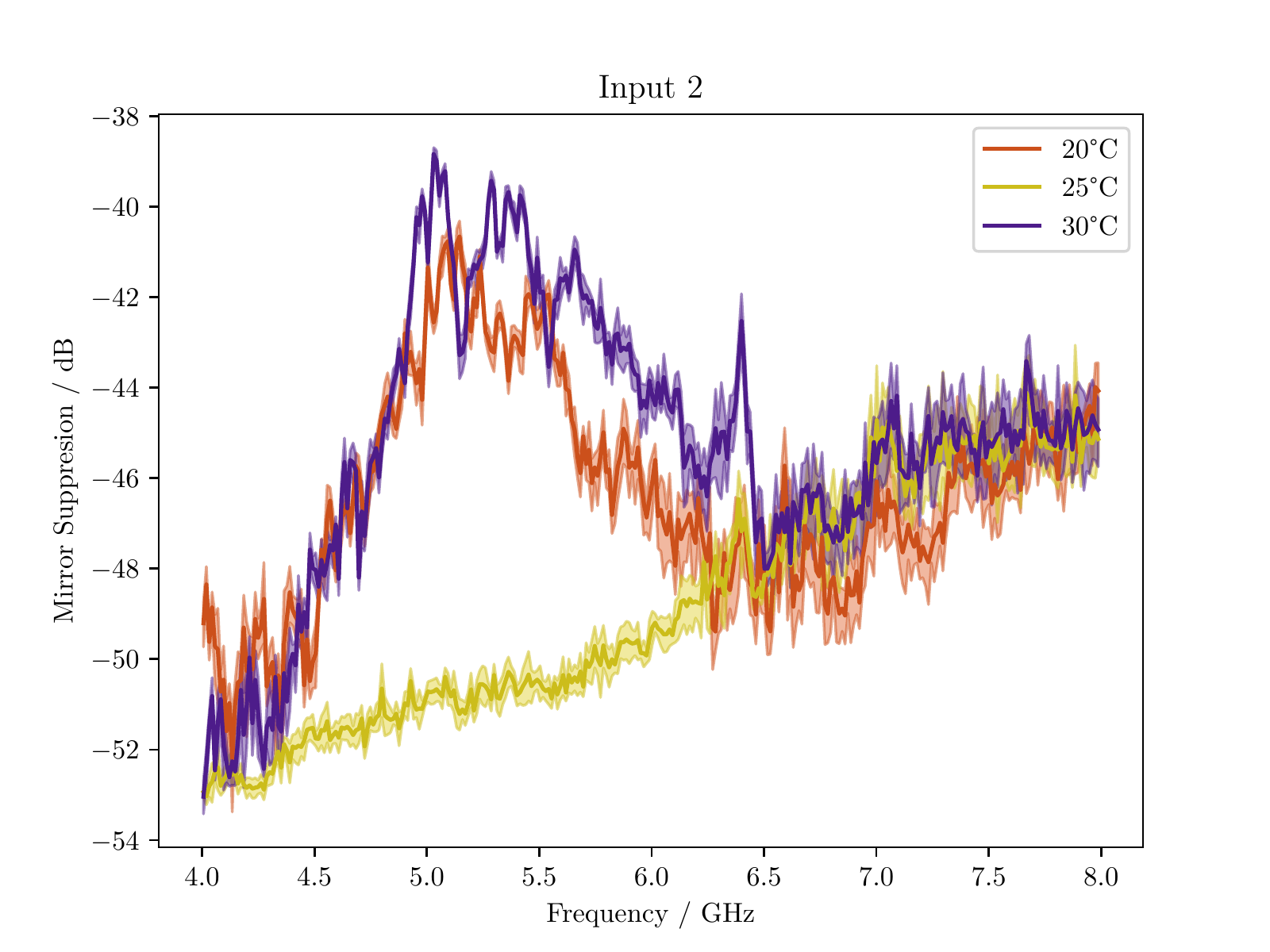}
	\caption{The temperature stability of the \gls{MirrorSuppression} in the second Nyquist-band for input \texttt{I} of the second \texttt{dFFTS-4G} board. The calibration is performed at \SI{25}{\celsius}. To qualify drifts due to time, a measurement at \SI{25}{\celsius} is taken at the beginning and end of the measurement series.}
	\label{fig:temperature_adc2}
\end{figure}
\begin{figure}[htbp]
	\centering
	\includegraphics[width=0.95\textwidth]{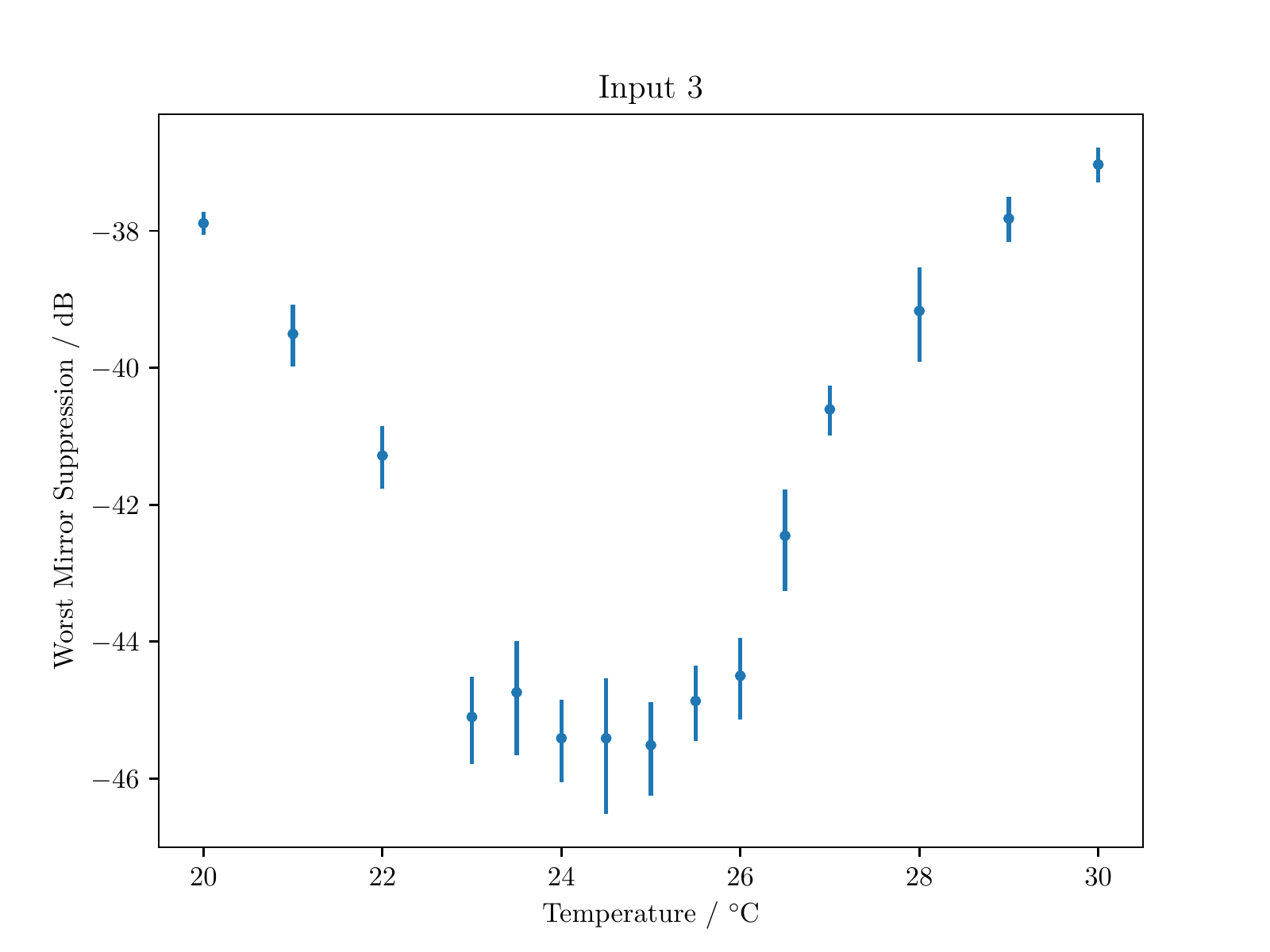}
	\caption{The temperature stability of the worst \gls{MirrorSuppression} in the second Nyquist-band for input \texttt{Q} of the second \texttt{dFFTS-4G} board. The calibration is performed at \SI{25}{\celsius}. The uncertainties of the worst \gls{MirrorSuppression} are given by the error bars.}
	\label{fig:temperature_adc3_worst}
\end{figure}
\begin{figure}[htbp]
	\centering
	\includegraphics[width=0.95\textwidth]{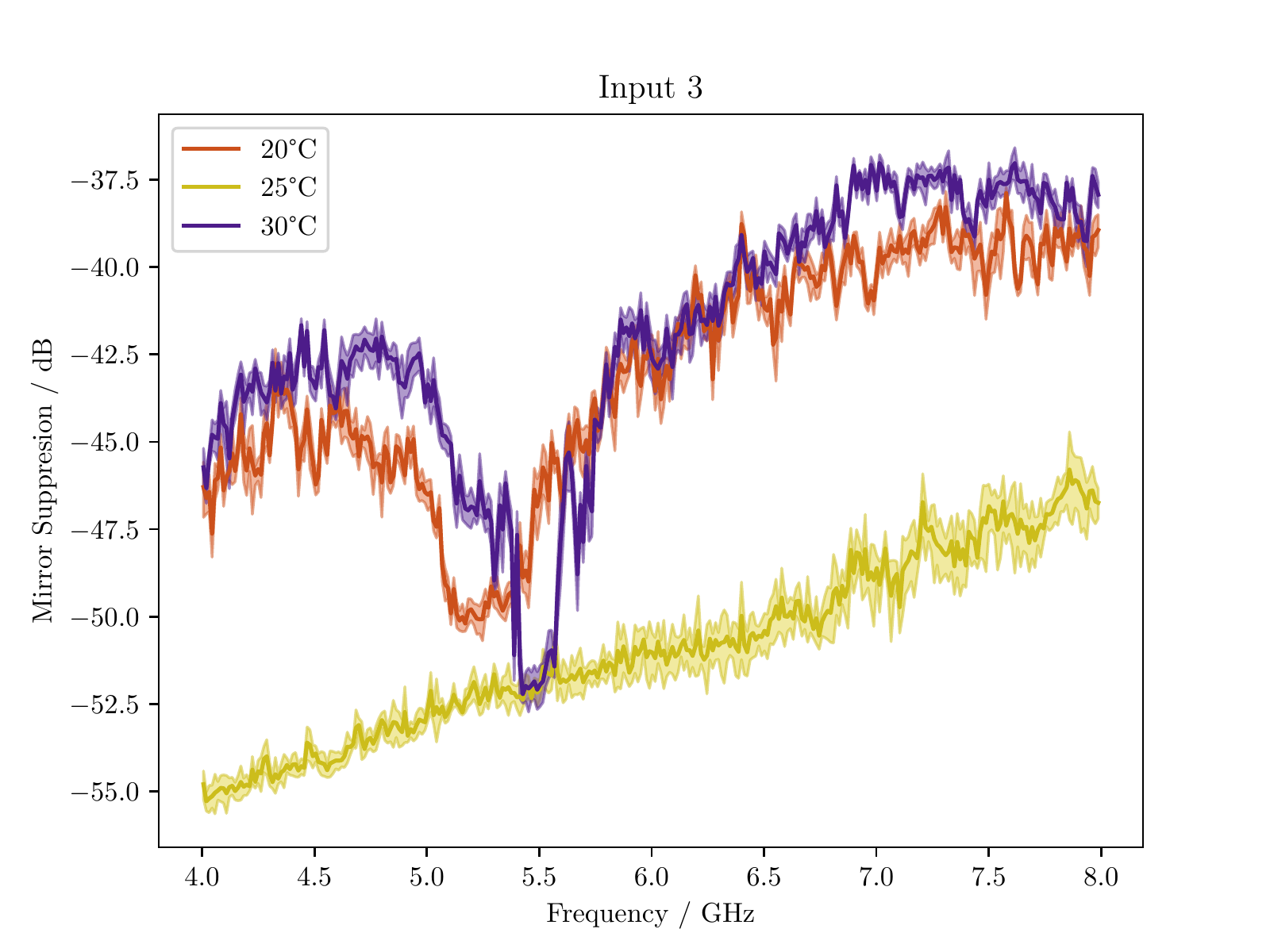}
	\caption{The temperature stability of the \gls{MirrorSuppression} in the second Nyquist-band for input \texttt{Q} of the second \texttt{dFFTS-4G} board. The calibration is performed at \SI{25}{\celsius}. To qualify drifts due to time, a measurement at \SI{25}{\celsius} is taken at the beginning and end of the measurement series.}
	\label{fig:temperature_adc3}
\end{figure}

\FloatBarrier
\section{Stability over power cycles}
\begin{figure}[htbp]
	\centering
	\includegraphics[width=0.95\textwidth]{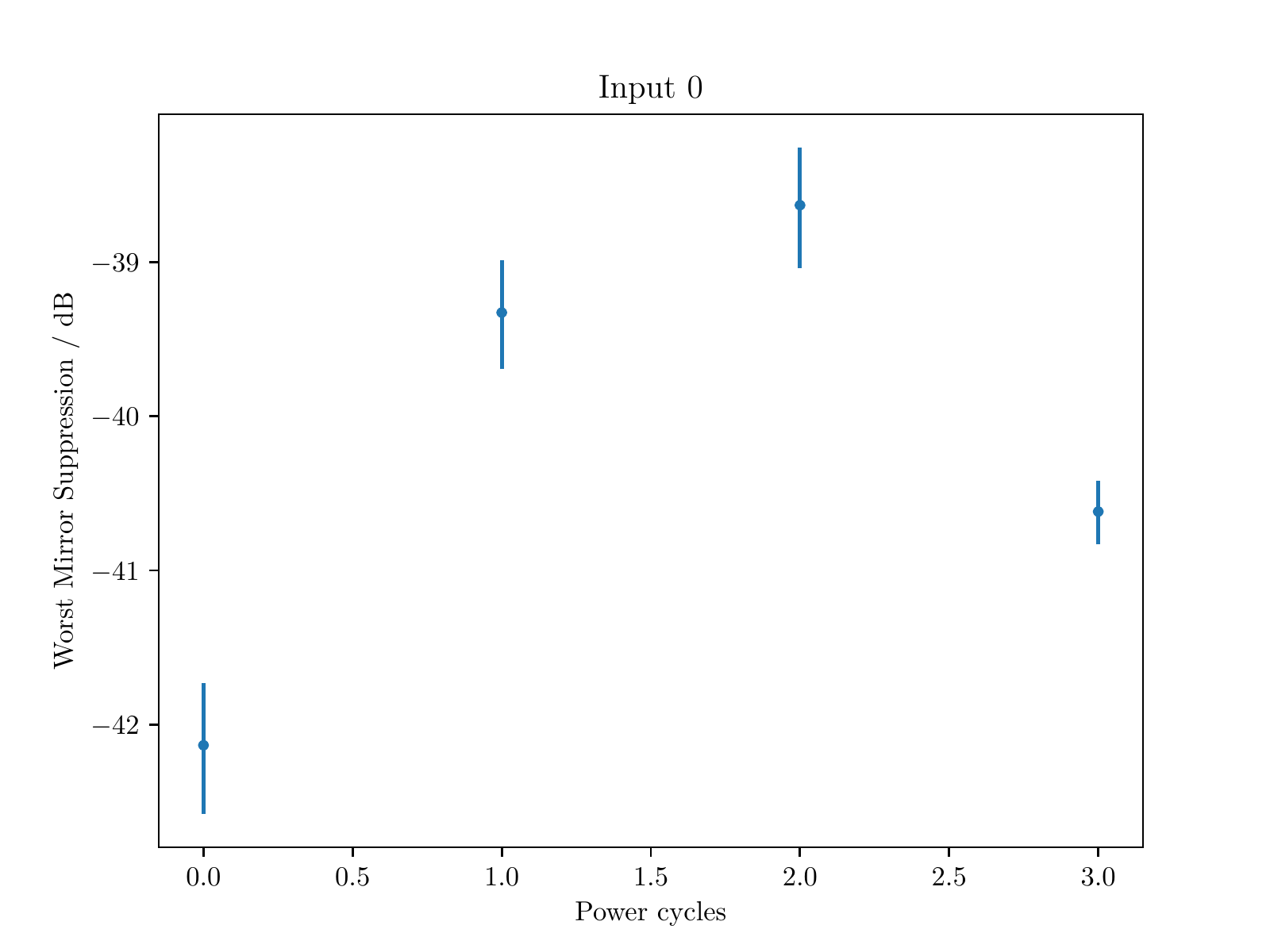}
	\caption{The stability of the worst \gls{MirrorSuppression} over power cycles in the second Nyquist-band for input \texttt{I} of the first \texttt{dFFTS-4G} board. The calibration and measurements are performed at an ambient temperature of \SI{25}{\celsius}. The uncertainties of the worst \gls{MirrorSuppression} are given by the error bars.}
	\label{fig:power_cycles_worst0}
\end{figure}
\begin{figure}[htbp]
	\centering
	\includegraphics[width=0.95\textwidth]{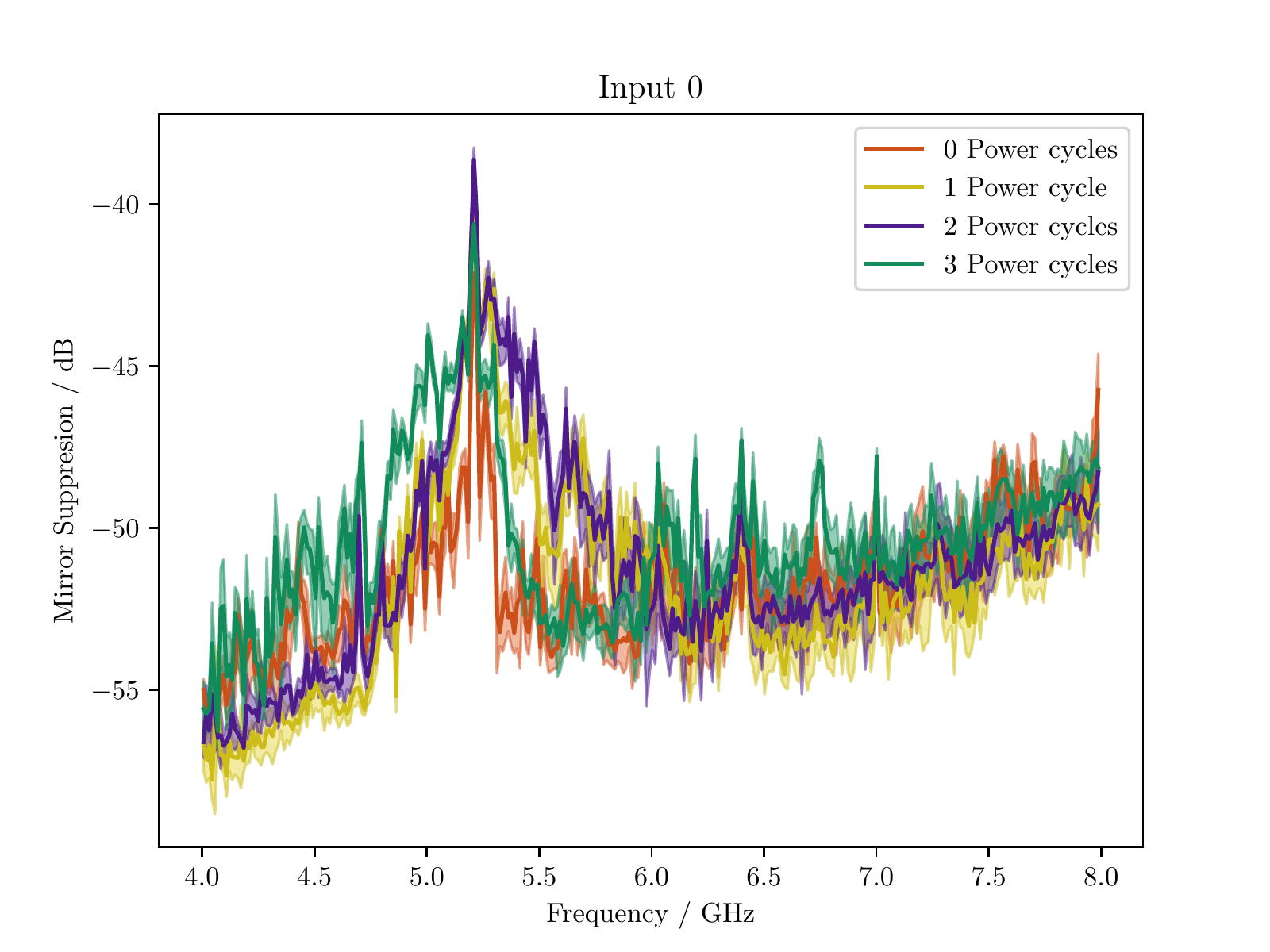}
	\caption{The stability of the \gls{MirrorSuppression} over power cycles in the second Nyquist-band for input \texttt{I} of the first \texttt{dFFTS-4G} board. The calibration and measurements are performed at an ambient temperature of \SI{25}{\celsius}.}
	\label{fig:power_cycles_adc0}
\end{figure}
\begin{figure}[htbp]
	\centering
	\includegraphics[width=0.95\textwidth]{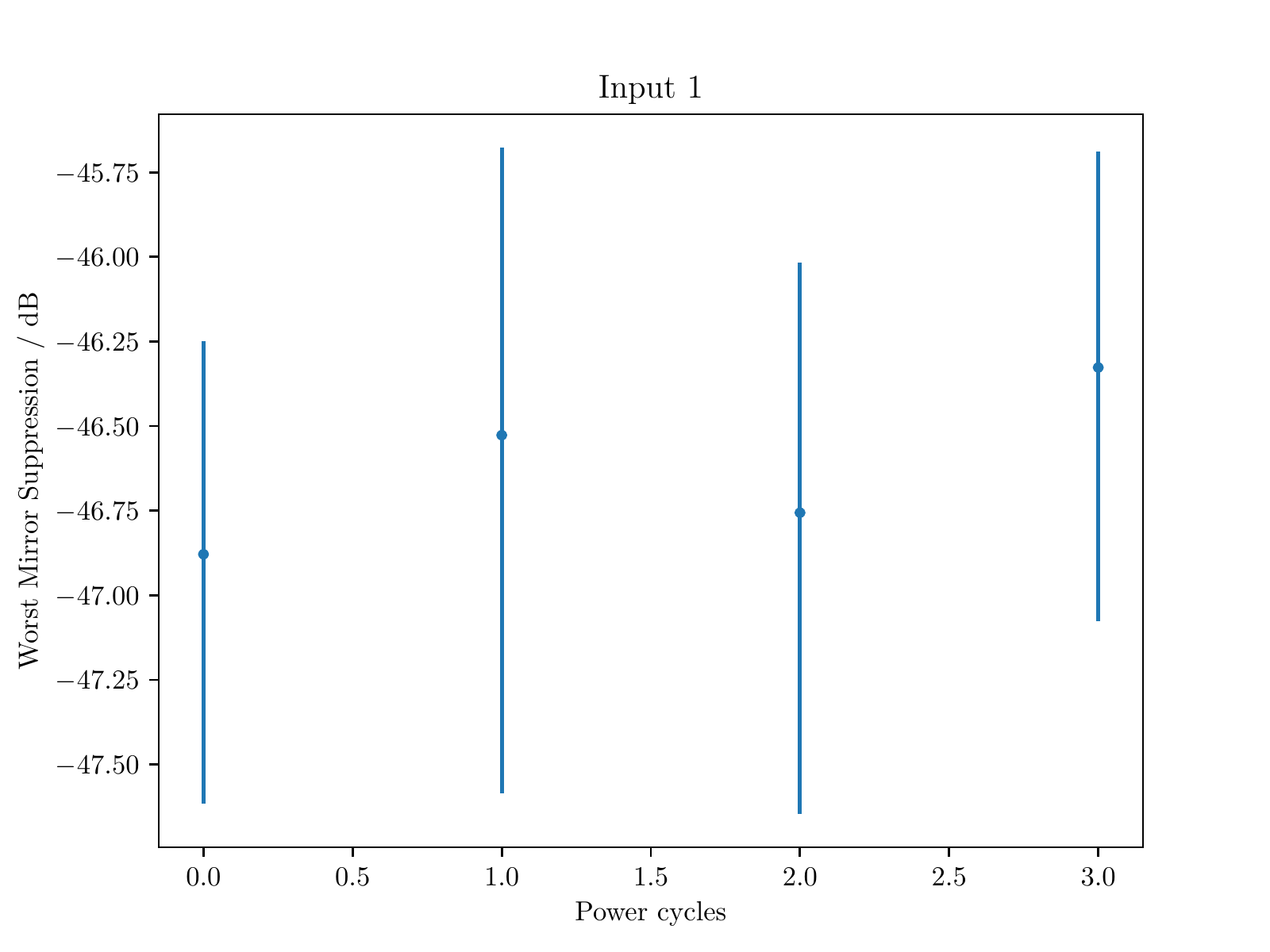}
	\caption{The stability of the worst \gls{MirrorSuppression} over power cycles in the second Nyquist-band for input \texttt{Q} of the first \texttt{dFFTS-4G} board. The calibration and measurements are performed at an ambient temperature of \SI{25}{\celsius}. The uncertainties of the worst \gls{MirrorSuppression} are given by the error bars.}
	\label{fig:power_cycles_worst1}
\end{figure}
\begin{figure}[htbp]
	\centering
	\includegraphics[width=0.95\textwidth]{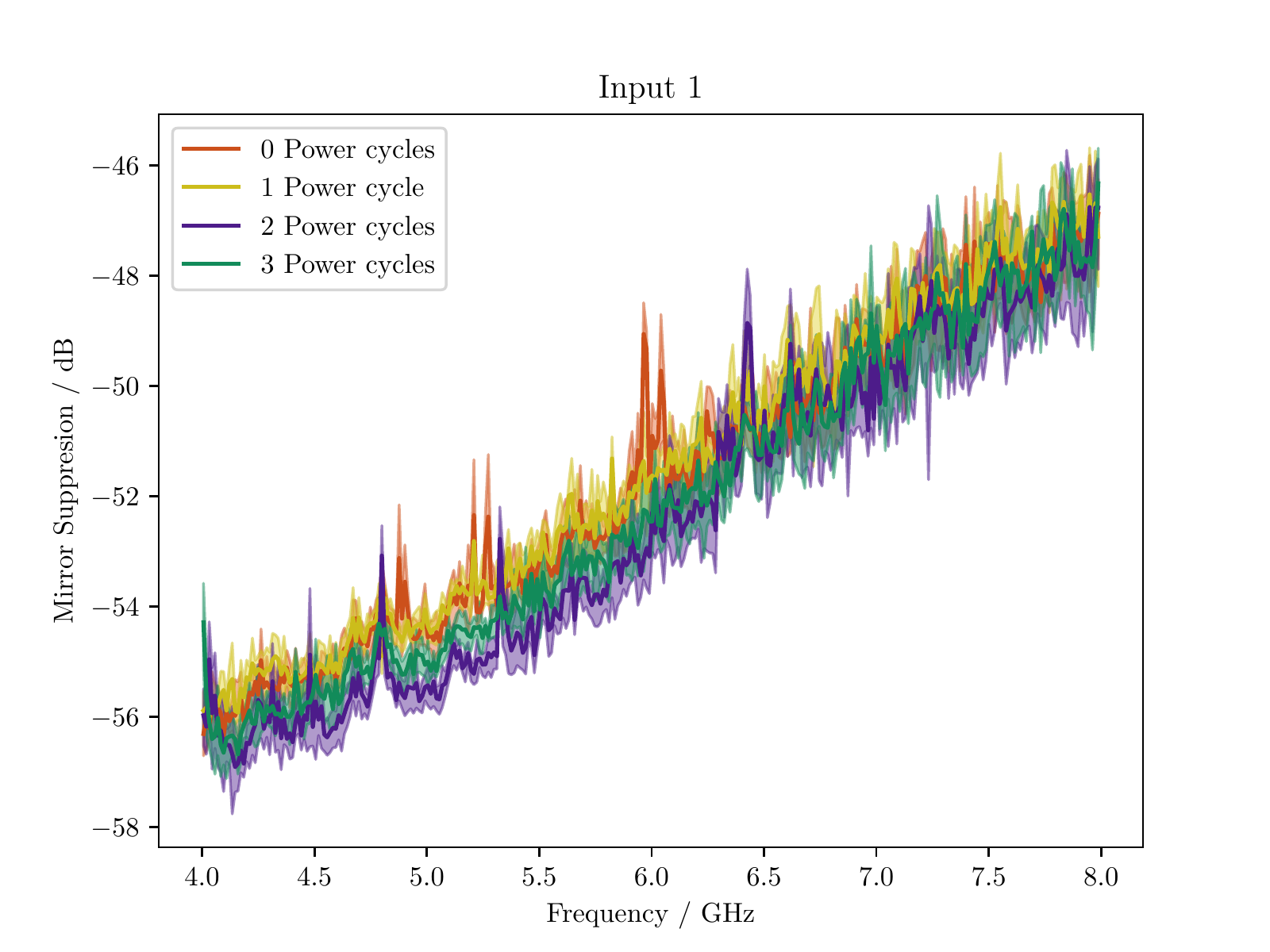}
	\caption{The stability of the \gls{MirrorSuppression} over power cycles in the second Nyquist-band for input \texttt{Q} of the first \texttt{dFFTS-4G} board. The calibration and measurements are performed at an ambient temperature of \SI{25}{\celsius}.}
	\label{fig:power_cycles_adc1}
\end{figure}
\begin{figure}[htbp]
	\centering
	\includegraphics[width=0.95\textwidth]{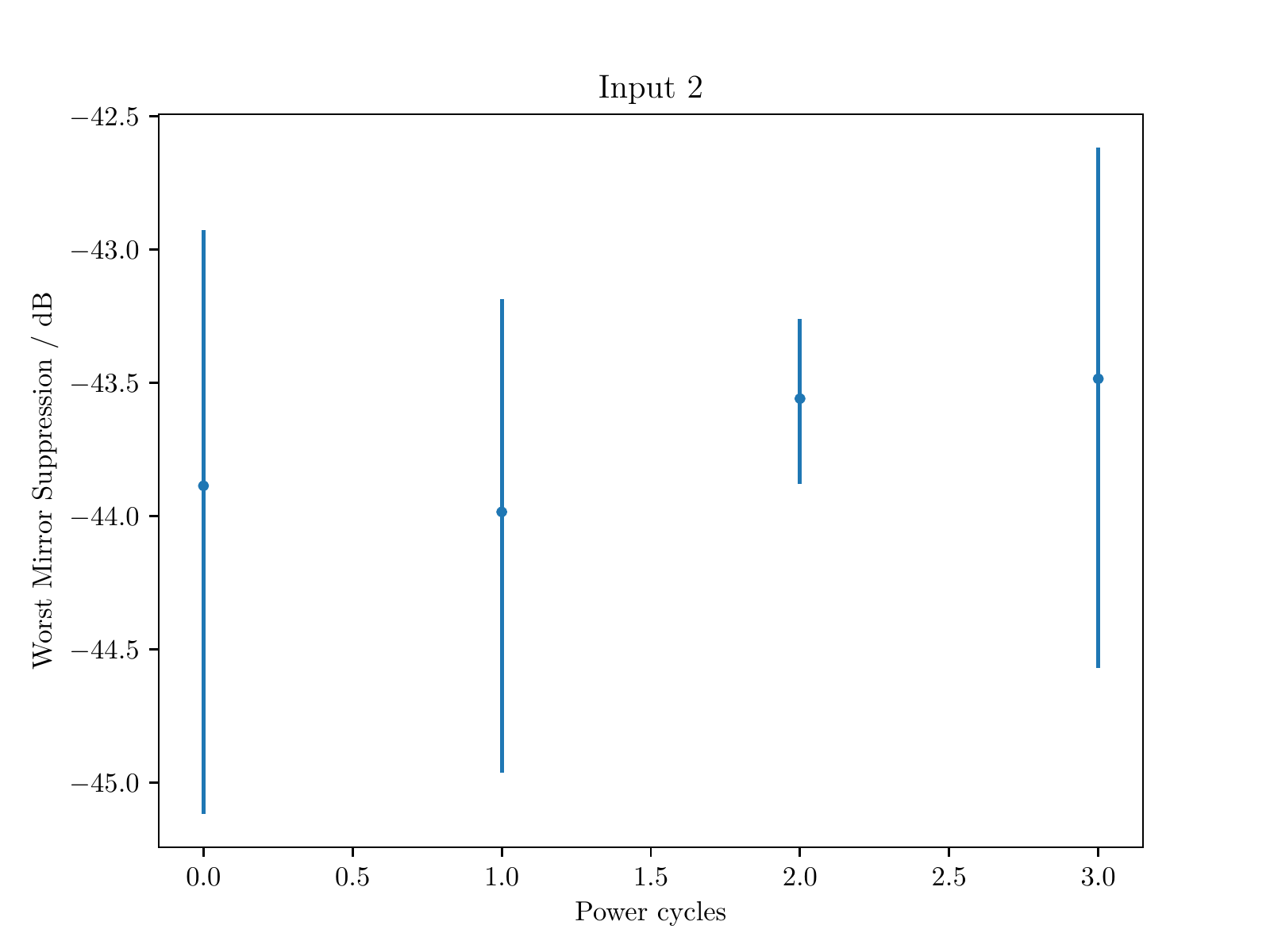}
	\caption{The stability of the worst \gls{MirrorSuppression} over power cycles in the second Nyquist-band for input \texttt{I} of the second \texttt{dFFTS-4G} board. The calibration and measurements are performed at an ambient temperature of \SI{25}{\celsius}. The uncertainties of the worst \gls{MirrorSuppression} are given by the error bars.}
	\label{fig:power_cycles_worst2}
\end{figure}
\begin{figure}[htbp]
	\centering
	\includegraphics[width=0.95\textwidth]{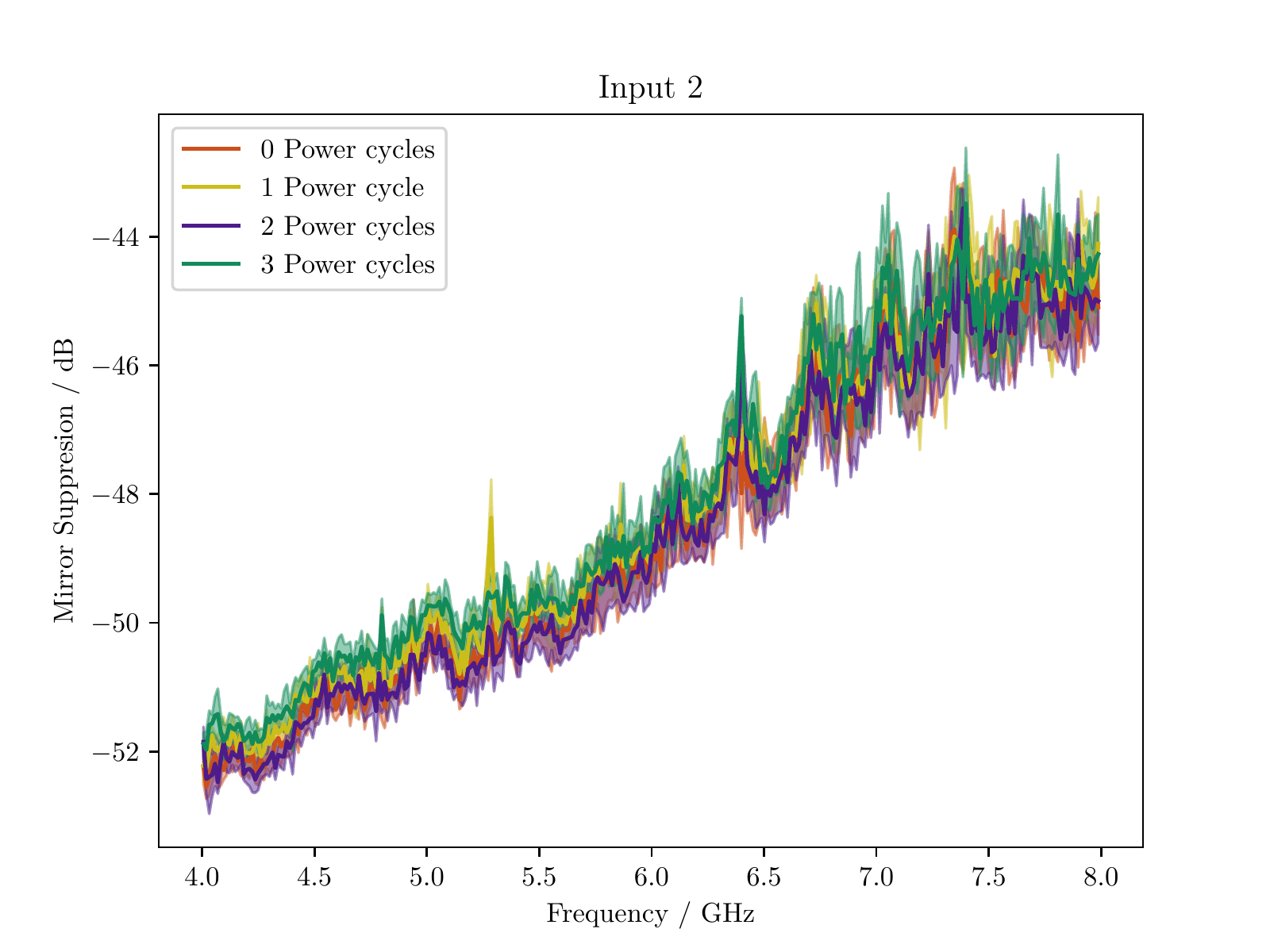}
	\caption{The stability of the \gls{MirrorSuppression} over power cycles in the second Nyquist-band for input \texttt{I} of the second \texttt{dFFTS-4G} board. The calibration and measurements are performed at an ambient temperature of \SI{25}{\celsius}.}
	\label{fig:power_cycles_adc2}
\end{figure}
\begin{figure}[htbp]
	\centering
	\includegraphics[width=0.95\textwidth]{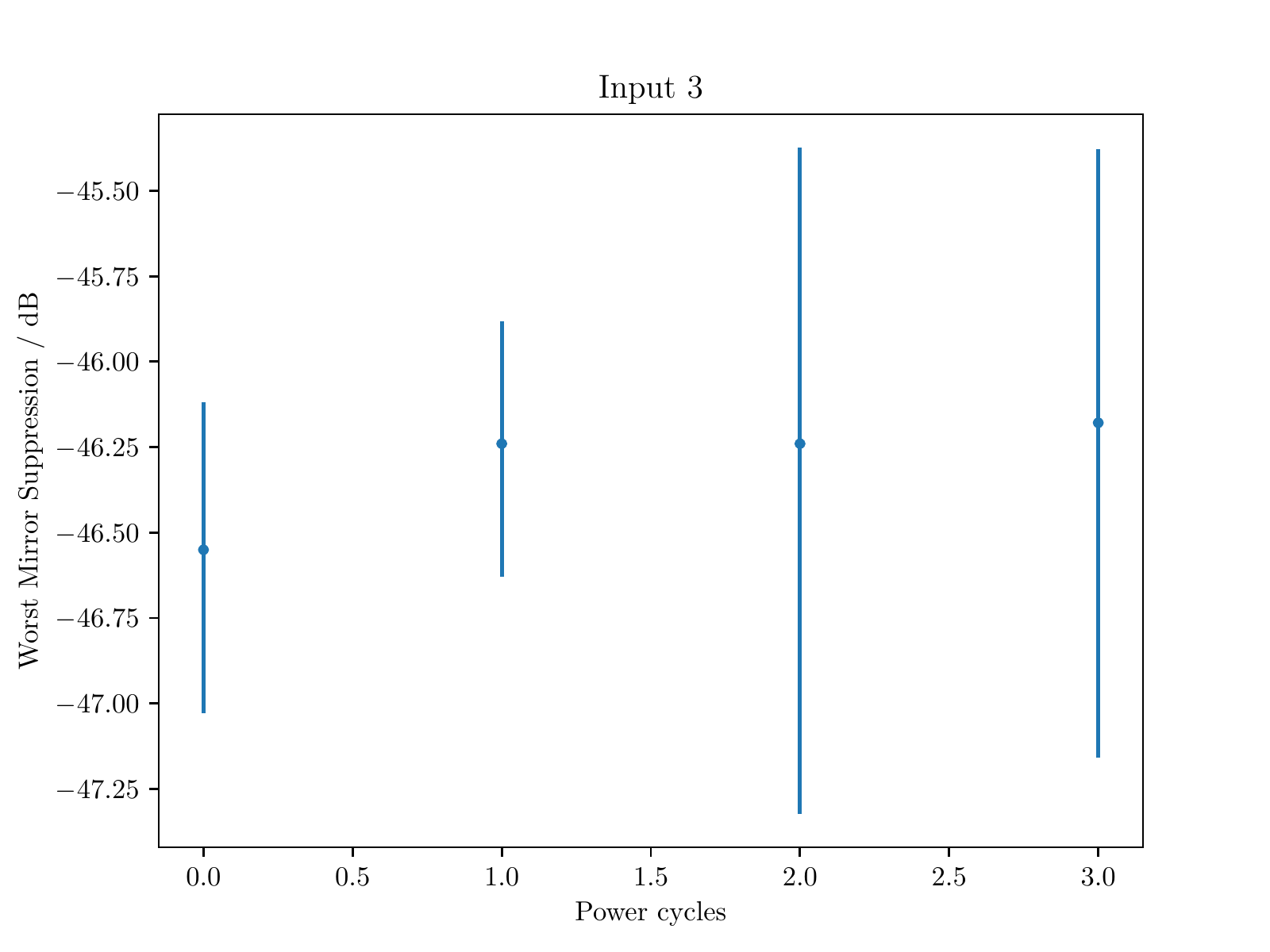}
	\caption{The stability of the worst \gls{MirrorSuppression} over power cycles in the second Nyquist-band for input \texttt{Q} of the second \texttt{dFFTS-4G} board. The calibration and measurements are performed at an ambient temperature of \SI{25}{\celsius}. The uncertainties of the worst \gls{MirrorSuppression} are given by the error bars.}
	\label{fig:power_cycles_worst3}
\end{figure}
\begin{figure}[htbp]
	\centering
	\includegraphics[width=0.95\textwidth]{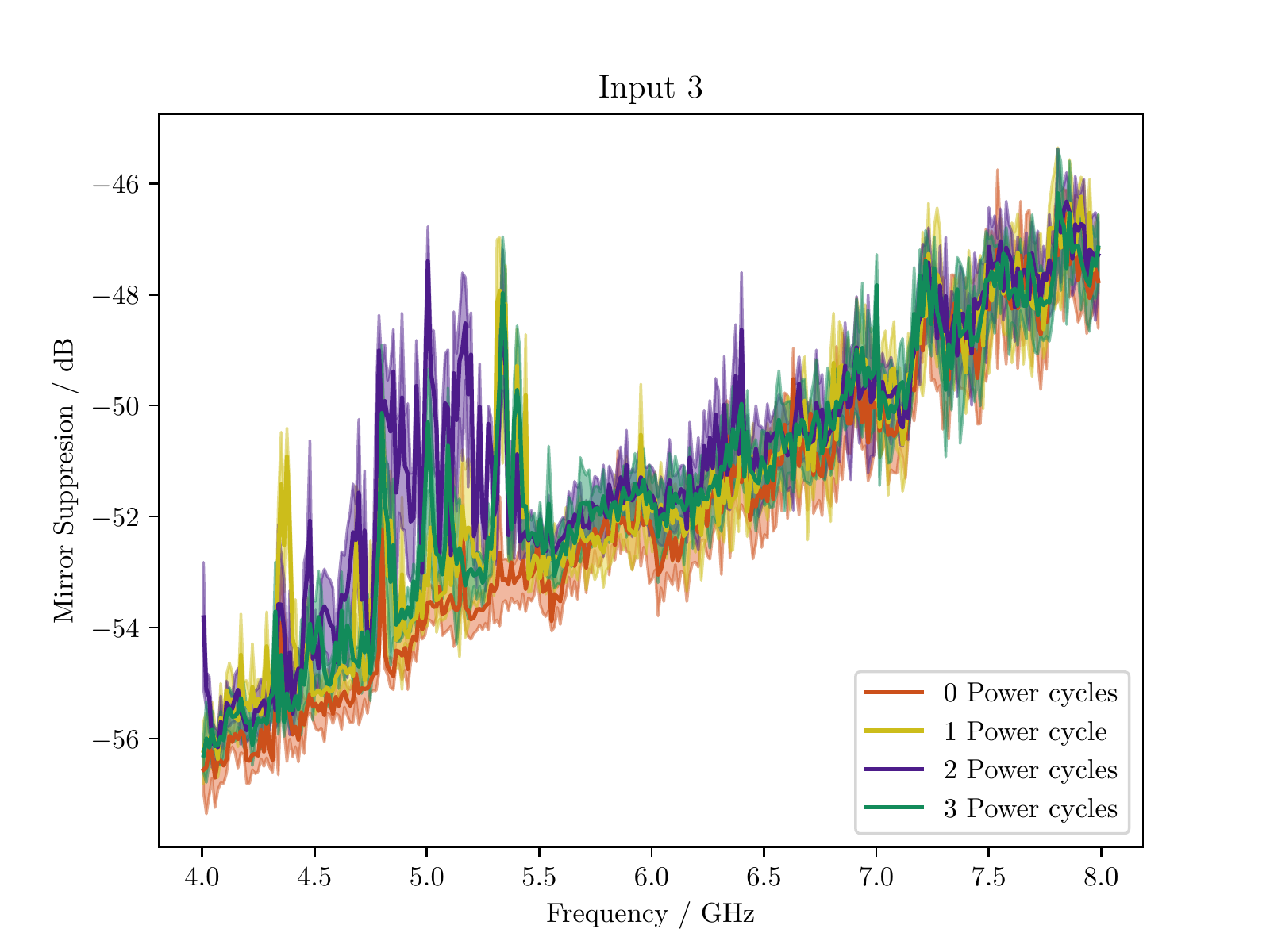}
	\caption{The stability of the \gls{MirrorSuppression} over power cycles in the second Nyquist-band for input \texttt{Q} of the second \texttt{dFFTS-4G} board. The calibration and measurements are performed at an ambient temperature of \SI{25}{\celsius}.}
	\label{fig:power_cycles_adc3}
\end{figure}

\FloatBarrier
\section{Impact of Interpolation}
\begin{figure}[htbp]
	\centering
	\includegraphics[width=0.95\textwidth]{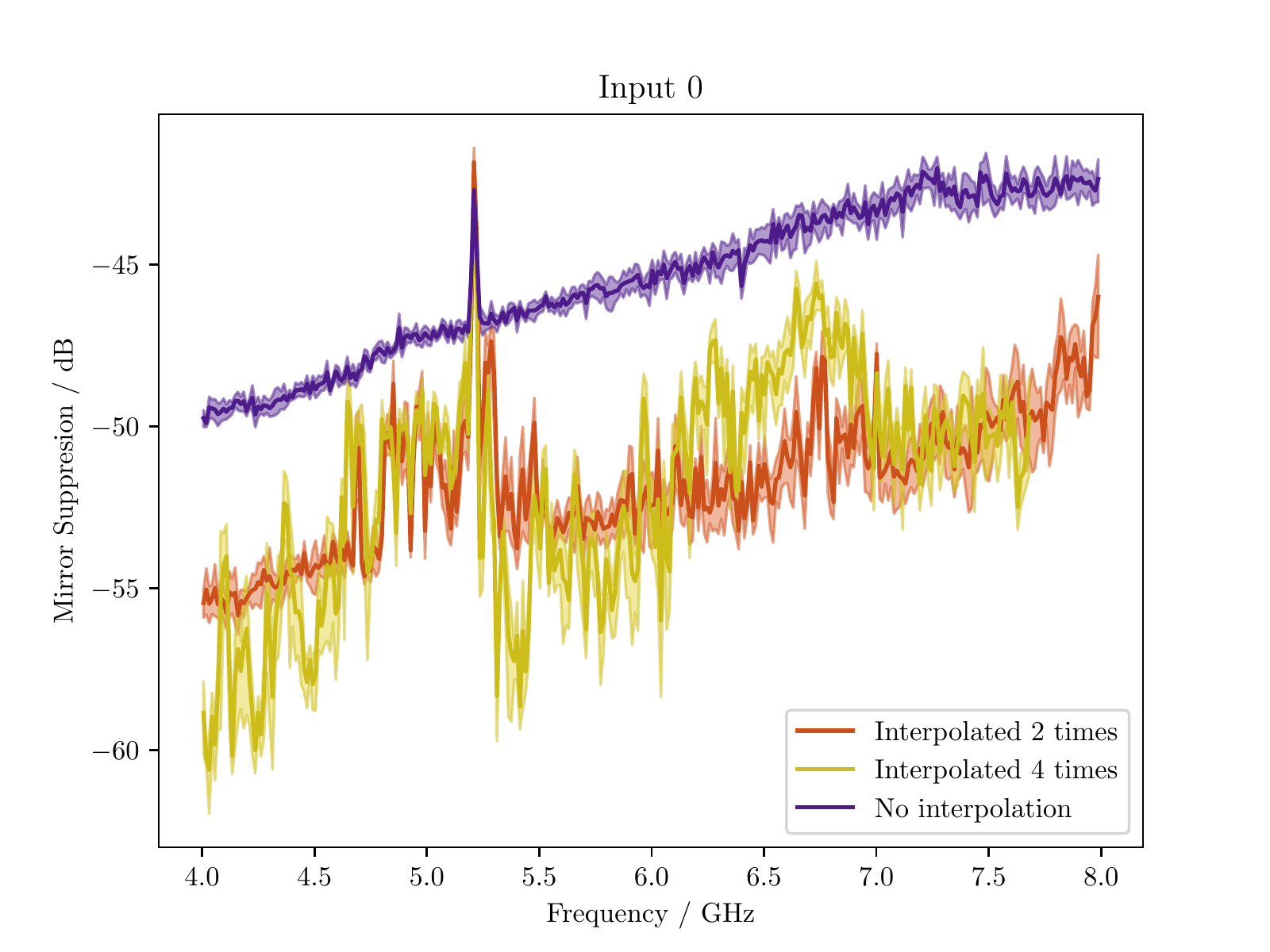}
	\caption{The \gls{MirrorSuppression} in the second Nyquist-band for different interpolations for input \texttt{I} of the first \texttt{dFFTS-4G} board. The calibration and measurements are performed at an ambient temperature of \SI{25}{\celsius}.}
	\label{fig:interpolation_adc0}
\end{figure}
\begin{figure}[htbp]
	\centering
	\includegraphics[width=0.95\textwidth]{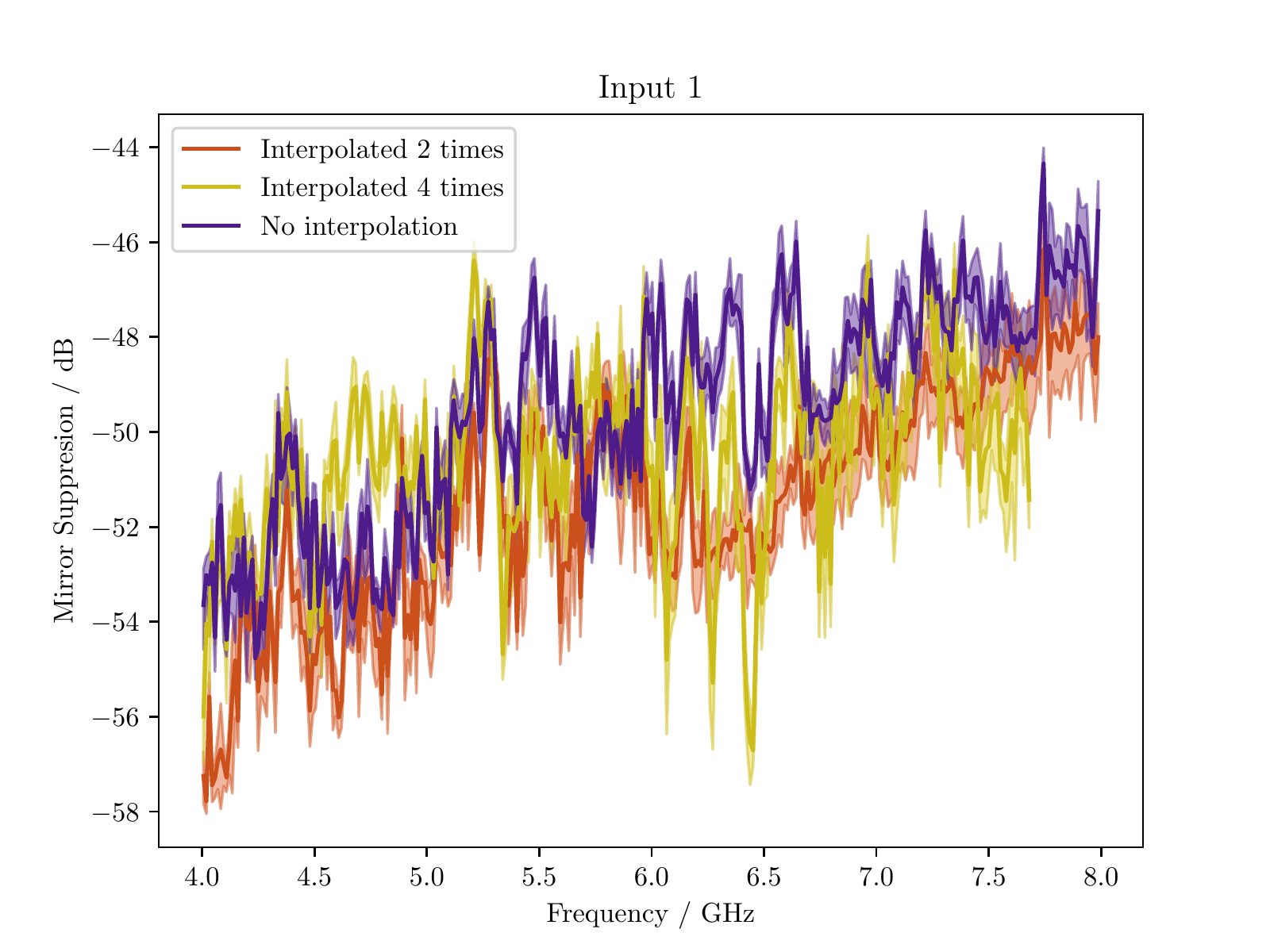}
	\caption{The \gls{MirrorSuppression} in the second Nyquist-band for different interpolations for input \texttt{Q} of the first \texttt{dFFTS-4G} board. The calibration and measurements are performed at an ambient temperature of \SI{25}{\celsius}.}
	\label{fig:interpolation_adc1}
\end{figure}
\begin{figure}[htbp]
	\centering
	\includegraphics[width=0.95\textwidth]{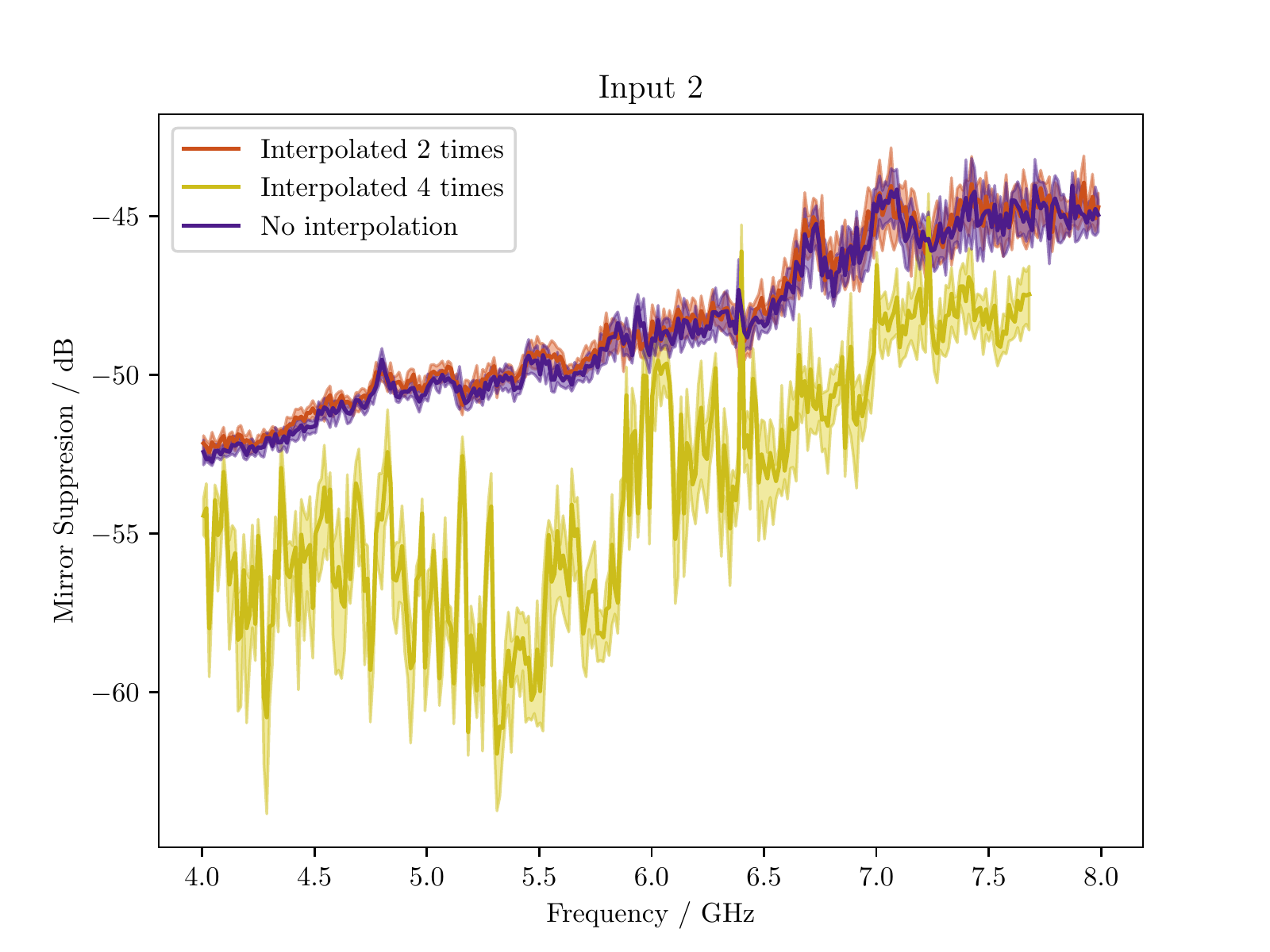}
	\caption{The \gls{MirrorSuppression} in the second Nyquist-band for different interpolations for input \texttt{I} of the second \texttt{dFFTS-4G} board. The calibration and measurements are performed at an ambient temperature of \SI{25}{\celsius}.}
	\label{fig:interpolation_adc2}
\end{figure}
\begin{figure}[htbp]
	\centering
	\includegraphics[width=0.95\textwidth]{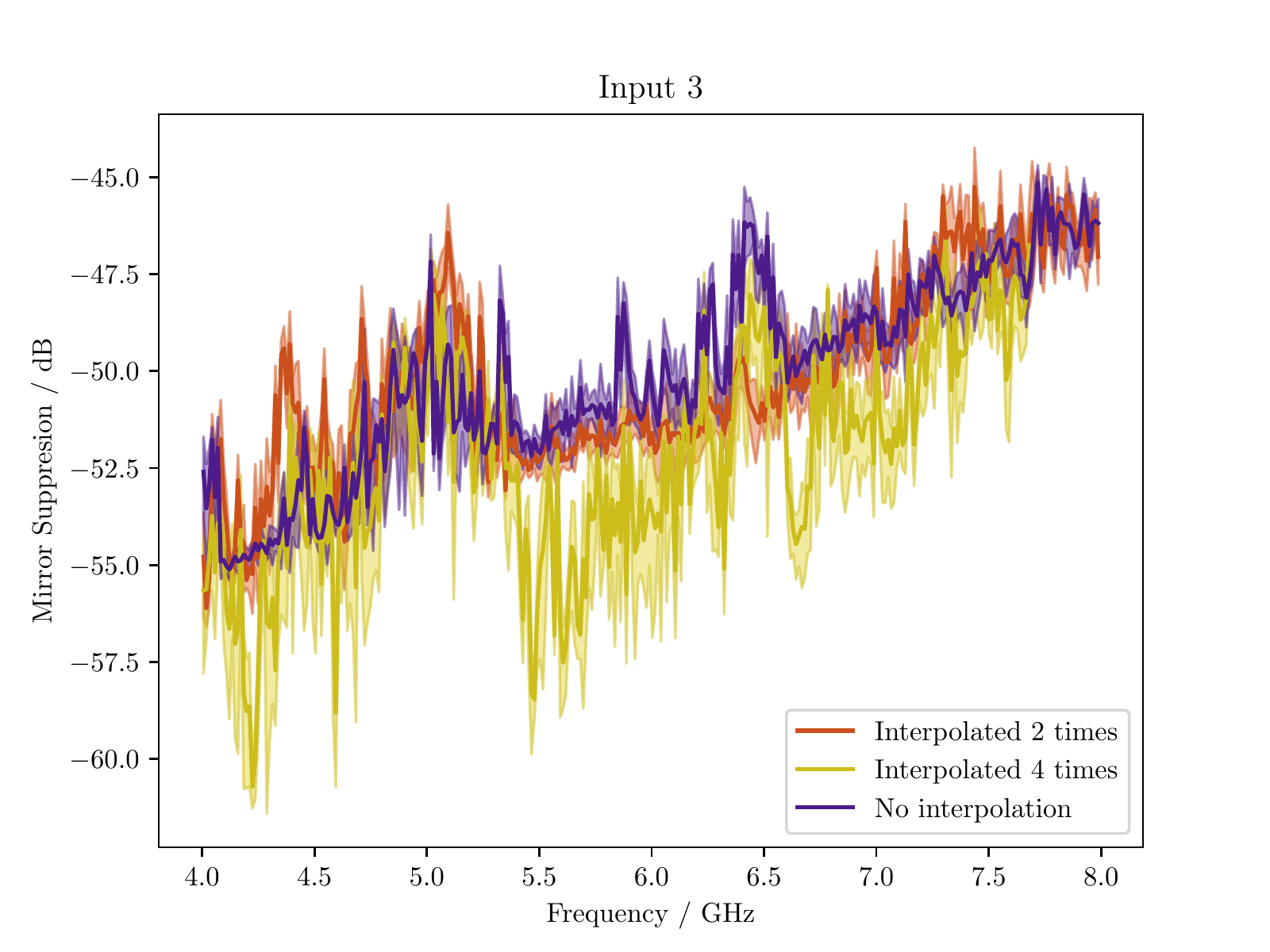}
	\caption{The \gls{MirrorSuppression} in the second Nyquist-band for different interpolations for input \texttt{Q} of the second \texttt{dFFTS-4G} board. The calibration and measurements are performed at an ambient temperature of \SI{25}{\celsius}.}
	\label{fig:interpolation_adc3}
\end{figure}

\FloatBarrier
\section{Power consumption}
\begin{figure}[htbp!]
	\centering
	\begin{subfigure}[t]{0.74\textwidth}
		\centering
		\includegraphics[width=\textwidth]{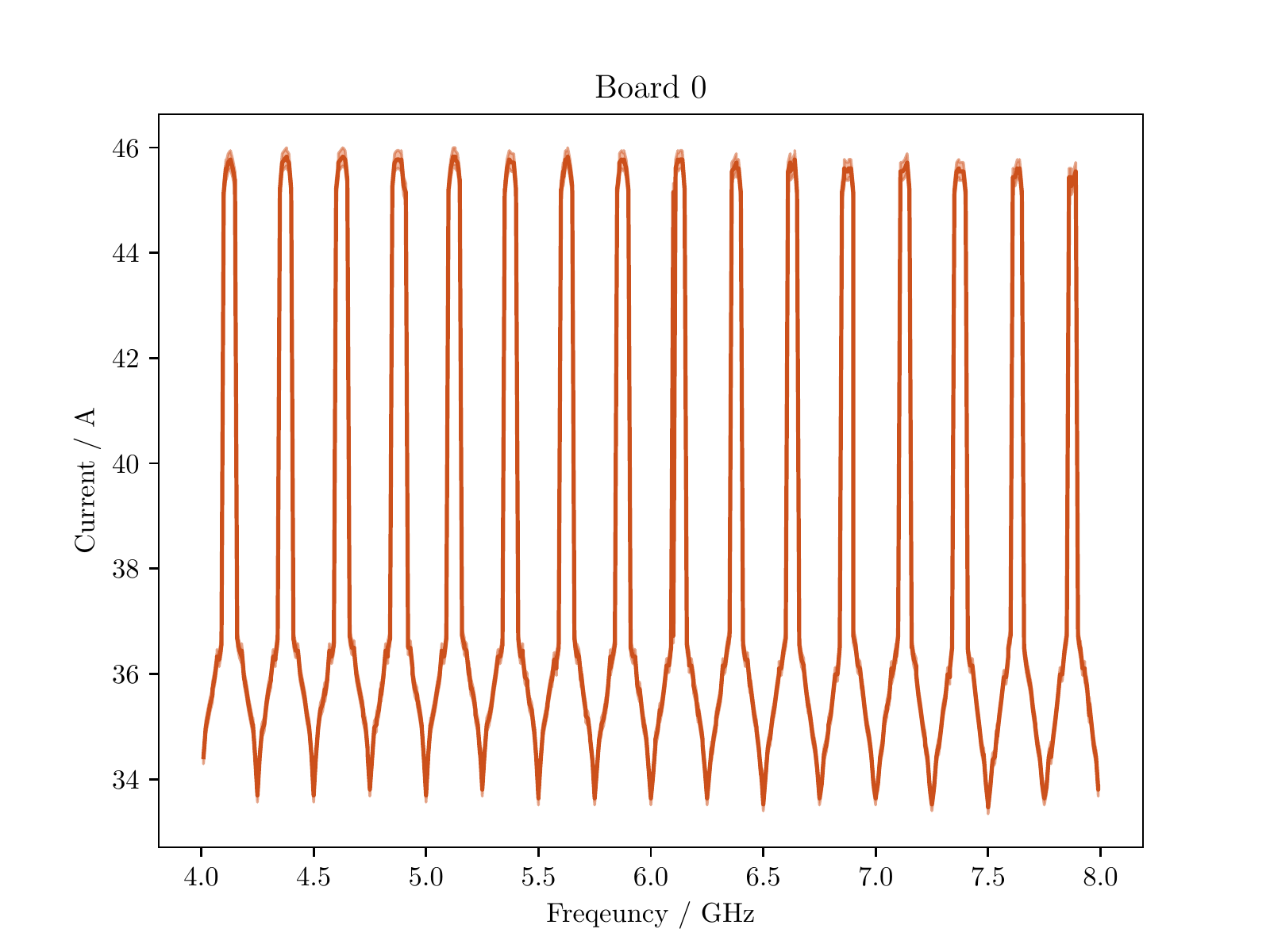}
		\caption{Board 0}
		\label{fig:current_noInterpolation_board0}
	\end{subfigure}
	\quad
	\begin{subfigure}[t]{0.74\textwidth}
		\centering
		\includegraphics[width=\textwidth]{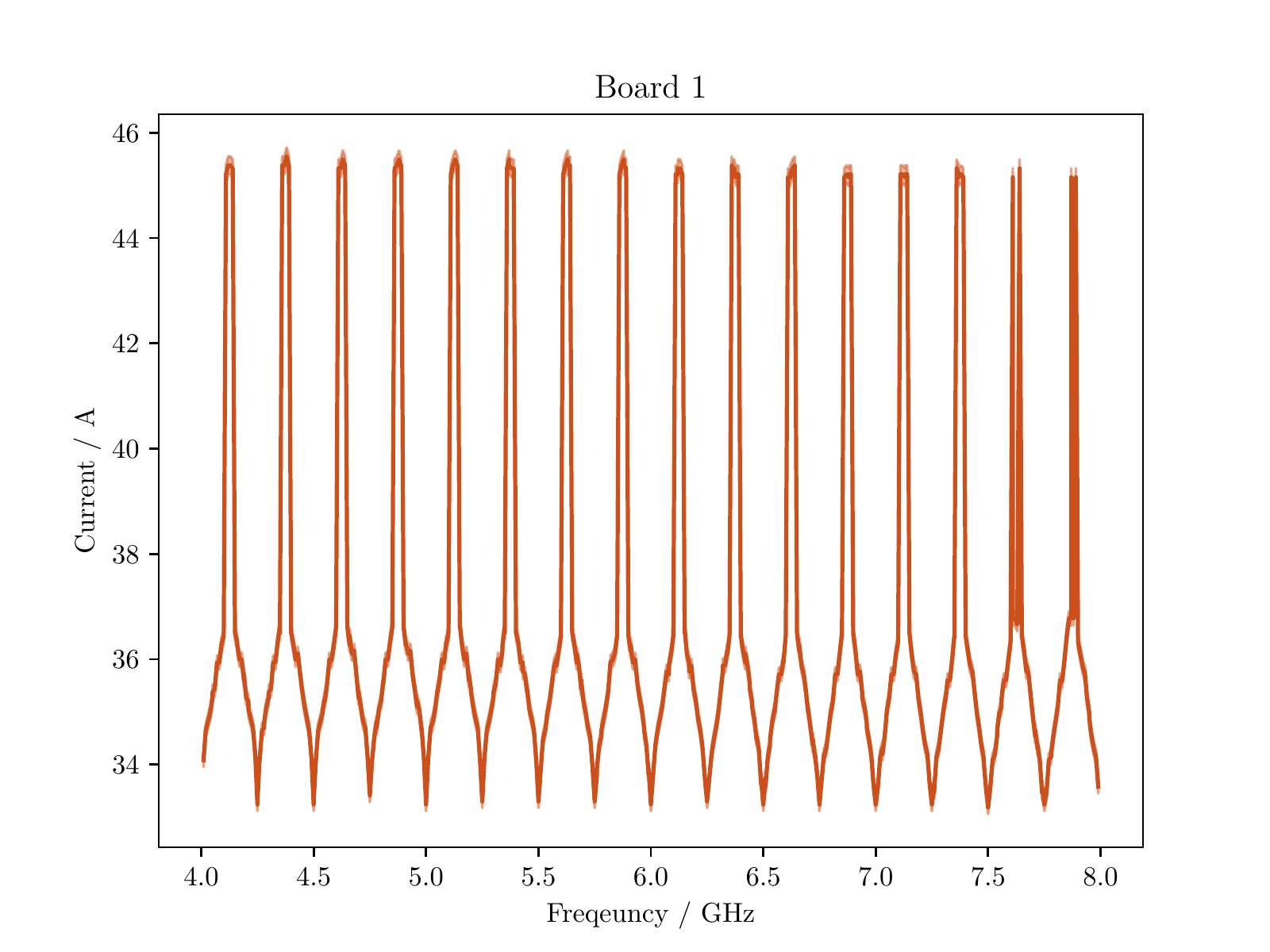}
		\caption{Board 1}
		\label{fig:current_noInterpolation_board1}
	\end{subfigure}
	\caption{The power consumption for the \SI{1}{\volt} power rail of the \texttt{dFFTS-4G} spectrometers with a \gls{gateware} doing two \glspl{FFT} with each $16$k spectral channel including a frequency independent calibration without interpolation. The frequency of the applied sinusoidal signal varies and has an amplitude of \SI{170}{\milli\volt}. The shaded array represents the error of the power measurements.}
	\label{fig:current_noInterpolation}
\end{figure}

\begin{figure}[htbp]
	\centering
	\begin{subfigure}[t]{0.74\textwidth}
		\centering
		\includegraphics[width=\textwidth]{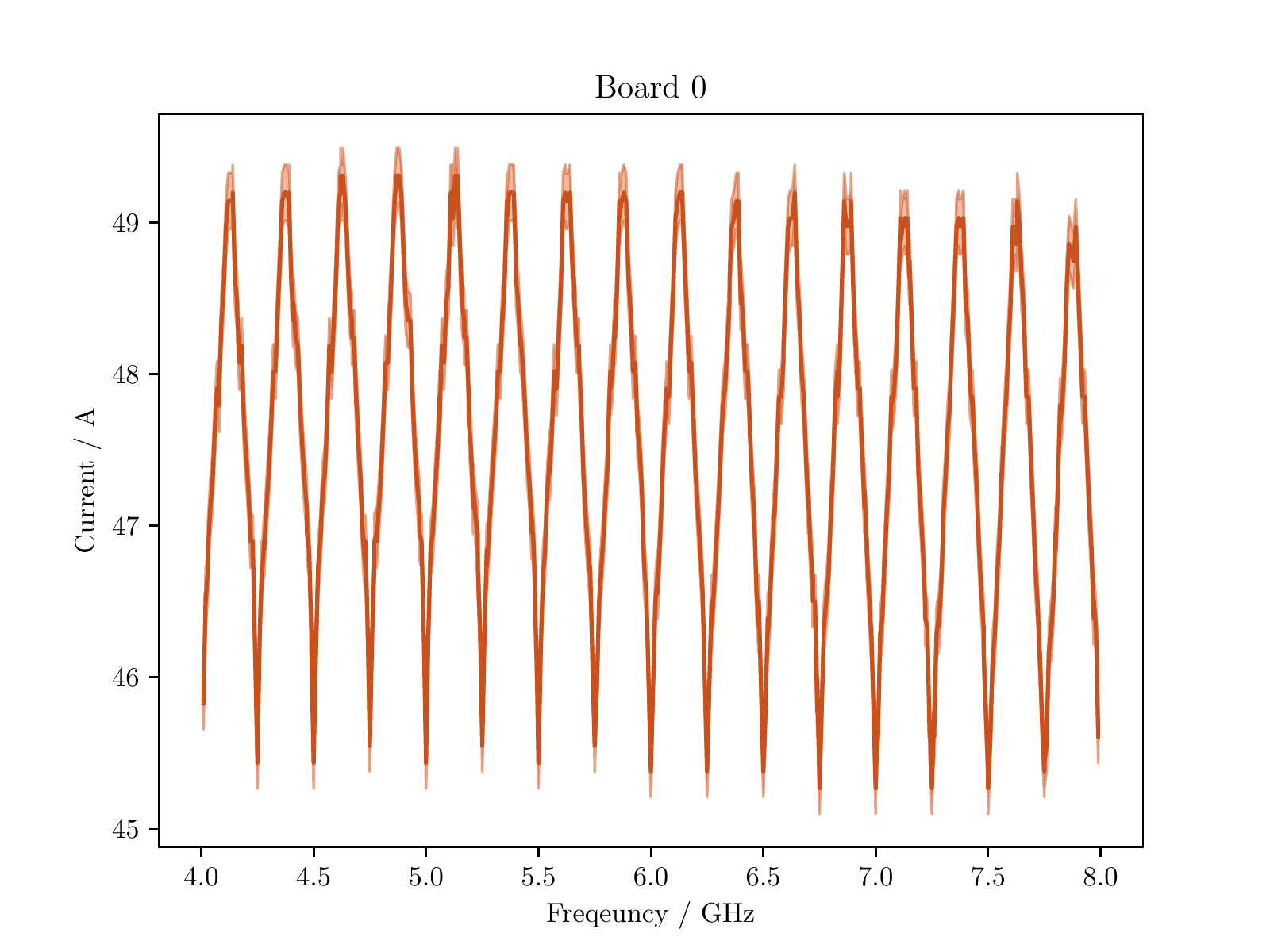}
		\caption{Board 0}
		\label{fig:current_2interpolation_board0}
	\end{subfigure}
	\quad
	\begin{subfigure}[t]{0.74\textwidth}
		\centering
		\includegraphics[width=\textwidth]{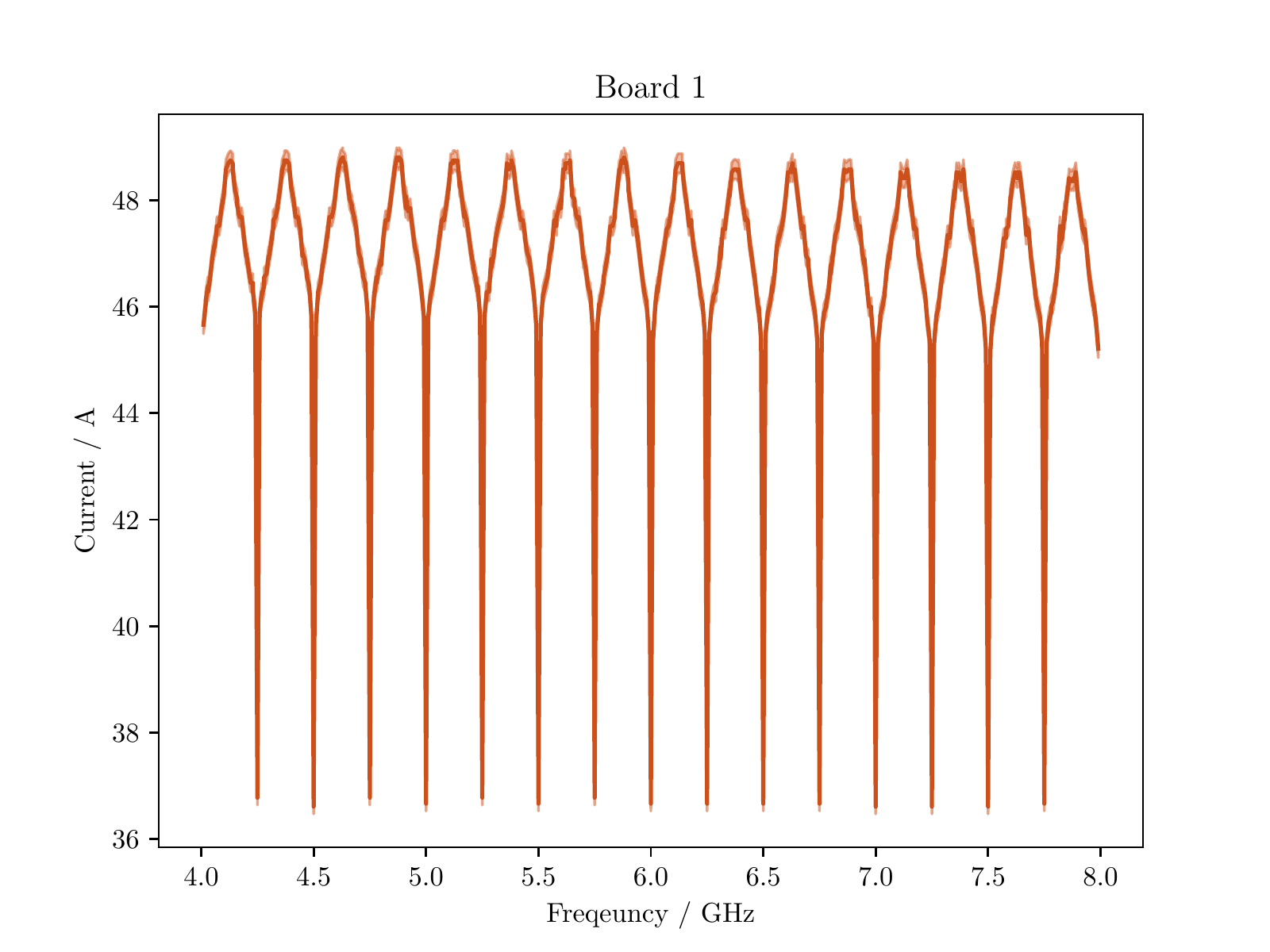}
		\caption{Board 1}
		\label{fig:current_2interpolation_board1}
	\end{subfigure}
	\caption{The power consumption for the \SI{1}{\volt} power rail of the \texttt{dFFTS-4G} spectrometers with a \gls{gateware} doing two \glspl{FFT} with each $16$k spectral channel  including a frequency independent calibration with an 2 times interpolation of the coefficients. The frequency of the applied sinusoidal signal varies and has an amplitude of \SI{170}{\milli\volt}. The shaded array represents the error of the power measurements.}
	\label{fig:current_2Interpolation}
\end{figure}

\begin{figure}[htbp]
	\centering
	\begin{subfigure}[t]{0.74\textwidth}
		\centering
		\includegraphics[width=\textwidth]{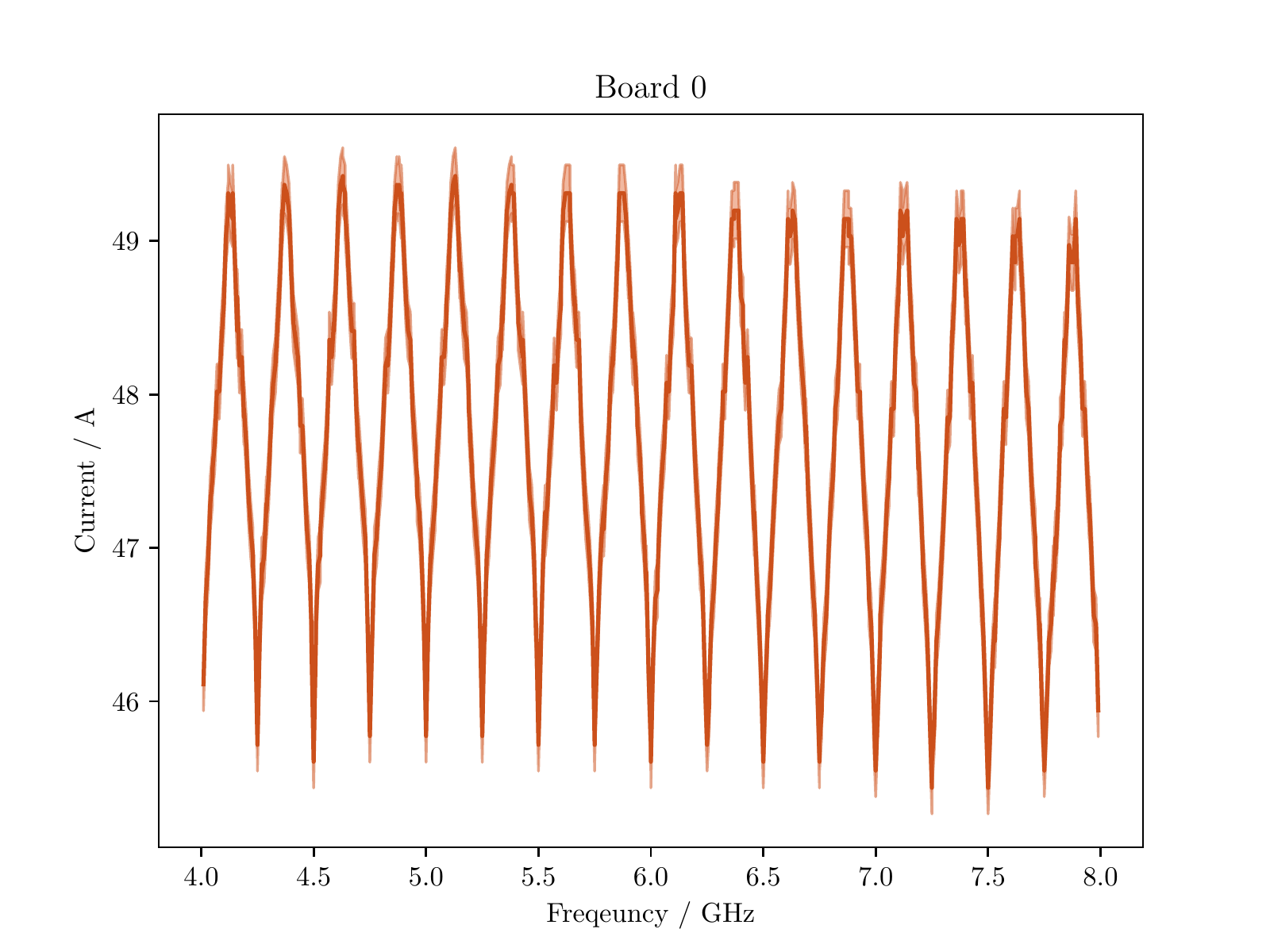}
		\caption{Board 0}
		\label{fig:current_4interpolation_board0}
	\end{subfigure}
	\quad
	\begin{subfigure}[t]{0.74\textwidth}
		\centering
		\includegraphics[width=\textwidth]{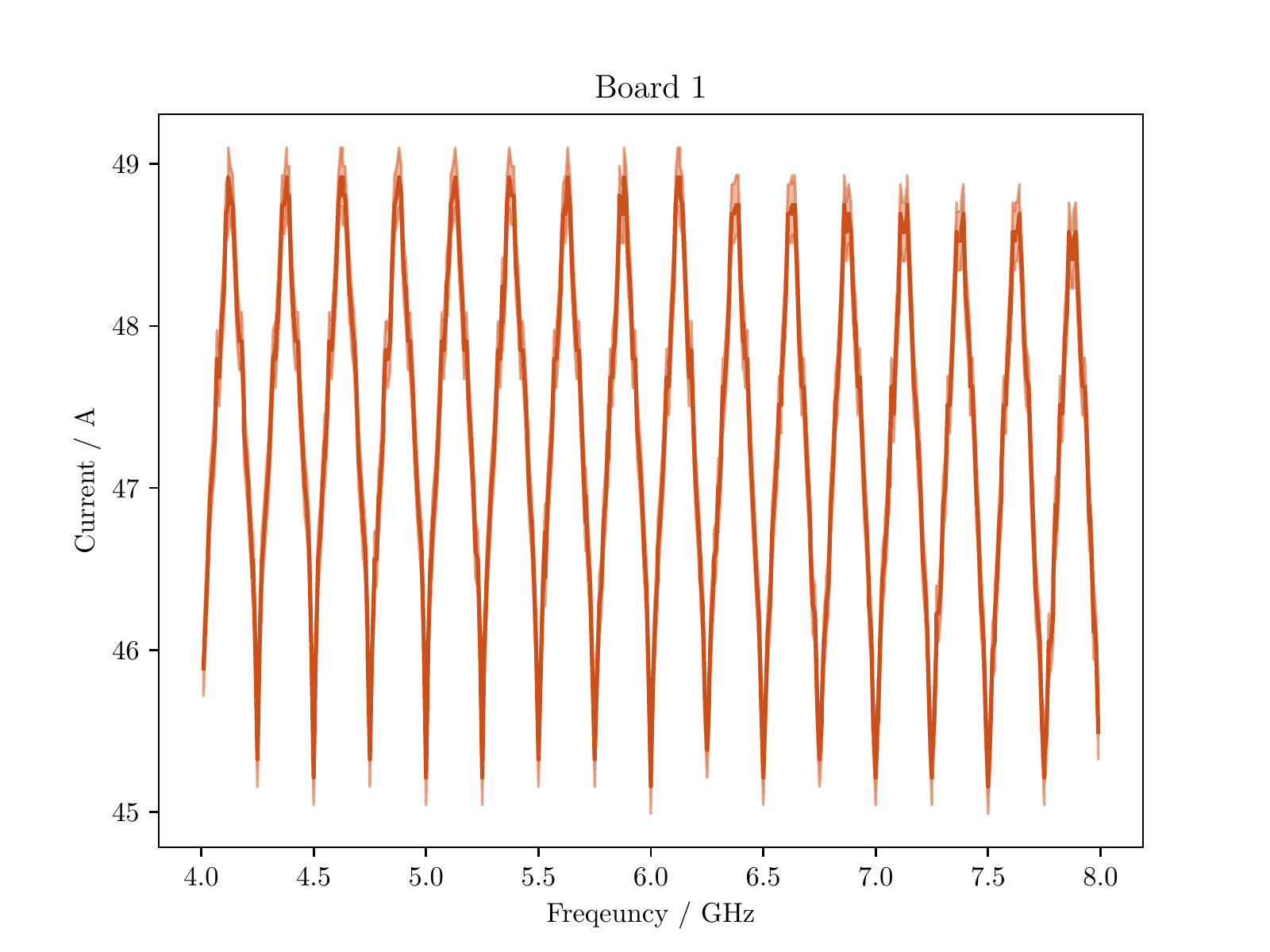}
		\caption{Board 1}
		\label{fig:current_4interpolation_board1}
	\end{subfigure}
	\caption{The power consumption for the \SI{1}{\volt} power rail of the \texttt{dFFTS-4G} spectrometers with a \gls{gateware} doing two \glspl{FFT} with each $16$k spectral channel  including a frequency independent calibration with an 4 times interpolation of the coefficients. The frequency of the applied sinusoidal signal varies and has an amplitude of \SI{170}{\milli\volt}. The shaded array represents the error of the power measurements.}
	\label{fig:current_4Interpolation}
\end{figure}
%%% Local Variables: 
%%% mode: latex
%%% TeX-master: "../mythesis"
%%% End: 

%------------------------------------------------------------------------------
% Declare lists of figures and tables and acknowledgements as backmatter
% Chapter/section numbers are turned off
\backmatter

\listoffigures
\listoftables

%------------------------------------------------------------------------------
% Print the glossary and list of acronyms
\printglossaries

%------------------------------------------------------------------------------
% You could instead add your acknowledgements here - don't forget to
% also add them to \includeonly above
% \include{thesis_acknowledge}

%------------------------------------------------------------------------------
% CV needed when you submit your PhD thesis
% \input{thesis_cv}

\end{document}